# HabEx

Exploring planetary systems around our neighboring sunlike stars and
enabling a broad range of observatory science in the UV through near-IR

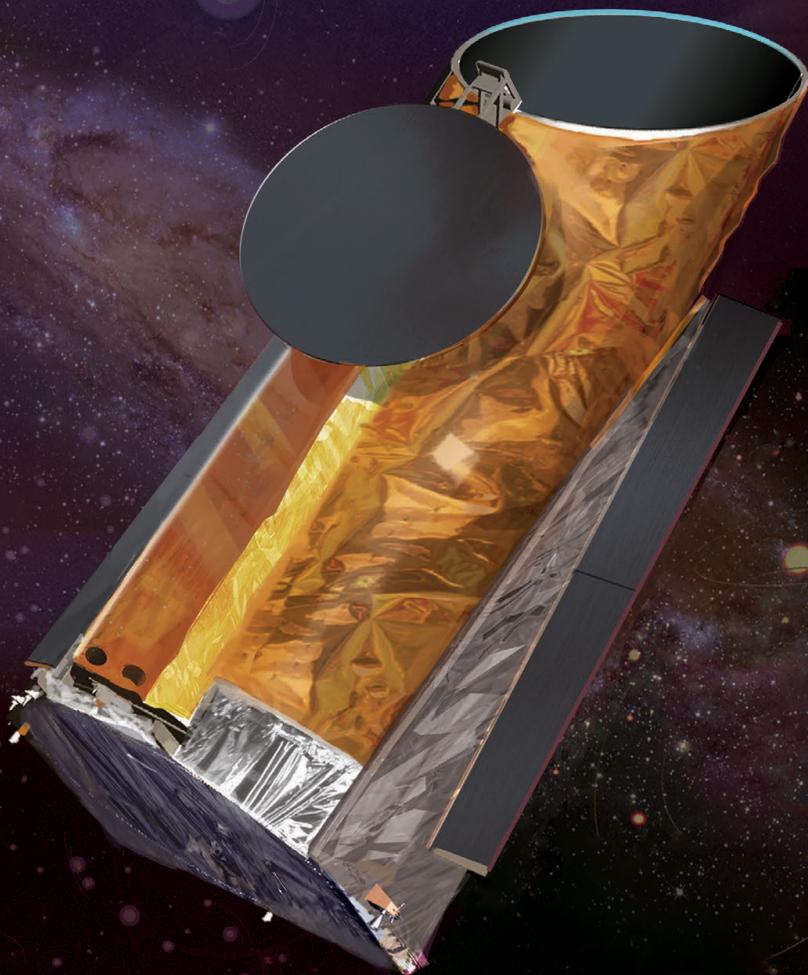

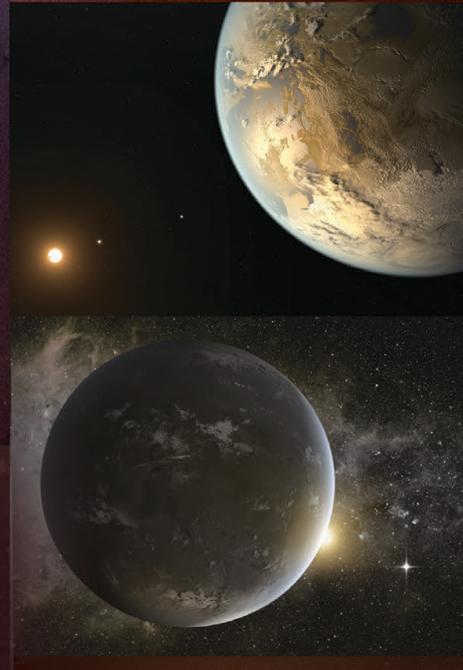

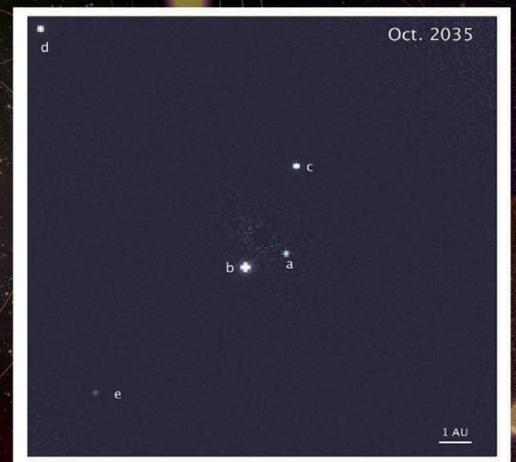

Oct. 2035

d

c

b a

e

1 AU

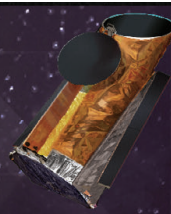
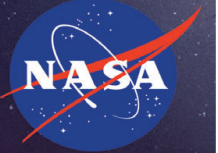

# HabEx

**EXPLORING PLANETARY SYSTEMS** AROUND NEARBY SUNLIKE STARS
AND ENABLING OBSERVATORY SCIENCE FROM THE UV THROUGH **NEAR-IR**

*www.jpl.nasa.gov/habex/*

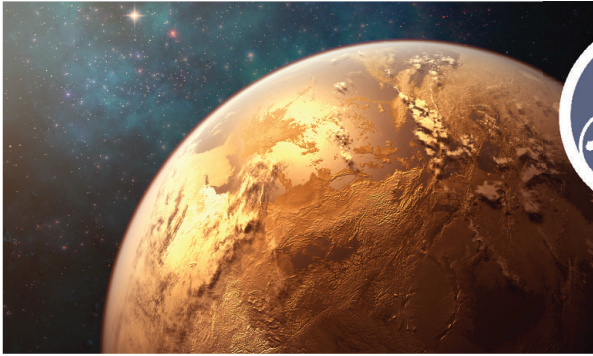

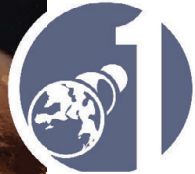

### GOAL 1

**To seek out nearby worlds and explore their habitability,** *HabEx* would search for habitable zone Earth-like planets around sunlike stars using direct imaging and spectrally characterize promising candidates for signs of habitability and life.

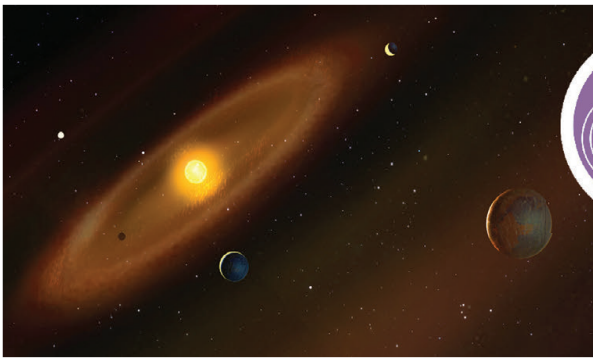

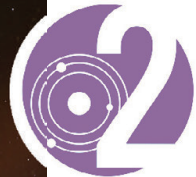

### GOAL 2

**To map out nearby planetary systems and understand the diversity of the worlds they contain,** *HabEx* would take the first "family portraits" of nearby planetary systems, detecting and characterizing both inner and outer planets, as well as searching for dust and debris disks.

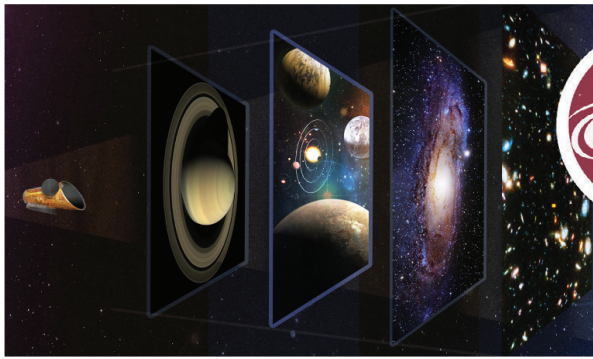

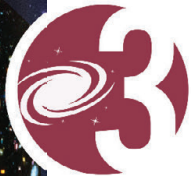

### GOAL 3

**To enable new explorations of astrophysical systems from our solar system to galaxies and the universe by extending our reach in the UV through near-IR,** *HabEx* would have a community-driven, competed Guest Observer program to undertake revolutionary science with a large-aperture, ultra-stable UV through near-IR space telescope.

**The HabEx concept design** relies on demonstrated, yet cutting edge, technologies wherever possible, which enables world-leading science in the 2030s while limiting risk and cost.

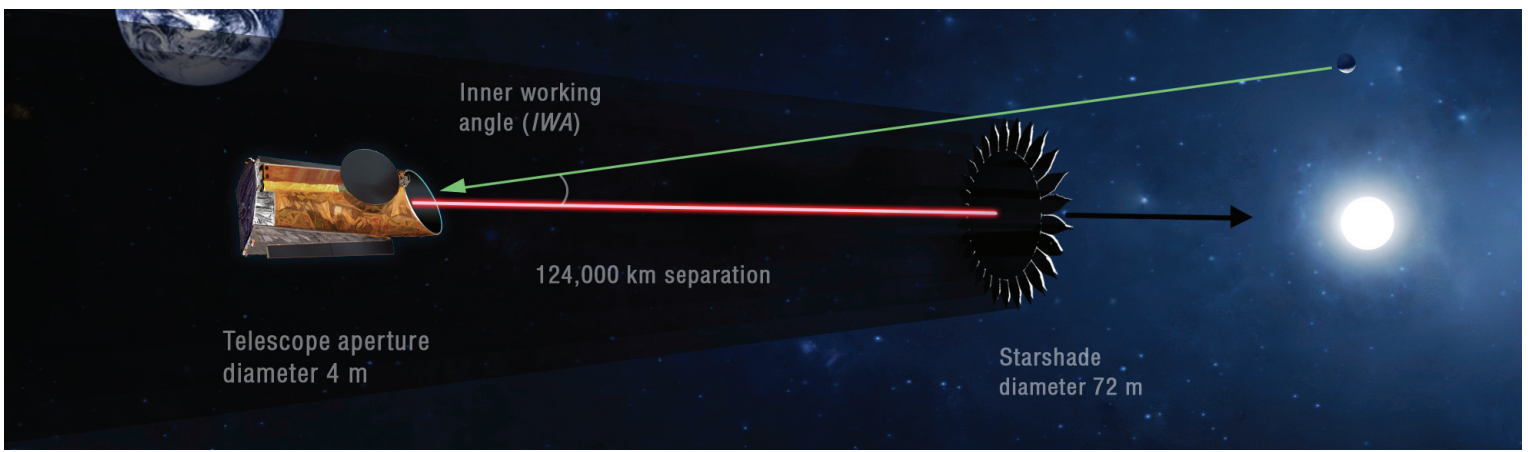

Inner working angle (*IWA*)

124,000 km separation

Telescope aperture diameter 4 m

Starshade diameter 72 m

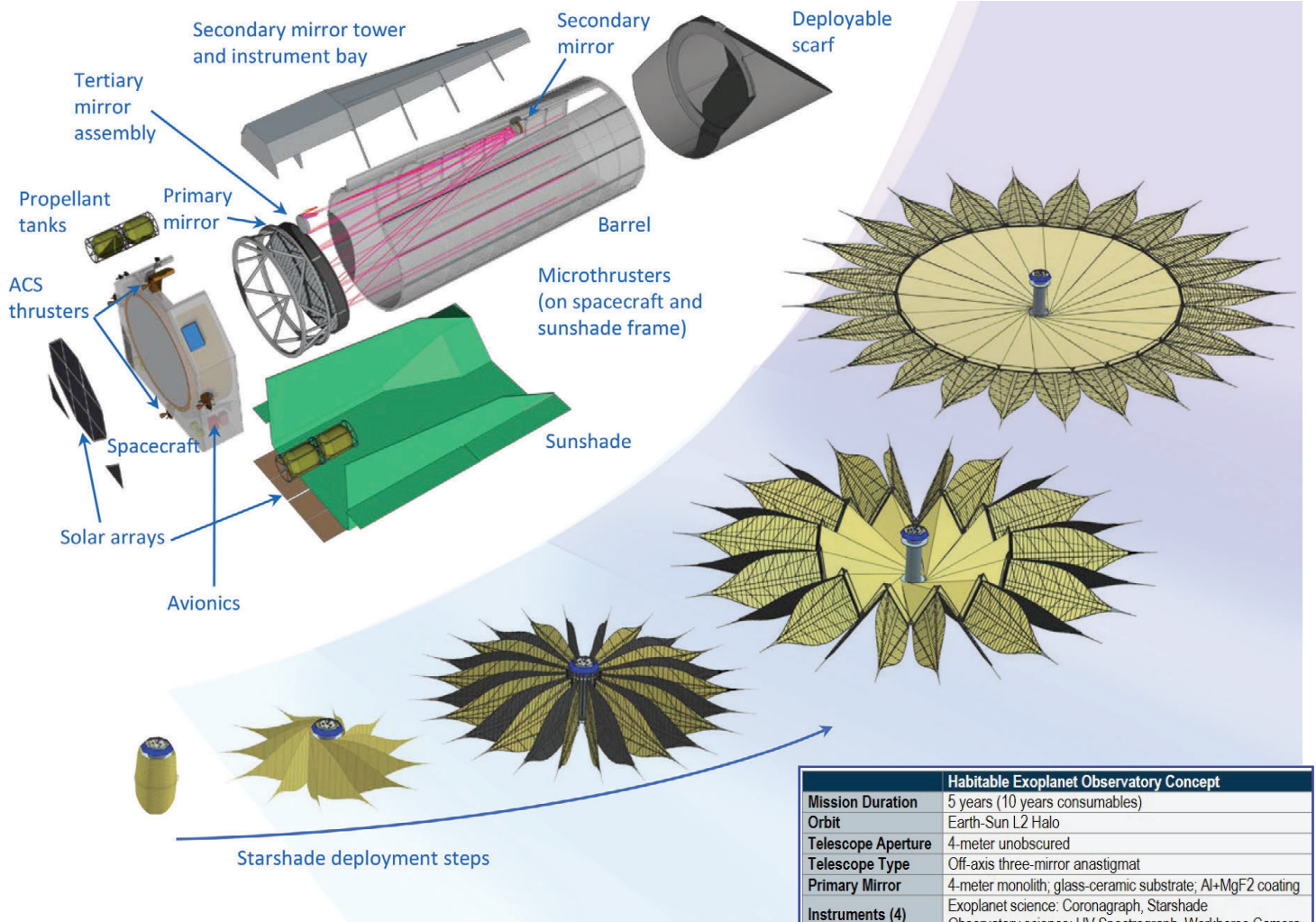

| | Habitable Exoplanet Observatory Concept |
|---|---|
| Mission Duration | 5 years (10 years consumables) |
| Orbit | Earth-Sun L2 Halo |
| Telescope Aperture | 4-meter unobscured |
| Telescope Type | Off-axis three-mirror anastigmat |
| Primary Mirror | 4-meter monolith; glass-ceramic substrate; Al+MgF2 coating |
| Instruments (4) | Exoplanet science: Coronagraph, Starshade<br>Observatory science: UV Spectrograph, Workhorse Camera |
| Attitude Control | Slewing: hydrazine thrusters; Pointing: microthrusters |

Starshade deployment steps

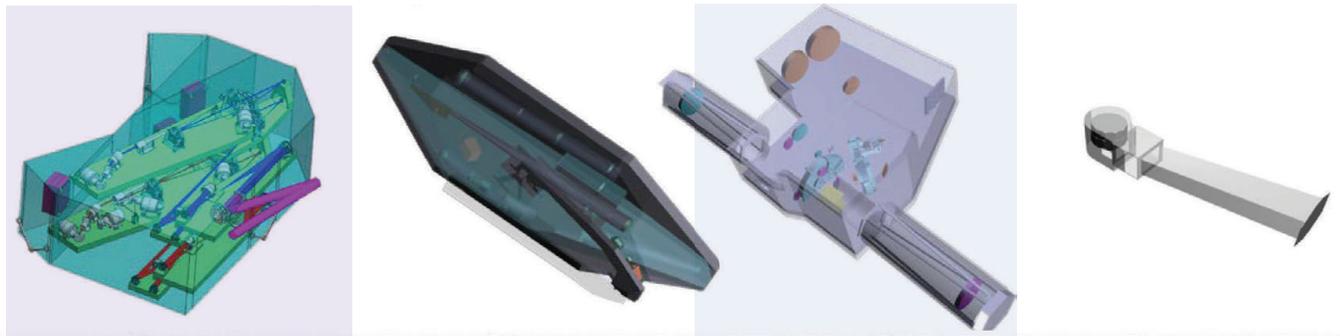

| | Coronagraph | Starshade | Workhorse Camera | UV Spectrograph |
|---|---|---|---|---|
| Purpose | Exoplanet imaging and characterization | Exoplanet imaging and characterization | Multipurpose, wide-field imaging camera and spectrograph for general astrophysics | High-resolution, UV spectroscopy for general astrophysics |
| Instrument Type | Vortex charge 6 coronagraph with:<br>- Raw contrast: 1×10⁻¹⁰ at IWA<br>- Δmag limit = 26.0<br>- 20% instantaneous bandwidth<br>Imager and spectrograph | 72 m dia starshade occulter with:<br>- 124,000 km separation<br>- Raw contrast: 1×10⁻¹⁰ at IWA<br>- Δmag limit = 26.0<br>- 107% instantaneous bandwidth<br>Imager and spectrograph | Imager and spectrograph | High-resolution spectrograph |
| Channels | Vis, Blue: 0.45–0.67 µm<br>  Imager + IFS with R = 140<br>Vis, Red: 0.67–1.0 µm<br>  Imager + IFS with R = 140<br>NIR: 0.95–1.8 µm, Imager + slit spectrograph with R = 40 | UV: 0.2–0.45 µm<br>  Imager + grism with R = 7<br>Vis: 0.45–1.0 µm<br>  Imager + IFS with R = 140<br>NIR: 0.975–1.8 µm<br>  Imager + IFS with R = 40 | UV/Vis: 0.15–0.95 µm<br>  Imager + grism with R = 2,000<br>NIR: 0.95–1.8 µm<br>  Imager + grism with R = 2,000 | UV: 0.115–0.3 µm (20 bands),<br>R = 60,000; 25,000; 12,000;<br>  6,000; 3,000; 1,000; 500 |
| Field of View | FOV: 1.5×1.5 arcsec² @ 0.5 µm<br>IWA: 2.4 λ/D = 62 mas @ 0.5 µm<br>OWA: 0.74 arcsec @ 0.5 µm | FOV: 11.9×11.9 arcsec² (Vis)<br>IWA: 60 mas (0.3–1.0 µm)<br>OWA: 6 arcsec (Vis) | 3×3 arcmin² | 3×3 arcmin² |
| Features | 64×64 deformable mirrors (2)<br>Low-order wavefront sensing & control | Formation flying sensing & control | Microshutter array for multi-object spectroscopy<br>2×2 array, 171×365 apertures | Microshutter array for multi-object spectroscopy<br>2×2 array, 171×365 apertures |

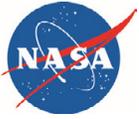

National Aeronautics and
Space Administration

**Jet Propulsion Laboratory**
California Institute of Technology
Pasadena, California

# HabEx: Habitable Exoplanet Observatory

# Interim Report

Astronomy, Physics and Space Technology Directorate
Jet Propulsion Laboratory
for
Astrophysics Division
Science Mission Directorate
NASA

August 2018

# Science and Technology Definition Team

| STDT Community Chairs | |
|---|---|
| Scott Gaudi, Ohio State University | |
| Sara Seager, Massachusetts Institute of Technology | |
| **Study Scientist** | |
| Bertrand Mennesson, NASA Jet Propulsion Laboratory | |
| **Deputy Study Scientist** | |
| Alina Kiessling, NASA Jet Propulsion Laboratory | |
| **Study Manager** | |
| Keith Warfield, NASA Jet Propulsion Laboratory | |
| **Science and Technology Definition Team Members** | |
| Kerri Cahoy, Massachusetts Institute of Technology | Tyler Robinson, Northern Arizona University |
| John T. Clarke, Boston University | Leslie Rogers, University of Chicago |
| Shawn Domagal-Goldman, NASA Goddard Space Flight Center | Paul Scowen, Arizona State University |
| Lee Feinberg, NASA Goddard Space Flight Center | Rachel Somerville, Rutgers University |
| Olivier Guyon, University of Arizona | Karl Stapelfeldt, NASA Jet Propulsion Laboratory |
| Jeremy Kasdin, Princeton University | Christopher Stark, Space Telescope Science Institute |
| Dimitri Mawet, California Institute of Technology | Daniel Stern, NASA Jet Propulsion Laboratory |
| Margaret Turnbull, SETI Institute | |
| **Design Team Members** | |
| Gary Kuan, Team Lead, NASA Jet Propulsion Laboratory | Ron Eng, NASA Marshall Space Flight Center |
| Stefan Martin, Payload Lead, NASA Jet Propulsion Laboratory | Jay Garcia, Jacobs Technology |
| Oscar Alvarez-Salazar, NASA Jet Propulsion Laboratory | Jonathan Gaskin, University of North Carolina |
| Rashied Amini, NASA Jet Propulsion Laboratory | Joby Harris, NASA Jet Propulsion Laboratory |
| William Arnold, a.i. solutions | Scott Howe, NASA Jet Propulsion Laboratory |
| Bala Balasubramanian, NASA Jet Propulsion Laboratory | Brent Knight, NASA Marshall Space Flight Center |
| Mike Baysinger, Jacobs Technology | John Krist, NASA Jet Propulsion Laboratory |
| Lindsey Blais, NASA Jet Propulsion Laboratory | David Levine, NASA Jet Propulsion Laboratory |
| Thomas Brooks, NASA Marshall Space Flight Center | Mary Li, NASA Goddard Space Flight Center |
| Rob Calvet, NASA Jet Propulsion Laboratory | Doug Lisman, NASA Jet Propulsion Laboratory |
| Velibor Cormarkovic, NASA Jet Propulsion Laboratory | Milan Mandic, NASA Jet Propulsion Laboratory |
| Charlie Cox, United Technologies Aerospace Systems | Luis Marchen, NASA Jet Propulsion Laboratory |
| Rolf Danner, NASA Jet Propulsion Laboratory | Colleen Marrese-Reading, NASA Jet Propulsion Laboratory |
| Jacqueline Davis, NASA Marshall Space Flight Center | Jim McGown, NASA Jet Propulsion Laboratory |
| Lisa Dorsett, ARCS | Andrew Miyaguchi, Northrop Grumman Corporation |
| Michael Effinger, NASA Marshall Space Flight Center | Rhonda Morgan, NASA Jet Propulsion Laboratory |

| Design Team Members, continued | |
| --- | --- |
| Bijan Nemati, University of Alabama | Scott Smith, NASA Marshall Space Flight Center |
| Shouleh Nikzad, NASA Jet Propulsion Laboratory | Mark Stahl, NASA Marshall Space Flight Center |
| Joel Nissen, NASA Jet Propulsion Laboratory | Phil Stahl, NASA Marshall Space Flight Center |
| Megan Novicki, Northrop Grumman Corporation | Hao Tang, University of Michigan |
| Todd Perrine, NASA Jet Propulsion Laboratory | David Van Buren, NASA Jet Propulsion Laboratory |
| David Redding, NASA Jet Propulsion Laboratory | Juan Villalvazo, NASA Jet Propulsion Laboratory |
| Michael Richards, Northrop Grumman Corporation | Steve Warwick, Northrop Grumman Corporation |
| Mike Rud, NASA Jet Propulsion Laboratory | David Webb, NASA Jet Propulsion Laboratory |
| Dan Scharf, NASA Jet Propulsion Laboratory | Rush Wofford, Northrop Grumman Corporation |
| Gene Serabyn, NASA Jet Propulsion Laboratory | Jahning Woo, NASA Jet Propulsion Laboratory |
| Stuart Shaklan, NASA Jet Propulsion Laboratory | Milana Wood, NASA Jet Propulsion Laboratory |
| John Ziemer, NASA Jet Propulsion Laboratory | |
| Additional Contributing Scientists | |
| Ewan Douglas, Massachusetts Institute of Technology | Tiffany Meshkat, California Institute of Technology/IPAC |
| Virginie Faramaz, NASA Jet Propulsion Laboratory | Peter Plavchan, George Mason University |
| Sergi Hildebrandt, NASA Jet Propulsion Laboratory | Garreth Ruane, California Institute of Technology |
| Neal Turner, NASA Jet Propulsion Laboratory | |
| Ex-Officio Non-Voting Members | |
| Martin Still, NASA Headquarters | |
| Douglas Hudgins, NASA Headquarters | |
| International Ex-Officio Non-Voting Members | |
| Christian Marois, NRC Canada (Canadian Space Agency, CSA, Observer) | |
| David Mouillet, IPAG Grenoble (Centre National d'Etudes Spatiales, CNES, Observer) | |
| Timo Prusti, ESA (European Space Agency, ESA, Observer) | |
| Andreas Quirrenbach, Heidelberg University (Deutschen Zentrums für Luft- und Raumfahrt, DLR, Observer) | |
| Motohide Tamura, University of Tokyo (Japanese Aerospace Exploration Agency, JAXA, Observer) | |
| Pietr de Wisser, Netherlands Institute for Space Research (SRON) | |

## Acknowledgments

This research was carried out at the Jet Propulsion Laboratory, California Institute of Technology, under a contract with the National Aeronautics and Space Administration.

## Disclaimer

The cost information contained in this document is of a budgetary and planning nature and is intended for informational purposes only. It does not constitute a commitment on the part of JPL and Caltech.

# Table of Contents

## Appendices

# EXECUTIVE SUMMARY

For the first time in human history, technologies have matured sufficiently to enable a mission capable of discovering and characterizing habitable planets like Earth orbiting sunlike stars other than the Sun. At the same time, such a platform would enable unique science not possible from ground-based facilities. This science is broad and exciting, ranging from new investigations of our own solar system to a full range of astrophysics disciplines.

The Habitable Exoplanet Observatory, or HabEx, has been designed to be the Great Observatory of the 2030s, with community involvement through a competed and funded Guest Observer (GO) program. HabEx is a space-based 4-meter diameter telescope mission concept with ultraviolet (UV), optical, and near-infrared (near-IR) imaging and spectroscopy capabilities. HabEx has three driving science goals during its five-year primary mission (**Figure ES-1**):

1. To seek out nearby worlds and explore their habitability.

2. To map out nearby planetary systems and understand the diversity of the worlds they contain.

3. To enable new explorations of astrophysical systems from our solar system to galaxies and the universe by extending our reach in the UV through near-IR.

## HabEx Science

**HabEx would seek out nearby worlds and explore their habitability.** A pervasive and fundamental human question is: Are we alone? Astronomy has recast this elemental inquiry into a series of questions: Are there other Earths? Are they common? Do any have signs of life? Space-based direct imaging above the blurring effects of Earth's atmosphere is the only way to discover and study exo-Earths—Earth-sized planets in Earth-like orbits about sunlike (F, G, and K-type) stars.

With unparalleled high-contrast direct imaging and spectroscopy, HabEx would find dozens of rocky worlds, including a dozen exo-Earths, and hundreds of larger planets around mature stars (**Figure ES-2**). HabEx would characterize exoplanets by determining orbital parameters and obtaining multi-epoch broadband spectra. Of particular interest for investigations of Earth-like exoplanets, HabEx would be sensitive to water vapor, molecular oxygen, ozone, and Rayleigh scattering, detecting these features if they have the same column density as modern Earth or greater. In addition, HabEx would detect other potential biosignature molecules, such as methane and carbon dioxide, if they have concentrations higher than modern Earth. For our nearest neighbors, HabEx would also search for evidence of surface liquid water oceans on exo-Earth candidates.

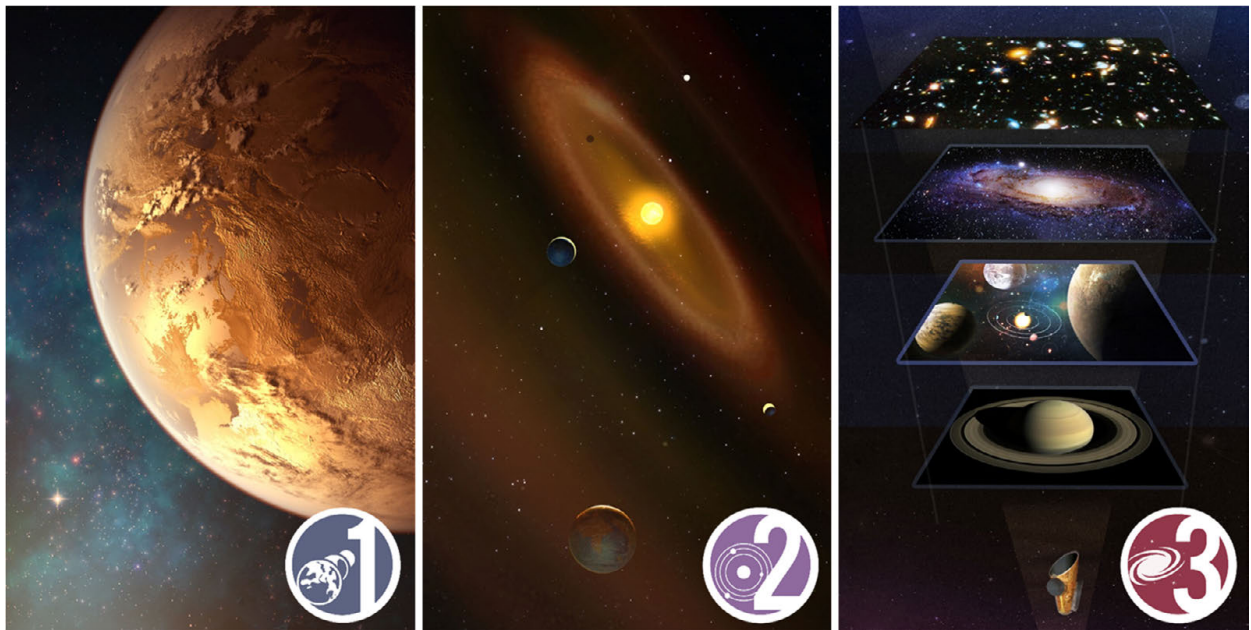

**Figure ES-1.** The HabEx Observatory has three science goals: 1. To seek out nearby worlds and explore their habitability, 2. To map out nearby planetary systems and understand the diversity of the worlds they contain, and 3. To enable new explorations of astrophysical systems from our solar system to galaxies and the Universe by extending our reach in the UV through near-IR.

**HabEx would map out nearby planetary systems and understand the diversity of the worlds they contain.** With high-contrast 11.9×11.9 arcsec$^2$ (equivalent to ~36×36 AU$^2$ at a distance of 3 pc) observations using the starshade, HabEx would be the first observatory capable of providing complete "family portraits" of our nearest neighbors. HabEx would characterize full planetary systems, including exoplanet analogs to Earth and Jupiter (**Figure ES-3**), and exodisk analogs to zodiacal dust and the Kuiper belt. HabEx is also expected to find and characterize a diversity of worlds that have no analogs in our solar system, including super-Earths and sub-Neptunes. These discoveries would provide detailed planetary system architectures, addressing open topics ranging from planetary system formation, to planetary migration, to the role of gas giants in the delivery of water to inner system rocky worlds. HabEx would test theories on planetary diversity, investigate planet-disk interactions, and place our solar system into detailed context for the first time.

**HabEx would carry out observations that enable new explorations of astrophysical systems from the UV through near-IR.** HabEx would be NASA's Great Observatory in the 2030s. Observing with a large aperture from above the Earth's atmosphere in an era

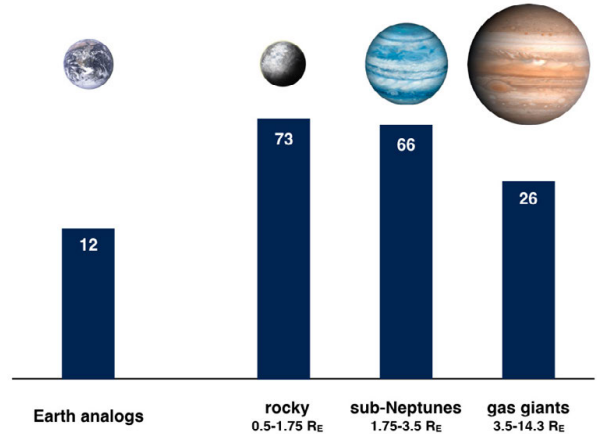

**Figure ES-2.** HabEx would characterize hundreds of exoplanets with a diversity of types, including >10 Earth-like planets in the habitable zone. Overall counts indicated on the right for different planet sizes are restricted to "hot" and "warm" planets, i.e., close enough to their stars that H$_2$O would *not* condensate in their atmospheres. Results shown are the mean HabEx exoplanet yields from realistic Design Reference Mission (DRM) simulations, using the nominal occurrence rates derived by the SAG13 meta-analysis of Kepler data for different planet radii and stellar insulation levels (Belikov et al. 2017; Kopparapu et al. 2018). While many "cold" planets are also expected to be detected with HabEx, their occurrence rate is poorly constrained by the Kepler data analysis and they are not included here. Planet counts are indicated in log-scale.

when neither the Hubble Space Telescope (HST) nor the James Webb Space Telescope (JWST) are operational, HabEx would provide the highest-resolution images yet obtained at UV and optical wavelengths (**Figure ES-4**). HabEx would also provide an ultra-stable platform and access to wavelengths inaccessible from the ground. These capabilities allow for a broad suite of unique, compelling science that cuts across the entire NASA astrophysics portfolio, as well as enabling new views of our own solar system. This "observatory-class" science, which would account for at least 25% of the HabEx primary mission and likely 100% of any extended mission, would be selected through a competed GO program, taking advantage of the community's imagination and priorities to maximize the science return of the mission.

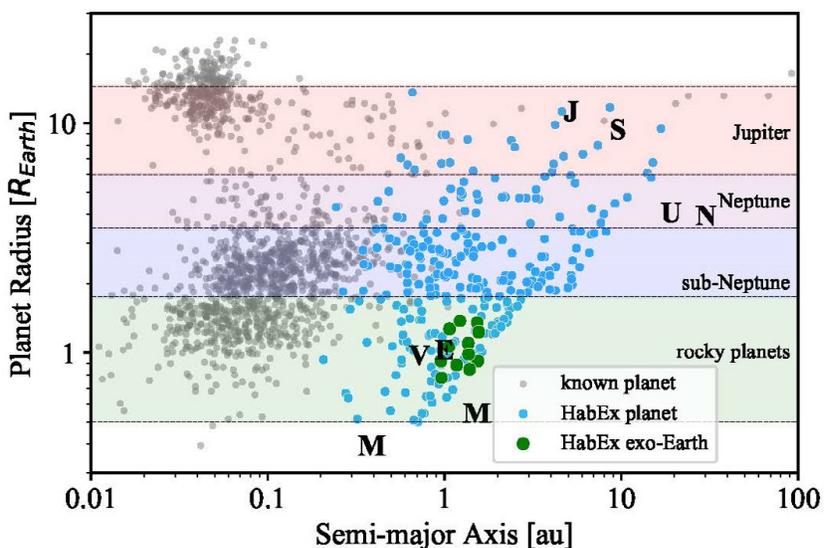

**Figure ES-3.** HabEx would discover and characterize hundreds of new exoplanets (cyan points), including exo-Earths (green points), populating previously unexplored regions of parameter space. Currently detected planets appear as grey dots.

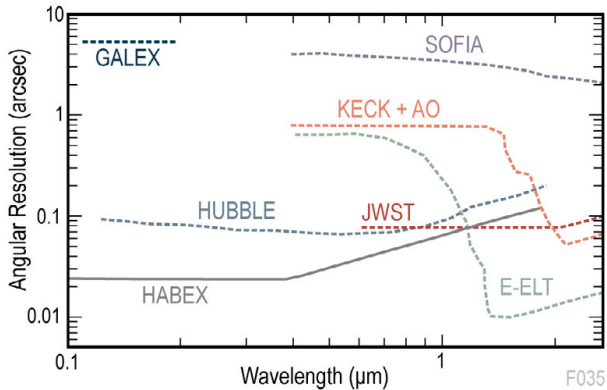

**Figure ES-4.** HabEx would provide the highest-resolution UV/optical images of any current or planned facility, enabling a broad suite of observatory science. Opportunities range from solar system, to stellar populations, to galaxies, to large-scale structure studies.

### HabEx Implementation

The HabEx Observatory baseline design is an off-axis, monolithic 4 m diameter telescope, diffraction-limited at 0.4 µm, in an Earth-Sun L2 orbit (**Figure ES-5**). HabEx has two starlight suppression systems: a coronagraph and a starshade, each with their own dedicated instruments for direct imaging and spectroscopy of exoplanets. HabEx also has two general purpose instruments: a UV spectrograph, and a UV through near-IR imaging spectrograph. The HabEx prime mission is five years, with up to 75% of the time dedicated to two ambitious exoplanet surveys, a *deep* survey of nine of our nearest sunlike stars, and a *broad* survey of 111 nearby mature stars. The primary difference between the surveys is that the deep survey would systematically search for fainter planets, integrating down to a planet-to-star flux ratio detection limit of $4\times10^{-11}$ at the inner working angle (IWA, defined as the closest detectable exoplanet separation), which corresponds to a Mars-sized planet around a sunlike star. In comparison, the individual exposure times for the broad survey are set to maximize the overall yield of Earth-like planets and the flux ratio detection limit will generally be higher than the deep survey ($\sim10^{-10}$).

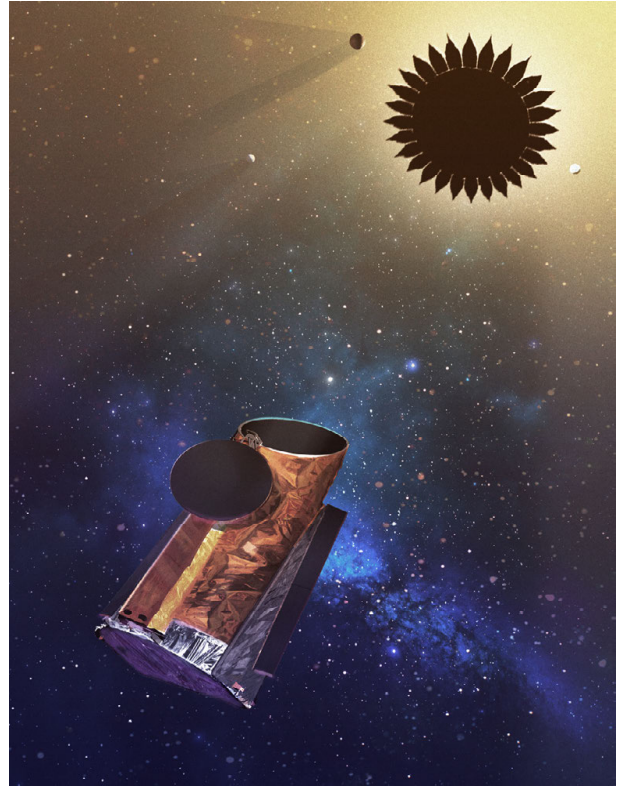

**Figure ES-5.** HabEx, consisting of a telescope and starshade flying in formation, is a proposed Great Observatory of the 2030s. The telescope would include two instruments for direct imaging and spectroscopy of exoplanets: the starshade and the coronagraph. The telescope baseline design includes two facility instruments: a UV spectrograph and a UV through near-IR camera and spectrograph.

The overall HabEx design has been optimized for high-contrast direct imaging and spectroscopy of Earth-sized and larger exoplanets. The off-axis monolithic primary mirror avoids the significant challenges faced by obscured and/or segmented mirrors in achieving both high contrast direct imaging *and* high planet light throughput with a coronagraph. The Earth-Sun L2 orbit provides a stable thermal and gravitational environment, ideal for high-contrast imaging. The dual starlight suppression capabilities provide a flexible approach for optimized exoplanet searches and detailed studies of exoplanets and their planetary systems and is more resilient to uncertainties. The coronagraph is nimble, residing inside the telescope, allowing for efficient multi-epoch surveying of multiple target stars to identify new exoplanet and exo-Earth candidates and also measure their orbits. However, the coronagraph has a narrow annular high-contrast field of view (FOV) with a bandpass limited to 20%. By contrast, the starshade provides a wider FOV and broader instantaneous wavelength coverage than the coronagraph but is fuel limited, rather than target limited, due to the relatively long slews needed to move the starshade between target stars. Importantly, this hybrid approach to direct exoplanet detection and characterization is a powerful combination, taking advantage of the strengths of each instrument and significantly increasing the resultant planetary yields over what is achievable by either instrument alone.

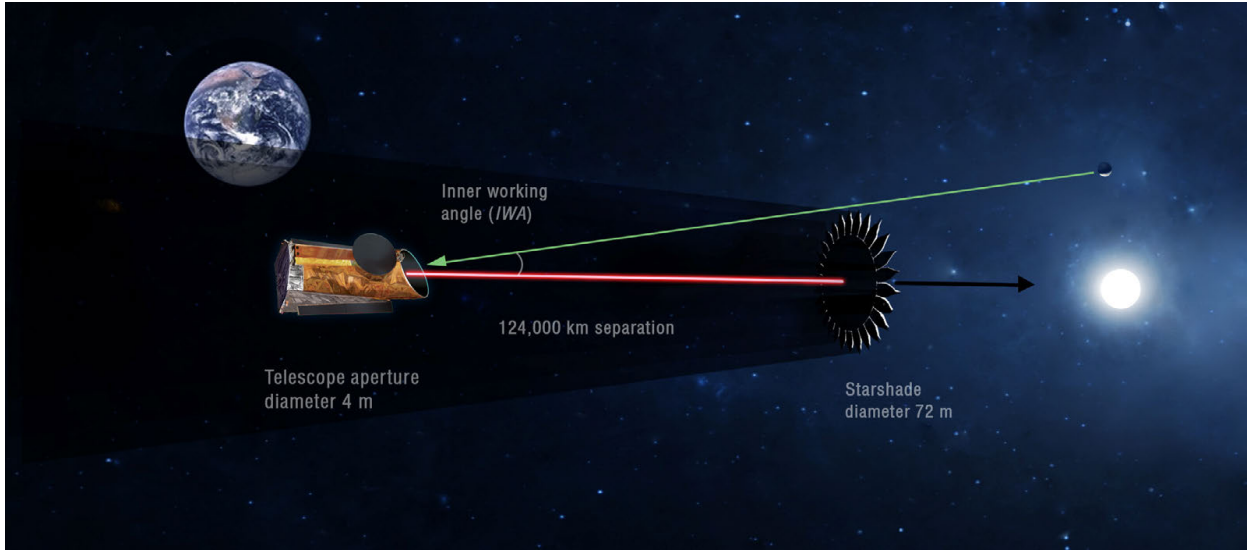

**Figure ES-6.** The HabEx telescope flying in formation with the starshade. HabEx is sensitive to exoplanets outside the shadow of the starshade, which defines the inner working angle.

The four baseline HabEx instruments are briefly described here:

**Starshade.** The starshade blocks starlight before it enters the telescope, allowing light from the exoplanet to be observed. The HabEx 72 m diameter starshade would fly in formation with the telescope at a nominal separation of 124,000 km (**Figure ES-6**). The starshade advantages include a high throughput, small IWA, with an outer working angle (OWA) limited only by the instrument FOV. The HabEx starshade has a 60 milliarcsecond (mas) IWA at 1 μm and a 6 arcsec OWA (for broadband imaging), with deep starlight suppression over an instantaneous bandwidth of 0.3–1.0 μm. The starshade may also operate at two additional separations from the telescope, a larger separation of 186,000 km that covers bluer wavelengths and a smaller separation of 69,000 km that covers redder wavelengths. The former covers an instantaneous bandwidth of 0.2–0.67 μm with a constant IWA of 40 mas, while the latter covers an instantaneous bandwidth of 0.54–1.8 μm with an IWA of 108 mas at 1.8 μm. The starshade instrument has three channels: a near-UV/blue channel covering 0.2–0.45 μm with a grism, a visible channel covering 0.45–1.0 μm with an integral field spectrograph (IFS) and camera, and a near-IR channel covering 0.975–1.8 μm with an IFS and camera.

**Coronagraph.** The coronagraph mask suppresses starlight from within the telescope to reveal the light from the exoplanets. HabEx is baselining a vortex charge-6 coronagraph (**Figure ES-7**) because of its high resilience to common low-order wavefront aberrations, which translates into significantly less stringent requirements on telescope thermal and mechanical stability than other coronagraph designs. The HabEx Observatory coronagraph has a 62 mas IWA at 0.5 μm with a 20% bandpass. The coronagraph has a blue channel with a camera and IFS covering 0.45–0.67 μm, a red channel with a camera and IFS covering 0.67–1.0 μm, and an IR imaging spectrograph that covers 0.95–1.8 μm.

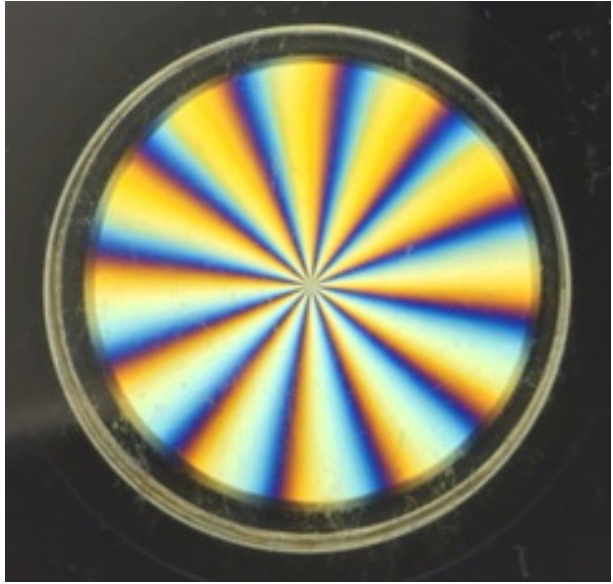

**Figure ES-7.** The vortex charge-6 coronagraph mask, baselined for HabEx, provides high resilience to telescope pointing errors and other common low-order optical wavefront aberrations. Credit: E. Serabyn.

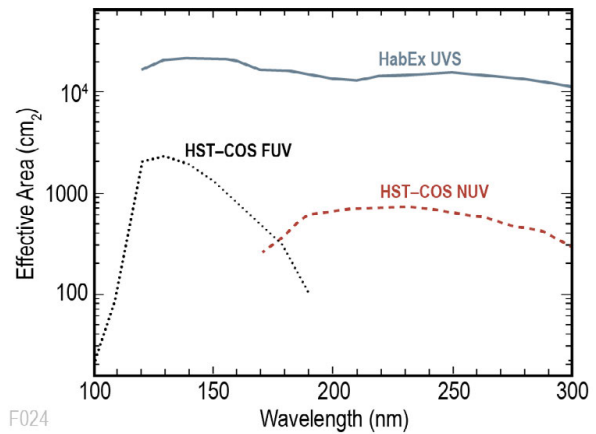

**Figure ES-8.** With more than ten times the effective area of HST-COS, combined with a microshutter array, the HabEx UVS provides several orders of magnitude improved efficiency for UV spectroscopic studies. This would enable the first multiplexed observations of multiple sightlines to a single galaxy, allowing a new probe of the baryon cycle in galaxies.

**UV Spectrograph (UVS).** The UVS has more than 10 times the effective area of HST's Cosmic Origins Spectrograph (COS; **Figure ES-8**). The UVS will be several orders of magnitude more capable than COS. Not only does the UVS provide improved angular resolution and throughput relative to HST, it also includes a microshutter array, allowing multiplexed UV slit spectroscopy for the first time in space. The UVS covers 0.115–0.3 μm with a FOV of 3×3 arcmin² and multiple spectroscopic settings up to resolutions of 60,000.

**HabEx Workhorse Camera (HWC).** The HWC is an imaging multi-object slit spectrograph with two channels covering wavelengths from the UV through near-IR and a spectral resolution of 2,000. The UV/visible channel covers 0.15–0.95 μm and the near-IR channel covers 0.95–1.8 μm. The HWC, with its larger 3×3 arcmin² FOV and higher resolution, will provide capabilities similar to, but significantly more sensitive than, HST's Wide-Field Camera 3 (WFC3) or Advanced Camera for Surveys (ACS).

### The HabEx Observational Strategy

The HabEx exoplanet observational strategy takes advantage of the dual starlight suppression instruments. A broad survey of 111 stars would be undertaken primarily for discovery of small exoplanets. This survey utilizes the coronagraph's pointing agility to revisit the target stars over multiple epochs for discovery, confirmation of physical association with the host star, and measurement of orbits for all detected planets with periods shorter than 10 years. Spectra of these planetary systems are obtained by the starshade. A deep survey utilizing the starshade for multi-epoch broad bandwidth observations of nine of the nearest sunlike stars would provide even more detailed information about our nearest neighbors, with access to even smaller planets and star-planet separations than in the broad survey. Overall, HabEx's hybrid coronagraph/starshade architecture enables a nimble and optimized

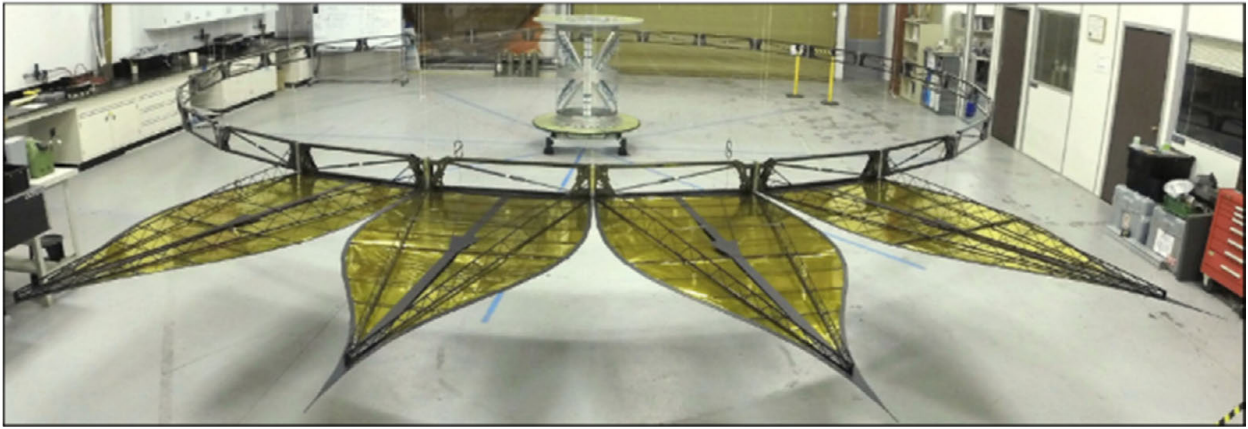

**Figure ES-9.** Prototype starshade truss with petals.

approach to exoplanet discovery and characterization.

The Guest Observer program would be community driven and competitively selected and would likely include solar system, exoplanet, galactic, and extragalactic studies. Both the UVS and HWC are designed for parallel observations of two separate 3×3 arcmin² FOV during observations by the starshade and coronagraph, providing two HST-like ultra-deep fields in the vicinity of the exoplanet target stars and greatly improving the scientific productivity of the HabEx mission concept.

HabEx would start this journey of exploration, providing the first detailed images and spectra of the full range of exoplanets orbiting nearby mature stars, and searching for signs of habitability and life on all of the small rocky worlds detected.

Over the last two decades, dramatic progress has also occurred in four key areas that make HabEx possible today: high-contrast imaging at small angular separations using broadband coronagraphs, starshade-specific modeling developments and technology demonstrations (**Figure ES-9**), manufacturing of large aperture

## Why Now? Scientific and Technological Readiness

There have been tremendous achievements in the discovery of exoplanets over the last 20 years. In particular, astronomers have discovered that small rocky planets around main sequence stars are common. This key result, and the already planned near-term atmospheric characterization of rocky planets orbiting M dwarf stars by missions like NASA's Transiting Exoplanet Survey Satellite (TESS), points to the next logical step: the detailed characterization of Earth-like worlds and complete planetary systems around sunlike stars.

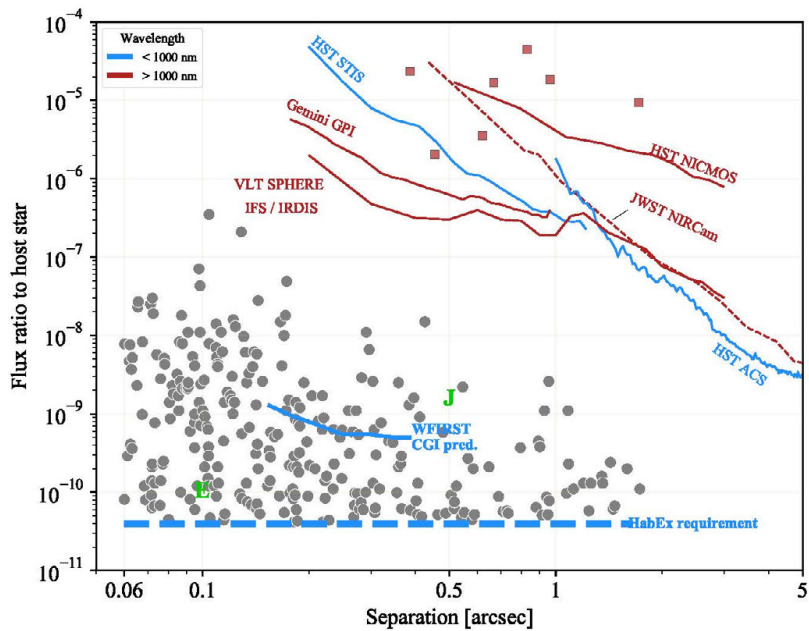

**Figure ES-10.** HabEx would detect and characterize newly discovered exoplanets at high contrast ratios (grey points), enabling the first detailed studies of Earth-like planets in the habitable zone.

monolithic mirrors, and vibration control using microthrusters for fine spacecraft pointing. In particular, steady progress in high contrast direct imaging technology has been very impressive, with the first direct detection of bright self-luminous exoplanets announced in 2008, and the characterization of closer-in self-luminous planets since then (**Figure ES-10**). Through careful design choices, lessons learned from past studies (particularly the Exo-Starshade and Exo-Coronagraph probe studies), and utilization of past and ongoing investments into these technologies, HabEx is able to present a design that minimizes cost and risk, while maximizing scientific return.

## Summary

HabEx is a cost-effective, low-risk, high-impact science mission concept. HabEx would leverage recent advancements in starlight suppression technologies to utilize both a coronagraph and starshade to seek new worlds and explore their habitability and map our nearest neighbor planetary systems to understand the diversity of the worlds they contain. While the HabEx mission architecture is optimized for direct imaging and spectral characterization of a broad range of exoplanets, HabEx also provides unique capabilities for UV through near-IR astrophysics and solar system science from the vantage of space, moving UV capabilities to the next level after HST retires. HabEx is a worthy UV/optical successor to HST in the 2030s with significantly improved sensitivity and spatial resolution stemming from HabEx's significantly larger 4 m diameter aperture, improved detector technology, exquisite wavefront control, and a more thermally stable orbit.

# 1  HABEX: A GREAT OBSERVATORY FOR EXOPLANETARY SCIENCE AND ASTROPHYSICS

Due largely to rapid technological advances, as well as strategic investments in ever more capable ground- and space-based observatories, the fields of astronomy and planetary science have been witness to numerous scientific breakthroughs over the past three decades. Despite this enormous progress in our understanding of our own solar system, other solar systems, and indeed the entire universe and its history, many essential questions remain to be answered, including:

- What fraction of planetary systems look like ours? Are there planets like the Earth, and if so, are these planets potentially habitable and indeed host life of some form?

- What is the complete life cycle of baryons and where are the missing baryons in the local universe? What are the formation histories of the nearest galaxies, including their past accretion, star formation, and dynamical evolutions? What is the nature of dark matter?

- What are the underlying mechanisms of atmospheric escape from bodies in our solar system, as well as exoplanets? What is the origin of Earth's water, and, by extension, the origin of water on other potentially habitable planets?

The Habitable Exoplanet Observatory (HabEx) would provide the crucial capabilities needed to address these questions. Although many new facilities will begin to explore these questions, only HabEx would have the capabilities required to fully answer many of them. Indeed, by identifying the gaps in capabilities that will *not* be filled by existing or planned facilities, the design of HabEx has been optimized to be a uniquely capable, powerful, versatile, and yet attainable observatory, particularly with regards to extreme starlight suppression and ultraviolet (UV) sensitivity.

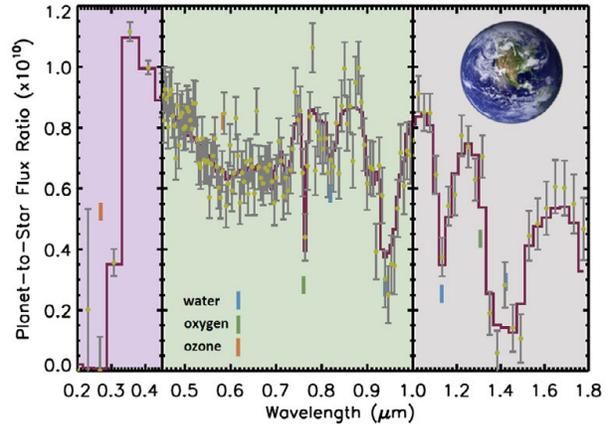

**Figure 1.1-1.** HabEx would be the first mission capable of detecting blue skies, oxygen, and water vapor on habitable planets around nearby stars, as shown in this simulated HabEx spectrum of an Earth-like planet in the habitable zone (see Section 2 for details).

As the Great Observatory of the 2030s and beyond, HabEx would play a major role not only in the next era of astrophysics and planetary science, but will also be the first observatory capable of directly imaging and characterizing Earth-like planets around sunlike stars (**Figure 1.1-1**). This introduction briefly reviews the progress in astrophysics, planetary science, and technology development over the past few decades that now enables, and indeed demands, an observatory with the proposed capabilities of HabEx, in order to address the questions above and many others.

## 1.1  Detecting and Characterizing Exoplanets with HabEx

What fraction of planetary systems look like ours? Are there planets like the Earth, i.e., rocky planets with thin atmospheres located at the right distances from their parent stars that they might have liquid water on their surfaces and so host life as we know it? If so, are these planets potentially habitable, and do these planets indeed host life of some form?

Just over three decades ago, whether other stars hosted planetary systems at all, let alone solar-like systems or Earth-like planets, was not known. It was not until technological developments in the 1980s achieved the capability to detect planets around other stars, that the quest to answer these questions in a scientific manner began.

The difficulty, of course, is that planets like the Earth are very small compared to their parent stars. The radius of the Earth is only roughly 1/100[th] of that of the Sun, the mass of the Earth is only about 3 millionths the mass of the Sun, and the luminosity of the Earth is only about 1 one 10 billionth that of the Sun. These comparisons demonstrate why detecting planets orbiting other stars, particularly Earth-like planets orbiting sunlike stars, is exceptionally challenging.

The first detections of exoplanets came in the late 1980s using radial velocity (RV) and pulsar timing techniques (Campbell, Walker, and Yang 1988, Latham et al. 1989, Wolszczan and Frail 1992). However, it was not until the discovery of the Jovian companion to 51 Pegasi with a period of only 4.2 days by Michel Mayor and Didier Queloz (1995) that the field of exoplanets suddenly 'took off.'

The first broad demographic survey of exoplanets to provide statistics of a large number of planetary systems over a substantial region of parameter space was NASA's Kepler mission, which used ultra-precise photometry to find small transiting planets around their host stars. Originally designed to find Earth-like planets—rocky planets with thin atmospheres in the habitable zones of sunlike stars—Kepler far exceeded expectations. It is now known from Kepler that small planets (with radii less than that of Neptune) on short periods (less than roughly 100 days) are very common, including the class of "super-Earths" and "sub-Neptunes" (planets with radii between that of the Earth and that of Neptune, which have no analogue in our solar system). It is also now know that most, if not all, stars host a planet.

With regards to Kepler's primary goal, quantifying the frequency of Earth-like planets orbiting sunlike stars in their habitable zones, $\eta_{Earth}$, there is a range of results from the different groups and studies that have attempted to answer this question using the Kepler data. This is due to several facts. First, the stellar variability of sunlike

## Why Direct Imaging of Sunlike Stars from Space?

Direct imaging from space is the only way to systematically spectroscopically observe atmospheres of rocky planets orbiting in the habitable zones of nearby sunlike stars. Such planetary atmospheres are inaccessible to ground-based direct imaging where instrumentation is unable to reach the necessary contrast ratios at the required close separations (e.g., Figure ES-10) and transit spectroscopy is not viable because of the low likelihood of transit.

In general, there are two primary methods of characterizing the atmospheres of exoplanets: direct imaging with starlight suppression, and transit spectroscopy of favorably aligned systems where the planetary orbit crosses our line of sight to the host star. For direct imaging, either in thermal emission or reflected light, one gets a spectrum of the host star that has been filtered through the planetary atmosphere once or twice, thus imprinting the constituents of that atmosphere on the stellar spectrum. Studying the brightness of the planet as a function of phase also reveals aspects of the planetary atmosphere, and potentially can detect the presence of a surface ocean (Section 2.1.4). Transit spectroscopy reveals information about the thermal emission of the planet via eclipse spectroscopy, atmospheric constituents via transmission spectroscopy, and the brightness of the planet as a function of longitude via phase curves.

Transit spectroscopy is an attractive approach for low-mass (i.e., M-class) stars for several reasons. First, the larger size of the orbiting exoplanets relative to the star both increases the likelihood of a transit, and increases the amplitude of the transit signal. Second, the habitable zones of low-mass stars are closer to their parent star than for sunlike stars, further increasing the likelihood and amplitude of transits. This, the so-called "small star opportunity," is the motivation behind ground-based surveys such as MEarth (Charbonneau et al. 2009) and Search for habitable Planets EClipsing ULtra-cOOl Stars (SPECULOOS; Gillon et al. 2017), and NASA's space-based Transiting Exoplanet Survey Satellite (TESS; Ricker et al. 2015). However, these same arguments imply that transit spectroscopy is not effective for small planets orbiting larger, sunlike stars.

stars turned out to be larger than was originally assumed. Second, transit surveys are subject to severe selection biases; careful modeling is required to correct for this and uncover the underlying truth. Finally, it is only recently that a thorough end-to-end quantification of the survey sensitivity has been performed (Burke et al. 2015).

Regardless, the net result, based partially on a small number of detected candidates and modest extrapolation, appears to be that $\eta_{Earth}$ is not unity (i.e., not every sunlike star hosts a potentially habitable planet), but nor is it very small. Our best estimates are that $\eta_{Earth}$ is in the range of 8–70% (1-sigma confidence range; Belikov 2017).

This has profound implications for missions whose primary purpose is to directly image and characterize potentially Earth-like planets and search for life. The yield of these missions is primarily contingent on three things: the ability to resolve the planet from the host star (also parameterized by the inner working angle, IWA), the ability to collect enough photons to be able to obtain a spectrum that can robustly identify biosignatures, and the frequency of potentially habitable planets (i.e., $\eta_{Earth}$), which sets (statistically) how far one must look to find a potentially habitable planet. The first two are essentially proportional to the aperture of the telescope. Thus, the larger the value $\eta_{Earth}$, the smaller the aperture required to have access to a given number of targets. Because $\eta_{Earth}$ is in the regime of tens of percent, it is possible to achieve the goal of reliably detecting and characterizing ~10 potentially habitable planets with telescope apertures as small as 4 m. This is the primary motivation for the HabEx architecture considered here.

While Kepler has revolutionized the field of exoplanets, and has provided our first estimates of $\eta_{Earth}$, it has not answered the second question above: How common are planetary systems like our own? This is because Kepler is basically insensitive to planets beyond 1 astronomical unit (AU; the mean distance between the Earth and the Sun). Yet, Kepler has found nearly 4,500 planetary candidates with orbital semi-major axes less than ~1 AU and masses from roughly that of the Earth to that of Jupiter and greater, indicating that most planetary systems *do not* look like our own. Other methods, such as radial velocity, are sensitive to Jupiter analogs, but in order to complete the statistical census of planetary systems, a method that is sensitive to more distant, lower-mass planets is required. NASA's

Wide-Field Infrared Survey Telescope (WFIRST), the next flagship mission after the James Webb Space Telescope (JWST), will perform a microlensing survey that will be sensitive to planets with mass greater than that of Earth and orbital separations from 1 AU to infinity, i.e., including free-floating planets. However, WFIRST microlensing will rarely detect multiple planets in a given system and will not inform the complete architecture of individual systems. The results from Kepler and WFIRST will be combined to provide a more complete census of planetary systems containing planets with mass greater than the Earth and separations from zero to free floating planets, including analogs to all those in our solar system except Jupiter.

However, it is important to emphasize that this compendium will be statistical in nature. As such, it will not be clear whether or not specific architectures like our own, with rocky planets within 2.5 AU concurrent with giant planets beyond 5 AU, are common or rare. For example, there is considerable controversy over the origin of the water on the Earth. It is somewhat of a cosmic paradox that the material in the regions of protoplanetary disks that correspond to the habitable zone in mature systems is well inside the 'snow line,' the distance from the star where water ice is stable in the near-vacuum of the protoplanetary disk. Thus, planets that form in the habitable zones of their parent stars are almost certainly largely devoid of water. Therefore, the liquid water that is now on Earth, which is requisite for life as we know it, was likely delivered from beyond the 'snow line.' While the details of the physics are still an area of active research, most researchers agree that giant planets beyond the snow are essential for delivering the water to the Earth. HabEx would provide data to test these theories and models and untangle the role of outer giant planets in delivering water to inner rocky worlds.

Barring dramatic improvements in radial velocity or astrometric techniques of detecting exoplanets, spaced-based direct imaging is likely the only method that can provide a nearly complete portrait of planetary systems, from potentially habitable inner planets to giant planets beyond the snow line. This is essentially the motivation for one of the primary components of the HabEx's exoplanet survey, deep observations of a handful of the most promising nearby sunlike stars. The goal of this survey is to understand the architectures, orbits, and atmospheres of the planets orbiting our nearest and best-understood neighbors, as well as to study the interaction of these planets with any extant dust disks and their parent stars.

## 1.2 HabEx Observatory Science: Capabilities and Example Applications

In addition to its ability to detect and characterize exoplanets using advanced starlight suppression technologies, as a Great Observatory with unique and unprecedented capabilities and instrumentation (e.g., **Figure 1.2-1**), HabEx would be capable of addressing a broad range of general astrophysics and solar system science questions, including many that cannot be anticipated today. In particular, (1) HabEx would provide the highest-resolution astronomical images ever obtained at UV and optical wavelengths; (2) HabEx would access wavelengths inaccessible from the ground; and, (3) observing from Earth-Sun L2, HabEx's ultra-stable platform would provide extreme precision measurements for a range of science. Highlighted here are a few compelling science themes that HabEx would be uniquely capable of addressing. These are a subset of the themes used to define the functional requirements of the HabEx observatory science instruments, and HabEx would achieve significantly more science than what is introduced here. In particular, this aspect of the HabEx Observatory would be community-led through a competitive, funded Guest Observer (GO) program, taking advantage of advances in our understanding and the community's imagination to produce the most scientifically productive mission.

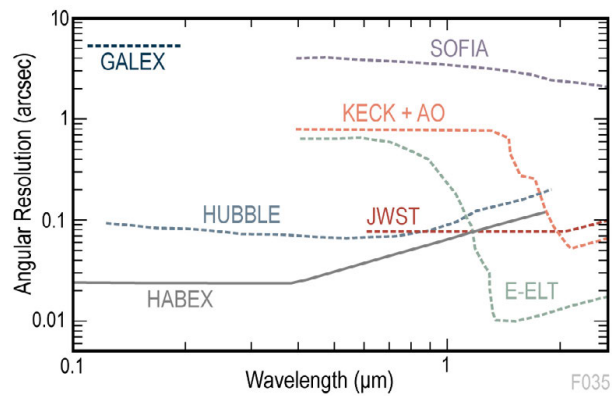

**Figure 1.2-1.** HabEx's 4 m, unobstructed, ultra-stable aperture takes provides unprecedented spatial resolution and effective area at UV and optical wavelengths. These capabilities are not replicated by any currently planned ground- or space-based observatory. This will enable not only the direct detection and characterization of a broad range of exoplanets, including Earth-like planets around sunlike stars, but also unique and exciting astrophysical research, including solar system, galactic, and large-scale structure science. See Sections 2 and 3 for details.

### 1.2.1 Observational Cosmology

Where are the missing baryons in the local universe? What is the complete life cycle of baryons? What is the local value of the Hubble constant and can we use its measurement to elucidate the nature of dark energy? What are the archeologies of the nearest galaxies, including their past accretion, star formation, and dynamical histories? What is the nature of dark matter?

HabEx would build upon the past three decades of research in observational cosmology and extragalactic astronomy, both of which have witnessed revolutions. Although the existence of dark matter has been known for decades (e.g., Zwicky 1933, Rubin & Ford 1970), and successive ambitious experiments and observations have refined our understanding of the properties of dark matter, our physical understanding of dark matter is still shockingly immature. What fraction of dark matter is baryonic? Is the non-baryonic dark matter composed of a single particle, or a whole periodic table of particles? Is dark matter self-interacting? Furthermore, the overall geometry of the universe—e.g., whether it is positively curved, negatively curved, or flat—was not known until the mid-1990s. Even one of the most fundamental properties of the universe, its age, was poorly constrained until the mid-1990s.

The situation is radically different today, with fairly precise measurements on many of the key cosmological parameters. Indeed, astronomy is now in the era of 'precision cosmology.' This is due to many breakthroughs, made possible by highly successful observational campaigns, space-based missions, and the enormous efforts of many astronomers.

These attempts to understand the contents and geometry of the universe were ultimately revolutionized by the discovery that the expansion rate of universe, at the present time, is accelerating (Riess et al. 1998, Perlmutter et al. 1999). One of the most remarkable discoveries of the past few decades, this was made possible by using Type Ia supernovae, which are standard candles, to measure the acceleration rate of the universe. Though theorized by Einstein, observational evidence that the universe is accelerating today was a surprise. Cosmic acceleration indicated that the majority (~70%) of the energy density universe today is composed of a mysterious component with negative pressure, known as dark energy.

Later space-based missions ushered in the era of precision cosmology by measuring the cosmic microwave background (CMB), an echo of the Big Bang, to exquisite precision. These missions include NASA's Wilkinson Microwave Anisotropy Probe (WMAP; Spergel et al. 2003) and the European Space Agency's (ESA) Planck mission (Ade et al. 2014). Meanwhile, wide-field ground-based surveys, such as the Sloan Digital Sky Survey (SDSS), provided complementary constraints, such as the measurement of baryon acoustic oscillations (Eisenstein 2005) and weak lensing measurements of the mass density of the universe and growth of structure. Because of these observations, the overall geometry of the universe in now known to unprecedented precision. Precise measurements of the fraction of its total energy density associated with its key components (i.e., relativistic particles, baryons, dark matter, and dark energy) have also been achieved.

However, this era of precision cosmology leaves at least as many questions as it provides answers. For example, now that a precise measurement of the mass density of baryons in the universe exists, it has become clear that a detailed census of baryons in the local universe falls short of this total by some ~30%. Where are the missing baryons? The most recent measurements of the local Hubble constant (which is the current expansion rate of the universe) appear to be inconsistent with those derived from the CMB. Is this an exciting sign of new physics, or simply evidence for systematics in the CMB and/or local Hubble constant measurements? Is the fact that many nearby low-mass, dark-matter dominated galaxies appear to have flattened cores, rather than having the cuspy cores predicted by the most vanilla flavors of cold dark matter, providing us clues as to the nature of dark matter, or is it simply due to missing physics in galaxy formation models? Finally, although we now appear to understand the contents and geometry of the universe to excellent precision, we still do not understand how the $\sim 10^{-6}$ fluctuations in the matter density at the epoch of the CMB led to the formation of clusters, galaxies, stars, planetary systems, and, ultimately, life.

### 1.2.2    Solar System Science

What is the basic physics underpinning the aurorae of the giant planets? Can we understand the physical mechanisms that determine the properties of the tenuous outer edges of the atmospheres of planets, and can we use this to inform our understanding atmospheric escape from bodies in our solar system, as well as exoplanets? What is the origin of Earth's water, and, by extension, the origin of water on other potentially habitable planets?

While the primary goal of HabEx is the study and characterization of exoplanets, particularly potentially habitable planets, the planets that can be studied in the most detail are those in our solar system. The bodies in our solar system exhibit diverse and complex behaviors, which are difficult to interpret from first principles. Therefore, in order to uncover the fundamental physics driving these phenomena, such as planetary aurorae, atmospheric escape, and water delivery, more detailed and incisive observations are needed to study them. HabEx's instrumentation would enable the detailed studies of solar system bodies.

Importantly, our improved understanding of these phenomena can then be used to complement and enhance the knowledge gleaned from HabEx's exoplanet exploration and characterization surveys. The representative questions posed above are exemplars of this complementarity between the exoplanet and solar system science enabled by HabEx, but they are certainly only a subset of the solar system applications of HabEx.

To place this into context, it is important to recognize that, concurrent with the dramatic progress in the fields of exoplanets and cosmology, our understanding of the contents of the solar system, and our models for its formation and evolution, have undergone dramatic revision, due to a combination of ground- and space-based observations and surveys, as well as an impressive fleet of missions that performed in situ explorations of many of the planets in our solar system and their satellites, as well as small bodies such as Ceres and Pluto.

In many ways, the discovery of the trans-Neptunian belt in 1992 (Jewitt, Luu, and Marsden 1992) heralded the beginning of a transformative era in the study of the solar system. This transformation was further fueled by the discovery of exoplanetary systems, and the recognition that an improved understanding of our solar system will inform our understanding of exoplanetary systems. In turn, the sheer number and diversity of exoplanetary systems places our solar system in context, and informs our understanding of its formation and evolution.

**Giant Planet Exospheres.** The details of the physical interaction of the outer atmospheres of planets (their "exospheres") with their local environments are essential for understanding atmospheric mass loss and the habitability of rocky planets. Planetary aurorae trace the interaction of stellar winds, which can erode atmospheres, with planetary magnetic fields, which can serve to protect those atmospheres from such winds. Thus, by studying these aurorae, the physics of the interactions between planets and their host stars, and how these interactions may affect the habitability of their planets, can begin to be understood. Observations of the tenuous outer atmospheres (exospheres) of planets in our solar system will help elucidate the physics of atmospheric escape, which can then be applied to better understand the observations of evaporating planets in other planetary systems.

**Origin of Earth's Water.** There is considerable controversy over the origin of water on Earth. The raw materials from which Earth was assembled were likely essentially devoid of water. Thus, the water content of Earth almost certainly originated from a population of water-rich bodies beyond the 'snow line,' which is estimated to be at ~2.7 AU for the solar system. There are three leading hypotheses for the primary reservoir from which the water of the Earth was delivered: the outer asteroid belt, which is known to be rich in volatiles; Jupiter-family comets, which likely originated from the trans-Neptunian belt; and long-period comets, which originated from much further out in the solar system. These three sources can be distinguished by their deuterium to hydrogen ratio, which is well measured for the Earth's oceans and is known to be quite distinct for asteroids, Jupiter-family and long-period comets. However, these measurements are quite challenging, and there have been only a handful of objects in each population for which this measurement has been made. The results are currently ambiguous, and it is not clear whether asteroids, Jupiter-family, or long-period comets contributed the bulk of the water on the Earth.

Different models for the formation and evolution of the solar system, which are now much better constrained by a more complete census of the constituents and properties of our solar system, make different predictions for which of these three primary reservoirs the majority of our water was delivered (e.g., Nesvorny et al. 2010). Furthermore, by comparing the architecture of our own solar system to that of nearby planetary systems obtained by the deep direct imaging survey, it will be possible to piece together the ingredients that are required to build not just our own inhabited planet, but potentially habitable planets in general.

## 1.3 Technological Advances that Enable the HabEx Mission

Dramatic technological progress in four key areas, accomplished over the last three decades, make HabEx possible today: high-contrast imaging with coronagraphs, starshade-specific technology developments, manufacturing of large monolithic mirrors, and vibration control using microthrusters for fine spacecraft pointing.

### 1.3.1 High-Contrast Coronagraphy

**History of Coronagraphy.** The idea of using an optical device to suppress the glare of a central bright object to study fainter surrounding structures dates back to French astronomer Bernard Lyot, who invented the coronagraph in 1930 to observe the hot gas ("corona") surrounding the Sun. While coronagraphic observations remained mostly limited to solar corona and solar system's objects until the 1980s (e.g., Trauger 1984), the benefits of using coronagraphy for the study of exoplanetary systems became obvious with the first optical images of beta Pictoris's extended edge-on circumstellar disk obtained by Smith and Terrile (1984). Following the discovery by the Infrared Astronomical Satellite (IRAS) of a large infrared excess around this star, these optical coronagraphic observations provided the first direct confirmation of planet formation and resolved images of dusty debris disks in another system. With the access to space provided by Hubble Space Telescope (HST) and the advent of adaptive optics (AO), wavefront sensing and control to correct the turbulent effects of the Earth atmosphere, increasingly powerful optical/near-infrared coronagraphs came in operation in the 1990s. Many more circumstellar disks were spatially resolved since (e.g., Krist et al. 1998, Clampin et al. 2000, Boccaletti et al. 2003, Ardila et al. 2005, Krist et al. 2005), and bright self-luminous exoplanets were directly imaged and characterized shortly afterwards for the first time using coronagraphy (e.g., Marois et al. 2008, Lagrange et al. 2008). This impressive progress was made possible through many parallel advances in wavefront sensing and control using extreme adaptive optics systems (e.g., Thomas et al. 2011, Langlois et al. 2012, Sauvage et al. 2016), deformable mirror technology, coronagraph designs (e.g., Trauger et al. 2011, Mawet et al. 2009, Kasdin et al. 2003, Guyon et al. 2005), detector technology and data reduction algorithms (e.g., Lafreniere et al. 2007, Soummer et al. 2012). Ground-based instruments can now detect exoplanets $10^6$ times fainter than their host star at separations of ~0.5 arcsec, in the near-IR (Macintosh et al. 2015). Recently, the Exo-C report (2015) presented a probe-class mission concept using 1.4 m unobscured telescope and a coronagraph to obtain direct imaging of exoplanets.

**Recent Technological Advances.** The required improvement over current state of the art to optically detect $10^3$ to $10^4$ times fainter Jovian and rocky planets orbiting solar-type stars has been the topic of a vigorous research and development effort at NASA and through the science community since the early 2000s. In particular, JPL laboratory testbeds have demonstrated adequate levels of narrow-band starlight suppression (~$6\times10^{-10}$, with ten times better stability, Trauger & Traub 2007) at relevant angular separations using unobscured apertures such as the one considered for the HabEx 4 m design. The latest laboratory demonstrations, funded under NASA's Strategic Astrophysics Technology/Technology Development for Exoplanet Missions (SAT/TDEM) program have since concentrated on suppressing starlight over broader wavelength ranges and wider regions of the science image, in particular using multiple deformable mirrors for simultaneous correction of amplitude and phase corrugations, and improved coronagraph designs to reach IWAs closer to the diffraction limit (Trauger et al. 2015, Guyon et al. 2014, Serabyn et al. 2014).

Finally, work completed by the WFIRST Coronagraph Instrument (CGI) also helped move the HabEx coronagraph technologies forward (**Figure 1.3-1**). Large (48×48) deformable mirrors have now been shown to work in the expected thermal environment and a more demanding vibrational environment. Notably, the CGI wavefront sensing and correction system has also already been shown to provide remarkable performance in the laboratory, with pointing and low-order wavefront drift residual errors adequate for reaching contrasts of ~$10^{-9}$ at $3\,\lambda/D$ with WFIRST. Interestingly, because the HabEx 4 m telescope is optimized for coronagraphy with its off-axis primary mirror and slower beam, it is more tolerant to aberrations than the 2.4 m WFIRST. As a result, the level of low-order wavefront control demonstrated by the WFIRST CGI may already be adequate for reaching contrasts of ~$10^{-10}$ at $2.5\,\lambda/D$ with HabEx (see Section 5).

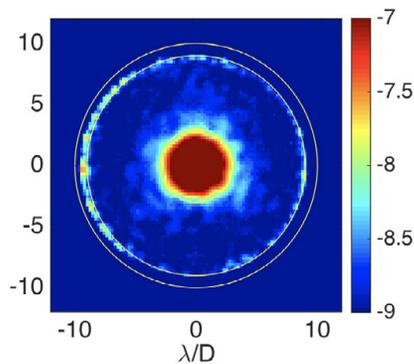

**Figure 1.3-1.** The WFIRST coronagraph testbed has achieved a 360 deg dark hole, with a contrast ratio of $10^{-9}$ from 3 λ/D to 8 λ/D at wavelength λ, where D is the telescope diameter. Results shown are unpolarized light with a 10% bandwidth centered at 0.55 µm. (Jun-Byoung Seo et al., private communication, Dec. 2016).

### 1.3.2 High Contrast with a Starshade

**History of Starshades.** The idea of using a starshade to image planets was first proposed in 1962 by Lyman Spitzer at Princeton (Spitzer 1962). In this landmark paper—in which he also suggested that NASA build and fly what would later become the HST and the Chandra X-ray Observatory—he proposed that an external occulting disk could be used to block most of the starlight from reaching the telescope, thus enabling the direct imaging of planets around nearby stars. He realized that diffraction from a circular disk would be problematic for imaging an Earth-like planet due to an insufficient level of light suppression across the telescope's pupil. He posited that a different edge shape could be used instead, foreshadowing today's approach. In 1974, the idea was revived by G.R. Woodcock of Goddard Space Flight Center. In 1985, Marchal (1985) discussed the use of an opaque disk surrounded by shaped petals, foreshadowing the modern design.

In 1995, the floodgates of exoplanet discovery were opened and interest in external occulters grew. Several mission concepts were proposed using apodized starshades. Copi and Starkman (2000) revisited the apodized starshade and found transmissive solutions defined by polynomials; their proposed mission was called the Big Occulting Steerable Satellite (BOSS). A few years later, Schultz et al. (2003) proposed a similar mission dubbed the Umbral Mission Blocking Radiating Astronomical Sources (UMBRAS). However, these suggestions were hampered by the difficulty in manufacturing a transmissive surface within the tight tolerances necessary. Simmons (2004, 2005) again looked at using starshades based on shaped pupil designs and suggested that the star-shaped design (Vanderbei, Spergel, and Kasdin 2003) was promising.

Shortly thereafter, Cash (2006) showed that an occulter consisting of an opaque solid inner disk surrounded by petals forming an offset hypergaussian function, tip-to-tip about 60 m in diameter, created a broadband, deep shadow. With a small IWA and reasonable manufacturing tolerances, this design finally allowed for the possibility of an affordable solution.

Designs based on a solid inner disk and shaped petals form the basis of several variations in the apodization function. Vanderbei (2007) developed a nonparametric, numerically generated approach to petal shape design. The resulting numerical designs allow for optimization considering engineering constraints, such as petal tip and valley width, petal length, and overall diameter, while preserving desired science performance.

In 2008, two teams were selected under the Astrophysics Strategic Mission Concept Study (ASMCS) to study starshades. Cash et al. (2009) developed the New Worlds Observer mission concept, while Kasdin et al. (2009) developed the Telescope for Habitable Exoplanets and Intergalactic/Galactic Astronomy (THEIA) concept. Both missions used 4 m aperture telescopes coupled with a starshade to achieve the sensitivity required to characterize Earth-like planets in the habitable zones of their parent stars. More recently, the Exo-S report (2015) presents two probe-class exoplanet direct imaging mission concepts, a rendezvous mission designed to work with the WFIRST 2.4 m telescope, and a dedicated mission with the co-launch of a 1.1 m telescope and a starshade.

**Recent Technological Advances.** A number of key starshade technologies have already been demonstrated to a high level through the TDEM component of NASA's SAT program since 2009, including manufacturing starshade petals (Kasdin et al. 2012) and verifying deployment mechanisms (Kasdin et al. 2014) at the required precision, the development of stray light mitigation techniques through modeling and sharp-edge materials development (Casement et al. 2016), and starlight suppression demonstration and model validation through field experiments (Glassman et al. 2016).

Further starshade technology work has been advanced through the "Starshade to Technology 5" (S5) project. Under that project, formation flying sensing will reach Technology Readiness Level (TRL) 5 before the HabEx final report is submitted, and starshade petal edge scatter (Martin et al. 2013) and performance modeling will reach TRL 5 before the National Academies issue their Decadal Survey report. Subscale deployment and shape stability testing will bring the overall starshade to TRL 5 by 2022, before the start of the HabEx project as currently planned.

### 1.3.3 Large Mirror Technology Advances

Schott now routinely makes 4 m blanks for the microlithography industry and has just cast a 4.2 m secondary mirror for the European Extremely Large Telescope (E-ELT) earlier this year. The primary challenge for the Zerodur® mirror had been its open back design, making it less stiff than closed-back mirrors. This added difficulty in achieving tight tolerances on surface figure during manufacture and increased susceptibility to environmental vibrations. This first issue was addressed by an experienced space telescope manufacturer, UTC Aerospace Systems (UTAS), with their method for measuring surface figure during manufacturing while in Earth's gravity. UTAS has been able to produce precision surfaces using this approach. The second issue was addressed by eliminating the primary source of vibrational disturbance, the reaction wheels.

### 1.3.4 Spacecraft Pointing and Vibration Control Advances

Like ESA's highly successful Gaia astrometry mission and the ESA/NASA Laser Interferometer Space Antenna (LISA) Pathfinder mission, HabEx has replaced reaction wheels with microthrusters for tight pointing control. The microthrusters only offset the effects of solar pressure; Earth-Sun L2 station keeping and slewing will be handled by a conventional monoprop hydrazine propulsion system. A microthruster system has been launched on four missions to date, and microthrusters are also baselined for ESA's upcoming Euclid and LISA missions. Gaia has already reached four years of successful L2 operations and will have completed its five-year baseline mission by the time the HabEx final report is submitted. Like Gaia, HabEx is using a phased array antenna, further reducing environmental vibration and enabling continuous science downlink, even during observations.

### 1.4 HabEx: A Great Observatory for the Next Era of Astronomy and Solar System Science

Enormous progress over the past three decades has vastly improved our understanding of our own solar system, other solar systems, and indeed the entire universe and its history. However, many elemental questions remain to be answered. Between now and the expected launch of HabEx, many missions and facilities will come online that will begin to answer these questions. By building upon newfound understanding in these fields, leveraging recent technological advances, and by identifying gaps in these areas of science inquiry that will *not* be filled by existing or planned facilities, HabEx will play a unique and crucial role in addressing many key scientific questions that cut across the full range of the NASA astrophysics and solar system portfolios. In particular, by optimizing the design of HabEx to provide unique and powerful, yet practically attainable, extreme starlight suppression and UV sensitivity, the HabEx Observatory will play a critical role in the next era of discovery, characterization, and understanding in these vast array of topics in astronomy, from the study of solar systems, ours and others, to cosmology.

# 2 CHARACTERIZATION OF EXOPLANETARY SYSTEMS

Humanity has reached an era where the long-standing scientific desire to seek and investigate new worlds and diverse planetary systems is now matched by the technological ability to fulfil that desire. There exists a wide variety of techniques to detect and characterize exoplanets, each with their own strengths, and each with their own limitations and biases (e.g., Seager 2010, Perryman 2014, Wright and Gaudi 2013).

Broadly conceived, detection techniques can be subdivided into direct detection methods, which are directly sensitive to the exoplanet's mass or light, and indirect detection methods, which detect the influence of the planet on its host star. The primary detection techniques are radial velocity (RV), transits, astrometry, microlensing, timing, and direct imaging. Planets that transit or are detected by direct imaging can also generally be characterized in detail as well, e.g., their spectra can be obtained. The uniqueness of space-based direct imaging of planets in reflected light, even in the mid-2030s (Appendix B), resides in the unmatched capability to obtain near-ultraviolet (UV) to near-infrared (IR) spectra of temperate rocky planets around sunlike stars (FGK dwarfs), and search for atmospheric biosignatures and the presence of surface liquid water (Section 2.1). At the same time, such observations will bring detailed family portraits (images and spectra) of most planets with a wide range of sizes and semi-major axes, as well as extended dust structures in nearby exoplanetary systems, thereby putting our own solar system in context for the first time (Section 2.2). The HabEx exoplanet science objectives, surveys strategy, and starlight suppression instruments are all designed to take full advantage of the vantage of space.

Over a nominal prime mission of five years, the HabEx Observatory will directly image and spectrally characterize 120 planetary systems within 17 pc of the Sun. Assuming no prior knowledge of planets in these systems—a worst-case scenario—HabEx direct imaging exoplanet surveys are expected to detect and characterize over 200 planets with a wide range of surface

| Science Goals and Objectives |
|---|
| 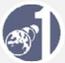 **Goal 1: To seek out nearby worlds and explore their habitability.** |
| **O1:** To determine if small (0.6–2.0 $R_{Earth}$) planets, continuously orbiting within the HZ exist around sunlike stars, surveying enough stars to reach a cumulative HZ completeness >40 for exo-Earths (0.6–1.4 $R_{Earth}$). |
| **O2:** To determine if planets identified in O1 have potentially habitable conditions (an atmosphere containing water vapor). |
| **O3:** To determine if any planets identified in O1 contain biosignature gases (signs of life) and to identify gases associated with known false positive mechanisms. |
| **O4:** To determine if any planets identified in O1 also contain water oceans. |
| 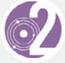 **Goal 2: To map out nearby planetary systems and understand the diversity of the worlds they contain.** |
| **O5:** To determine the architectures of planetary systems around sunlike stars within 12 pc. |
| **O6:** To determine the interplay between planets and dust in planetary systems around sunlike stars within 12 pc. |
| **O7:** To determine if the presence of giant planets is related to the presence of water vapor in the atmospheres of small planets detected in O2. |
| **O8:** To determine the diversity of planetary atmospheric compositions in planetary systems around sunlike stars within 12 pc. |

temperatures and planetary radii, from rocky planets to sub-Neptunes and gas giants. Remarkably, with a search completeness >50% for rocky planets in the habitable zone (HZ) of the vast majority of stars observed, HabEx will detect, measure the orbits, and spectrally characterize a dozen Earth-like planets around sunlike stars. The wavelength coverage from UV (0.2 µm) to near-IR (1.8 µm) and the spectral resolution (R = 140 in the 0.45–1 µm range) capture the absorption bands of key molecular species, which can be used to distinguish between different types of exoplanets. Strong water vapor bands, oxygen and ozone features, carbon dioxide and methane bands, and more are part of the HabEx exoplanet surveys wavelength range.

## 2.1    Goal 1: To seek out nearby worlds and explore their habitability

### 2.1.1    Objective 1: Are there Earth-sized planets orbiting in the habitable zone?

The main observational parameter that controls the number of detected exo-Earths over the 5-year primary mission life is the "completeness" of the search. Here, completeness refers to the probability of detection if every target star hosted an exo-Earth in its HZ. For a given sample of stars, the "cumulative completeness" is the sum of these probabilities over all target stars, and HabEx has chosen a cumulative completeness goal of >40. The cumulative completeness is a function of many mission parameters, with telescope diameter, inner working angle (IWA; the inner bound of the high contrast search area for starlight suppression instruments), point source detection limits, and overall survey time driving the calculation. The parameters are degenerate, but yield simulations demonstrate that a cumulative completeness of >40 is achievable with a 3.7 m aperture telescope and a survey time of ≥2 years using high-contrast imaging instruments with a planet-to-star flux ratio detection limit of $\leq 10^{-10}$ at an IWA ≤ 74 mas in the visible band. Adopting for instance a value of 0.24 for the occurrence rate of exo-Earths around sunlike stars (Belikov et al. 2017) corresponds to detecting at least 10 exo-Earths over the HabEx primary mission. Detection yields for exo-Earths—and other planet types—depend on actual occurrence rates around the HabEx target stars (see Section 2.3 for HabEx exoplanet yield mean estimates and Appendix B for detailed assumptions and yield uncertainties).

However, it is important to go well beyond simple exoplanet detection. HabEx will confirm physical association of any detected point sources with the host star, determine the orbital parameters of detected exoplanets, and place constraints on the exoplanet radii.

**Confirming Physical Association**

Following the detection of a point source, revisits are required to determine whether the point source is an orbiting planet or a distant background object. For the closest stars (<5 pc) the stellar parallax and proper motion will generally be sufficient to confirm common proper motion in two epochs. Two epochs will also suffice for stars between 5–10 pc, except in rare cases where the candidate exoplanet orbital motion is along the background track, where a third epoch will be required to break the degeneracy.

**Orbit Determination**

The planet's orbital semi-major axis, combined with the host star's luminosity, determines the stellar irradiance incident on the planet. The stellar irradiance on the planet, in turn:

- Provides an estimate as to whether or not the planet is inside the HZ (e.g., Kopparapu et al. 2013);
- Is needed to infer the reflectivity of the planet from its apparent brightness relative to the star; and
- Is a main input in atmospheric modeling and spectral retrieval.

For spectral retrieval studies, the incident flux of a planet should be measured to better than 10%.

Orbital eccentricity affects both the instantaneous and orbit-averaged stellar irradiance incident on a planet. Eccentricity may also provide constraints on the formation and dynamical evolution of the planet's orbit.

Measuring the orbit of potentially habitable planets is also important to place the planet in the context of the overall architecture of the planetary system (including inner and outer planets and dust/debris disks). In some multi-planet systems, dynamical stability considerations may refine the range of possible orbital solutions and constrain the planet masses (e.g., HR8799, Fabrycky & Murray-Clay 2010).

Orbit determination via direct imaging requires more revisits to the target star than simply confirming a planet's physical association with its host star. Based on simulations of orbit fitting, at least four well-spaced detections, with position uncertainty ≤ 5 mas rms (**Figure 2.1-1**), are required to achieve 10% precision measurements on the three, key planetary orbital parameters: semi-major axis; eccentricity; and inclination.

Details of the simulations are as follows. The fiducial cases were chosen to span a variety of planets of interest, with semi-major axis of 0.5, 1, 2, and 5 AU, inclination angles of 30, 50, and 80 degrees, and planet radius of 0.5, 1, 2, 4, and 11 Earth radii (for a total of 60 cases). Planets were all assigned the same eccentricity (e = 0), time of periastron, position angle of nodes, and argument of periastron, and are all taken to orbit a solar mass star at 10 pc. Planet position measurements were generated with observations spaced evenly 6 times an orbit, with Gaussian uncertainty added with sigma of 5 mas. At each epoch, the simulation calculates whether the planet is detected using a coronagraph in V-band with an IWA = 62 mas. Finally, utilizing the rejection sampling algorithm Orbits for the Impatient (OFTI; Blunt et al. 2017), each orbit is fit, progressively adding more epochs. **Figure 2.1-1** shows results for a representative set of orbits at 1 AU with an inclination angle of 80 degrees. A precision of 10% is achieved on the key parameters of semi-major axis, eccentricity, and inclination in this particular case after three detections (not counting the two non-detections when the planet was 1) within the IWA and 2) was too faint to be detected because it was in a crescent phase).

For planets discovered during precursor observations using the RV technique, fewer direct imaging epochs are needed to recover the orbital parameters. If the RV orbital parameters are well constrained at the time of the HabEx observation (in particular, assuming that the argument of periastron and time of periastron are well-constrained via, e.g., RV measurements taken during the HabEx mission), a single well-timed observation can determine the last missing parameter, the inclination angle, to within 10 deg.

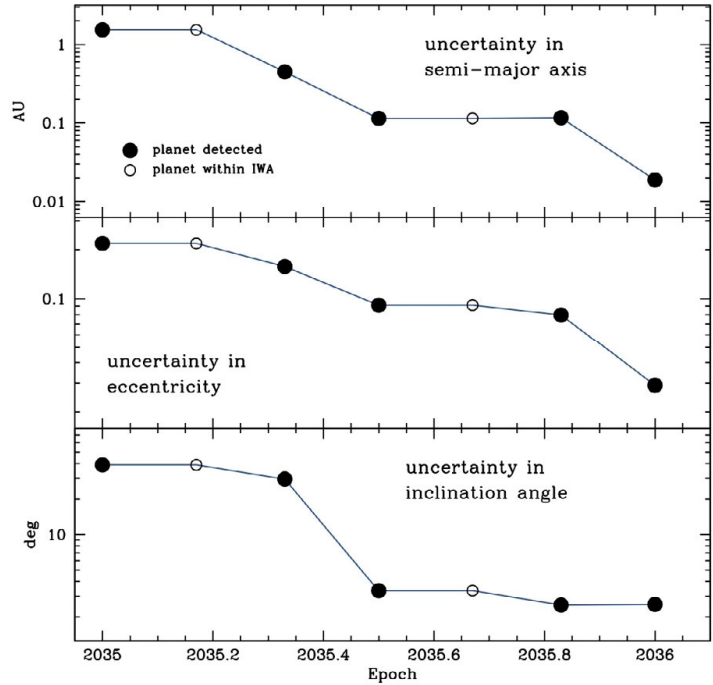

**Figure 2.1-1.** Four well-spaced coronagraph detections are needed to constrain the semi-major axis, eccentricity, and inclination angle of an exo-Earth in a 1 AU orbit to better than 10% (1σ) in this simulation. Credit: E. Douglas

Even in unfavorable cases where the RV orbital parameters are poorly constrained, or the HabEx measurements are not optimally placed, two to three direct detection epochs will suffice to recover the orbital phase and constrain inclination to within 10 deg.

**Planet Radius Constraints**

Photometry alone cannot determine planet size due to the degeneracy between planetary radius and geometric albedo (hereafter simply referred to as albedo), e.g., a given planet could be either small with high reflectivity or large with low reflectivity. However, spectroscopic data can constrain planet radii to a factor of two. In the case of a high signal-to-noise ratio (SNR) observation of a planet with a thin atmosphere, the spectrum provides even better radius constraints (Feng et al. 2018).

The fundamental measurements that HabEx will make are planet position at multiple epochs and the wavelength-dependent planet-to-star flux ratio, $F_p/F_s$ ($\lambda$), such that:

$$\frac{F_p}{F_s} = A_g(\lambda)\Phi(\lambda, \alpha)\left(\frac{R_p}{d}\right)^2,$$

where $\mathcal{A}_g$ is the albedo, $\lambda$ is wavelength, $\Phi$ is the scattering phase function, $\alpha$ is the phase angle (i.e., the star-planet-observer angle), $R_p$ is the planetary radius, and $d$ is the planet's distance from the host star.

To demonstrate that the planet size can be constrained, consider first that the planet illumination phase and orbital semi-major axis will be well known. The primary remaining degeneracy is the $\mathcal{A}_g R_p^2$ product. Based on solar system analogs and transiting exoplanets, $\mathcal{A}_g$ can reasonably be assumed to be between 0.06 and 0.96. With this albedo range, the corresponding values for $R_p$ fall within a factor of 4. For an Earth-twin (same size as the Earth), the estimated radius would span 0.5 to 2 times Earth's radius.

The scattering phase function, $\Phi$, plays only a minor role in the planet-to-star flux ratio because it is dominated by the geometry of the illumination phase (for illumination phase angles less than 100 degrees). This statement is supported by both measured and modeled scattering phase functions, which show relatively little spread at a given phase (Sudarsky et al. 2005) and are typically slowly varying functions of phase angle. Thus, phase uncertainties in the planet-to-star flux ratio equation are unlikely to be a dominant noise source when constraining planetary size. The weak influence of the phase function is further highlighted by the planet-to-star flux ratio expression, which shows that $R_p \propto \Phi^{-1/2}$.

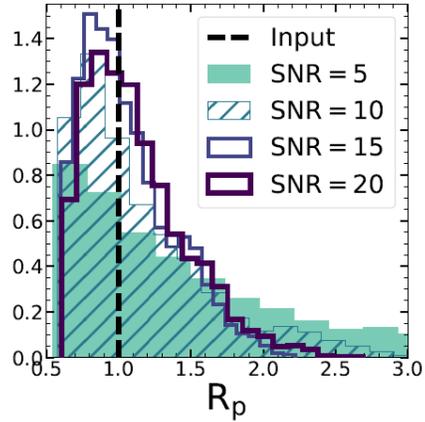

**Figure 2.1-3.** Moderate signal-to-noise (SNR = 10–15) spectroscopy constrains exoplanet radius to a factor of 2 in this simulation of the posterior probability distribution function of the radius of an exo-Earth. Credit: K. Feng

Spectroscopy breaks some of the degeneracy between albedo and exoplanet radius. The shape and strength of key reflected light atmospheric features (e.g., a Rayleigh scattering slope between 0.45–0.70 $\mu$m) encode information about the planet radius (**Figure 2.1-2**). Recently, Bayesian techniques, which originated from the Earth sciences (Rodgers 2000), have been adapted for use in interpreting exoplanet simulated direct imaging reflected light observations. Detailed atmospheric retrieval studies on simulated visible-wavelength spectra of rocky exoplanets (e.g., Lupu et al. 2016) quantify the ability to estimate planet radius based on reflected-light spectral observations. In the case of an Earth-twin, detailed atmospheric retrieval studies (Feng et al. 2018) of a visible spectrum from 0.4–1.0 $\mu$m, indicate that at R $\geq$ 140 and SNR $\geq$ 10 the planetary radius can be retrieved with a 1$\sigma$ uncertainty <60% (**Figure 2.1-3**). The information that constrains the planet radius is found between 0.45–0.70 $\mu$m (**Figure 2.1-2**).

**Objective 1 top-level requirements**

| Parameter | Requirement |
|---|---|
| Telescope diameter | 3.7 m |
| Survey time | ≥2-years |
| Inner working angle | ≤74 mas |
| Planet-to-star flux ratio | ≤10⁻¹⁰ |
| Star-planet separation measurement accuracy (1σ) | ≤5 mas rms |
| Wavelength range | ≤0.45 µm to ≥0.70 µm |
| Spectral resolution | R ≥ 140, SNR ≥ 10 |

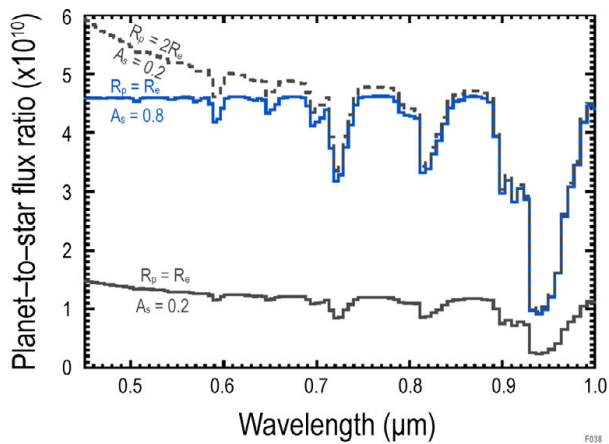

**Figure 2.1-2.** Spectroscopy constrains surface albedo and radius of exoplanets with the same optical brightness due to wavelength-dependent scattering in the atmosphere.

### 2.1.2 Objective 2: Are there Earth-like planets with atmospheres containing water vapor?

"Follow the water." Both in and beyond the solar system, this mantra is key to humanity's search for habitable environments beyond Earth. Water ($H_2O$) is central to life on Earth. Water's ability to act as a polarized solvent that undergoes hydrogen bonding gives it a unique role for all Earth-based life. As a result, water is one of the few requirements shared by all life on Earth, and life is ubiquitous on Earth where appropriate amounts of water can be found.

An important aspect of the HabEx science goal to explore the habitability of new worlds is to infer the presence of surface liquid water oceans on Earth-like exoplanets using several different methods. Objective 2 solely focuses on the detection of atmospheric water vapor (**Figure 2.1-4**), a key indicator of surface liquid water on a small planet. Small (temperate) planets do not hold onto atmospheric water vapor without a liquid ocean reservoir because stellar UV radiation dissociates water vapor in the atmosphere and the hydrogen then escapes into space. In contrast, sub-Neptunes are massive enough to hold onto hydrogen, so water vapor naturally occurs due to atmospheric equilibrium chemistry. Objective 4 (Section 2.1.4) discusses additional observations that can result in greater confidence that liquid surface water has been detected.

Atmospheric water vapor has five broad spectral absorption features from 0.7 µm to 1.5 µm, which can all be detected with a resolution of $R \geq 37$ (DesMarais et al. 2002). Detection of just one water vapor spectral feature is sufficient to securely identify water vapor in the planet's atmosphere. Moreover, detailed measurements of the shapes of multiple water vapor features will provide constraints on water vapor atmospheric abundances. In order to detect at least two water vapor absorption features in the atmospheres of Earth-like exoplanets, HabEx requires an IWA $\leq 74$ mas at 1.0 µm.

**Objective 2 top-level requirements**

| Parameter | Requirement |
|---|---|
| Inner working angle @ 1.0 µm | ≤74 mas |
| Planet-to-star flux ratio | ≤10⁻¹⁰ |
| Wavelength range | ≤0.7 µm to ≥1.5 µm |
| Spectral resolution | R ≥ 37, SNR ≥ 10 |

### 2.1.3 Objective 3: Are there Earth-like planets with signs of life?

The search for life fundamentally requires two related sets of investigations: a search for the gases attributable to life, and characterization of the environment in which those gases arose. This context is critical to support whether biological activity might be active at the surface of the planet, or whether nonbiological production of the signals is also a possibility.

One of the most significant and most detectable signals of life in modern Earth's atmosphere is the presence of large quantities of atmospheric molecular oxygen ($O_2$; **Figure 2.1-4**). Created as a byproduct of oxygenic photosynthesis, $O_2$ has accumulated to the level of 20% by volume of atmosphere on modern Earth. $O_2$ has a strong spectral feature at 0.76 µm. $O_2$ also leads to the accumulation of ozone ($O_3$) in the atmosphere, which has a strong cutoff feature at 0.33 μm and a broad, shallow feature at 0.55 μm (**Figure 2.1-4**). The spectral range required for HabEx to detect both $O_2$ and $O_3$ is 0.3–0.8 μm, with a spectral resolution R ≥ 5 for ozone < 0.35 μm, and R ≥ 70 for molecular oxygen at 0.76 μm.

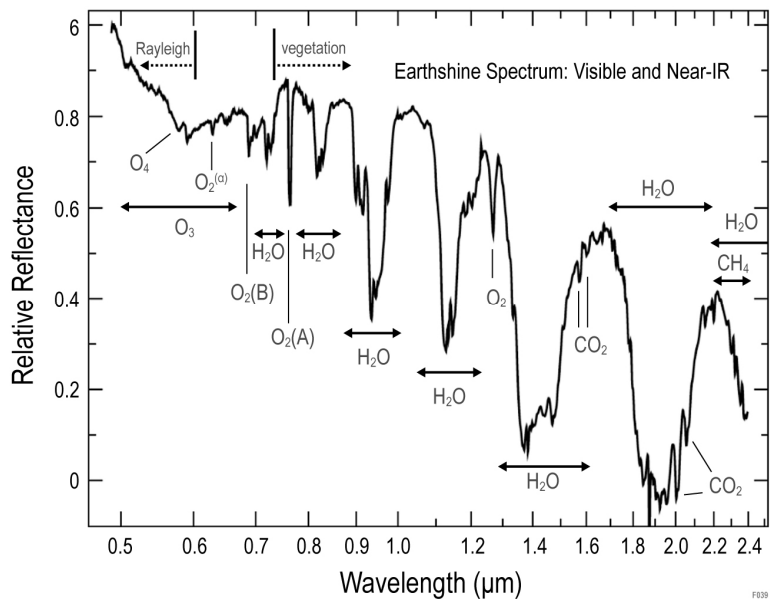

**Figure 2.1-4.** The visible and near-IR spectral range provides a wealth of key molecular features for Earth-like planets, as shown in this disk integrated, high-resolution spectrum of Earthshine. Credit: Turnbull et al. (2006)

Detecting biosignature gases—the byproducts of a biosphere—is much easier than conclusively demonstrating that the gases arise from biological activity. There are two major ways in which a "false positive" for life could arise. First, there could be spectral confusion, where a molecule other than the biosignature gas absorbs at the same wavelength. For a putative detection of molecular oxygen at 0.76 μm, a false positive can be eliminated by looking for additional $O_2$ features at other wavelengths (e.g., 0.69 μm and 1.27 μm) and by searching for absorption from $O_3$, which is a photochemical byproduct of $O_2$ (**Figure 2.1-4**).

The second source of false positives is where the gas in question exists in the exoplanet's atmosphere but has a nonbiological source. Much recent work on confirming biosignatures has focused on how planets can produce $O_2$ and $O_3$ via nonbiological means, and on how to identify such worlds. Many of these scenarios involve very different surface environment conditions, such as a lack of water vapor (Gao et al. 2015), a lack of water clouds due to a sparsity of non-condensable background gases (Wordsworth and Pierrehumbert 2014), or photochemical processes on planets with high carbon dioxide ($CO_2$) concentrations. Notably, some of these mechanisms result in $O_2$ dominated atmospheres with multiple bars of $O_2$ (Luger and Barnes 2015), which appear as wide and deep absorption features from $O_2$, $O_3$, and oxygen dimer ($O_4$), which are all detectable between 0.3–0.76 μm. Fundamentally, the same bulk atmospheric phenomenon is at the heart of all these known false positive mechanisms for $O_2$ and $O_3$: a high oxygen to hydrogen ratio in the planet's atmosphere. If that ratio can be constrained, then all of the known false positive mechanisms can be eliminated. There are three molecules that may provide such constraints: water vapor, methane ($CH_4$), and $CO_2$.

- Photo-dissociation of atmospheric water vapor, followed by massive loss of hydrogen through the top of a planet's atmosphere, may be driven by high-energy radiation from the host star or by a lack of a cold trap that would otherwise keep hydrogen-containing water molecules low in the atmosphere (Wordsworth and Pierrehumbert 2014). This nonbiological process generates a high oxygen to hydrogen ratio and presents as an absence of atmospheric water vapor (Gao et al. 2015). It can be ruled out as an abiotic mechanism for producing $O_2$ if water vapor abundances can be measured through multiple features. Water vapor features become broader and deeper with increasing wavelength, between 0.7–1.8 μm.

- Photochemical processes on planets with $CO_2$ concentrations orders of magnitude greater than modern-day Earth can generate high concentrations of nonbiological $O_2$. This false positive mechanism can be revealed through detection of a deep and broad $CO_2$ feature at 1.59 μm.

- $CH_4$ at modern-Earth abundance level is challenging to detect in the atmosphere of an Earth-like exoplanet. However, detection of $CH_4$ at any abundance level is indicative of an atmosphere that is not $O_2$ dominated. The broadest near-IR $CH_4$ features are at 1.00 μm and 1.69 μm, and can be detected with a resolution of R ≥ 20 (DesMarais et al. 2002).

Another key parameter for constraining false positive mechanisms is the UV flux from the host star. Abiotic $O_2$ and $O_3$ generation mechanisms are believed to be more prevalent for planets around late M dwarf stars with high UV flux rates. This makes exoplanets around M dwarfs a priori less attractive for biosignature searches than F-, G-, and K-type stars. Regardless of spectral type, flare activity from the host star can create a short-lived stellar environment where strong UV radiation can trigger $O_2$ and $O_3$ production. This should be constrained through UV observations of the host star prior to exoplanet direct imaging observations. This combination of stellar UV spectrum and optical to near-IR exoplanet spectra will eliminate all known nonbiological $O_2$ production mechanisms and increase confidence in the identification of exoplanet biospheres.

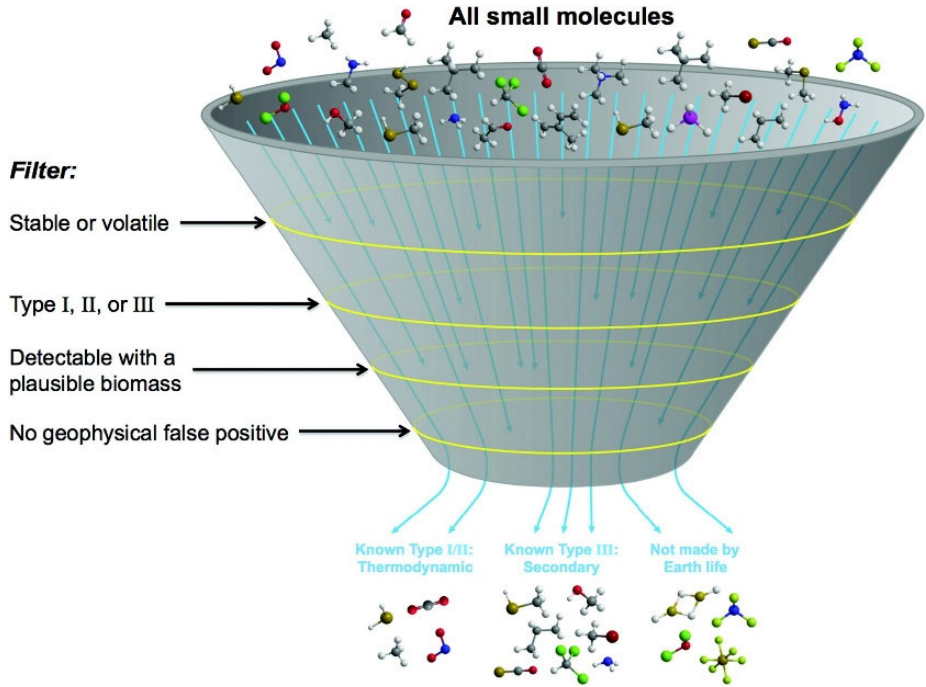

**Figure 2.1-5.** Gases can be ranked in their utility as biosignatures following a progressive filtering process (Seager, Bains, and Petkowski 2016).

Beyond modern Earth's major biosignatures—chiefly $O_2$, $O_3$, and $CH_4$—there surely are other biosignatures for other kinds of exoplanets. The community has an ongoing effort to comprehensively investigate biosignature gases and their false positives (**Figure 2.1-5**). Some, such as methyl chloride ($CH_3Cl$) and dimethyl sulfide ($C_2H_6S$), have been studied in the literature, and others are also starting to be explored (Seager, Bains, and Petkowski 2016).

**Objective 3 top-level requirements**

| Parameter | Requirement |
|---|---|
| **Wavelength range** | ≤0.3 µm to ≥1.7 µm |
| **Spectral resolution** | $O_3$: R ≥ 5, 0.30–0.35 µm with SNR ≥ 5 per spectral bin<br><br>$O_2$: R ≥ 70, at 0.76 µm and $CO_2$, $CH_4$: R ≥ 20, 1.0–1.7 µm with SNR ≥ 10 per spectral bin |

### 2.1.4 Objective 4: Are there Earth-like planets with water oceans?

Water vapor (Objective 2; Section 2.1.2) is strongly suggestive evidence for, yet does not conclusively demonstrate the presence of, liquid water oceans on the surface of the planet. There are two ways that have been proposed to directly detect surface liquid water.

Specular reflection (glint), a disproportionate increase in the brightness of a planet in a crescent phase, indicates a liquid surface (**Figure 2.1-6**; Robinson et al. 2014). The main challenge with detecting glint is that its signal occurs when the planet is at a high illumination phase, resulting in a small apparent separation from the host star. Ocean glint is observed on the Earth at illumination phases >120 deg and also results in an apparent reddening of the planet (**Figure 2.1-7**). To detect glint on an Earth-like exoplanet requires at least two broadband observations at different epochs. One observation is required at an illumination phase > 120 deg for which the continuum planet-to-star flux ratio is a constant $7\times10^{-11}$ for an Earth-twin (**Figure 2.1-6**).

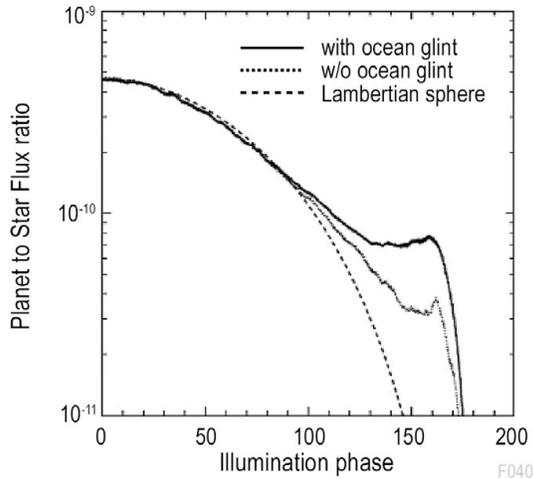

**Figure 2.1-6.** Ocean-glint brightness enhancement is prominent for illumination phases > 120 deg. Shown here is the simulated broadband 1.0-1.1 µm flux ratio of an Earth-twin orbiting a sunlike star as a function of illumination phase Credit: Robinson et al. (2014).

Glint detection is a powerful observational technique to confirm liquid water on a planet surface. Observations combining evidence for the presence of a liquid surface with the detection of water vapor in the atmosphere could reliably point to oceans on other worlds.

The second, related method for detecting oceans is through the polarization that liquid surfaces imprint on reflected light. Depending on the exact nature of the planet observed, ranging from no atmosphere and covered by a calm ocean, to a thick Rayleigh scattering atmosphere with or without clouds, simulations show that the polarization fraction and peak will be reached at different illumination phases (Zugger et al. 2010). Consequently, polarization adds another dimension to the search for surface liquid water oceans, which can be used in combination with unpolarized light curves and planet-to-star flux ratio measurements at different illumination phases to more robustly confirm extrasolar oceans than glint detection alone. The full potential of polarimetry for water ocean studies will be further explored in the HabEx final report.

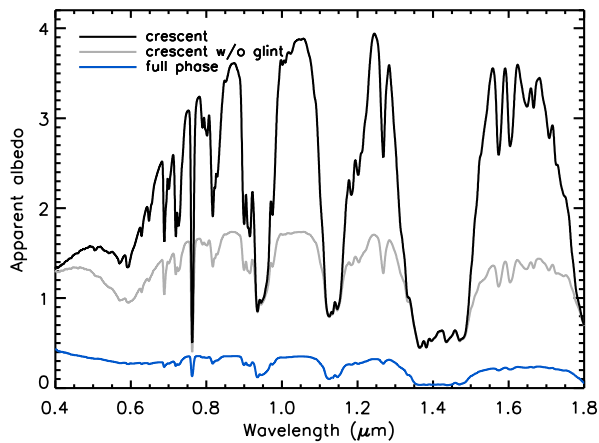

**Figure 2.1-7.** Apparent albedo of Earth at full phase (blue), and at crescent phase (150 deg) both with glint (black). Ocean glint causes Earth-like planets to appear brighter and redder at crescent phase than predicted without oceans (adapted from Robinson (2017) based on the validated model described in Robinson et al. (2011). Apparent albedo is defined as the albedo a Lambert sphere (with radius equal to the planetary radius) would need to reproduce the observed brightness of the planet, and values larger than unity imply forward scattering.

**Objective 4 top-level requirements**

| Parameter | Requirement |
|---|---|
| **Planet-to-star flux ratio detection limit** | ≤7×10⁻¹¹ with SNR > 7 |
| **Phase Coverage** | Broadband photometric measurements at >120 deg<br>Exo-system apparent inclination > 30 deg |

## 2.2 Goal 2: To map out nearby planetary systems and understand the diversity of the worlds they contain

### 2.2.1 Objective 5: What are the architectures of nearby planetary systems?

The solar system has a surprising architecture compared to other planetary systems that have been glimpsed so far. The Sun hosts four terrestrial planets within 1.5 AU, followed by the asteroid belt at 2–3 AU. Beyond the 'snow line' at roughly 2.7 AU, where water ice was stable in the Sun's protoplanetary disk in a near-vacuum, is Jupiter at 5.2 AU, the most massive and most dynamically dominant planet in the solar system. Beyond Jupiter, the remaining three outer planets become less massive and more metal rich. Notably, all eight planets in our solar system orbit in the same direction as the rotation of the Sun, and all have nearly circular orbits. Beyond the orbit of Neptune lies another belt of relatively small objects, the Kuiper belt, which is estimated to be substantially more massive than the asteroid belt.

This architecture was fairly clearly explained in a number of papers before the discovery of exoplanets (e.g., Kokuba and Ida 1998, Pollack et al. 1996). These models invoked in situ formation of the planets within the Sun's protoplanetary disk to explain the features of our solar system, with little-to-no migration of the planets from their birth sites. Granted, theorists had identified a number of problems with this simplistic model (e.g., Weidenschilling 1995), but they all seemed surmountable.

The discovery of 51 Peg b (Mayor and Queloz 1995), a giant exoplanet with a mass similar to Jupiter's but with an orbital period of only ~4 days (the first example of a what is now known as a "hot Jupiter"), led to the questioning of the entire paradigm of solar system formation models. Models generically predict that gas giant planets (primarily made of hydrogen and helium) must form beyond the snow line (e.g., Lin et al. 1996). Thus, the discovery of the planet 51 Peg b implied that large-scale planet migration must be considered, at least for some planetary systems. Further discoveries identified planets that have no analogs in our solar system, such as super-Earths and sub-Neptunes, with masses between that of the Earth and Neptune, as well as planets in highly eccentric and/or non-coplanar orbits.

Despite all this, it still appears as though our solar system followed the basic formation paradigm developed before the discovery of exoplanetary systems, with no large-scale planetary migration. Comparing the architecture and properties of the planets in our solar system to existing knowledge of the properties of other planetary systems, several natural questions arise: What are the architectures of our neighboring planetary systems? Why did the solar system follow one path of formation history, but so many other systems did not? Is our solar system architecture rare?

The solar system is fortuitous because it has an 'inverted' architecture, such that the largest planets are further from the Sun. As a result, the visible planet-to-star flux contrast is roughly constant for Venus, Earth, Jupiter, and Saturn. In addition to the hot Jupiter systems, Kepler has identified many closely packed systems of super-Earths and sub-Neptunes with periods of less than ~100 days (e.g., Lissauer et al. 2011). However, it is unknown whether such systems contain giant planets on more distant orbits.

Characterizing the architectures of full planetary systems requires a deep integration over a wide range of angular separations. The nearest systems form the most favorable targets for these observations, since nearby targets require shorter integration times. However, detecting outer planets in these systems drives outer working angle (OWA; the outer bound of the high-contrast search area) requirements for high-contrast imaging and spectroscopy. In the fiducial case of a nearby solar system analog at 5 pc, detecting the equivalent of Neptune at 30 AU requires HabEx to have an OWA $\geq$ 6 arcsec. For an 0.6 $R_{Earth}$ planet at 1 AU seen at quadrature in the same system, HabEx requires a detection limit for the planet-to-star flux ratio of $\leq 4 \times 10^{-11}$ at V-band, which is equivalent to a $\Delta$mag detection limit of 26.0.

**Objective 5 top-level requirements**

| Parameter | Requirement |
|---|---|
| Planet-to-star flux ratio detection limit | $\leq 4 \times 10^{-11}$ with SNR > 7 at distances <5pc |
| OWA | $\geq 6$ arcsec |

### 2.2.2   Objective 6: How do planets and dust interact?

In addition to planets, circumstellar dust is a key component of an exoplanetary system that can be directly imaged. This section concentrates on the faint exozodiacal dust and exo-Kuiper structures that will be imaged and characterized as part of the HabEx observations focused on addressing Goals 1 and 2, particularly the planetary system observations described in Objective 5 (Section 2.2.1). Optically thick protoplanetary disks and bright extended debris disks are covered in Section 3.7. To detect and characterize the full extent of dusty debris disks in nearby systems, whether exo-Kuiper belt analogs or inner HZ dust structures at solar dust density levels, HabEx requires the capability to image to $\leq 0.3$ times the solar dust level.

**Is the solar system's two-belt architecture common?** The solar system's planetesimal belts are located in regions where the density of solid material was insufficient to form a dominant planetary body to accrete the remaining planetesimals. In this standard solar system formation paradigm, the regions of low planetesimal density are just inside the snow line and at the outer edge of the solar system. Alternatively, stochastic processes could leave planetesimal belts in stable regions between any pair of planets, disconnecting the belt locations from the ice line.

Observations of a large sample of debris disks to solar dust density levels and below have the potential to distinguish between these two formation processes. A comprehensive understanding of planetary system architectures requires measuring the location, density, and extent of dust and planetesimal belts around nearby mature stars. The measurements can also reveal shepherding planets that maintain the shapes of the dust rings, and azimuthal asymmetries in the disk that can be used to infer the mass of the perturbing planet. Additionally, these dust disk observations may also be used to infer the presence of planets too small or faint to image directly, as demonstrated in the case of the bright asymmetric debris disk around Beta Pic (Burrows et al. 1995, Mouillet et al. 1997), and its later imaged planet (Lagrange et al. 2008).

**How is dust produced and transported in debris disks?** Imaging much fainter dust structures with respect to the central star than currently possible crosses an important threshold in disk physics. Brighter disks—all those currently imaged—are collision dominated; the dust grains observed were mainly destroyed by collisions with other grains. Disks with optical depths less than $\sim v_{Keplerian}/c_s$ (where $c_s$ is the speed of sound) are predicted to be transport dominated (Krivov, Mann, and Krivova 2000), meaning that grain-grain collisions are rare enough that grains can flow throughout the planetary system under the influence of radiation drag forces until they are sublimated in the star's corona or ejected from the system by an encounter with a planet.

Dust transport critically depends on the grain size, which defines the surface area to mass ratio and thus the strength of radiative forces on the grain. Grain sizes can be constrained by observing disk colors, scattering phase functions, and the strength of optical polarization. Adding spatially resolved spectroscopy of the disk over a broad-range of wavelengths would also help to constrain grain size distribution, chemical composition, and mineralogy all the way into the HZ. As such, HabEx requires the capabilities for polarimetry and spatially resolved spectroscopy.

Dust disk grains may be too large to produce narrow spectral features at optical and near-IR wavelengths. However, spatially resolved spectroscopy over a 0.8 µm to 1.5 µm band, even at low spectral resolution, R ≥ 20, will enable faint exoplanet broad molecular absorption features to be disentangled from bright dust resonance structures.

**Objective 6 top-level requirements**

| Parameter | Requirement |
|---|---|
| Polarimetric capability | Broadband images in at least 2 polarization states |
| OWA | ≥6 arcsec |
| Spatially resolved spectroscopy | From IWA to OWA ≤0.8 µm to ≥1.5 µm at R ≥ 20 |

### 2.2.3 Objective 7: Do giant planets impact the atmospheric water content of small planets inside the snow line?

The prevailing opinion in planetary science is that the architecture of our solar system had a substantial influence on the habitability of Earth. In particular, Jupiter may have regulated the dynamical delivery of water to Earth by perturbing the orbits of small bodies from beyond the snow line (Raymond, Quinn, and Lunine 2004).

Despite the fact that roughly 70% of the surface of Earth is covered by water, it is important to recognize that the Earth is relatively dry. Ganymede, Europa, Titan, and even Triton, although 30–40 times less massive than Earth, are all believed to have more water by volume than Earth. Earth's surface water is actually just a "thin veneer," although there are debates about how much water is stored in the Earth's interior. This suggests a fine-tuning issue: if the Earth were significantly drier, life might not have been able to thrive; if it were significantly wetter, then it might have been a water world, loosely defined as planets of 50% or more water by mass, planets with deep oceans and no continents. Water worlds may be poor places for life to originate since they suppress the carbon cycle, thought to be essential to maintaining the Earth's temperature and therefore habitability, although this conclusion is controversial (Kite and Ford 2018).

Raymond et al. (2004) argue that if Jupiter had a significantly higher eccentricity, little water would have been delivered to the inner solar system, leaving Earth too dry for habitability. Additionally, if there were no giant planets beyond the orbit of the Earth, Earth would have been bombarded by migrating, water-rich planetesimals from beyond the snow line, leaving a water world. Conversely, Faramaz et al. (2017) argue that mean-motion resonances with exterior planets on moderately eccentric (e ≳ 0.1) orbits scatter planetesimals onto cometary orbits with delays of the order of several 100 Myr, resulting in continuous delivery of water to Earth over Gyr time-scales. The presence of outer Jovian planets in eccentric orbits may then correlate with higher concentrations of water in the atmosphere of rocky worlds found in these systems (Morbidelli and Raymond 2016).

High-contrast observations with HabEx will be able to address this controversy by correlating the abundance of atmospheric water vapor found on detected rocky planets in Objective 2 (Section 2.1.2) with the presence/absence of a Jovian planet in an eccentric orbit. This requires an IWA small enough to image the inner rocky planets and an OWA large enough to detect outer Jovian planets at 5–10 AU distances, between the snow-line and an outer dust belt. For stars at 3 pc, this corresponds to an OWA greater than 3.3 arcsec.

**Objective 7 top-level requirements**

| Parameter | Requirement |
|---|---|
| IWA | ≤74 mas |
| OWA | ≥3.3 arcsec |
| Wavelength range | ≤0.7 μm to ≥1.5 μm |

### 2.2.4 Objective 8: How diverse are planetary atmospheres?

Clement terrestrial and sub-Neptune exoplanets are expected to present a wider range of atmospheres than giant planets. Giant exoplanets have 'primary' atmospheres, formed by accretion during the formation of the planetary system, and are predominantly hydrogen and helium with some degree of metal-enrichment. In contrast, the atmospheres of super-Earths may be 'secondary,' formed by active geological and/or biological processes as is the case for Earth, or may be a primary atmosphere so metal-enriched that the primary species is water.

**Figure 2.2-1** shows examples of the anticipated spectral diversity of a range of sub-giant exoplanets. Cool Neptune-size exoplanets are expected to show strong signatures of $CH_4$ and water vapor—key chemical species that are the dominant forms of carbon and oxygen, respectively. Complex photochemistry in the chemically reduced atmospheres of Neptunes and sub-Neptunes may lead to haze formation, reddening the spectra of these planets. Some terrestrial planets, including Earth and super-Earth sized exoplanets, may possess atmospheres substantially thicker than that of our Earth, which would be indicated by strong Rayleigh scattering features. Alternatively, rocky worlds like Mars, which have experienced atmospheric loss or erosion may appear as barren rock, presenting few spectral signatures beyond their red color. The types of planets and their

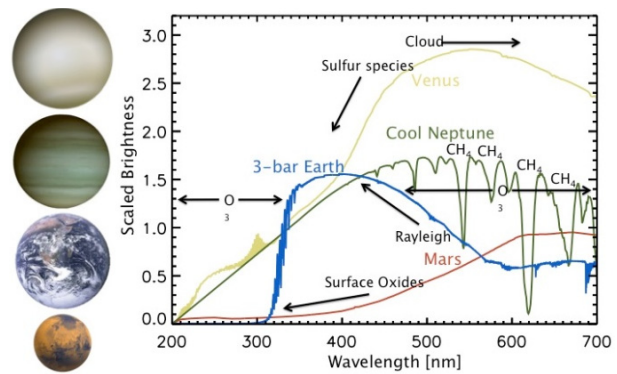

**Figure 2.2-1.** HabEx will begin to map out the true diversity of exoplanets as terrestrial through Neptune-like planets are expected to show a wide diversity in their atmospheric spectral features.

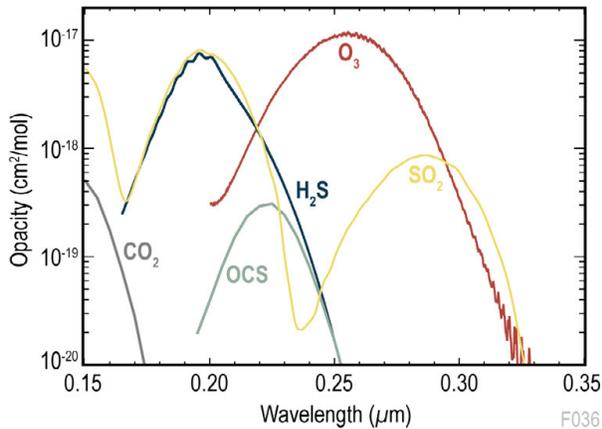

**Figure 2.2-2.** Key molecular absorption features appear in the UV/optical (0.2-0.45 μm), including $O_3$. UV opacities of key species shown are assembled from the Virtual Planetary Laboratory Molecular Spectroscopic Database.

atmospheres are likely to surpass our models and imaginations. By detecting and characterizing terrestrial through Neptune-like exoplanets orbiting nearby stars, it will be possible for HabEx to begin mapping the true diversity of exoplanets.

While the precise mix of atmospheric species to expect is unknown, the absorption bands of key species are well known. Strong water vapor bands, $O_2$ and $O_3$ features, $CO_2$ bands, and more are found in the 0.3 μm to 1.7 μm wavelength range, as shown in **Figures 2.2-2** and **2.2-3**. This wavelength coverage is broad enough to detect and distinguish between deep $CO_2$ atmospheres, water-rich steam atmospheres, the $CH_4$-rich atmospheres of Neptunes and sub-Neptunes, and, critically, the atmospheres of $O_2$-containing Earth-like planets. Furthermore, at a spectral resolution of $\geq 70$, the sharpest features in the visible (i.e., the $O_2$ bands) are well resolved, and a near-IR resolution of $\geq 20$ will reach numerous broad infrared bands for $CO_2$, $CH_4$ and $H_2O$ in particular.

**Objective 8 top-level requirements**

| Parameter | Requirement |
|---|---|
| **Wavelength range** | $\leq 0.3$ μm to $\geq 1.7$ μm |
| **Spectral resolution** | $O_3$: R $\geq 5$, $0.30 - 0.35$ μm with SNR $\geq 5$ per spectral bin<br><br>$O_2$: R $\geq 70$, at 0.76 μm and $CO_2$, $CH_4$: R $\geq 20$, 1.0–1.7 μm with SNR $\geq 10$ per spectral bin |

## 2.3 Exoplanet Science Yield Estimate

HabEx will revolutionize exoplanet science by searching 120 nearby stars for potentially Earth-like exoplanets during the primary mission, discovering hundreds of diverse exoplanets in the process. Precisely estimating the expected exoplanet science yield necessitates modeling the execution of such a mission, which in turn requires constraints on several key astrophysical parameters, such as exoplanet occurrence rates, as well as a high-fidelity simulator of exoplanet imaging missions. A decade ago, such modeling was not possible. Now, the Kepler Mission has constrained the frequency of Earth-sized potentially habitable planets around sunlike stars (e.g., Burke et al. 2015), the Keck Interferometer Nuller (Mennesson et al. 2014), and the Large Binocular Telescope Interferometer (Ertel et al. 2018) have placed constraints on the presence of warm dust around nearby stars, and mission

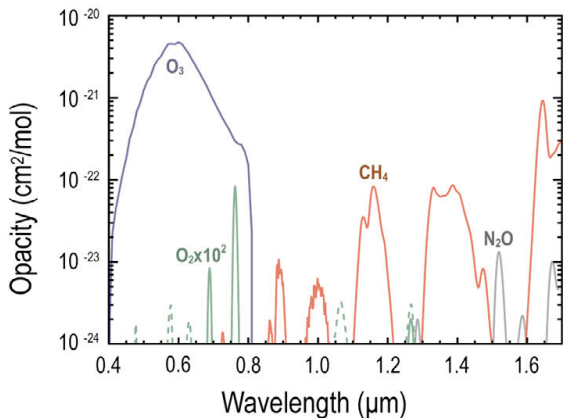

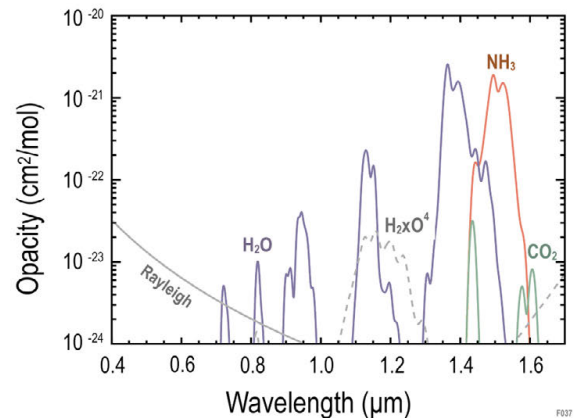

**Figure 2.2-3.** Optical to near-IR coverage 0.4–1.7 μm ensures sensitivity to key molecular absorption features including $O_2$, $H_2O$, and $CH_4$. Opacities of key molecular species are assembled from the HITRAN 2012 database, following Meadows and Crisp (1996).

simulators have advanced dramatically (e.g., Stark et al. 2014, 2015; Savransky et al. 2016).

This section assumes the HabEx architecture described in Section 5, which traces directly from the science requirements described earlier in Sections 2.1 and 2.2. The HabEx exoplanet direct observations make use of a 4 m, off-axis primary mirror and of a hybrid coronagraph and starshade suppression system (**Table 2.3-1**) enabling direct imaging and spectroscopy of Earth-sized and larger exoplanets. HabEx direct imaging spectroscopic capabilities are designed to search for atmospheric biosignature gases and the presence of surface liquid water on HZ rocky planets. The wavelength coverage from UV (0.2 μm) to near-IR (1.8 μm) and the spectral resolution (R = 140 in the 0.45–1.0 μm range) captures the absorption bands of key molecular species, which can be used to distinguish between different types of exoplanets. Strong water vapor bands, oxygen and ozone features, carbon dioxide and methane bands, and more are detectable in the HabEx UV wavelength range. Both the coronagraph and starshade instruments have through near-IR capabilities, including imaging and integral field spectroscopy (IFS). The starshade also includes a UV grism for low-resolution spectroscopy down to 0.2 μm. The high-contrast imaging field of view extends from an IWA of 60 mas and a (maximum) OWA of 6 arcsec. This enables detection of planets over a broad range of orbital semi-major axes.

*One key characteristic of the HabEx dual starlight suppression system is that the starshade is designed to provide the same IWA over the whole 0.3–1.0 μm region as the coronagraph for broadband detection at 0.5 μm.* This means that all of the planets detected by the coronagraph can be characterized spectroscopically by the starshade from 0.3–1.0 μm (within the 100 starshade slews available).

**Table 2.3-1**. Top-level specifications of HabEx direct imaging systems.

| | Coronagraph | Starshade |
|---|---|---|
| **Purpose** | Exoplanet imaging and characterization | Exoplanet imaging and characterization |
| **Instrument Type** | Vortex charge 6 coronagraph with:<br>Raw contrast: 1×10$^{-10}$ at IWA<br>Dmag limit = 26.0<br>20% instantaneous bandwidth<br>Imager and spectrograph | 72 m dia starshade occulter with:<br>124,000 km separation<br>Raw contrast: 1×10$^{-10}$ at IWA<br>Dmag limit = 26.0<br>107% instantaneous bandwidth<br>Imager and spectrograph |
| **Channels** | Vis, Blue: 0.45–0.67 μm<br>Imager + IFS with R = 140<br>Vis, Red: 0.67–1.0 μm<br>Imager + IFS with R = 140<br>NIR: 0.95–1.8 μm,<br>Imager + slit spectrograph with R = 40 | UV: 0.2–0.45 mm<br>Imager + grism with R = 7<br>Vis: 0.45–1.0 μm<br>Imager + IFS with R = 140<br>NIR: 0.975–1.8 μm<br>Imager + IFS with R = 40 |
| **Field of View** | FOV: 1.5×1.5 arcsec$^2$ @ 0.5 μm<br>IWA: 2.4 λ/D = 62 mas @ 0.5 μm<br>OWA: 0.74 arcsec @ 0.5 μm | FOV: 11.9×11.9 arcsec$^2$ (Vis)<br>IWA: 60 mas (0.3–1.0 μm)<br>OWA: 6 arcsec (Vis) |
| **Features** | 64×64 deformable mirrors (2)<br>Low-order wavefront sensing & control | Formation flying sensing & control |

The quantity and quality of exoplanet science that the HabEx mission concept can produce was estimated using established exoplanet yield calculation and target prioritization methods (Stark et al. 2014). The science yield expected from the baseline 4 m HabEx concept and exoplanet surveys–nominally 3.75 years long—is summarized in this section. A complete description of the exoplanet science yield estimates, the techniques used, assumptions made, and justification for the adopted exoplanet observing strategy can be found in Appendix B.

Only the science yield of the high-contrast imaging instruments is presented here. In addition, the HabEx Observatory has two broad-purpose instruments dedicated to a Guest Observer (GO) program: the HabEx Workhorse Camera (HWC) and the Ultraviolet Spectrograph (UVS; see Section 3).

### 2.3.1 Exoplanet Observing Programs and Operations Concepts

HabEx is designed to obtain three primary exoplanet data products: multiband photometry to detect planets and dust disks, spectra to assess chemical compositions, and precise astrometry to measure exoplanet orbits and determine if a planet resides in the HZ. HabEx will obtain these measurements on all planetary systems observed, via two primary observation programs.

### 2.3.1.1 HabEx Broad Exoplanet Survey

HabEx will devote 3.5 years of wall clock time to conduct a search optimized for the detection of exo-Earth candidates around 111 target stars (Appendices B and C). This broad survey will have two components:

1. First, HabEx will obtain multi-epoch coronagraph images of all 111 target stars in the broad survey, with the goal of detecting exo-Earth exoplanets, measuring their colors, and constraining their orbits. Each star will be observed ~6 times on average, with low-priority stars observed at less than a few epochs and high-priority targets observed up to ~10 times. The yield simulations predict that if all target stars within 12 pc had an exo-Earth in their HZ, 57% of them would have such planets detected and their orbits determined using this scenario. In other words, this initial phase is characterized by a >50% "HZ search completeness." These initial multi-epoch coronagraph observations (**Figure 2.3-1**) will allow prioritization of targets and enable the next step of the survey to be optimized with respect to the phases of the detected planets, especially for the smaller inner planets in fast orbits.

2. The next step in this survey uses the starshade to spectrally characterize the most interesting systems identified in the previous step.

Detailed simulations of starshade slews and consumables (Appendix B) indicate that all of the systems observed with the HabEx coronagraph that have an exo-Earth candidate identified, and at least 50% of the systems that did not show an exo-Earth can be imaged and spectrally characterized with the HabEx starshade. Starshade observations will begin with a 11.9"×11.9" broadband visible image, possibly revealing additional outer planets as well as outer dust belts inaccessible to the coronagraph. **Figure 2.3-2** shows a starshade simulated visible image of a putative five-planet system around Bet Hyi, a sunlike target star at 7.5 pc. The whole exoplanetary system can be imaged, with its Earth, sub-Neptune, Saturn, Jupiter, and Neptune analogs, together with its exozodi and exo-Kuiper belts.

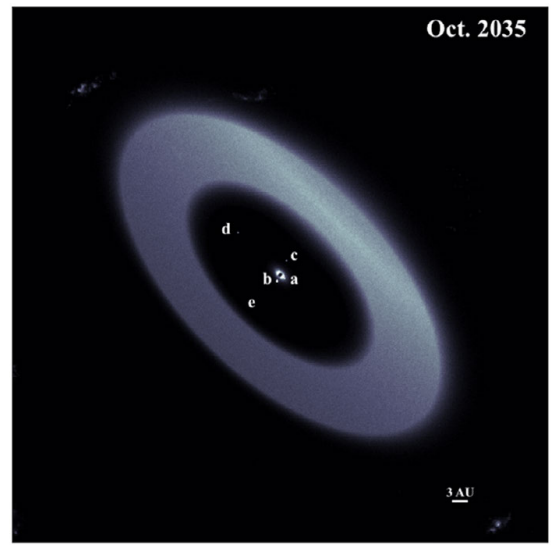

**Figure 2.3-2.** The HabEx starshade instrument detects an exo-Earth (a), sub-Neptune (b), Jupiter (c), Saturn (d), and Neptune (e) analogs around Bet Hyi, a sunlike star 7.5 pc away. The inner dust belt (zodiacal dust analog within 1 AU) and outer dust belt (Kuiper belt analog extending from 20 to 33 AU), both with three times the density of solar system level, are clearly visible together with some background galaxies. This is the same system as in Figure 2.3-1 but now with a field of view: 11.9"×11.9" revealing the outer planets and dust belt. Credit: S. Hildebrandt

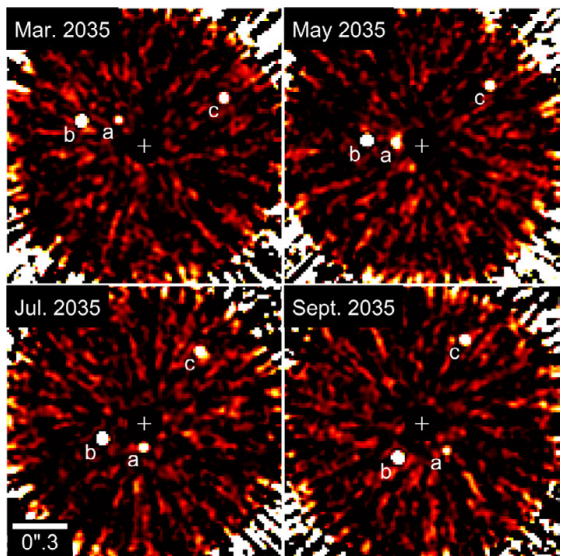

**Figure 2.3-1.** The HabEx coronagraph instrument detects an exo-Earth (a), sub-Neptune (b) and Jupiter analog (c) around a sunlike star in this simulation of four observing epochs. Distance: 7.5 pc; orbital inclination 60 deg; exo-Earth semi-major axis: 1 AU; wavelength range: 0.45–0.55 μm. Credit: G. Ruane.

For all systems with exo-Earth candidates detected, the starshade visible IFS will obtain planetary spectra from 0.3–1.0 μm *in a single observation*, with R = 7 from 0.3–0.45 μm and R = 140 from 0.45–1.0 μm, with an SNR of 10 or higher per spectral channel. This 0.3–1.0 μm portion of the spectrum is obtained by placing the starshade at its nominal distance of 124,000 km. For a few select high-priority exo-Earths, multi-epoch visible spectra will be obtained, and further UV and near-IR characterization will be performed by slewing the starshade to two different distances: 182,000 km to cover 0.2–0.67 μm and 69,000 km to cover 0.54-1.8 μm. **Figure 2.3-3** illustrates the case of a planetary system 7.5 pc away, with an Earth analog at 1 AU, a sub-Neptune at 2 AU, Jupiter, Saturn and Neptune analogs at 5, 10, and 15 AU, respectively.

Planetary systems with no exo-Earth candidate found by the coronagraph will be prioritized for spectral characterization depending on the types and number of planets detected during the previous step of the broad survey.

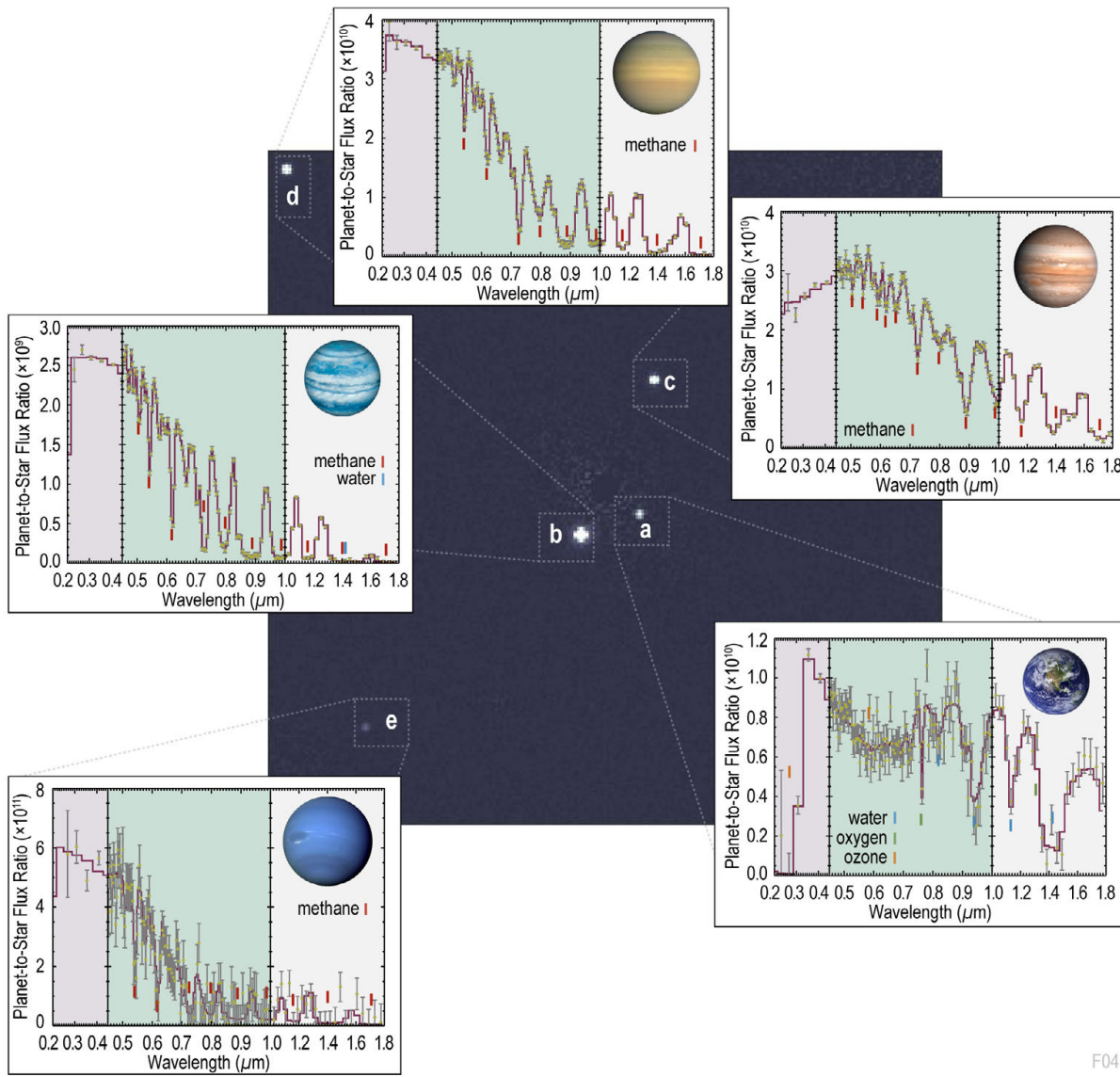

F041

**Figure 2.3-3.** The starshade instrument collects broadband images then near-UV to near-IR spectra of all planets found (2"×2" inner area of simulation in Figure 2.3-1 after fitting and subtracting the extended dust components). This is the same hypothetical planetary system around Bet Hyi shown in the simulations of Figures 2.3-1 and 2.3-2. The assumed exposure is 180h, and covers the 0.3–1.0 μm region of all planetary spectra in a single observation.

As shown by Stark et al. (2016), coronagraphs excel at orbit determination, but take longer to provide a spectrum with broad wavelength coverage. Starshades on the other hand, excel at quickly providing spectra, but can only constrain the orbits for a handful of targets due to the cost of slewing the starshade. The HabEx broad exoplanet survey is designed to fully capitalize on the complementary strengths of both instruments, combining them to provide higher yield and better characterization than either one alone (Appendix B).

### 2.3.1.2   HabEx Deep Exoplanet Survey

For the remaining 3 months of available exoplanet observations, HabEx will perform a "**deep survey**" of nine nearby (3–6 pc) high-priority sunlike stars (**Table 2.3-2**). These stars have been selected based on the very high search completeness that can be achieved through observations at even a single epoch with relatively short exposure times, for a broad range of planet types and physical separations (Appendix B). For this program, HabEx will exclusively use the starshade to observe each star an average of three times. During each deep survey observation, HabEx will:

1. Obtain a deep broadband image down to the assumed systematic detection floor of $\Delta$mag =

26 to search for faint objects. This corresponds to a planet-to-star flux ratio of $4 \times 10^{-11}$, similar to a Mars size planet seen at a gibbous phase in the HZ of a sunlike star. These deep broadband searches can be made quickly given the relative closeness of the target stars.

2. Obtain an R = 7 (grism) spectrum from 0.3–0.45 μm using the starshade UV channel and an R = 140 spectrum from 0.45–1.0 μm using the starshade visible channel IFS. The exposure times will be determined to enable detection of an Earth-twin at quadrature with an SNR = 10 per spectral channel, regardless of whether an exo-Earth candidate exists in the planetary system. Once again, these spectra can be obtained relatively quickly given the targets distance (Appendix B).

These multi-epoch deep exposures and spectra will provide an unprecedented reconnaissance of nine of our closest neighbors, revealing the characteristics of their planetary systems and interplanetary dust structures in exquisite detail. The current list of deep survey stars remains illustrative: the exact number and identity of these high-priority targets may be revised based on additional knowledge about specific systems available by the time of the HabEx observations.

**Table 2.3-2** Many of the nine deep survey targets, nearby sunlike stars, have captured the public's imagination for centuries.

| Star | Type | Dist. (pc) | V-mag | Age (Gyr) | Notes |
|------|------|-----------|-------|-----------|-------|
| τ Ceti | G8V | 3.7 | 3.5 | 5.8 | **Astronomy:** closest solitary G-star, 4 confirmed planets (2 in HZ) plus debris disk<br>**Popular culture:** homeport of *Kobayashi Maru* in *Star Trek* and location of *Barbarella* (1968) |
| 82 Eridani | G8V | 6.0 | 4.3 | 6.1–12.7 | **Astronomy:** 3 confirmed planets (all super-Earths) plus dusk disk |
| ε Eridani | K2V | 3.2 | 3.7 | 0.4–0.7 | **Astronomy:** 1 unconfirmed planet (Ægir) plus dust disk<br>**Common name:** Ran<br>**Popular culture:** location of *Babylon 5* (1994–1999) |
| 40 Eridani | K1V | 5.0 | 4.4 | | **Astronomy:** triple-system, with white dwarf and M-dwarf<br>**Common name:** Keid<br>**Popular culture:** in *Star Trek*, host star to Vulcan |
| GJ 570 | K4V | 5.8 | 5.6 | | **Astronomy:** quadruple-system, with 2 red dwarfs and brown dwarf |
| σ Draconis | K0V | 5.8 | 4.7 | 3.0 ± 0.6 | **Astronomy:** 1 unconfirmed planet (Uranus-mass)<br>**Common name:** Alfasi<br>**Popular culture:** visited in *Star Trek* episode "*Spock's Brain*" (1966) |
| 61 Cygni A | K5V | 3.5 | 5.2 | 6.1 | **Astronomy:** wide-separation binary<br>**Common name:** Bessel's star |
| 61 Cygni B | K7V | 3.5 | 6.1 | | **Popular culture:** home system of humans in Asimov's *Foundation* series |
| ε Indi | K5V | 3.6 | 4.8 | 1.3 | **Astronomy:** triple-system, with 2 brown dwarfs<br>1 unconfirmed planet (Jupiter-mass) |

**Figure 2.3-4** shows the time allocation assumed in our initial DRM between the deep and broad exoplanet surveys, and between the two starlight suppression instruments. Latest DRM results indicate that a ~2 times higher time fraction, i.e., 50%, could be devoted to GO with only minor (~15%) impact on the overall exoplanet surveys yield. Further DRM and time allocation optimization will take place by the final study report. Time fractions shown include all wavefront control and pointing overheads, but not starshade slew times, since coronagraphic and GO observations are conducted while the starshade is slewing from target to target. GO observations would cover a much larger fraction of an extended mission. The pie chart does not include the anticipated parallel observations (e.g., deep field imaging and spectroscopy with the HWC and UVS instruments during long exoplanet direct imaging exposures).

### 2.3.2   Exoplanet Yields for the Baseline 4-Meter Concept

For the *combined deep and broad surveys*, the expected yield of detected and spectrally characterized exo-Earth candidates for the baseline HabEx mission is $12^{+18}_{-8}$, where the range is set by uncertainties in the frequency of exo-Earths ($\eta_{Earth}$) and finite sampling uncertainties (see Appendix B). For each exo-Earth candidate characterized, the spectra will reveal the presence of water vapor, molecular oxygen, ozone, and Rayleigh scattering in the planet's atmospheres, if present with the same column density of modern Earth. **Figure 2.3-5** shows a simulated HabEx starshade spectrum of an Earth-like planet at quadrature around a sunlike star located at 10 pc. The starshade provides a broad instantaneous spectral range from 0.3–1.0 μm with a single exposure and instrument setting; spectral coverage reaching bluer (down to 0.2 μm) or redder (up to 1.8 μm) wavelengths requires additional observations with the starshade, more distant or closer to the telescope, respectively.

While searching for and characterizing exo-Earth candidates, HabEx will detect hundreds of other planets, from hot rocky worlds to cold gaseous planets. **Figure 2.3-6** shows the nominal number and types of exoplanets expected to be detected during the HabEx broad coronagraphic survey (1.95 years), using the default occurrence rates derived from Kepler data (Belikov 2017), a constant exozodi level of 3 zodis per star, and the planet size and temperature classification scheme recently proposed by Kopparapu et al. (2018). Red, blue, and cyan bars indicate hot, warm, and cold planets, respectively. The inset green bar shows the predicted yield of exo-Earth candidates, which is a subset of the warm rocky planets. Using these assumptions and instrument performance models consistent with its detailed telescope and coronagraph design specifications (Appendix B and Section 5), it is estimated that HabEx will detect and characterize the orbits of 92 rocky planets (radii between 0.5–1.75 Re), among which ~12 Earth analogs, 116 sub-Neptunes (1.75–3.5 Re) and 65 gas giants (3.5–14.3 Re). The yields are based on optimizing the observation plan for the detection and characterization of exo-Earth candidates.

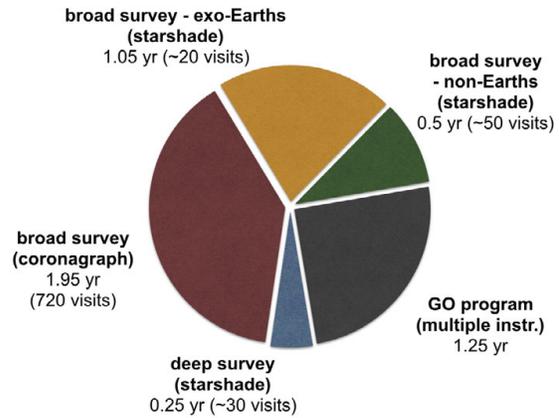

**Figure 2.3-4.** Notional HabEx time allocation for a 5-year primary mission. The broad-survey uses both the coronagraph (for multi-epoch imaging) and the starshade (for spectroscopy). The deep survey only uses the starshade. Latest DRM results indicate that a significantly higher (50%) time fraction could be devoted to GO with minor (~15%) impact on the exoplanet surveys yield.

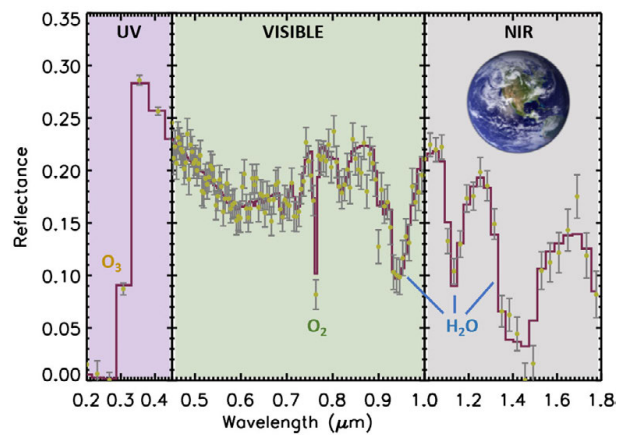

**Figure 2.3-5.** HabEx will clearly detect the ozone cutoff below 0.33 μm, atmospheric Rayleigh scattering at blue wavelengths, and multiple signatures of molecular oxygen and water vapor. Shown is a simulated 300-h starshade observation of an exo-Earth at quadrature around a sunlike star at 10 pc.

Both the broad and deep surveys rely on multiple visits to individual stars, and HabEx will hence obtain multiband, multi-epoch photometry for all planets discovered. For the vast majority of planets detected on orbits shorter than ~15 years, HabEx will also measure orbital periods and phase-dependent color variations.

HabEx will use the starshade to perform a total of ~100 slews for spectral characterization observations. The deep survey program will use ~30 of these starshade observations on nine targets. The remaining ~70 slews will be used to spectrally characterize all systems with Earth-analogs discovered, and >~50% of the other planetary systems. As can be seen in **Figure 2.3-7**, all of the planets discovered by HabEx occupy a region currently unexplored in the radius vs. separation parameter space. Additionally, most of the rocky planets found and spectrally characterized by HabEx will be orbiting around sunlike stars (FGK dwarfs). Transit measurements from space (e.g., Transiting Exoplanet Survey Satellite [TESS], JSWT, PLAnetary Transits and Oscillations of stars [PLATO]) and direct imaging with ground-based ELTs should be able to spectrally characterize such

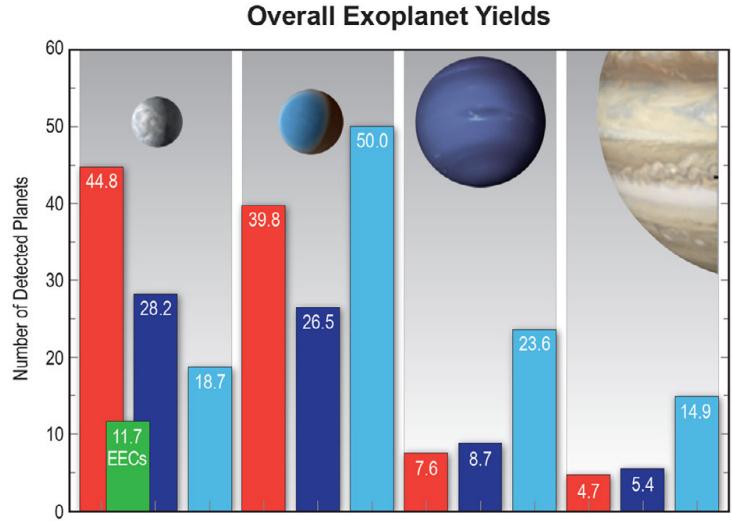

**Overall Exoplanet Yields**

Number of Detected Planets

44.8 · 28.2 · 18.7 · 11.7 EECs · 39.8 · 26.5 · 50.0 · 7.6 · 8.7 · 23.6 · 4.7 · 5.4 · 14.9

**Figure 2.3-6.** HabEx is expected to detect more than 200 exoplanets over a wide range of surface temperatures and planetary radii: 92 rocky planets, among which ~12 Earth analogs, 116 sub-Neptunes, and 65 gas giants. Counts are based on the nominal SAG13 occurrence rates for each planet size and stellar insulation level, following the classification established by Kopparapu et al. 2018 (red bars: "hot" planets; dark blue bars: "warm" planets; ice blue bars: "cold" enough planets that $H_2O$ would condensate in their atmosphere). Occurrence rates estimates are highly uncertain for cold planets. See Appendix B for details on occurrence rates uncertainties.

planets around M dwarfs in the near to mid-term future. But only a mission like HabEx could provide detailed near UV to near-IR spectra of temperate rocky planets and true Earth analogs around sunlike stars, detect biosignatures in their atmosphere, and take complete family portraits of individual exoplanetary systems and dust structures. The yield estimation results highlight

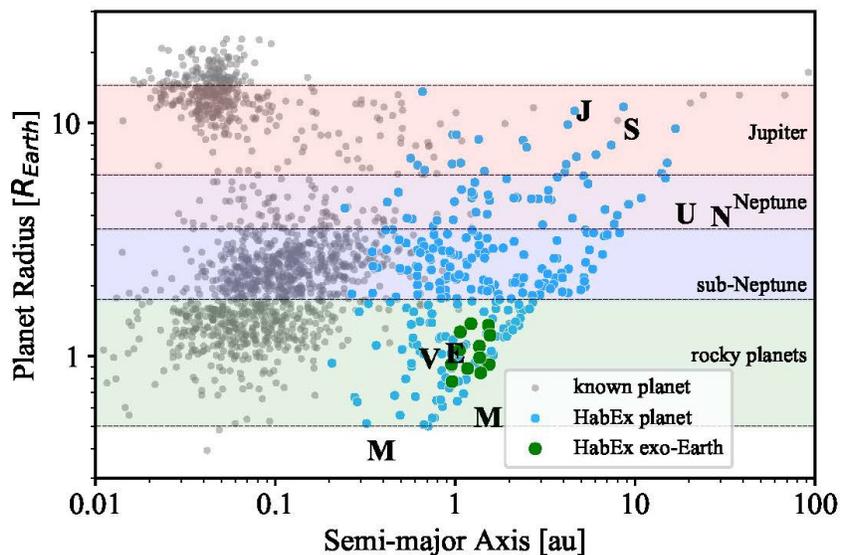

**Figure 2.3-7.** Expected distribution of HabEx discovered planets as a function of physical separation (at quadrature) and radius. Credit: T. Meshkat

both the uniqueness and complementarity of HabEx's exoplanet science grasp.

The baseline target star list characteristics for the combined deep and broad surveys are summarized in **Figure 2.3-8**. HabEx will observe a total of 120 stars covering a wide variety of spectral types, among which 109 are FGK stars. Because this list is prioritized for high exo-Earth search completeness, nearly all A stars are discarded (because of prohibitive flux-ratio requirements), and only 9 nearby early-type M-stars are selected (due to prohibitive IWA requirements). For the same reasons, the highest completeness for HZ planets is obtained for G and K stars, which provide the best trade-off between contrast and angular separation requirements. HabEx will achieve >50% HZ completeness on the majority of stars observed, with 57% average HZ completeness for stars closer than 12 pc (meeting Objective 1, see Section 2.1.1).

Over 90% of the stars are brighter than $7^{th}$ magnitude and all are closer than 17 pc. The full target list can be found in Appendix C.

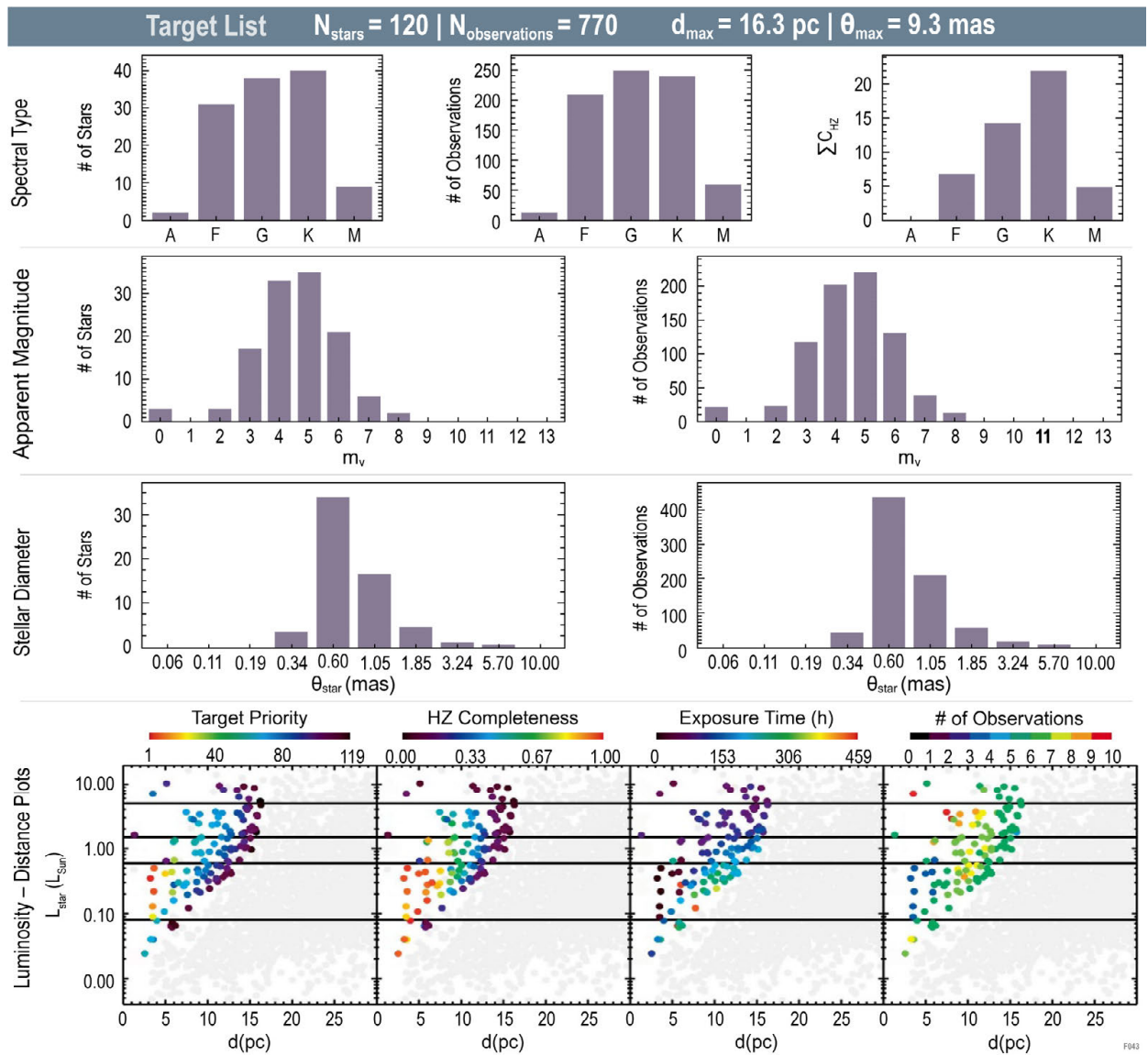

**Figure 2.3-8. HabEx target stars**. A total of 120 stars will be surveyed: 9 during the deep survey (starshade only) and 111 during the broad survey (multi-epoch coronagraphic searches and planet orbit determination, followed by planet spectral characterization with the starshade). Based on the HabEx survey strategy, the upper-right panel shows the number of HZ Earth-like planets that would be characterized around stars of different types, assuming each star had one such planet.

### 2.3.3 Science Yield Dependencies to Astrophysical and Instrument Parameters

The exoplanet yield results presented in the previous section were derived under the nominal astrophysical and engineering parameter assumptions listed in Appendix B and hence represent the most likely values. In reality, however, yields may vary from the expected values shown in **Figure 2.3-6** due to astrophysical uncertainties and the actual distribution of planets around nearby stars. The yield uncertainties *for all planet types* are given in Appendix B. They are estimated as an RMS combination of the SAG13 1-sigma exoplanet occurrence rate uncertainties (Belikov 2017) and the uncertainty due to the random distribution of planets around individual stars. The latter was estimated by assuming that planets are randomly assigned to stars, such that multiplicity is governed by a Poisson distribution, and that each observation represents an independent event with probability of success given by that observation's completeness. The error bars listed in Appendix B do not include uncertainty in the median exozodi level, any uncertainties in mission performance parameters, or uncertainty in observational efficiency. The uncertainty on the cold planet yields is underestimated, as the SAG13 occurrence rates are derived from Kepler data and purely extrapolations in this temperature regime (Belikov 2017).

Of particular interest is the dependence of exo-Earths yield on telescope diameter, assumed occurrence rate, and exozodi level. The top panel of **Figure 2.3-9** shows the number of exo-Earths expected to be detected and spectrally characterized between 0.3 and 1.0 µm as a function of telescope diameter, using nominal HabEx assumptions for all other parameters (including an exo-Earth occurrence rate of 0.24 and a constant 3 zodi level for all stars) and the same yield optimization algorithm used for the 4 m nominal HabEx architecture. The lower panel of **Figure 2.3-9** shows the probability of having zero exo-Earths characterized during the HabEx exoplanet surveys, under pessimistic, nominal, and optimistic assumptions for the exo-Earth occurrence rate value, and folding in the current results derived from the LBTI exozodi characterization survey (Ertel et al. 2018). For the latter, the calculation is made assuming that all stars have the same exozodi level, but that this common level follows a probability distribution consistent with the data gathered to date. At the nominal exo-Earth occurrence rate of 0.24 and for a 4 m diameter HabEx mission, the nominal number of exo-Earth spectra obtained is 12, and the probability of obtaining none is less than 0.2%. While the telescope diameter selection was essentially driven by independent technical considerations (Section 5), we find here that a ~4 m telescope provides a reasonable yield of exo-Earths and a high probability of success.

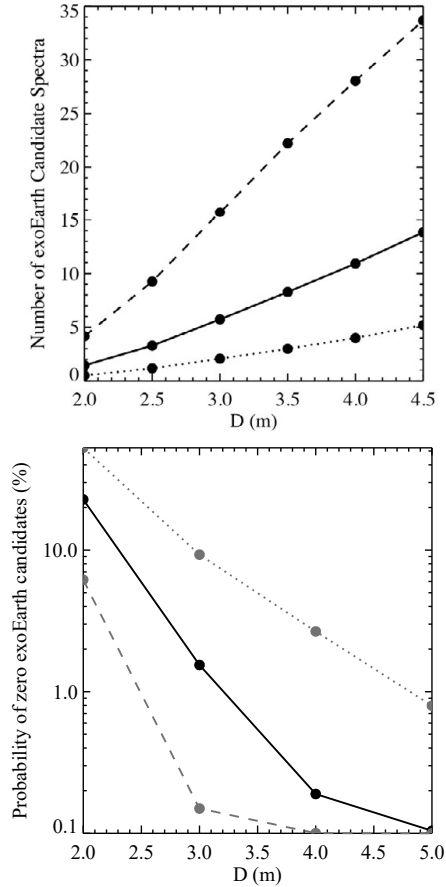

**Figure 2.3-9.** *Top panel:* Number of exo-Earths detected and spectroscopically characterized over the full 0.3–1.0 µm range as a function of telescope diameter, assuming different exo-Earth occurrence rates: 0.08 (dotted curve), 0.24 (plain), and 0.70 (dashed). *Bottom panel:* Probability of not detecting an exo-Earth candidate as function of telescope diameter, for these same 3 occurrence rates.

## 2.4    Value of Additional Observations

The science outlined in this section can be accomplished by a direct imaging mission without relying on any prior knowledge provided by other facilities, whether ground- or space-based. However, new observatories are expected to be operational by the time HabEx launches (see Appendix A), providing additional data on the target systems, and enabling more robust HabEx target prioritization and scheduling. For example, contemporaneous radial velocity or astrometric observations, at precisions not achievable today, may confirm a small planet's location in the HZ. This could reduce the required number of HabEx direct imaging visits. Similarly, simultaneous precision astrometry observations of the host star and HabEx direct imaging observations are expected to improve the planet mass and orbit determination precision (Guyon et al. 2013). A higher-precision mass estimation would improve the characterization of the observed exoplanet atmosphere. An assessment of the potential for precursor and contemporaneous observations to enhance the HabEx observations will be presented in the final report.

# 3  OBSERVATORY SCIENCE

Following in the tradition of NASA astrophysics flagships, such as Hubble Space Telescope (HST), James Webb Space Telescope (JWST), and Wide Field Infrared Survey Telescope (WFIRST), HabEx would be a Great Observatory with at least 25% of the primary mission and most of an extended mission reserved for guest observers. With the largest aperture ultraviolet (UV)/optical mirror ever deployed in space and two powerful instruments designed for this aspect of the mission, HabEx would enable unique science, not possible from ground- or space-based facilities in the 2030s, when HabEx would launch. This science would be broad and exciting, addressing the full range of primary NASA science disciplines, from solar system investigations to Cosmic Origins (COR) science to Physics of the Cosmos (PCOS). Furthermore, in addition to the exoplanet direct imaging science discussed in Section 2, HabEx would enable significant additional exoplanet exploration science, including transit spectroscopy and imaging of protoplanetary disks. The Guest Observer (GO) time encompasses everything beyond the planned exoplanet direct imaging surveys, and these competed programs are referred to as "observatory science." It is expected that HabEx would serve a very similar role to that played by HST in the astronomical community and the world at large for decades: a flexible and powerful tool producing an extremely broad range of exciting astrophysics, and fueling the public's interest in science, the cosmos, and NASA.

HabEx observatory science relies on three unique capabilities that define its discovery space. First, HabEx would provide the highest resolution UV/optical images ever obtained (**Figure 3-1**). Diffraction limited at 0.4 µm, HabEx would outperform all current and approved facilities, including the 30 m class ground-based extremely large telescopes (ELTs), which will achieve ~0.01 arcsecond resolution at near-infrared (near-IR) wavelengths with adaptive optics, but will be seeing-limited at optical wavelengths. Second, HabEx would observe wavelengths inaccessible from the ground, including the UV and in optical/near-IR atmospheric absorption bands. Finally, operating at L2, far above the Earth's atmosphere and free from the large thermal swings inherent to HST's low-Earth orbit, HabEx would provide an ultra-stable platform that would enable science ranging from precision astrometry to the most sensitive weak lensing maps ever obtained.

Two capable instruments are included in the HabEx design specifically to enable observatory science programs: the UV Spectrograph (UVS) and the HabEx Workhorse Camera (HWC). Both instruments rely on low-risk, flight-proven technology, and neither leverages strong additional requirements on the observatory design. The UVS is an evolved version of HST's Cosmic Origins Spectrograph (COS), taking advantage of several decades of improvement in detector and optics technology, as well as the larger aperture of HabEx relative to HST. The HWC is an evolved version of the dual-beam Wide-Field Camera 3 (WFC3) on HST. The HWC would provide imaging and multi-slit spectroscopy in two channels: a UV/optical channel and a near-IR channel. The designs and requirements of these instruments flow down from a few specific science cases detailed below, though with this aspect of the mission competed, some of the example science cases might not be executed. There would also be a multitude of additional applications for the UVS and HWC instruments on HabEx, many not yet anticipated; the science cases presented here are only meant to be exemplary and to span a breadth of compelling science with which to define the instrument requirements and capabilities.

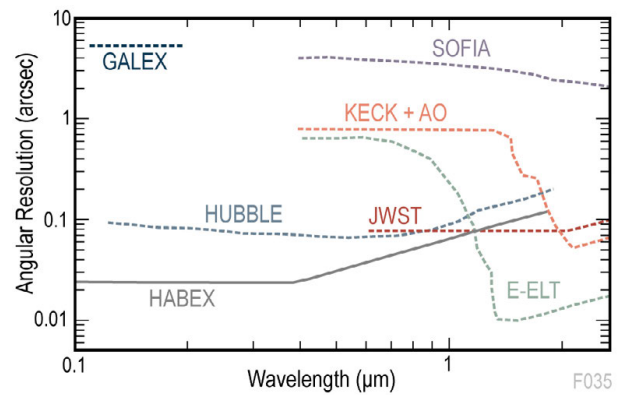

**Figure 3-1.** HabEx would provide the highest resolution UV/optical images of any current or planned facility, enabling a broad suite of unprecedented observatory science.

The remainder of this section is structured as follows. Section 3.1 presents the anticipated scientific and technological landscape in the 2030s, emphasizing expected key open questions in several disciplines and detailing the discovery space for a UV-to-near-IR space-based facility. Several science programs enabled by HabEx are then detailed, distinct from the direct imaging exoplanet studies described in Section 2. These science programs form the basis for the general astrophysics instrument suite and design requirements.

### 3.1    The Astrophysics Landscape in the 2030s

Between now and the 2030s, astronomers will commission a wide array of impressive facilities and instruments. These include both surveys that will image large swaths of the sky with unprecedented sensitivity at optical (e.g., Large Synoptic Survey Telescope [LSST]), near-IR (e.g., Euclid, WFIRST), and X-ray energies (e.g., eROSITA), as well as facilities with more limited fields of view, optimized for detailed follow-up studies (e.g., JWST, ELTs). Below is a brief discussion of some of the key questions expected to be left partially or wholly unanswered in the 2030s, and how the unique discovery space afforded by a space-based 4 m class UV-to-near-IR telescope will address these questions.

### 3.1.1    Key Science Questions for the 2030s

Several billions of dollars are currently being spent on ground- and space-based facilities with a primary goal of mapping large swaths of the universe in order to study the history of cosmic expansion and address fundamental questions of cosmology. As a byproduct of these studies, many classes of rare, exciting astronomical sources are expected to be found, from dwarf galaxies in the nearby universe, to a hundred-fold increase in the census of strong gravitational lenses, to quasars at redshift z ~ 10 and beyond. These discoveries will demand a range of follow-up studies, some of which will be amenable to ground-based facilities available in that era, but many of which will require space-based follow-up. Besides the exoplanet characterization questions addressed in the other portions of this report, a multitude of key science questions are expected to remain unanswered into the 2030s, including, but not limited to, the missing baryon problem, the nature of dark matter, the history of cosmic acceleration, the history of cosmic reionization, the nature of the seeds of supermassive black holes, the sources and physics of gravitational wave events, detailed understanding of solar system analogs to exoplanets, and a detailed understanding of the formation and evolution of galaxies. HabEx would provide a unique and important platform for these studies, and many more.

**HabEx Discovery Space**
- Highest angular resolution UV/optical images
- Access to wavelengths inaccessible from the ground
- Ultra-stable platform

### 3.1.2    Discovery Space for the 2030s

With no more servicing missions planned, HST is expected to degrade into disservice sometime in the 2020s, thereby shutting off access to the UV portions of the electromagnetic spectrum (e.g., 0.115–0.32 μm), as these wavelengths are absorbed by the Earth's atmosphere. Many key diagnostic features are in this energy range, particularly from highly ionized species in hot plasmas. This energy range is essential for studying the hot phase of the interstellar medium (ISM), intergalactic absorption, as well as for understanding the physics of a range of objects, from planets to galaxies to gravitational wave sources. Access to the UV will be essential to the astronomical community for studying the wide array of sources to be found between now and the 2030s. Furthermore, with marked improvements in technology, the grasp of a UV instrument built in the 2030s will greatly exceed a simple scaling with aperture size, thereby providing important, new information on targets already observed by HST. With no large-aperture UV satellite currently planned, there will be a gap in future capabilities. With its next-generation UV instrument and large-aperture in space, HabEx would fill this gap, thereby realizing tremendous discovery potential.

There is a similarly large discovery potential for a next-generation optical/near-IR satellite. First light is expected to occur for several 30 m class, ground-based ELTs by the 2030s—specifically the Giant Magellan Telescope (GMT), the Thirty Meter Telescope (TMT), and the European ELT (E-ELT). Since it is widely recognized that adaptive optics (AO) will remain infeasible at optical wavelengths for the foreseeable future (i.e., well past the 2030s), the greatest gains for these facilities will occur at longer wavelengths, where diffraction-limited AO-assisted observations of point sources provide gains that scale as aperture diameter, D, to the fourth power (i.e., $D^4$) rather than the simple seeing-limited $D^2$ gains provided by the larger aperture. Accordingly, significant effort is going into designing the AO systems for these telescopes, which will allow the full gains from these large apertures to be realized. Indeed, all the first-light instruments for the E-ELT are diffraction-limited, AO-fed infrared instruments, while GMT and TMT include first-light plans for both diffraction-limited, AO-fed infrared instruments and seeing-limited optical instruments. Therefore, the sharpest imaging at optical wavelengths will remain a domain best achieved from space for the foreseeable future.

Finally, space-based observations provide a platform significantly more stable than ground-based observatories, which is essential for a range of science applications, from sensitive weak lensing studies, which require an exceptionally stable, well-characterized point spread function (PSF), to astrometric studies that require a stable, well-characterized focal plane, to studies that require extremely accurate and stable photometry or spectrophotometry.

Much of the extraordinary progress in astrophysics over the past 20 years has been enabled by combining HST's exquisite resolution and stability, with the light-gathering power of larger-aperture 10 m-class telescopes, such as Keck and the Very Large Telescopes (VLTs). Often these resources were employed in tandem, with HST providing high-resolution imaging and the ground-based facilities providing spectroscopy (e.g., the Hubble Deep Field). We expect the 2030s to witness similar, but considerably more powerful synergies between HabEx and the ELTs.

In the following sections, several science cases are detailed that take advantage of a 4-meter UV-to-near-IR mirror in space to uniquely address pressing open questions in astrophysics. These science cases then form the basis to define the functional requirements of the UVS and HWC instruments on HabEx.

## 3.2 Tracing the Life Cycle of Baryons

Despite decades of efforts, approximately one-third of the baryons in the local universe remain unaccounted for (**Figure 3.2-1**). Notably, stars only account for <10% of the baryons in a typical galaxy. The "missing baryons" are thought to be predominantly in the form of diffuse, hot gas around and between galaxies, but many fundamental questions remain open about this gas, even within the very local universe. This material, the intergalactic medium (IGM; i.e., the gas between galaxies) and the circumgalactic medium (CGM; i.e., the gas external to, but near galaxies), is the fuel from which stars ultimately form, and, later in their lives, the material that galaxies redistribute and enhance through supernovae and violent mergers. Studying and understanding this gas is key towards understanding the life cycle of baryons in the cosmos. However, this presents observational challenges since the bulk (60%) of the IGM is predicted to be extremely hot, with the key diagnostic transitions at UV and X-ray energies and thus inaccessible to the ground.

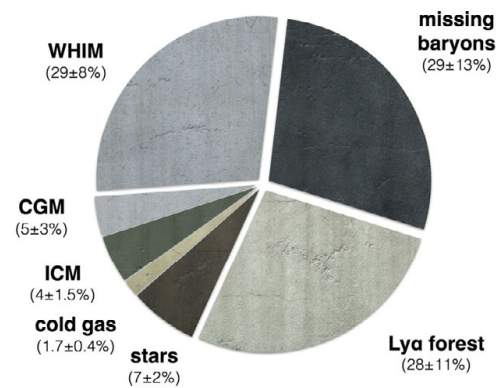

**Figure 3.2-1.** Approximately one-third of the baryons in the local universe are unaccounted for, likely tied up in a hot gas phase (Shull, Smith, and Danforth 2012). WHIM: warm/hot interstellar medium, CGM: circumgalactic medium, ICM: intracluster medium.

Outlined below is a broad observational program with the goal of better understanding the nature of the IGM, and probing the baryon cycle (**Figure 3.2-2**)—i.e., how stars are formed in galaxies, material is ejected from galaxies in the late stages of stellar evolution (i.e., supernovae), and then this enriched material is subsequently returned to galaxies. Specifically, to constrain the cosmic baryon cycle over the past 10 Gyr, it is necessary to:

- Measure the amount of gas and heavy elements around $z < 1$ galaxies;
- Complete the census of baryons in the local universe; and

- Determine the dynamical state and origin of the various components of the IGM, i.e., determine what fraction of the IGM is primordial, and what fraction is due to outflowing material, recycled accretion, or other physical causes.

The most promising observational approach to study the IGM is to use bright distant sources, such as quasars (QSOs), as backlights for absorption line studies of the intervening IGM. UV observations trace the "invisible" baryons— diffuse, extremely hot ($> 10^6$ K) material that is too energetic to strongly participate in star formation or galactic assembly. This hot IGM is therefore left out of most assays of the material between the galaxies.

Sensitive studies of the hot IGM present specific observational challenges. First, at least in the local universe, the observations must be obtained at wavelengths inaccessible from the ground since the Earth's atmosphere absorbs and scatters photons blue-wards of ~0.32 μm.

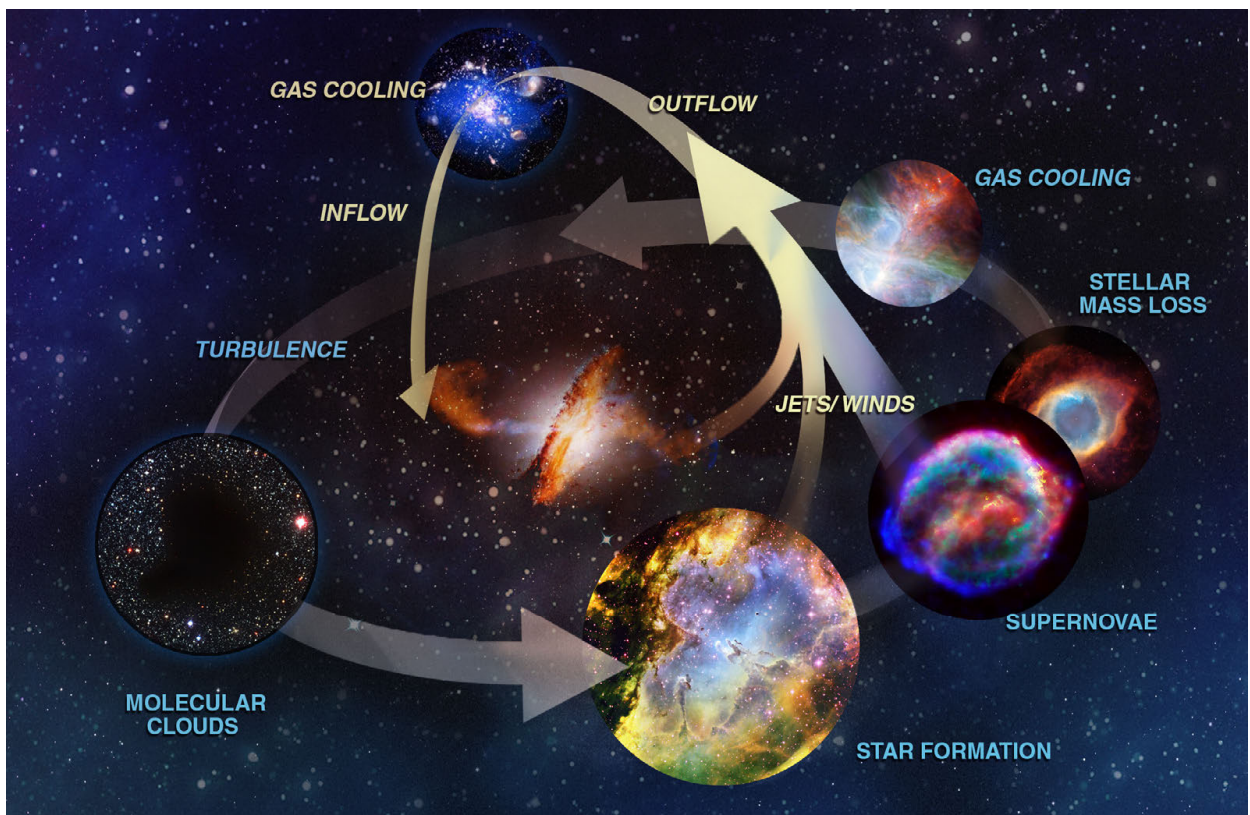

**Figure 3.2-2.** With its improved UV sensitivity and multiplexing capabilities, HabEx would be two orders of magnitude more efficient for investigations of the lifecycle of baryons.

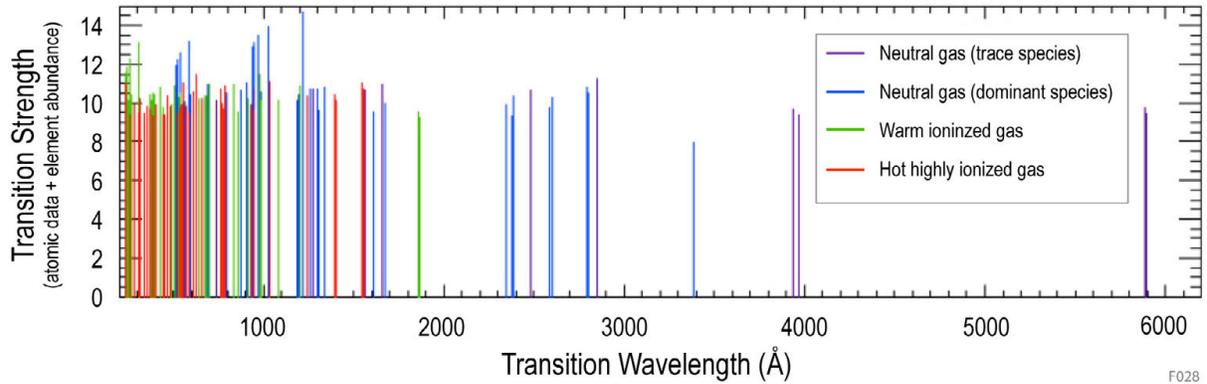

**Figure 3.2-3.** While optical and near-UV observations offer access to neutral gas, observations of near-UV and far-UV features are required to probe the more dominant warm and hot components of the IGM. This graphical representation shows the wealth of diagnostic lines the far-UV and near-UV offer to astrophysical investigation, comparing transition strength to rest-frame transition wavelength (Tumlinson, personal communication).

UV astronomers distinguish near-UV observations (0.12–0.36 μm) and far-UV observations (0.09–0.12 μm). As shown in **Figure 3.2-3**, the density of diagnostic features is highest at the blue end of the far-UV, blueward of 0.1 μm. While an instrument sensitive to these energies in the observed frame would be provide strong benefits, cosmological redshifting allows us to access these rest-frame wavelengths and this science with a slightly redder blue cut-off; an instrument requirement of access to a minimum wavelength of 0.115 μm was set, identical to the blue cutoff of HST. To provide significant gains relative to current studies, this work would require sensitive UV absorption line studies on a flagship-class (i.e., 4 m or larger) UV-optimized space telescope.

Another way to address the flow of material between the IGM and galaxies is to map and track the metallicity evolution of the IGM, and through that data the physics and the contents of galactic haloes, as well as the evolution of UV-irradiated environments. Such work requires UV imaging to determine the centers of massive star formation, and a very large number of spectroscopic lines of sight across a single galaxy. The latter requires a large collecting area, ~4 m diameter or larger, in order to have a sufficient surface density of background sources amenable to spectroscopic study in reasonable exposure times. Large-format photon-counting detectors would significantly enhance the efficiency of these observations by allowing simultaneous observations of multiple sightlines. Recent improvements in UV coatings and UV optics materials should also provide significant enhancements in throughput relative to current technology (e.g., Scowen et al. SPIE, submitted, and references therein), providing gains well beyond the simple scaling due to the primary mirror aperture. The required capabilities are broad, involving a field of view (FOV) of at least a square arcminute (to enable multiplexing), and subarcsecond angular resolution (to resolve both star formation in the host galaxy and filamentary structure in the IGM and CGM). In order to enable imaging and spectroscopy of the full extent of typical nearby galaxies, the field of view requirement for UV imaging was set to be 2.5 arcmin on a side, with a required capability to obtain multi-object slit spectroscopy.

Also, focus would be needed on understanding how gaseous material, and, in particular, the chemical elements, are distributed and dispersed into the CGM and the IGM. Specifically, how does baryonic matter flow from the IGM into galaxies and from there into stars and planets, and ultimately life? As our nearest galactic neighbors, the Magellanic Clouds and their constituent HII regions provide the best extragalactic targets for a high-resolution multiband UV-through-near-IR imaging survey, including narrowband imaging. A complementary high-resolution spectroscopic far-UV survey of a statistically significant subset of the ~1,300 early-type (OB) stars catalogued in the Magellanic Clouds would simultaneously provide insight into the stellar atmospheres, and probe the conditions of the ISM and CGM near those stars. Similar studies of galactic halo stars would also be scientifically important, as illustrated in **Figure 3.2-4** which shows absorption lines from r-process nucleosynthesis, i.e., elements synthesized by rapid neutron capture in neutron star mergers or supernovae. Approximately half of the atoms heavier than iron were created through this process. This work would provide stringent tests of stellar models of hot stars. As stated above, this program requires a reasonable field of view (>5 arcmin$^2$) with high-resolution, subarcsecond imaging at 0.3 μm, a telescope aperture of at least 1.5 m, and a large suite of filters. The far-UV spectroscopy would require spectral resolution $R = 60,000$, requiring next generation reflective coatings combined with new microchannel plate (MCP) technology.

In more general terms, a case can be made for a program that addresses the large number of diagnostic lines available. Such work can be done with a 4-meter aperture and would allow additional science such as determination of chemical abundances in star-forming galaxies, the effect of UV irradiation on exoplanets with potential biosignatures, and the nature of reionization and the escape fraction of ionizing radiation from star-forming galaxies. By targeting galaxies at modest redshifts, such work would not require instrument sensitivity below observed 0.1 μm (e.g., **Figure 3.2-3**). For this science, spectral resolutions as high as 30,000 are required, with a sensitivity 10× better than HST-COS, and a >10× multiplexing capability.

Looking at the requirements to achieve the scientific investigations described above, the following requirements flow down for UV spectroscopy:

- Blue-end cutoff ≤ 0.115 μm;
- Multiple spectral resolutions, ranging from $R = 10,000$ to $R \geq 60,000$;
- Telescope aperture ≥ 4 m;

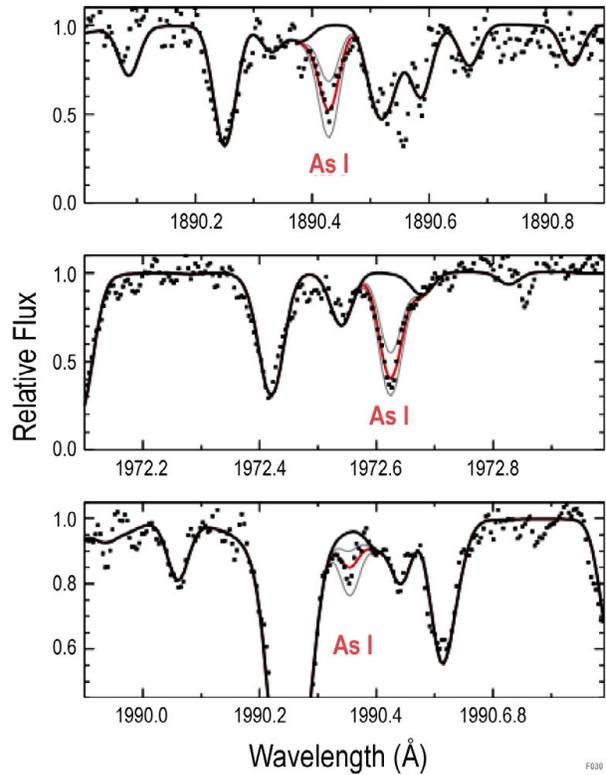

**Figure 3.2-4.** The HabEx UV Spectrograph's sensitivity makes observations of spectral lines from elements synthesized by r-process nucleosynthesis in neutron star mergers or supernovae (black points) ~50 times more efficient than with HST/STIS. The red curve in each panel fits detections of arsenic (As) in a metal-poor subgiant (Roederer and Lawler 2012), while the thin black curves show abundance variations by a factor of ±2, and the bold black curve shows a model spectrum with no arsenic.

- Field of view ≥ 2.5×2.5 arcmin$^2$; and
- Multi-object slit spectroscopy, with >10 multiplexing.

The scientific investigations described here are quite extensive and broadly describe much of the UV science that HabEx may perform over the course of a several-year mission. However, it is estimated that extremely significant gains relative to current knowledge would be enabled with just several weeks of observations, encompassing both QSO absorption line studies and studies of UV-bright stars in our galaxy and the Magellanic Clouds.

## 3.3    Local Value of the Hubble Constant

Recent measurements of the local value of the Hubble constant, $H_0$ (i.e., the local expansion rate of the universe), have been controversial, and hint at possible new physics. One set of observations is based on an extensive HST/WFC3 program of imaging nearby galaxies at optical and near-IR wavelengths (Riess et al. 2016). This study finds a local value of the Hubble constant that is $3.4\sigma$ higher than the latest value measured by the Planck satellite, based on measurements of the cosmic microwave background (CMB). With the HST program reporting a value of $H_0 = 73.24 \pm 1.74$ km s$^{-1}$ Mpc$^{-1}$ and Planck reporting $H_0 = 66.93 \pm 0.62$ km s$^{-1}$ Mpc$^{-1}$, the era of *precision* cosmology has certainly arrived, but, at first glance, perhaps not yet fully the era of *accurate* cosmology. Importantly, the HST program measures the local value of the Hubble constant, while Planck observes the surface of last scattering of the CMB at high redshift ($z \sim 1,100$) and infers the local value of the Hubble constant based on an assumed cosmology. Potentially, the discrepancy arises from the assumption of a "vanilla" $\Lambda$CDM cosmology (i.e., the simplest dark energy equation of state, with a temporally invariant cosmological constant, $\Lambda$). One plausible explanation for the apparent discrepancy could involve an additional source of dark radiation in the early universe.

Riess et al. (2016), which highlighted this tension in recent measurements, reduced the uncertainty in the local value of the Hubble constant from 3.3% to 2.4%, with the bulk of the improvement coming from near-IR observations of Cepheid variables in 11 galaxies that hosted recent type Ia supernovae (SNe Ia). This work more than doubled the sample of reliable SNe Ia with Cepheid-calibrated distances to a total of 19 and improved the local measurement of the Hubble constant by improving calibrations for the lowest rungs on the cosmic distance ladder. WFIRST, with the same aperture as HST, will only be able to improve upon HST if more SNe Ia occur within the small volume of the local universe in which Cepheid variables are accessible to a 2.4 m class telescope. The near-IR channel on HabEx/HWC would vastly increase the volume accessible to such measurements, allowing precision Cepheid-based measurements to dozens of galaxies that have hosted SNe Ia identified between now and when HabEx launches, thereby significantly reducing the uncertainty in the local value of the Hubble constant. The required precision photometry is not achievable from the ground. JWST will be able to achieve some of this science, but fewer accessible SN Ia will have been identified when JWST launches, and JWST is highly inefficient for cadenced observations given its slow slew and settle times.

The time required for such a program is estimated based on Riess et al. (2016), which used HST to identify and measure Cepheids in 20 nearby galaxies with a mean exposure time of ~15 kiloseconds per galaxy. Assuming similar exposure times, but reaching to much greater volumes given the eight-fold improvement in HabEx sensitivity relative to HST (i.e., taking advantage of the D$^4$–scaling for unresolved sources), a survey of a few dozen galaxies could be accomplished in a few weeks of observations. This would increase the number of well-calibrated Cepheid distances to galaxies known to host type-Ia supernovae by a factor of several, thereby decreasing the uncertainties in the local value of the Hubble constant. Such data would also be valuable for a range of nearby galaxy science, such as resolved studies of their stellar populations. These observations would require a 4 m class telescope (or larger), with a field-of-view comparable to nearby galaxies (i.e., $\geq 2.5 \times 2.5$ arcmin$^2$), and multiple filters options for imaging from the optical ($\geq 0.4$ μm) to the near-IR ($\leq 1.7$ μm).

### 3.4 Measuring the Star Formation Histories of Nearby Galaxies from Stellar Archaeology

One of the primary goals of studies of galaxy formation and evolution is to map how galaxies formed their stars and produced heavy elements over cosmic time. This is essential for understanding the life cycle of baryons in a cosmological context, as well as how and when the conditions fertile for forming planets and life arise. However, the current picture of galactic star formation has many open questions. How does the distribution of stellar types formed out of gaseous clouds—i.e., the stellar initial mass function (IMF)—vary with the metallicity of these clouds? What role does environment play? For example, in denser regions, an important impact from UV photons emitted by nearby stars, stellar remnants, and potentially active galactic nuclei may be expected.

The formation history of stars can be probed in a statistical way by studying galaxies at different redshifts, providing snapshots at different cosmic epochs. However, a complementary and very powerful technique identifies individual stars within nearby galaxies. Applying knowledge of how stars evolve in color and brightness as they age, the ages and chemical abundances of these stars can be determined. This allows a "fossil record" of when the stars formed to be extracted. HST can resolve individual stars down to stars like our Sun only for the very nearest galaxies. This means that HST can directly detect sunlike stars in only one other large galaxy, the Andromeda spiral galaxy (M31). HabEx would enable the mapping of star formation histories for a much larger and more diverse sample of galaxies, probing galactic environments beyond the Local Group. By pushing out to larger distances, HabEx would enable studies of the diversity of galaxy formation histories as a function of mass, environment, and other properties.

This science goal requires high-resolution, wide-field precision photometry of crowded fields down to the stellar main sequence in multiple UV-optical bands. As shown in **Figure 3-1**, HabEx provides the highest resolution UV and optical images of any facility currently in development. A resolution of 0.1 arcsec or better is required to minimize stellar blending, while a field of view of at least a few arcminutes on a side is required to obtain a sufficient source density to study the properties of the population (e.g., age, metallicity). Multiple bands are required to determine stellar colors. This work requires an extremely stable PSF over arcmin-scale fields, which will not be possible with ground-based telescopes, particularly at optical wavelengths. Stellar archaeology has been pioneered and demonstrated with HST for very nearby galaxies, mostly dwarf galaxies within the Local Group (Tolstoy, Hill, and Tosi 2009, Weisz et al. 2014). Even if HST's lifetime were extended, few galaxies are sufficiently close to resolve their stellar populations in this way with a 2.4 m telescope (specifically, there are only two: M31 and M33). JWST will be able to push somewhat further, but UV-optical measurements are critical for breaking the well-known degeneracy between dust, metallicity, and age (Brown et al. 2008). Similarly, although WFIRST will have a large FOV, roughly two orders of magnitude larger than what HST, JWST, or HabEx provide, it has an HST-class aperture, and so will have a similar angular resolution to HST, and thus will not be able to do these studies beyond the Local Group.

In terms of the time required to do such studies, exposure of hours to tens of hours per galaxy will be required, implying that this program could be implemented in a moderate, several day program, easily accommodated as a HabEx GO program. In terms of instrument requirements, the capabilities demanded by Section 3.3 would suffice, with the additional requirements of UV imaging capabilities ($\geq 0.25$ µm) and multi-object slit spectroscopy with a minimal multiplexing factor of 20.

## 3.5 Probing the Nature of Dark Matter with Dwarf Galaxies

One of the most fundamental unanswered questions in physics regards the nature of dark matter. We know that dark matter comprises most (~85%) of the matter and about a third (~30%) of the total energy density in the universe (Planck Collaboration et al. 2016), but beyond that, little is known. Is dark matter a single particle, or is there a whole dark periodic table of particles? Standard or "vanilla" dark matter only interacts with itself and with normal matter (i.e., baryons) through gravity (and perhaps through the weak force). However, particle physics allows for many other possibilities. For example, it is possible that dark matter could be "self-interacting."

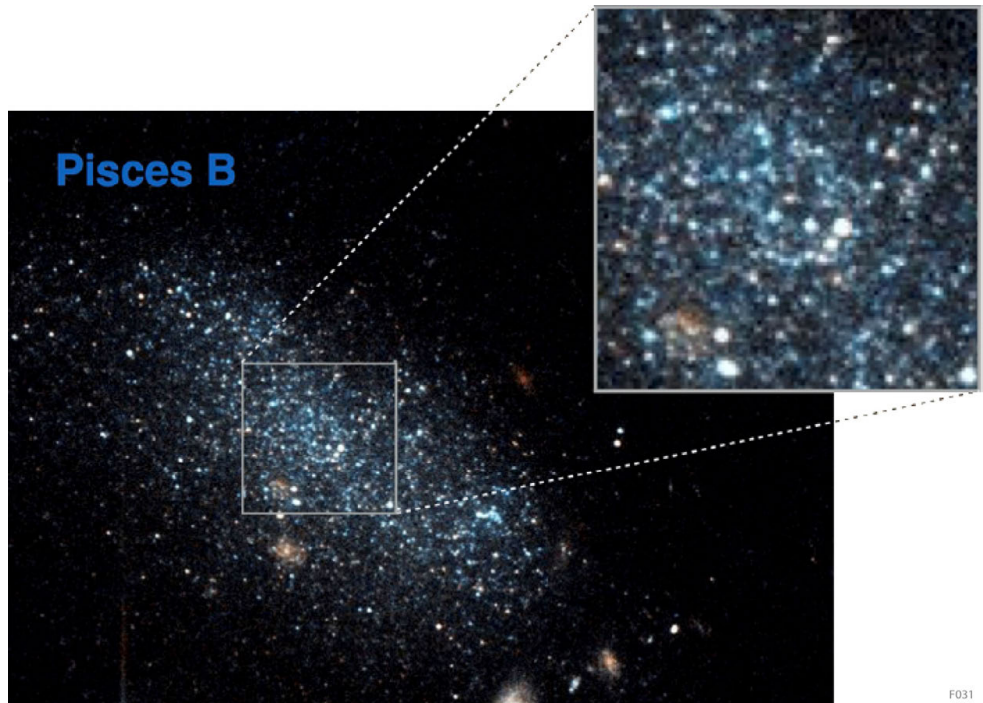

**Figure 3.5-1.** With high-resolution images of dwarf galaxies in the local universe, HabEx would measure the distances and star formation histories of local analogs of the first galaxies. Shown here is a Hubble image of the recently discovered nearby dwarf galaxy Pisces B at a distance of 8.9 Mpc (Tollerud et al. 2016). Compared to HST, HabEx would resolve fainter stars in galaxies like Pisces B, and obtain images like the one shown here for galaxies over a ~10× larger volume.

Dwarf galaxies in the Local Group (e.g., **Figure 3.5-1**) provide promising laboratories for probing the nature of dark matter because, unlike larger galaxies, which are mostly comprised of 'normal' matter (e.g., stars and gas) near their centers, dwarf galaxies are overwhelmingly dominated by dark matter all the way to the center. If galaxies formed out of pure standard dark matter (i.e., with no stars or gas, as well as no additional dark matter self-interactions), theory robustly predicts that their density profiles should monotonically increase all the way to their centers—i.e., that their density profiles should have "cusps" at their center. However, there has been much debate and consternation over the fact that many observed dwarf galaxies instead have "cores"—i.e., their density profiles plateau to a constant value at the center. **Figure 3.5-2** illustrates how the density profiles of dwarf galaxies depend on the nature of dark matter.

There are two main proposed solutions to explain these observations. Either (1) dark matter is not "vanilla," or (2) the large amounts of energy created by massive stars as they explode in supernovae removes the dark matter from the cusps, thus flattening them out into cores. There is a very large parameter space of "non-vanilla" dark matter models that are considered equally plausible, or natural, to particle physics theorists, and astrophysical observations are likely the most efficient way to narrow down this large parameter space. Theory groups largely agree on one clean prediction: if the flat density profile galaxy cores are created by non-vanilla dark matter, they should be seen universally in all galaxies. On the other hand, if the flat core profiles are created by stars and supernovae, then pristine "cusps" should be seen surviving in galaxies with truncated star formation histories (Read, Agertz, and Collins 2016), because they did not have vigorous enough star formation to produce the requisite energy to remove the dark matter and therefore destroy the cusps. It should also be possible to see correlations of the central galactic density profiles with galaxy properties, such as the ratio of the mass of stars to the mass of dark matter.

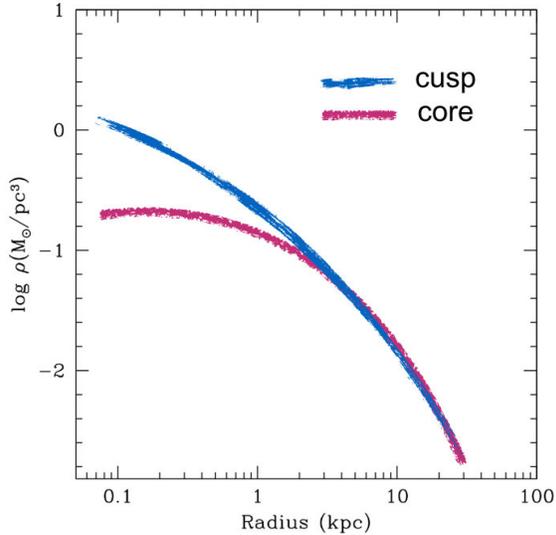

**Figure 3.5-2.** By spatially resolving the inner kpc of a sample of dwarf galaxies with a range of star formation histories, HabEx would determine if the flattened "core" profiles seen in many galaxies are indicative of self-interacting dark matter, or simply due to supernovae feedback clearing out the inner baryons in galaxies with standard dark matter, which has a "cuspy" density profile.

In order to test these predictions, high-resolution photometry is needed to probe galactic light profiles, as well as spectroscopy and proper motions to probe the stellar velocities, which act as tracers of the overall gravitational potential. Such data must be collected for a sample that spans a range of masses and star formation histories. Ultra-faint dwarf galaxies have total luminosities of $100$–$10^5$ L$_{sun}$, while the more luminous "classical" dwarf galaxies have total luminosities of $10^5$–$10^7$ L$_{sun}$ (i.e., $M_V$~-2 to -15) and physical sizes ranging from 100 pc to 1 kpc (half-light radii). Roughly 100 dwarf galaxies are known currently within ~3 Mpc (e.g., McConnachie 2012), and many more are expected to be discovered by Euclid, LSST, and WFIRST.

HabEx is essential for multiple parts of this study. First, the high spatial resolution and photometric precision of HabEx is required to obtain accurate star formation histories using optical colors for individually resolved stars. Line-of-sight (LOS) velocity measurements are probably best obtained by ELTs from the ground. However, there is a well-known degeneracy between the velocity anisotropy of the stars and the density profile. Therefore, constraints of the velocity anisotropy can be obtained by measuring the proper motions of stars, requiring astrometric accuracy of better than 40 mas yr$^{-1}$ assuming a fiducial distance of 60 kpc (Postman et al. 2009). For the ultra-faint dwarfs that possess the largest dark-to-baryonic matter ratios, main sequence stars are needed to measure these proper motions. This is likely to be infeasible even with ground-based 30 m telescopes.

In terms of the time required to do these HabEx dark matter investigations, exposure times of hours to tens of hours per galaxy would be required for the photometric studies, while the proper motion studies would require multiple observations, ideally with large temporal baselines. Assuming such studies are done on a few dozen galaxies, sampling a range of ages, masses, and morphologies, an ambitious version of this program should be executable within a few weeks of observation time, while a more limited version observing a smaller sample of galaxies could be done more economically, in a few days of observing time. Either way, this unique and fundamental investigation into the nature of dark matter with HabEx could easily be accommodated as a GO program. The instrument requirements levied by the two previous sections would suffice for this program.

## 3.6 Exoplanet Transit Spectroscopy

Through its GO program, HabEx would also enable important new exoplanet science. As one example, exoplanet transit spectroscopy is a rapidly evolving area of study and is currently the primary technique used to study the composition and structure of exoplanet atmospheres. Here, the transit or occultation depth is measured as a function of wavelength. When the exoplanet passes in front of the star, it will appear slightly larger at wavelengths where the atmosphere is more strongly attenuating (e.g., within molecular absorption bands). Thus, the wavelength-dependent transit depth can be used to measure a low-resolution spectrum of the planet's atmosphere, and thereby detect features that are diagnostic of its physical properties and constituents, at least at low pressures. Similarly, when the planet passes behind the star, the drop in flux from the combined system is indicative of the planet's brightness temperature at that wavelength, and so can be used to produce a low-resolution emission spectrum of the planet.

The magnitude of the impact that JWST will have on the understanding of exoplanet atmospheres (e.g., Greene et al. 2016) depends on how well the systematics can be controlled, which is currently unknown. For JWST to significantly push the boundaries of knowledge about cool, rocky exoplanet atmospheres, the mission would need to achieve systematic noise floors with its instruments that are comparable to the photon noise floor, as well as devoting a large fraction of the mission lifetime to exoplanet transit science (e.g., Deming et al. 2009, Cowan et al. 2015, Greene et al. 2016). In addition, the UV and visible capabilities of HabEx provide access to key molecular bands (e.g., the extremely strong ozone Hartley and Huggins bands) and gas/haze scattering features (e.g., molecular Rayleigh scattering), which JWST will not be able to observe given its infrared spectral range.

HabEx provides an opportunity to build a telescope that, by design, achieves the requisite precision to study transiting rocky exoplanet atmospheres, and is complementary to proposed exoplanet-themed missions. For example, while the Atmospheric Remote-sensing Exoplanet Large-survey (ARIEL) mission, recently selected as ESA's next medium-class (M4) mission, would provide a first-of-its-kind dedicated survey of hundreds of exoplanet transit spectra, this relatively small (0.64 m²) space-based telescope would not achieve the precision needed to study rocky exoplanet atmospheres.

The key to studying rocky exoplanet atmospheres with transit spectroscopy is to focus on worlds around late-type stars, as these exoplanets provide the most favorable planet-to-star size ratios (and thus, the largest transit depths). Such worlds are known to be common (Dressing and Charbonneau 2015) and key targets have already been discovered to be transiting nearby ultra-cool dwarfs (e.g., Gillon et al. 2017, Bonfils et al. 2017). Rocky exoplanets transiting mid-type M dwarfs are expected to present transit features throughout the UV, visible, and near-IR with characteristic depths of 10–100 ppm (Meadows 2017). The depth of these features can easily exceed 100 ppm for late-type M dwarfs, although the overall lower luminosity of these stars will require longer integration times (or stacked transits; Barstow & Irwin 2016) to achieve the signal-to-noise ratios (SNRs) needed to detect molecular or atomic features in either the transmission or emission spectra of these systems.

**Figure 3.6-1** demonstrates the capabilities of HabEx for transit spectroscopy, where a simulated spectrum of TRAPPIST-1e is shown, assuming it has an Earth-like atmosphere. The error bars are for stacking 10 transits, corresponding to ~10 hrs of in-transit integration. Such an observation would provide robust detections of several key atmospheric molecules in the optical/near-IR wavelength range, including ozone and water. In particular, the ozone feature is critical. Though it occurs at a wavelength accessible from the ground, having an observation from above the Earth's atmosphere will be essential for a robust and reliable detection.

Simulations were also undertaken of HabEx observations of target stars of type M5V and M8V placed at a distance of 10 pc (e.g., similar to the distance of the TRAPPIST-1 system). For the mid-type M dwarf, precisions of 10 ppm in the atmospheric composition are achieved with less than 10 hours of observations throughout most of the visible and near-IR. Atmospheric features in the blue and UV are generally broad, implying spectral elements at these wavelengths could be combined to increase SNRs (e.g., Greene et al. 2016). Similarly, precisions of 10–100 ppm in the atmospheric composition of an HZ Earth-like planet orbiting the M8V target are typically achieved in less than 10 hrs. As transit durations for habitable zone planets around M dwarf hosts are of order 1 hour, stacking only a handful of transit observations would enable HabEx to acquire transit spectra of potentially habitable exoplanets. Ultimately, the ability of HabEx to characterize terrestrial planet atmospheres using transmission and eclipse spectroscopy would be determined by the systematic noise floors achievable by the HabEx instruments, as well as systematic noise floors set by the host stars themselves. JWST will help to characterize these levels. Barring limitations by unknown systematic noise floors, the instrument requirements motivated by the earlier science cases would suffice for this science.

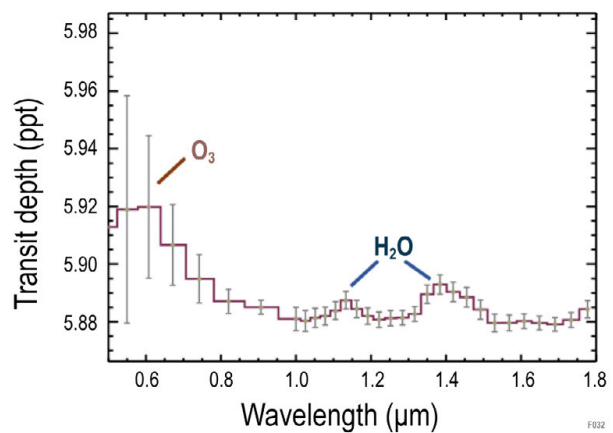

**Figure 3.6-1.** HabEx would identify molecules in the atmospheres of Earth-like planets from transit spectroscopy of eclipsing planets. Shown above is a simulated 10 transit spectrum of TRAPPIST-1e assuming an Earth-like atmosphere (ppt = parts per thousand).

## 3.7    HabEx GO Observations of Bright Circumstellar Disks

While >50 faint exozodi and exo-Kuiper belt analogs would already be characterized as part of the two exoplanet surveys (Section 2.6), the HabEx GO program provides the opportunity for additional dedicated observations of circumstellar disks. Of special interest are HabEx high-contrast images and IFS spectra of known optically thick protoplanetary disks and bright extended debris disks. Both types would benefit from HabEx's much improved contrast and spatial resolution over current facilities in the optical and near-IR.

### 3.7.1    Protoplanetary Disks

These targets are intrinsically interesting as the first stage in the evolution of planetary systems. Thousands are known from infrared surveys, but their Herbig Ae and T Tauri star hosts are generally faint (7 < V < 15) and more distant (~140 pc) than the targets of the primary HabEx exoplanets surveys. Young disks are optically thick in the optical and near-IR. This substantially eases the planet/star contrast required to detect them (~$10^{-6}$), but also can lead to them being self-shadowed. Generally, only the disk's upper and lower surfaces (or a cleared inner region if present) are detectable in reflected light; their midplanes are completely inaccessible. HST and ground-based AO have had some success imaging protoplanetary disks with inner holes, but the majority of protoplanetary disks are not detected in the presence of the central star's direct light. Indeed, edge-on star-occulting disks account for a substantial fraction of the scattered light detections to date. The faintness of the host stars limits ground-based AO to accessing only a small, brighter subset of stars with protoplanetary disks. However, these disks are sufficiently massive that they are bright at submillimeter wavelengths and thus ALMA is extremely capable of studying them. HabEx optical/near-IR imaging would reveal the distribution of small grains relative to the large grains emitting in the submillimeter, and the presence of shocked emission from accretion and outflows onto the star and protoplanets. Protoplanetary disks still retain substantial amounts of primordial gas and thus giant planet formation could still be ongoing and detected with HabEx high-spatial resolution, high-contrast imaging.

**Constraining Formation Mechanisms of Planets.** Of specific interest will be HabEx IFS images of transition disks, i.e., protoplanetary disks with inner clearings inferred by broad infrared spectral measurements, in which embedded planets may be caught in formation and best revealed through high-contrast imaging in accretion lines. The HabEx starshade visible IFS would, for instance, image Taurus-Aurigae disks over spatial scales ranging from 8–250 AU at $10^{-10}$ contrast levels, extending current disk images to the inner parts, and at contrast levels far below what is currently achieved from the ground: ~$10^{-3}$ at 20 AU, e.g., for the planet-forming regions around LkCa15 (Sallum et al. 2015) or AB Aurigae (Hashimoto et al. 2011). HabEx deep images of protoplanetary disks would also be able to detect fainter planets accreting at much lower rates than currently possible from the ground and hence explore the correlation between the observed disk structures and the presence of planets down to a Jupiter mass or less (**Figure 3.7-1**). Since the core accretion and disk instability models predict different formation efficiencies and timescales at a given separation, the direct detection of a statistical number of these newly formed exoplanets would help determine which process dominates. Similarly, since predictions for the luminosity of planet at a given mass and age differ by several orders of magnitude at very early ages, depending on the models used (e.g., Marley et al. 2007), measuring the brightness of very young giant exoplanets would provide crucial information.

### 3.7.2 Bright Debris Disks

Nearby mature stars often have bright debris dust disks, well-suited to a GO program. These disks contain small dust grains, continuously generated by the collisions of small bodies and the sublimation of comets. Depending on their size, these dust grains can be continuously swept out of the system by stellar wind and radiation pressure forces or can spiral into the star due to Poynting-Robertson drag. Debris disks are generally optically thin, so the brightness of a disk can be used to infer the total dust mass, given some assumptions. As little as a lunar mass of small grains will have a large surface area and thus will be readily observable. These small dust grains can be detected as a circumstellar reflection nebulosity at optical and near-infrared wavelengths, or through the starlight they absorb and reradiate at thermal infrared wavelengths.

Finally, HabEx's inner and outer working angles would enable observations of at least 60 RV planets known today, a number that can be expected to grow by the time the mission would launch. Five bright debris disks are known in this RV planet sample and would provide a clear guaranteed opportunity to study disk/planet interactions. Deep imaging of the others may similarly discover previously unknown dust populations whose internal structures could be used to calibrate theories of disk/planet interactions. The instrument requirements motivated by the earlier science cases would suffice for the disk science.

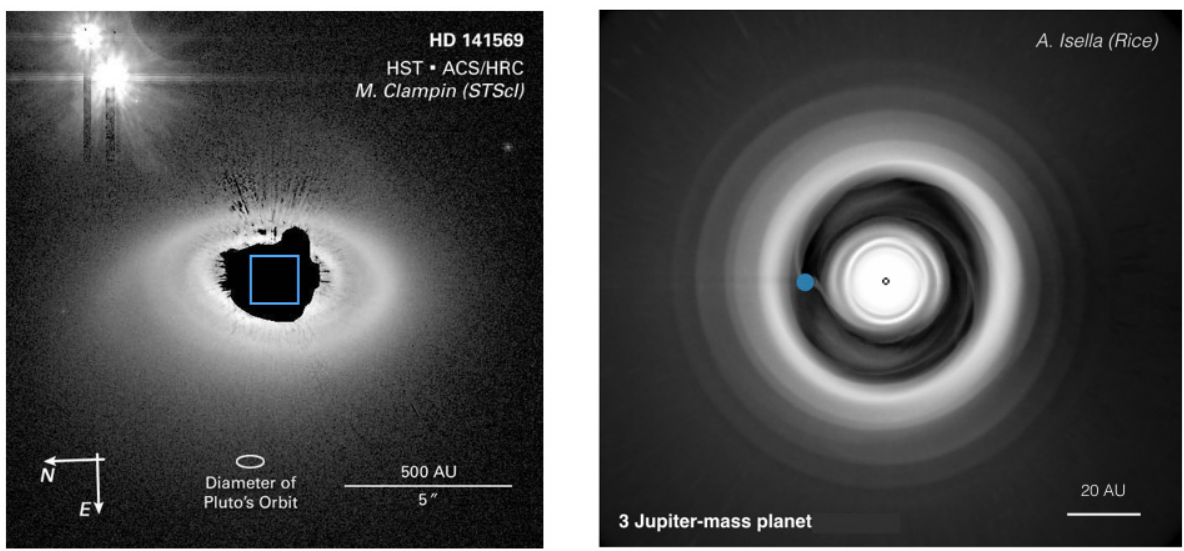

**Figure 3.7-1.** The HabEx coronagraph would reveal the inner-most structure of protoplanetary disks, unobtainable by HST (left panel). The right panel shows a theoretical model protoplanetary disk with a planet three times more massive than Jupiter (blue dot), illustrating the rich structure expected within an area corresponding to the box in the left panel. For this system, the box corresponds to approximately twice the size of the HabEx coronagraph field of view.

## 3.8    Solar System Observations

The discovery of thousands of exoplanets orbiting nearby stars is a historic advance, with incredibly broad implications ranging from fundamental questions about the development of life, to detailed astrophysics questions to understand this new scientific terrain. In terms of the latter, there is a strong desire to characterize and understand these exoplanets, how they formed, and how they interact with their host stars. To inform such studies, the planets within our own solar system are the ones that can be studied most closely, thereby providing important and unique laboratories to understand the basic physical principles of how planets form and evolve. In this section, several unique and important solar system studies enabled by HabEx are discussed, which would enhance understanding of solar system planets, and thereby exoplanets. Specifically, studies of planetary aurora, exospheres, cryovolcanism, atmospheric composition, and high-contrast imaging are discussed.

**Planetary Aurora.** Aurora, such as the northern lights (*aurora borealis*) and southern lights (*aurora australis*) when viewed in the Earth's northern or southern sky, respectively, are the results of a planet's atmosphere being strongly perturbed by its host star's stellar wind, causing charged particles in both the wind and the planet's magnetospheric plasma to precipitate into the planet's upper atmosphere. Besides being seen on Earth, this phenomenon has also been seen on all the gas giants in our solar system (e.g., **Figure 3.8-1**). Planetary aurorae are best studied at UV wavelengths, where the bulk of the emission is produced and the level of reflected sunlight is low, i.e., the highest contrast is obtained when observing the sunlit face of a planet.

With its improved UV sensitivity relative to HST, HabEx would probe the basic principles of planetary aurora, which are one of the key examples of star-planet interactions. The solar system provides examples of aurorae that cover a wide range of physical scales and conditions, thereby providing an important testing ground for probing star-planet interactions in exoplanetary systems. For example, what controls auroral processes on different scales of time and planet size, different levels of stellar winds, different planetary rotation rates, and different magnetic field strengths? On the Earth, the solar wind flow time past the planet is a few minutes, and auroral storms develop in a complex interaction with the southward-pointing interplanetary magnetic field. On Jupiter and Saturn, the flow time is hours to days. Jupiter sometimes responds to changes in the solar wind, other times not at all, while Saturn's auroral activity responds to every solar wind pressure front. Is auroral activity at Saturn controlled just by solar wind pressure, or is the interplanetary magnetic field direction important? An open question is whether Saturn's aurora is similar to the Earth's, or does it have a different interaction with the solar wind?

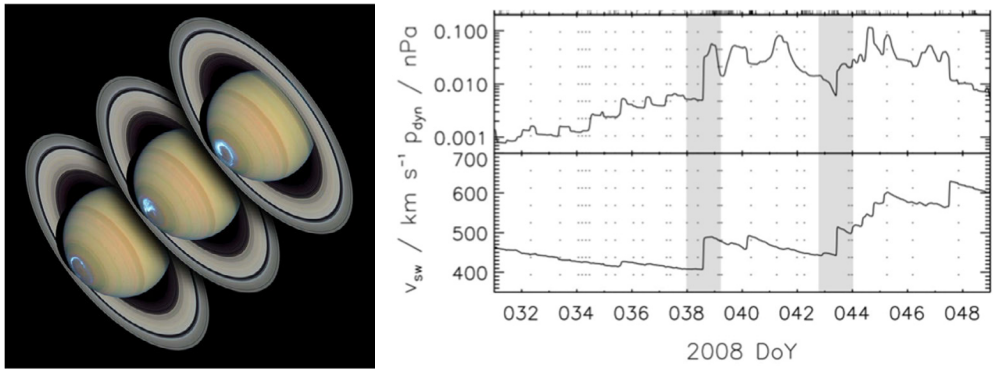

**Figure 3.8-1.** HabEx would enable the studies of solar system aurorae, such as in this HST far-UV image of Saturn's aurora and changes during an auroral storm (left), and total auroral power at Saturn vs. arriving solar wind speed (right). The shaded regions indicate the arrival of solar wind shocks at Saturn.

Besides monitoring and investigations of Jupiter's and Saturn's aurora, HabEx can extend UV auroral imaging to Uranus and Neptune with the sensitivity and resolution to detect patterns in faint auroral emissions. This would provide solar system analogs to the large number of recently discovered 2–5 Earth-radii exoplanets.

**Planetary Exospheres.** The outermost atmosphere-like, gravitationally bound, low-density gas around a planet is referred to as the planetary exosphere. By studying exospheres in our own solar system, HabEx would further our understanding of exoplanetary exospheres, which form the interaction region of a planetary atmosphere with the space environment. Exospheres are best observed in the vacuum UV, where the strongest transitions occur and reflected solar continuum is weak.

What physical principles govern the loss of an atmosphere into space? In the solar system, this process varies strongly from planet to planet. For the Earth, atmospheric loss is predominantly due to the high-energy tail of the atmosphere's Maxwell-Boltzmann velocity distribution exceeding the escape speed, and thereby being lost into space. This process is referred to as Jeans escape. At Mars and Venus, hot hydrogen gas populations are likely to dominate the exosphere loss. Also, for Mars, large annual variations exist, implying a strong seasonal control of the escape flux

(**Figure 3.8-2**). At Mercury, solar radiation pressure and solar wind proton charge exchange may dominate. At Uranus, a high-temperature hydrogen gas corona affects ring particle lifetimes. At Pluto, there is the potential for hydrodynamic flow of escaping hydrogen, which could entrain heavier species. Many of these phenomena are not observable with HST but would be observable with HabEx thanks to its significantly improved sensitivity at UV wavelengths.

**Cryovolcanism.** Cryovolcanism is an analog of the volcanism commonly observed on Earth, except that rather than molten rock (magma) being spewed by a volcano, the eruptions consist of volatiles such as water, ammonia, or methane. From HST and the Voyager flybys, several examples of cryovolcanism and cryoventing have been observed in the solar system, including on moons of gas giant planets, such as Jupiter's Europa (**Figure 3.8-3**), Saturn's Enceladus, and Neptune's Triton. With the improved UV sensitivity and resolution of HabEx as compared to HST, cryovolcanism may be discovered on many other small bodies in the solar system. Establishing statistics on the conditions in which cryovolcanism occurs, and what sets off the eruptions, is key to understanding the principles of volcanism in general. Current observations of eruptive plumes on Europa by HST are critical to the design and planning of the Europa flyby mission, and address the important question of extant life on Europa. HST only sees evidence for plumes ~10% of the time, and always observes in the vacuum UV for high sensitivity to small columns of gas. With 5 to 10 times greater sensitivity than HST, HabEx would vastly improve understanding of cryovolcanism on Europa, probing the duty cycle and distribution of activity. HabEx would also extend studies to Enceladus and other gas giant moons. Even Pluto would be in the range of observations by HabEx.

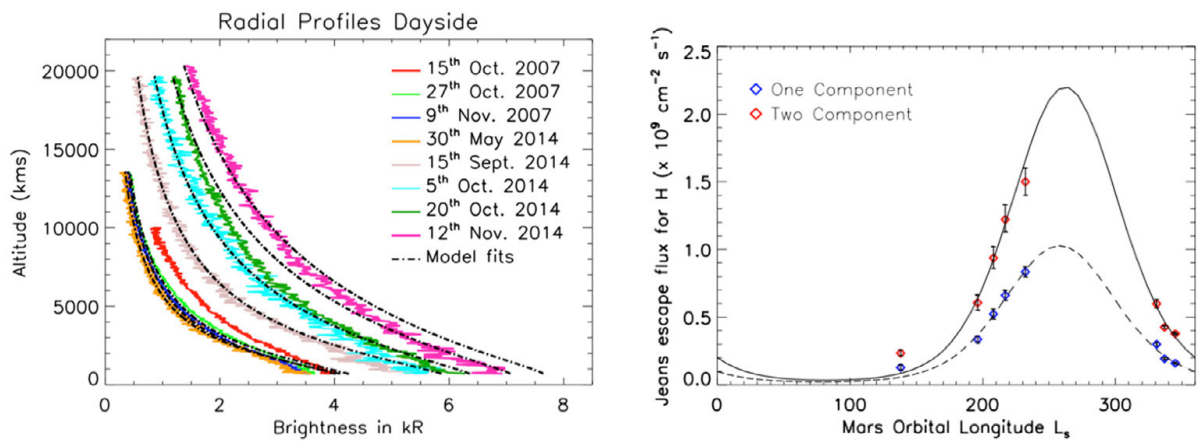

**Figure 3.8-2.** With access to hydrogen Lyman-α emission, HabEx would expand the study of solar system exospheres. Shown here are HST altitude profiles of Lyman-α emission from the martian exosphere (left) showing large changes over time, and hydrogen escape flux versus solar longitude (right) derived from the observations using a radiative transfer model. Solar longitude corresponds to martian season: the broad increase around 270 degrees roughly corresponds to perihelion and southern summer.

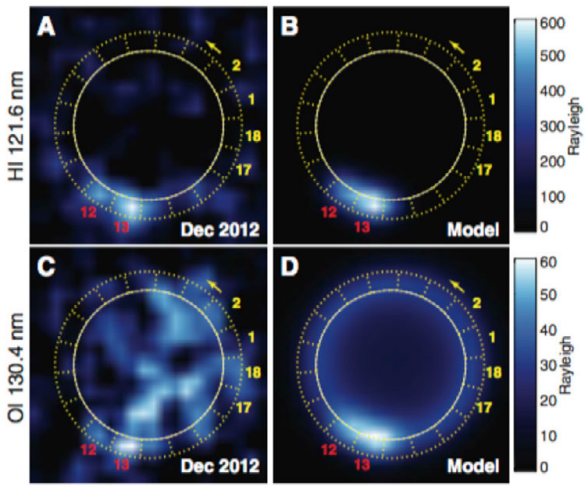

**Figure 3.8-3.** The HabEx UV Spectrograph would investigate cryoplumes on bodies throughout the solar system. Shown here are HST far-UV images of oxygen airglow emission from cryoplumes on Europa.

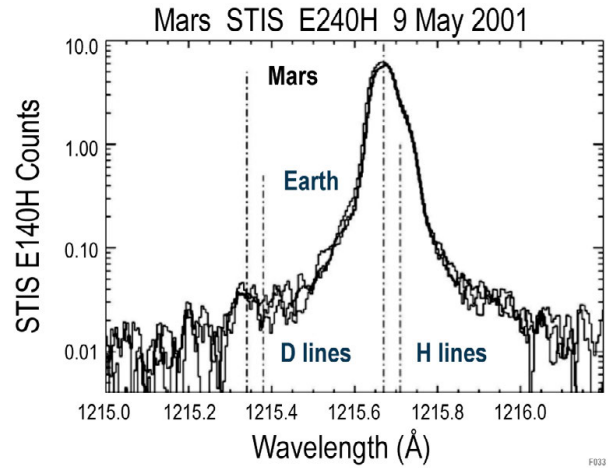

**Figure 3.8-4.** Outside the Earth's geocorona, HabEx would have a far lower background for far-UV observations than HST. Shown here is an HST spectra of deuterium (D) and hydrogen (H) Lyman-α emission from Mars.

**High-Contrast Imaging.** With its unprecedented high-contrast imaging capabilities, HabEx would search for faint companions to objects in our solar system, including small satellites around outer solar system objects such as Pluto and Kuiper belt objects. Such satellites provide key information about formation and evolution of the solar system, including tidal interactions with larger bodies.

These unique and important solar system studies made possible by HabEx primarily rely upon system capabilities and requirements already designed into the mission. The one new requirement is a capability for non-sidereal tracking. Roughly speaking, the studies outlined here could be accomplished in a Legacy-scale, or several-week observing program (or several efforts adding up to that scale program).

### 3.9 Science Traceability Matrix

The science cases presented above have been used to define the HabEx observatory science instrument capabilities using the HabEx Observatory Science Traceability Matrix (STM-2; Section 4). These programs both define a very exciting set of unique observatory science that HabEx would enable, as well as create a very flexible, capable instrument suite that would enable a broad suite of additional science. In particular, the three key capabilities demanded by these science programs are:

1. High-resolution ($R \leq 60{,}000$), multi-object UV slit spectroscopy to $\leq 0.115$ µm;

2. Imaging and multi-object slit spectroscopy from the near-UV to the near-IR over a field of view of at least 2.5 arcmin on a side; and

3. Ability to track solar system objects.

### 3.10 Parallel Observations

The UVS and HWC instruments have been designed to operate in parallel, both with each other and with the exoplanet direct imaging observations. Since much of the exoplanet imaging would involve long, multiday exposures of nearby stars, this would provide opportunities for ultra-deep parallel observations and increase the observatory scientific efficiency. Notably, since the HabEx exoplanet direct imaging surveys would observe nearby stars, these targets are distributed roughly isotropically across the sky, and thus are amenable to both galactic and extragalactic science goals.

Parallel observations with the exoplanet direct imaging program would enable multiple ultra-deep imaging fields, similar to the extremely successful Hubble Deep Fields and Hubble Ultra-Deep Survey. In addition, ultra-deep spectroscopic surveys done in parallel with the exoplanet direct imaging observations would enable a range of science, from ultra-deep probes of the IGM with the UVS, to ultra-deep and/or highly complete spectroscopic surveys with the HWC. The UVS and HWC fields of view are sufficiently offset from the direct imaging instrument fields of view that scattered light is not expected to be an issue, and the HWC instrument design incorporates a fine steering mirror, which would enable dithering during the deep exoplanet stares.

# 4 TOWARDS THE HABEX SCIENCE TRACEABILITY MATRIX

The confirmation of oxygen, water vapor, and other signs of habitability, or even life, will certainly make headline news not only in the scientific literature, but in the news media around the world. Yet the design of a large space mission calls for a more rigorous approach to formulating goals than to "measure exoplanetary spectra," or to "obtain measurements such as luminosities, polarization, or redshifts." Design requirements flow from mission goals, and these goals should lead to clearly articulated objectives. An established framework to capture science goals and objectives and the derived instrument and mission requirements is the Science Traceability Matrix (STM). STM-1 in this section gives an example of the requirement flow down for the HabEx exoplanet direct imaging and characterization science.

While detecting and characterizing the reflected light from exoplanets is the initial motivation for the telescope, coronagraph, and starshade, the telescope also enables a wide range of astrophysics and solar system science beyond the deep and broad exoplanet surveys. To illustrate these possibilities and to define the observatory instruments, a representative sample of challenging and interesting science goals have been selected. STM-2 in this section addresses observatory science with the HabEx Workhorse Camera (HWC) and the UV Spectrograph (UVS) described in Section 3.

In each row of the STM, goals and objectives trace through detailed calculations and simulations to physical parameters and observables, and to the instrument and mission requirements. This ensures that as long as HabEx meets these engineering requirements, the mission will be able to meet the science goals and objectives. The STMs included in this interim report are still under active development with some requirements still to be reviewed (TBR) or to be determined (TBD) and some of the trace to be completed. The science goals and science objectives may evolve prior to the HabEx final report through the current process of refinement and clarification.

## 4.1 HabEx Science Goals and Objectives

### 4.1.1 Science Goals

The HabEx science goals are framed around the desire to detect and characterize Earth-like exoplanets, undertake detailed investigations of our nearest neighbor planetary systems, and enable an exciting community led Guest Observer (GO) program that takes advantage of an ultra-stable, large-aperture, ultraviolet (UV) through near-infrared (near-IR) telescope in space. The goals have been articulated to spawn compelling science objectives and motivate a quantitative trace to measurement, instrument, and mission requirements. The HabEx goals are defined as:

**Goal 1:** To seek out nearby worlds and explore their habitability.

**Goal 2:** To map out nearby planetary systems and understand the diversity of the worlds they contain.

**Goal 3:** To enable new explorations of astrophysical systems from our solar system to galaxies and the universe by extending our reach in the UV through near-IR.

### 4.1.2 Science Objectives

The HabEx science objectives (see STM-1 and STM-2) address the HabEx science goals and can be supported or refuted directly with measurements by HabEx.

The best available data on the occurrence rates of Earth-like exoplanets located in orbits that are consistent with habitability (Belikov 2017) have been used to ensure that HabEx has a very high probability of detecting and characterizing at least one exo-Earth (with a high priority on many more than one). However, it is important to note that occurrence rates are still uncertain in the literature and ultimately a fixed property of the universe that HabEx would end up constraining directly. Regardless of the nature of the universe, the suite of science objectives that HabEx would address would lead to an enormous range of revolutionary science in the 2030s.

STM-1. HabEx Exoplanet Surveys Science Traceability Matrix.

| Science Goals | Science Objectives | Scientific Measurement Requirements | | Instrument Functional Requirements | Projected Performance | Mission Functional Requirements (Top Level) |
|---|---|---|---|---|---|---|
| | | Physical Parameters | Observables | | | |
| **Goal 1:**<br><br>**To seek out nearby worlds and explore their habitability** | O1: To determine if small (0.6–2.0 $R_{Earth}$) planets continuously orbiting within the habitable zone (HZ) exist around sunlike stars, surveying enough stars to reach a cumulative HZ search completeness > 40 for exo-Earths (0.6–1.4 $R_{Earth}$) | Planet position wrt the central star over time to determine the orbit semi-major axis, eccentricity, and inclination to within 10% accuracy | Star to planet separation measured at 4 (or more) different orbital positions to an angular positional accuracy of 5 mas rms each | IWA: ≤ 74 mas at 0.5 µm | Coronagraph IWA: 62 mas @ 0.5 µm for detection | Observe target stars ≥ 4 times, up to 10 times, a few months to a few years apart<br><br>Telescope aperture ≥ 3.7 m<br><br>Survey time ≥ 2 years |
| | | | | Shall be able to detect a point source ≥ $10^{10}$ times fainter than a sunlike star located at 12 pc (Vmag = 5.3) and 83 mas from it (exo-Earth at quadrature) with SNR ≥ 7 in < 60 hours using broadband ≤ 0.45 µm to ≥ 0.55 µm filter and coronagraph | Can detect a point source $10^{10}$ times fainter than a sunlike star at 12 pc and 83 mas from it at SNR = 7 in 40h using broadband 0.45–0.55 µm filter | |
| | | Planet radius within a factor of 2 at 95% confidence | Planetary spectrum, including Rayleigh scattering slope between 0.45 µm and 0.7 µm with R ≥ 140 | IWA: ≤ 74 mas at 0.7 µm | Starshade IWA: 60 mas for spectroscopy anywhere between 0.3 µm and 1.0 µm | Single epoch observation |
| | | | | Spectral range: ≤ 0.45 µm to ≥ 0.7 µm with R ≥ 140 | Starshade instantaneous spectral coverage: 0.3 µm to 1.0 µm: with R = 140 from 0.45–1.0 µm | |
| | | | | Shall be able to detect a point source ≥ $10^{10}$ times fainter than a sunlike star located at 12pc (Vmag = 5.3) and at 83 mas from it with SNR ≥ 10 in < 700 hours in a R ≥ 140 spectral bin located anywhere between 0.45 µm and 0.7 µm | Starshade can detect a point source $10^{10}$ times fainter than a sunlike star located at 12 pc (Vmag = 5.3) and at 60 mas from it with SNR = 10 in 400 hours in a R ≥ 140 spectral bin located anywhere between 0.45 µm and 0.7 µm | |
| | O2: To determine if any planets detected in O1 have potentially habitable conditions (an atmosphere containing water vapor) | Detect atmospheric water vapor ($H_2O$) if column density > (TBD) | Planetary spectrum, including water absorption features between 0.7 µm and 1.5 µm with R ≥ 37 | IWA: ≤ 74 mas at 1.0 µm<br>Spectral range: ≤ 0.7 µm to ≥1.5 µm with R ≥ 37<br>SNR ≥ 10 per spectral bin | Starshade IWA: 60 mas for spectroscopy anywhere between 0.3 µm and 1.0 µm<br><br>Starshade IWA: 90 mas at 1.5 µm | Starshade reposition to cover ≥ 1.0 µm |
| | O3: To determine if any planets identified in O1 contain biosignature gases (signs of life) and to identify gases associated with known false positive mechanisms | Detect molecular oxygen ($O_2$) if column density > 2 g cm$^{-2}$ and ozone ($O_3$) if column density > 1 g cm$^{-2}$ (i.e., Paleoproterozoic Earth, where oxygen at 1% modern) | Planetary spectrum:<br>$O_3$ cutoff around 0.3 µm to 0.35 µm with R ≥ 5<br>$O_2$ absorption feature between 0.75 µm and 0.78 µm with R ≥ 70 | Spectral range<br>$O_3$: ≤ 0.3 µm to ≥ 0.35 µm at R ≥ 5. Photometric SNR ≥ 5.<br><br>$O_2$: ≤ 0.75 µm to ≥ 0.78 µm with R ≥ 70. SNR ≥ 10, per spectral bin | Spectral range: Starshade<br>Near-UV/blue: 0.2 µm to 0.45 µm with R = 7. SNR = 10 per spectral bin<br><br>Red: 0.45 µm to 1.0 µm with R = 140. SNR = 10 per spectral bin | Pointing times of up to 1,440 hours (60 days) |
| | | Detect methane ($CH_4$) if concentration > 0.02% (TBR) and carbon dioxide ($CO_2$) if concentration > 1% (TBR) | Planetary spectrum:<br>$CO_2$ and $CH_4$ absorption features between 1.0 µm and 1.7 µm with R ≥ 20 | Spectral range: Starshade<br>≤ 1.0 µm to ≥ 1.7 µm with R ≥ 20 and SNR ≥ 10 per spectral bin | Spectral range: Starshade<br>Red: 0.45 µm to 1.0 µm with R = 140. SNR = 10 per spectral bin<br><br>Near-IR: 1.0 µm to 1.8 µm with R = 40. SNR = 10 per spectral bin | |
| | O4: To determine if any planets detected in O1 contain water oceans | Glint from surface oceans | Near-IR planetary broadband photometry measured at ≥ 2 illumination phases, with ≥ 1 measurement at illumination phase < 120 deg and at > 120 deg<br><br>Exoplanetary system inclination > 30 deg | Photometric range ≥ 0.9 µm<br><br>Planet-to-star flux ratio detection limit: ≤ 7×$10^{-11}$ with SNR ≥ 7 | Coronagraph and Starshade:<br>Photometric range up to 1.8 µm<br><br>Planet-to-star flux ratio detection limit: 4×$10^{-11}$ with SNR ≥ 7 | Multi-epoch planet photometry |
| | | Polarization from surface oceans | TBD | TBD | TBD | |

| Science Goals | Science Objectives | Scientific Measurement Requirements | | Instrument Functional Requirements | Projected Performance | Mission Functional Requirements (Top Level) |
|---|---|---|---|---|---|---|
| | | Physical Parameters | Observables | | | |
| **Goal 2:**<br><br>**To map out nearby planetary systems and understand the diversity of the worlds they contain** | O5: To determine the architectures of planetary systems around sunlike stars within 12 pc | Planetary orbits and radii. Detect molecular species in planetary atmospheres | Star-to-planet separation measured at ≥ 3 different orbital positions<br>Broad planetary spectra TBD | OWA: ≥ 6 arcsec<br>Planet-to-star flux ratio detection limit ≤ 4×10⁻¹¹ with SNR ≥ 7 at distances ≤ 5 pc | Starshade<br>OWA: 6 arcsec<br>Planet-to-star flux ratio detection limit: 4×10⁻¹¹ with SNR ≥ 7 at 5 pc | ≥ 3 observations a few months to a few years apart |
| | O6: To determine the interplay between planets and dust in planetary systems around sunlike stars within 12 pc | Disk colors<br>Disk degree of polarization<br>Disk morphology | Disk broadband images<br><br>Disk spectrally resolved images from 0.8 to 1.5 μm at R ≥ 20 | OWA: ≥ 6 arcsec<br>Spatially resolved spectroscopy from IWA to OWA with range ≤ 0.8 μm to ≥ 1.5 μm at R ≥ 20 | Starshade<br>OWA: 6 arcsec<br>Spatially resolved spectroscopy from IWA to OWA with R = 140 from 0.45 μm to 1.0 μm and R = 40 from 1.0 μm to 1.8 μm | Broadband images in at least two polarization states |
| | O7: To determine if the presence of giant planets is related to the presence of water vapor in the atmospheres of small planets detected in O2 | Same as O2 plus:<br>Planet orbit and radius for giant planet. | Star-to-planet separation measured at ≥ 3 different orbital positions<br><br>Broad planetary spectra TBD | IWA: ≤ 74 mas; OWA ≥ 3.3 arcsec<br>Spectral range: 0.45 μm to 1.0 μm | Starshade<br>IWA: 60 mas at 1.0 μm: OWA: 6 arcsec<br><br>Spectroscopy from IWA to OWA<br>R = 7 from 0.2 μm to 0.45 μm and<br>R = 140 from 0.45 μm to 1.0 μm and<br>R = 40 from 1.0 μm to 1.8 μm | Starshade repositioning for ≤ 0.3 μm and ≥ 1.0 μm observations |
| | O8: To determine the diversity of planetary atmospheric compositions in planetary systems around sunlike stars within 12 pc | Detect atmospheric molecular species H₂O, O₂, O₃, CO₂, CH₄ if concentration > (TBD) | Planetary spectra:<br>Molecular absorption features between 0.3 μm and 1.7 μm | Spectral range: ≤ 0.3 μm to ≥ 1.7 μm<br>O₃: R ≥ 5, 0.3–0.35 μm with SNR ≥ 5 per spectral bin<br>O₂: R ≥ 70, at 0.76 μm: and<br>CO₂, CH₄: R ≥ 2 at 1.0–1.7 μm with SNR ≥ 10 per spectral bin | Starshade<br>IWA: 60 mas: OWA: 6 arcsec<br><br>Spectroscopy from IWA to OWA<br>R = 7 from 0.2 μm to 0.45 μm and<br>R = 140 from 0.45 μm to 1.0 μm and<br>R = 40 from 1.0 μm to 1.8 μm | Starshade repositioning for ≤ 0.3 μm and ≥ 1.0 μm observations |

**STM-2. HabEx Observatory Science Traceability Matrix.**

| Science Goals | Investigation Themes | Investigation Science Objectives | Science Measurement + Technical Requirements | | Mission Requirements |
|---|---|---|---|---|---|
| | | | **Observables** | **Physical Parameters** | |
| **1. Trace the Life Cycle of Baryons** | Spectroscopically map the distribution of HI and metals in the IGM and CGM, in order to understand the cosmic baryon cycle over the last 10 Gyr | Column densities and kinematics of intergalactic medium (IGM) and circumgalactic medium (CGM) gas | UV absorption lines | Physical parameter: UV continuum strength: Parameter: SNR ≥ 25 for AB = 20 source in 1,200s Physical parameter: Wavelength range: Parameter: ≤ 0.115 µm to ≥ 0.300 µm Physical parameter: Spectral resolution: Parameter: ≥60,000 | • UV throughput ≥10% at 0.115 µm --> UV sensitivity coatings |
| | | | Spectroscopic multiplexing | Physical parameter: ≥10 sightlines simultaneously Parameter: Field of view ≥ 2.5×2.5 arcmin² | |
| **2. Measure the Local Value of the Hubble Constant** | Investigate and constrain possible evolution in value of hubble constant over cosmic time, potentially indicative of new physics | Precise Cepheid distances to hosts of Type Ia supernovae in local universe | Multiepoch imaging of ≥20 Cepheids per galaxy | Physical parameter: Field of view Parameter: ≥ 2.5×2.5 arcmin² Physical parameter: Wavelength range Parameter: ≤ 1.5 µm to ≥ 1.7 µm Physical parameter: Photometric Precision Parameter: 1% absolute | • Near-IR imaging channel |
| **3. Measure the Star Formation Histories of Nearby Galaxies** | Investigate and constrain formation and evolution of nearby galaxies by directly studying their stellar populations | Resolved stellar populations of nearby galaxies, down to solar mass main sequence stars | Multiband imaging of nearby galaxy | Physical parameter: Field of view Parameter: ≥2.5×2.5 arcmin² Physical parameter: Wavelength range Parameter: Broadband imaging from ≤ 0.3 µm to ≥ 1.7 µm Physical parameter: Photometric precision Parameter: 3% absolute | • Near-UV and optical imaging channels |
| **4. Probe the Nature of Dark Matter with Dwarf Galaxies** | Investigate and constrain dark matter models through detailed studies of nearby dwarf galaxies | High-resolution photometry to measure galactic light profiles, including kinematic information from spectroscopy and proper motions | Multiband imaging of nearby galaxy | Physical parameter: Field of view Parameter: ≥ 2.5×2.5 arcmin² Physical parameter: Wavelength range Parameter: Broadband imaging from ≤ 0.4 µm to ≥ 0.8 µm Physical parameter: Photometric precision Parameter: 3% absolute | • Near-UV and optical imaging channels • Optical and near-IR multi-object spectroscopy |
| | | | Multi-object spectroscopy | Physical parameter: Field of view Parameter: ≥ 2.5×2.5 arcmin² Physical parameter: Wavelength range Parameter: ≤ 0.5 µm to ≥ 1.0 µm Physical parameter: Spectral resolution Parameter: R ≥ 2,000 | |
| **5. Exoplanet Transit Spectroscopy** | Transit spectroscopy of exoplanets to investigate atmospheric composition | Understand exoplanet atmospheres, search for interesting chemistry | Exoplanet transit spectrum | Physical parameter: Wavelength range Parameter: ≤ 0.35 µm to ≥1.4 µm | • Spectroscopic channel covering near-UV to near-IR |
| **6. Solar System Observations:** studying planetary aurora, planetary exospheres, cryovolcanism, atmospheric composition, and high-contrast imaging | Study planets within our solar system, to better inform our understanding of exoplanets | Study a range of planetary characteristics, from aurora to cryovolcanism to atmospheric and exospheric composition | Images of solar system planets | Physical parameter: Wavelength range Parameter: ≤ 0.15 µm to ≥ 0.30 µm Physical parameter: Field of view Parameter: ≥ 1.0×1.0 arcmin² | • Capability to track moving solar system moving targets |

# 5   HABEX 4-METER BASELINE DESIGN

The baseline HabEx 4 m observatory concept will be the largest, most stable, space telescope covering ultraviolet (UV), visible, and near-infrared (near-IR) wavelengths ever built. With an unobscured 4 m diameter aperture, it is capable of collecting three times as many photons (12.6 $m^2$ collecting area) as the 2.4 m Hubble Space Telescope (HST; 4.0 $m^2$ collecting area). Its diffraction resolution limit is 21 milliarcseconds (mas) at 0.4 $\mu$m, compared to HST's performance of 34.4 mas. In addition, HabEx is designed to be the most stable astronomical observing platform ever; capable of relative pointing stability to 0.7 mas. HST's best pointing stability is 2 mas (Nelan et al. 1998). Furthermore, this design is solely based on manufacturing capabilities and state-of-the-art telescope technologies presently available.

The HabEx 4 m architecture takes advantage of the significantly larger launch capability of NASA's new Space Launch System (SLS) Block 1B. The SLS Block 1B utilizing an 8.4 m diameter, 27.4 m tall payload fairing (PLF) can deliver greater than 38,000 kg to Earth-Sun L2. By comparison, the SpaceX Falcon Heavy has a 5.2 m diameter by 13.1 m tall fairing volume and can deliver 12,500 kg to L2. The greater launch capacity of the SLS is increasingly important for future large telescope missions. It shifts the design boundaries and enables the use of mass to reduce design complexity as space telescope apertures grow larger. The ability to select a monolithic primary mirror, as opposed to a segmented mirror requiring precision actuators and edge sensors, is an example of this paradigm change.

Section 5 begins with the system overview, followed by a discussion of requirements and error budgets. The mission design is covered in Section 5.4, with a detailed description of the payload in Section 5.5. Later subsections include the telescope bus, pointing control, starshade occulter, starshade bus, and system integration.

## 5.1   System Overview Description

The baseline HabEx mission is composed of two separate spacecraft flying in formation in a Earth-Sun L2 orbit. One spacecraft carries a 4 m off-axis telescope, and four science instruments—a coronagraph (CG) and a starshade instrument (SSI) for exoplanet direct imaging, and a wide-field workhorse camera (HWC) and a wide-field high-resolution ultraviolet spectrograph (UVS) for observatory science. The other spacecraft is a 72 m starshade. Together, while in formation, they form an externally occulting observatory for exoplanet imaging and spectral characterization. The starshade suppresses the light from the target star while the telescope's starshade imager observes the planetary system surrounding the target star. To form this observatory, the starshade positions into the line-of-sight (LOS) between the telescope and the target star at an approximate 124,000 km separation from the telescope, and maintains alignment using a positional control loop carried over a spacecraft-to-spacecraft radio link. Position sensing is carried out by instrumentation on the telescope spacecraft and position control is handled by the propulsion system on the starshade spacecraft. **Figure 5.1-1** illustrates the formation flying configuration.

The starshade and telescope spacecraft are co-launched on a SLS block 1B launch vehicle into a 780,000 km diameter L2 orbit. The primary mission will run for five years, but the telescope includes enough fuel to continue operations for five additional years. The starshade has fuel for five years of operations after which it can no longer slew to new targets and must hold the L2 orbit until serviced. Serviceability is a requirement for all large astrophysics observatories; both the starshade and the telescope are able to be refueled and upgraded, however, the starshade occulter cannot be replaced during servicing.

Exoplanet science observations are accomplished with the internally occulting coronagraph working in concert with the external starshade occulter. These two instruments are complementary in nature. While the starshade instrument is capable of very high-contrast imaging and spectroscopy over a large field-of-view (FOV), it is limited in the number of observations due to the large slew times of the starshade occulter. The coronagraph, on the other hand, is capable of faster slews, making many more observations possible, but has a narrower, high-contrast FOV with reduced spectrographic capability. Working together, the starshade and coronagraph can achieve the repeated planet detections required to determine orbits, and the high-resolution spectral profiles needed to characterize exoplanet atmospheric gases.

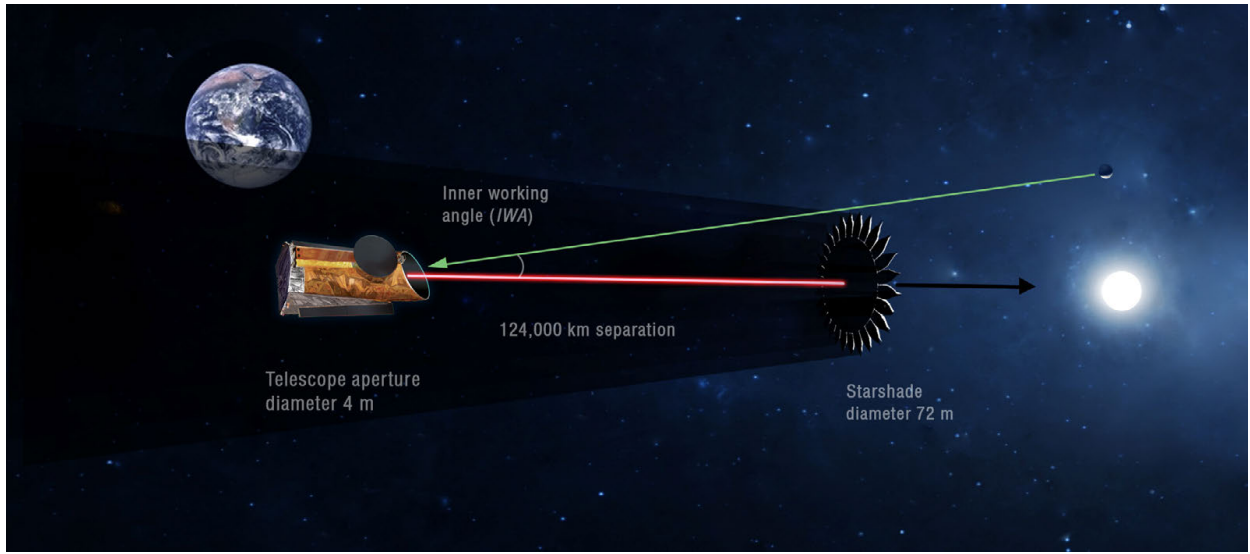

**Figure 5.1-1.** HabEx Observatory configuration.

Coronagraphy requirements drive most features of the HabEx telescope design. The unobscured aperture and monolithic primary mirror avoid light-diffracting mirror edges and obscurations that reduce coronagraph contrast performance and throughput. To ensure coronagraph-driven LOS pointing and wavefront stability, the telescope includes precision thermal control, mirror positional control using laser metrology, and a fine guidance sensor (FGS) paired with microthrusters for pointing control during observations. Some HWC and UVS needs also influenced the telescope design. Wide fields of view needed by the HWC and UVS instruments led to adoption of a three-mirror anastigmat (TMA) layout, and UV sensitivity to mirror contamination set the operating temperature for the primary mirror.

The starshade spacecraft includes the 72 m deployable starshade occulter payload, solar electric propulsion (SEP) to move the starshade from target to target, a bipropellant for the propulsion system to hold alignment when observing, and a formation flying beacon and communication link. The starshade is spin stabilized, rotating at a rate of 0.33 rpm.

In addition to the complementary nature of the HabEx baseline's joint coronagraph and starshade direct imaging measurements, the inclusion of both imaging and spectroscopy capabilities within both instruments adds resiliency against both technical and programmatic risks. Loss of one occulting instrument does not eliminate the exoplanet science return of the mission since both the coronagraph instrument and the starshade instrument carry imaging and spectroscopy channels with similar (but not identical) capabilities. While the starshade is better suited to spectroscopy and the coronagraph to searches, either can serve the purpose in the event of a failure in the other instrument. In addition, a delay in starshade development could be addressed with separate launches and the completion of coronagraph measurements before the starshade joins the telescope at L2.

## 5.2 Requirements

HabEx key baseline design requirements are derived from the concept's science goals and are traceable to those goals through the Science Traceability Matrix (STM; see Section 4) and the subsequent starshade and coronagraph error budgets (see Section 5.3). The requirements are organized into five areas: mission, coronagraph, starshade, observatory science, and telescope.

### 5.2.1 Mission Requirements

The science-driven key requirements for the HabEx mission are derived directly from STM-1:

- An orbit suitable for supporting formation flying with long periods of telescope/ starshade alignment on target
- A mission capable of revisiting the target star systems at least four times
- A baseline mission duration sufficiently long to allow at least four observations of each planetary system, separated by months or years

In addition to the science-driven requirements, the HabEx mission must also meet two programmatic requirements. The first is that the mission must be serviceable. This requirement was established by law, but given the significant investment required for a mission like HabEx, having the ability to extend and expand the facility's science return seems only practical. The second programmatic constraint is that the mission must be able to be launched by an American-built launch vehicle likely to be available at the time of the HabEx mission. This requirement will set overall launch mass and volume limits on the flight system.

#### 5.2.1.1 Formation Flying Requirements

To properly suppress the parent star's starlight to enable planet observations, the starshade must maintain its position relative to the telescope and its sight-line to the star. This, by definition, is formation flying. Tolerancing of the starshade design to maintain performance levies key requirements on the position of the starshade (see

**Table 5.2-1.** Key formation flying requirements.

| Formation Flying Requirements | |
|---|---|
| Separation: | 124,000 km ± 250 km |
| Lateral displacement: | ±1 m |
| Occulter tilt: | ≤1 deg |
| Distance sensing: | ≤25 km |
| Lateral displacement sensing: | ±24 cm |

**Figure 5.3-2** for the formation flying positional requirements in relation to the overall starshade contrast error budget). Maintaining the necessary contrast levels for direct imaging Earth-sized planets in the habitable zone (HZ) of nearby stars, requires that the starshade stay within 250 km of the nominal separation distance from the telescope, and within 1 m lateral displacement from the telescope-to-star sight-line (**Table 5.2-1**).

### 5.2.2 Coronagraph Requirements

Many requirements throughout the HabEx flight system stem from the coronagraph contrast requirements. While the coronagraph instrument can "dig a dark hole" and achieve $10^{-10}$ contrast using deformable mirrors (DMs) to overcome coronagraph and telescope quasi-static wavefront errors (WFEs), maintaining this high contrast throughout an observation sets requirements on the stability of the wavefront at the coronagraph diffraction mask. Contrast and contrast stability requirements are set in the coronagraph error budget in Section 5.3, and decompose into further requirements on the telescope, pointing control, thermal control, low-order wave front sensing and control, and laser metrology systems.

Contrast, signal-to-noise ratio (SNR), spectral resolution and inner working angle (IWA) requirements (see **Table 5.2-2**) are all established in STM-1.

### 5.2.3 Starshade Requirements

The starlight suppression capability of the starshade instrument is achieved in combination with the formation flying starshade occulter. The ability of the starshade instrument and occulter system to image an exoplanet is highly dependent on an accurately deployed and stable occulter shape, as well as proper telescope-starshade positioning. The former drives occulter fabrication and deployment tolerances, and thermal design (see Section 5.3 for starshade design requirements in relation to the contrast error budget). The latter sets the formation flying requirements, which were previously discussed in Section 5.2.1.1.

**Table 5.2-2.** Key coronagraph instrument requirements.

| Coronagraph Requirements | |
|---|---|
| Waveband:<br>• Imaging:<br>• Spectroscopy: | • ≤0.45 µm to ≥1.7 µm<br>• ≤0.45 µm to ≥1.7 µm |
| Spectroscopy resolution (R = λ/Δλ) | R ≥ 140 (0.45–1.0 µm)<br>R ≥ 40 (1.0–1.7 µm) |
| IWA of coronagraphic dark field: | ≤74 mas |
| Starlight suppression raw contrast in the dark field: | ≤$10^{-10}$ between IWA & OWA |
| Stability of starlight suppression raw contrast in the dark field: | ≤$2 \times 10^{-11}$ between IWA & OWA |

Instrument waveband, spectral resolution and signal-to-noise ratio requirements are directly linked to the HabEx science objectives through the Science Traceability Matrix (Section 4). Pointing requirements are set by the telescope's diffraction-limited point spread function.

Key starshade instrument and occulter requirements are summarized in **Table 5.2-3**.

### 5.2.4 Observatory Science Requirements

In addition to exoplanet direct imaging, HabEx would also support other astrophysics science through the UVS and the HWC instruments. The UVS would conduct high-resolution UV spectroscopic observations and requires high throughput at the FUV end of the band (0.115 μm) as well as high spectral resolution (**Table 5.2-4**). The UVS would also conduct intergalactic medium and circumgalactic medium emissions emission mapping. This requires the use of FUV multi-object spectroscopy (MOS) over a modest-sized field. For HabEx, the UVS has a 3×3 arcminute$^2$ FOV and a microshutter array to allow for MOS. All UVS requirements come from STM-2 and directly map to the baryon life cycle science objective.

**Table 5.2-3.** Key starshade requirements.

| Starshade Requirements | | |
|---|---|---|
| **Instrument** | Waveband:<br>• Imaging:<br>• Spectroscopy: | • ≤0.3 μm to ≥1.7 μm<br>• ≤0.3 μm to ≥1.7 μm |
| | Spectroscopy resolution<br>(R = λ/Δλ) | R ≥ 7 (0.3–0.45 μm)<br>R ≥ 140 (0.45–1.0 μm)<br>R ≥ 37 (1.0–1.7 μm) |
| | Signal-to-noise | ≥ 10 per spectral bin |
| | Pointing:<br>• Accuracy:<br>• Stability: | • <2 mas/axis<br>• <2 mas/axis |
| **Occulter** | Starlight suppression raw contrast in the dark field: | ≤10^{-10} from IWA |
| | IWA of full dark field: | <100 mas at 1.1 μm |
| | Petal position (manufacture): | Bias: ±500 μm (3σ)<br>Random: ±1,500 μm (3σ) |
| | Petal shape, quasi-static (manufacture) | Bias: ±115 μm (3σ)<br>Random: ±230 μm (3σ) |
| | Petal radial position (deployment): | Bias: ±500 μm (3σ)<br>Random: ±1,500 μm (3σ) |
| | Petal shape stability (thermal):<br>• Disk-petal differential strain (bias):<br>• Petal width (bias): | • 30 ppm<br><br>• 20 ppm |

**Table 5.2-4.** Key UV spectrograph requirements.

| UV Spectrograph Requirements | |
|---|---|
| **Waveband:** | ≤0.115 μm to ≥0.3 μm |
| **Field-of-view:** | ≥ 2.5 × 2.5 arcmin$^2$ |
| **Spectral resolution:** | R ≥ 60,000 (highest resolution band) |
| **Throughput:** | > 10% |
| **Pointing:**<br>• **Accuracy:**<br>• **Stability:** | • ≤2 mas rms/axis<br>• ≤2 mas rms/axis |

The HWC requirements also stem from the objectives in STM-2. Like the UVS, the HWC requires a fairly large FOV and a microshutter array to conduct MOS. The minimum spectral resolution is set by the dark matter science. Hubble constant science sets the photometric precision. Like the starshade instrument and the UVS, pointing for the HWC is driven by the telescope's diffraction-limited point spread function. **Table 5.2-5** identifies the HWC key requirements.

### 5.2.5 Optical Telescope Assembly Requirements

The HabEx telescope requirements are primarily driven by the need to direct image Earth-sized planets in the habitable zone with the coronagraph. Imaging these small planets orbiting close in to their host stars requires a telescope/coronagraph system that can produce a "dark hole" at $10^{-10}$ contrast while maximizing irradiance throughput. Additionally, the number of habitable zones that can be observed, and accordingly the science yield achievable with the system, will increase as the IWA decreases. Contrast and IWA requirements are specified in the Science Traceability Matrix (STM-1) and are directly linked to the HabEx science objectives. Throughput contributes to the SNR ratio; the requirement for which is also specified in STM-1.

**Table 5.2-5.** Key HWC requirements.

| Workhorse Camera Requirements | |
|---|---|
| **Waveband:**<br>• **Imaging:**<br>• **Spectroscopy:** | • ≤0.15 μm to ≥1.7 μm<br>• ≤0.35 μm to ≥1.4 μm |
| **Field of view:** | ≥2.5×2.5 arcmin$^2$ |
| **Spectral resolution:** | R ≥ 2000 |
| **Photometric precision:** | < 1% absolute |
| **Pointing:**<br>• **Accuracy:**<br>• **Stability:** | • ≤2 mas rms/axis<br>• ≤2 mas rms/axis |

A coronagraph works best with a clear aperture. Without obscurations, the system throughput is maximized and diffracting edges within the field of view are avoided, simplifying the coronagraph design and improving contrast performance. For these reasons, clear aperture coronagraph yields are comparable to significantly larger diameter on-axis and segmented aperture yields. Accordingly, HabEx adopted an off-axis, unobscured monolithic telescope architecture to maximize coronagraph science yield while minimizing aperture diameter and telescope size and cost.

The telescope primary mirror diameter was set at 4 meters early in the study. A telescope with such a mirror was seen by the HabEx team as within industry's current capabilities, and a significant advancement over HST and the Wide Field Infrared Survey Telescope (WFIRST). A subsequent aperture yield sensitivity study (see Section 2.3.3) supported the initial decision, indicating that a 3.7 m aperture, at minimum, is necessary to reduce the probability of not characterizing an exo-Earth to below 0.5%. The 4 m aperture provides additional performance margin with little impact on mirror fabrication readiness.

Another important telescope design parameter was the F number. A slower telescope (i.e., larger F#) is longer and consequently, heavier and more costly, but a faster telescope has a greater issue with polarization crosstalk, which impacts coronagraph contrast. An early trade study was conducted to evaluate the minimum acceptable telescope F# for HabEx (**Figure 5.2-1**). The trade indicated that f/2.0 would meet contrast requirement for the vector vortex coronagraph (VVC) charge 6.

**Table 5.2-6.** Optical Telescope Assembly (OTA) requirements.

| OTA Requirements | |
|---|---|
| Telescope architecture: | Off-axis, unobscured aperture |
| Aperture diameter: | ≥3.7 m |
| Primary mirror f/#: | ≤f/2.0 |
| Diffraction limit wavelength: | 0.4 μm |
| Bandpass: | ≤0.115 μm to ≥ 1.7 μm |
| Operating temperature, telescope optics: | ≥270K |
| Wavefront error, quasi-static: | ≤30 nm rms |
| Wavefront error, stability:<br>• "Correctable" <1 mHz:<br>• "Uncorrectable" >1 mHz: | <br>• ≤1 nm rms<br>• ≤5 pm rms |
| Internal line-of-sight stability:<br>• "Correctable" <200 Hz:<br>• "Uncorrectable" >200 Hz: | <br>• ≤2 mas rms/axis<br>• ≤0.7 mas rms/axis |

Telescope reflectivity bandpass requirements were set by the observational needs of the four science instruments. UVS set the blue end of the telescope band to a maximum wavelength of 0.115 µm and the exoplanet direct imaging instruments set the red end of the bandpass to a minimum of 1.7 µm. These requirements map to the HabEx science case through the STMs.

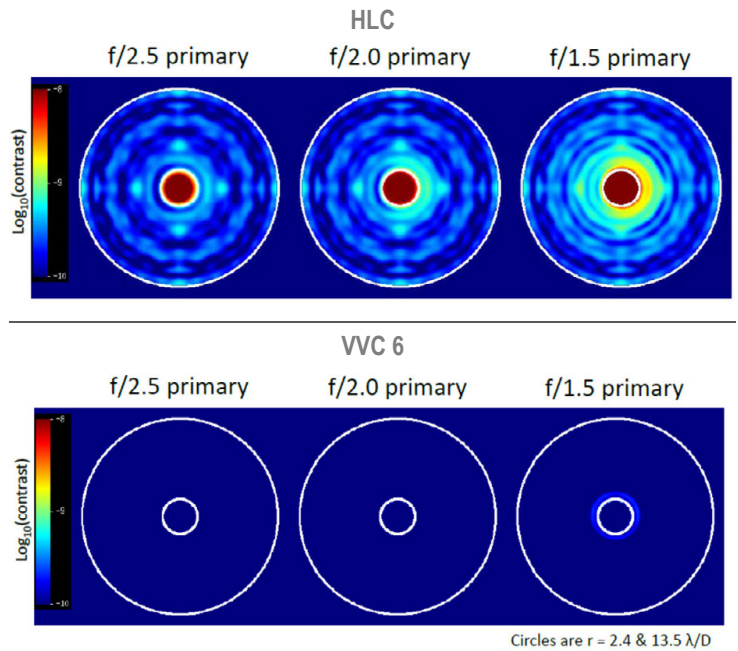

**Figure 5.2-1.** Simulated contrast vs. F# for the VVC 6 and hybrid Lyot coronagraph (HLC) )over the 0.4–0.49 μm band. All simulations assumed an HST-like aluminum coating with magnesium-fluoride overcoat on the primary and secondary mirrors. The effect of F# is more pronounced in the HLC.

The need to carry out UV science set the minimum operating temperature for the telescope optics at 270K. Telescopes at colder temperatures face significant contamination issues (Bolcar et al. 2016) Power considerations and bandpass red end performance will prevent the telescope from running much above the design minimum temperature.

Coronagraph contrast performance requires that the Optical Telescope Assembly (OTA) be ultra-stable in both internal LOS errors and output wavefront quality. These requirements fall out of the coronagraph error budget, which is specified in Section 5.3. To aid in meeting these requirements, laser metrology and control (MET) of the secondary mirror (SM) and tertiary mirror assembly (TMA) with respect to the primary mirror (PM) maintains the alignment of the OTA optics. The use of microthrusters also aids in achieving an ultra-stable telescope by not introducing high frequency jitter. Details of these telescope design features are given later in this design section.

## 5.3    Error Budgets for Exoplanet Instruments

The performance estimates of both the coronagraph and starshade are based on detailed error budgets that compute the scattered light level in the image plane as a function of instrument perturbations. Perturbation amplitudes are allocated using experience from laboratory tests, the results of dedicated technology development, and engineering models.

For the coronagraph, which has a wavefront control system that can compensate for static wavefront errors, the performance is driven by the stability of optical components, mainly the low-order bending modes of the primary mirror, motion of the secondary and tertiary relative to the primary, and pointing variations. The coronagraph error budget (**Figure 5.3-1**) is thus based on sensitivity studies that map minute perturbations to the light level in the dark hole. The error budget rolls up the effects of optics bending, rigid-body and flexible-body motions of optics, beam walk across imperfect optics, and both drift and vibrations during an observation. It includes the ability to control a set of low-order aberrations, and options for several coronagraph configurations. A detailed description of the coronagraph error budget can be found in Shaklan et al. (2005) and Marchen and Shaklan (2009).

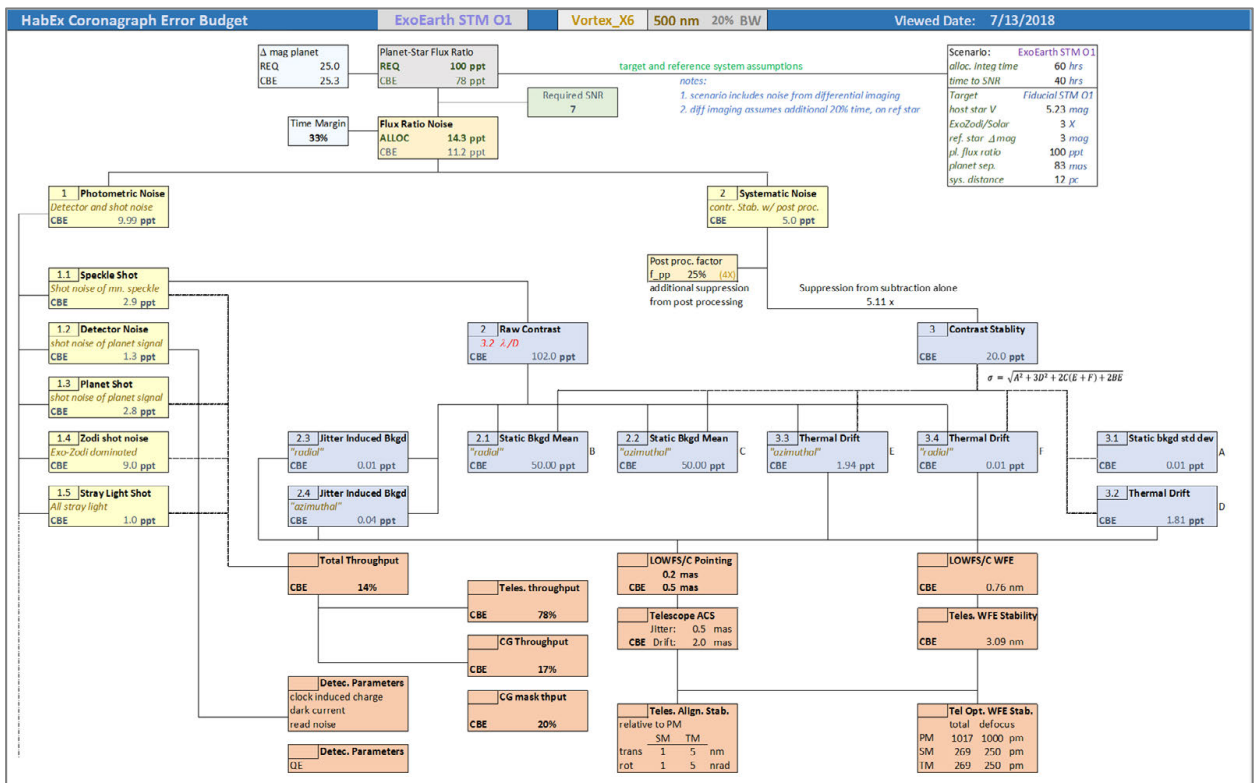

**Figure 5.3-1.** Coronagraph error budget for imaging an Earth-like planet around a sunlike star at 14 pc with a required delta magnitude of 25. A coronagraph bandwidth of 20% at a wavelength of 500 nm is assumed.

Likewise, the starshade error budget (**Figure 5.3-2**) also uses model-based optical sensitivities (Shaklan et al. 2010, 2011) to determine the effects of starshade shape errors on instrument light level in the telescope focal plane. The starshade has no active compensation mechanism so the error budget includes both manufacturing and time-varying terms.

Both the coronagraph and starshade error budgets assume that perturbations are completely uncorrelated, with two exceptions. The first is that the coronagraph optics can move as a unit.

The second is that for the starshade, every random term that appears on a petal (e.g., where is the petal located, and what shape does it have) also appears in global forms, perfectly correlated amongst all the petals. This provides a means of tolerancing manufacturing and thermal errors that repeat from petal to petal.

It is important to note that the error budgets contain only model-based optical sensitivities and a methodology for combining error terms. The budgets are not based on thermal or mechanical models. They provide a framework for allocating motions and manufacturing errors that either originated with, or can be compared to, thermal and mechanical models.

Both error budgets have two important and compensating assumptions. They both assume Model Uncertainty Factors (MUFs) of 2 in the expected raw (uncalibrated) performance, while also assuming the ability to calibrate the raw contrast by a factor of 2. The MUFs are mainly present to account for oversimplifications such as correlated errors and ideal optical masks that are not captured in the model. The calibration factor is conservative because better performance has been demonstrated (Cady and Shaklan 2014; albeit at lower contrast levels than needed for HabEx).

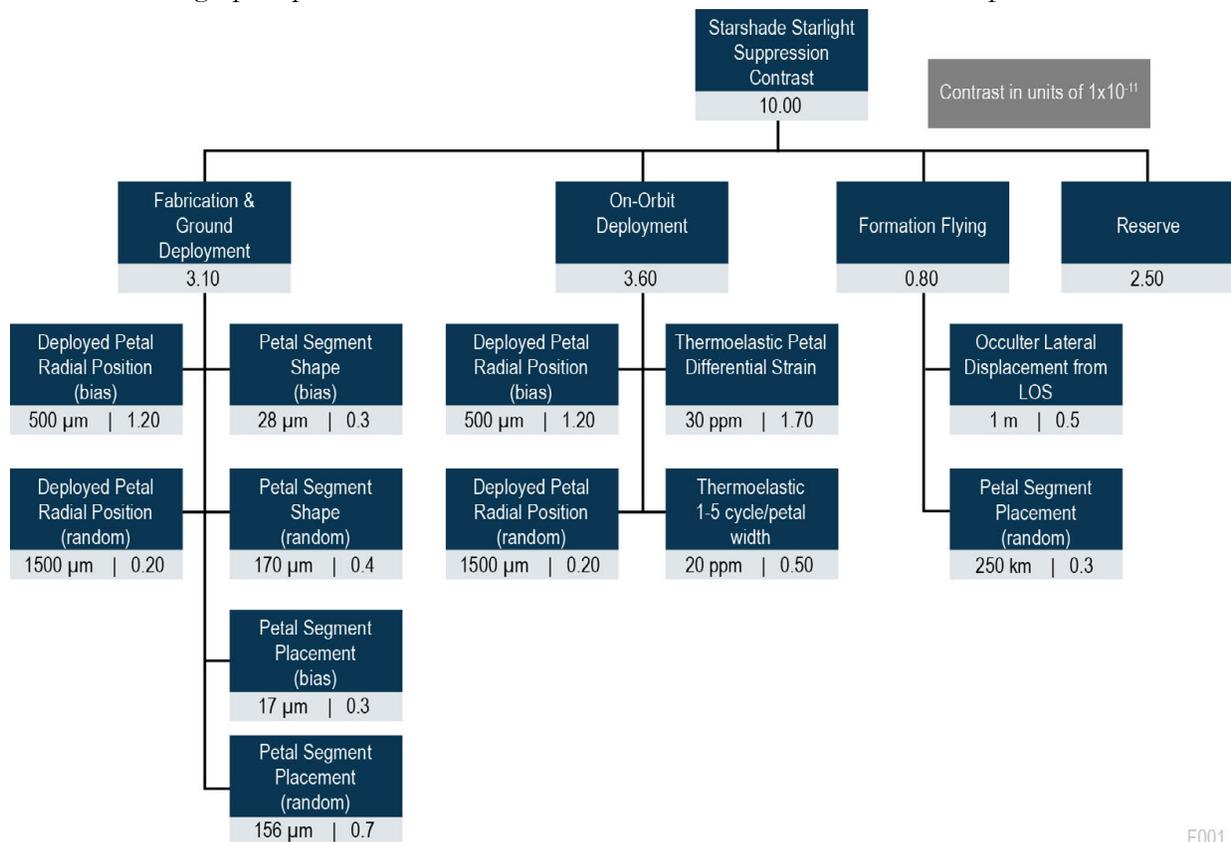

**Figure 5.3-2.** Starshade contrast error budget.

### 5.3.1 Coronagraph Error Budget

Coronagraph performance requirements drive many aspects for the telescope and telescope flight system designs so some additional discussion of the coronagraph error budget and these relationships is warranted. The high-level science requirement for the coronagraph is to image an Earth-like planet in the habitable zone around a sunlike star as far out as 12 pc. The instrument therefore must be able to distinguish a planet having a delta magnitude of 25 with respect to its parent star (magnitude 5) and a SNR of 7. The two main categories of error are photometric noise and systematic error. The sources of photometric error are shot noise from the planet, the residual speckle, and zodiacal dust (both local and at the exoplanetary system), along with detector noise. The detector currently baselined for HabEx, following the WFIRST approach, is an electron multiplication CCD (EMCCD) operated in photon counting mode. There are three sources of noise in an EMCCD: dark current, clock-induced charge, and read noise. The gain in an EMCCD effectively makes the read noise negligible. Clock-induced charge arises from impact ionization that contributes noise electrons as the signal is clocked out of a CCD. It is present in all CCDs but only noticeable when the read noise is near zero, as is the case in an EMCCD. For the most demanding spectroscopy, dark current will dominate and clock-induced charge is usually the second most important term.

The systematic noise can be thought of as the spatial standard deviation of the residual speckle after differential imaging. Such a residual can produce false positives and is hence a noise source. The contributions to background in a coronagraph dark hole can be broken into two categories: the incoherent background light (e.g., from zodiacal dust or background galaxies) and coherent scattered starlight. The former contributes only to the photometric noise, via the shot noise of the speckle pattern, but the latter, in the presence of optical changes (e.g., from thermal drifts), can produce a time-varying speckle. The amplitude of the change goes as the geometric mean of the coherent part of the original leakage field and the perturbation field from instability. This means that limiting the initial coherent field is important in suppressing the effect of the speckle instability. Any instability in the speckle results in a residual speckle after differential imaging. The residual speckle can produce false positives.

The source of systematic noise is primarily wavefront instability as a consequence of three disturbance modes: the initial static background noise of the dark hole, slowly changing thermal related drift, and fast changing jitter. Each of these noise modes produce radial (background) and azimuthal (speckle) terms that combine either coherently or incoherently to reduce contrast.

The physical events producing the two dynamic disturbance modes include rigid body motion of telescope and instrument optics, bending of optics, and beam walk over optical surfaces, all resulting from mechanical jitter, thermal drift, or LOS sensing errors.

Telescope and instrument requirements related to these disturbance sources include LOS pointing control to less than 2 milliarcseconds rms, rigid body alignment of the telescope optics relative to the primary mirror to within 5 nanometers in translation and 5 nanoradians in rotation, and telescope wavefront error stability to about 3 nanometers rms.

Architectural features of the HabEx telescope flight system limit the effect of two of the three disturbance sources. Mechanical jitter is largely nonexistent due to the low noise characteristics of microthrusters. Rigid body motion of the secondary mirror and tertiary mirror assembly as a result of thermoelastic effects on the telescope metering structure are tightly compensated with the laser metrology truss and rigid body actuators.

The remaining disturbance source expected to have the greatest effect on coronagraph performance is the thermoelastic bending of the telescope optics—specifically, the primary mirror.

However, given the very low coefficient of thermal expansion (CTE) and large thermal inertia of the glass-ceramic substrate and the insensitivity of the VVC 6 to low order wavefront aberrations within its null space (Mawet, Pueyo, et al. 2010), fine thermal control of the primary mirror assembly is expected to be well within current capabilities.

## 5.4 Mission Design

The HabEx mission concept has three distinct phases: launch, cruise, and science operations. During science operations, several different operational configurations exist to accomplish the science mission. Observations requiring joint positioning of the telescope and starshade necessitate precision formation flying.

### 5.4.1 Launch

The telescope and starshade were designed to be co-launched as a stack in an SLS Block-1B with an 8.4 m diameter fairing. As shown in **Figure 5.4-1**, the starshade is at the bottom of a stack, surrounded by a launch vehicle adapter that carries the telescope. The scarf shade required on the last 1.5 meters of the telescope is deployable in order for the stack to fit within the launch vehicle.

The HabEx baseline concept has selected an operations orbit at Earth-Sun L2. Coronagraphy requires a very low disturbance environment and L2 offers one of the most thermally stable orbits in close proximity to Earth. Heliocentric drift away orbits are also possible but the observatory could not be serviced in the future and the starshade would need to launch at close to the same time as the telescope. A two-launch option with a later starshade launch for programmatic reasons, could not be used.

To reach HabEx's halo orbit at L2, an excess energy (C3) of -0.6 km²/s² is required for HabEx's direct transfer. The maximum possible value (MPV) of the combined telescope, starshade, Petal Launch Restraint & Unfurler Subsystem (PLUS), and adapter mass is 34,869 kg and includes 29% average contingency plus an additional 12% system margin and another 3% of launch margin. The high-level mass budget of the HabEx mission is described in **Table 5.4-1**.

### 5.4.2 Cruise, Commissioning, and Checkout

Following launch, the telescope and the starshade separate. Three days after separation, both the telescope and the starshade perform their first trajectory correction maneuver (TCM-1), which begins their transfer to Earth-Sun L2. TCM-1's expected ΔV is approximately 50 m/s, depending on launch error and when it occurs following launch. This cruise phase lasts 6 to 8 months during which the spacecraft commissioning occurs. Two-way Doppler and ranging with the Deep Space Network (DSN) using X-band for ephemeris reconstruction and trajectory updates will confirm both trajectories. Delta Differential One-way Ranging (DDOR) may also be used closer to the halo orbit insertion (HOI) to refine knowledge of the trajectory. **Table 5.4-2** provides the key orbit parameters for the mission.

During cruise, both spacecraft undergo a range of commissioning and checkout activities.

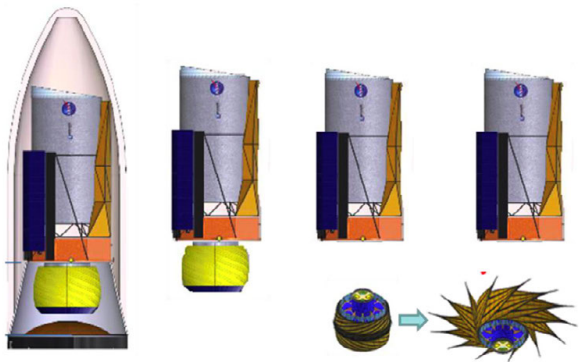

**Figure 5.4-1.** HabEx launch configuration accommodates for both the telescope and starshade. After launch, they separate and starshade later deploys.

**Table 5.4-1.** High-level mass breakdown of HabEx flight elements with LV margin.

| Element | Value |
|---|---|
| Telescope flight system wet mass (kg) | 19,385 |
| Starshade flight system wet mass (kg) | 11,274 |
| PLUS mass (kg) | 500 |
| Launch vehicle adaptor mass (kg) | 3,709 |
| Total launched mass (kg) | 34,869 |
| Launch vehicle capacity (kg) | 36,000 |
| LV margin (kg) | 1131 |
| LV margin (%) | 3% |

**Table 5.4-2.** HabEx key L2 halo orbit parameters

| Parameter | Value |
|---|---|
| Target/destination | Earth-Sun L2 |
| Trajectory type | Direct transfer |
| Cruise duration | 6–8 months from launch |
| Orbit diameter | ~780,000 km |
| Z amplitude | ~40,000 km |
| Orbit period | 175 days |
| Eclipse time | 0 minutes |
| Orbits/year | 2 |
| Max S/C-Sun distance | 1.012 AU |
| Max S/C-Earth distance | 1,800,000 km |

The telescope will open its cover 30 days after TCM-1. The telescope spacecraft commissions its subsystems and performs thruster calibrations for both the chemical propulsion system and the SEP microthruster system, which are described in Section 5.4.4. In parallel, the starshade commissions its own subsystems and unfurls its petals as soon as safely possible after TCM-1. Following unfurling, the starshade will then jettison the PLUS and the high-voltage SEP system will be brought online for calibration. At this point, the precise mass properties of the deployed system would be established. At launch +90 days, both the telescope and starshade perform a second TCM (TCM-2) of about 5 m/s ΔV, which prepares them for the HOI. Owing to the low-energy halo orbit about Earth-Sun L2, there is a window starting at launch +180 days lasting to about launch +240 days where HOI can occur. The HOI burn is about 5 m/s ΔV for both telescope and starshade.

### 5.4.3    Science Operations

Science operations can be divided into observations requiring only use of the telescope and those requiring both the telescope and starshade, with the latter taking about 20% of the total observing time.

During joint observations, both the telescope and starshade rely on propulsion for station keeping. The telescope must maintain orbit around the L2 point resulting in an annual station keeping budget of 10 m/s ΔV per year. While observing, the starshade requires 1.5 m/s ΔV per day for formation flying. However, the starshade will spend most of its time repositioning for new observations, where ~130 m/s ΔV is required for each repositioning. Additionally, the starshade must offset solar pressure and maintain the halo orbit using about 0.01 m/s to 0.2 m/s ΔV each day depending on the starshade's angle toward the Sun and the amount of propellant remaining on board.

### 5.5    Payload Overview

The telescope spacecraft payload consists of the telescope itself, four science instruments (**Figure 5.5-1**), plus ancillary equipment. Two of the instruments are designed for exoplanet direct imaging science. The first exoplanet science instrument contains a pair of coronagraphs, each with a variety of filters, capable of covering a broad spectral range from 0.45 to 1.0 μm in two observations. The second exoplanet science instrument is a starshade camera, able to cover 0.3 to 1.0 μm with a variety of spectral filters in conjunction with an external starshade at a single observational distance. These instruments are complementary; the coronagraph can make rapid survey observations while the starshade camera can observe angles closer to the star and once on target, enables more efficient spectroscopic observations. Both instruments offer capabilities into the infrared up to 1.8 μm and the starshade instrument works into the ultraviolet, down to 0.2 μm.

The other two optical instruments enable observatory science observations. The first observatory science instrument is a high-resolution ultraviolet spectrograph and camera (UVS) operating from 0.115 μm to 0.3 μm with resolution $R$ up to 60,000 on a 3'×3' field of view. The second observatory science instrument is also a camera and spectrograph (HWC) enabling imaging and spectroscopy, also on a 3'×3' field of view, in two bands stretching from the UV to the near-IR. In spectroscopy mode, the HWC operates as a multi-object spectrograph with resolution of 2,000.

Ancillary optical payload equipment consists of a laser metrology system to maintain the telescope alignment, and a fine guidance sensor.

### 5.5.1    Payload System/Optical Design

The presence of the coronagraph drives many aspects of the telescope design since it is sensitive to small changes in the incoming stellar wavefront and therefore requires a very stable telescope. Most ground- or space-based telescopes have an on-axis secondary mirror with its inevitable spider support, and as a result of the diffraction from this obstructed aperture, the coronagraph's optical throughput may become seriously degraded to as low as a few percent as shown in in the case of WFIRST-CGI (Krist, Nemati, and Mennesson 2016). In addition, the optical losses inherent in a long beam train brings the net throughput below 1%. For HabEx, the use of an off-axis secondary mirror dispenses with the obscuration losses and will allow at least ten times better throughput than CGI. An off-axis design leads to a smaller F# for the primary mirror parent compared to an equivalent on-axis design and this leads to worsened polarization effects that adversely affect coronagraph performance. To test this, telescopes with F#s ranging from 2.5 down to 1.5 were modeled and the effect of polarization on coronagraph contrast was calculated. From these simulations, the choice was made to set the F# of the primary to f/2.5. Such a slow telescope appeared to meet contrast requirements easily with the VVC 6 and marginally with the hybrid Lyot coronagraph (HLC), giving HabEx at least two different coronagraph options for further study. Furthermore, the length of the telescope could still be accommodated with the SLS fairing when co-launched with the starshade.

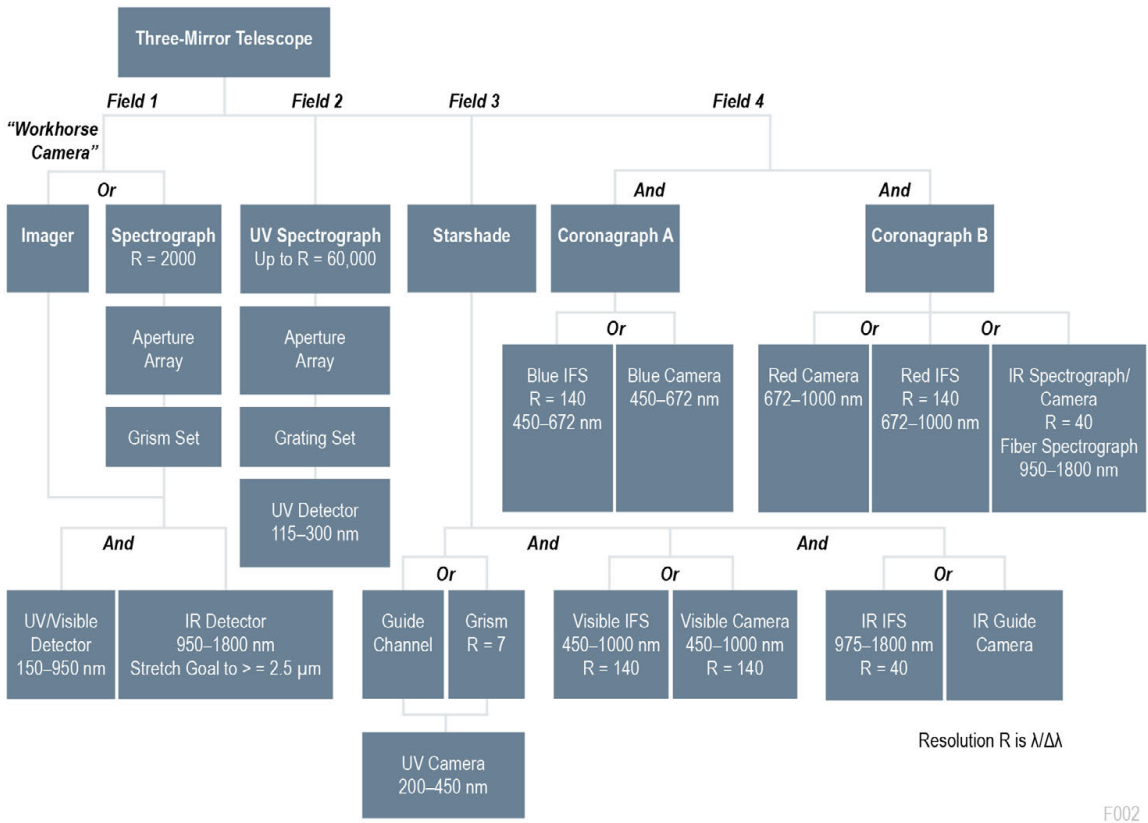

**Figure 5.5-1.** HabEx instruments, showing all instruments and modes (some of which can operate simultaneously).

For exoplanet observations, only a very narrow field of view is needed so that a simple Cassegrain or Ritchey-Chretien telescope could be considered, but HabEx employs a three-mirror telescope design to provide a larger field of view for each of the two observatory science instruments.

To achieve sufficient pointing stability when on target, a fine guidance sensor operating through the telescope aperture is needed. With the relatively large aperture, the point spread function (PSF) of the system has a small angular size (~21 mas full-width half-maximum at 0.4 μm) and to maintain the benefit of this high optical resolution on the cameras and starshade instrument, telescope pointing must be controlled to ~1/10th of the PSF size. To maintain maximum contrast on the coronagraph the pointing requirement is more stringent, so a fine-steering mirror (FSM) within the coronagraph beam train pushes the residual pointing error down to sub-milliarcsecond levels.

### 5.5.2  Telescope

The baseline HabEx optical telescope is a 4-meter, off-axis, three-mirror anastigmatic telescope with a scarfed straylight tube. It is essentially a scale-up of the Exo-Coronagraph (Exo-C) 1.4-meter telescope concept (Stapelfeldt et al. 2015) including an off-axis primary mirror to provide the coronagraph with an unobscured aperture, science instruments mounted on the side of the telescope (**Figure 5.5-2**, right) for mechanical and thermal isolation from the spacecraft, and improved polarization performance in the coronagraph. The primary is a 4-meter diameter, 400 mm thick, open-back Zerodur mirror. To minimize polarization anisotropy, the HabEx primary mirror focal length is f/2.5, creating a fairly long telescope configuration. Fortunately, the SLS can accommodate this length, including a scarfed straylight baffle, without the need for any physical deployments (**Figure 5.5-2**, left) or as a shared launch with the starshade using a deployed scarf (**Figure 5.5-3**). For this report, the co-launched configuration was selected. **Figure 5.5-4** shows the baseline HabEx telescope flight system concept including forward scarf, actuated baffle tube cover, solar panels, sun shade, and science instrument box.

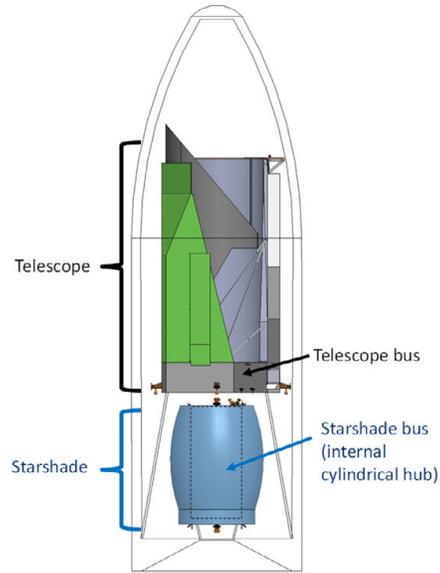

**Figure 5.5-3.** The HabEx telescope and starshade co-launch configuration.

#### 5.5.2.1  Telescope Optical Design

The HabEx telescope is a TMA with a 4 m diameter primary mirror, 2.5 m off axis. Robb's method (Robb 1978) was used to obtain the initial parameters for the design and the result was optimized in Zemax® to produce a collimated 50 mm beam at the output. **Table 5.5-1** shows the design parameters. The collimated output greatly simplifies the accommodation of the observatory science instruments, since otherwise, off-axis optics would be needed throughout those systems. **Figure 5.5-5** shows the telescope's optical layout.

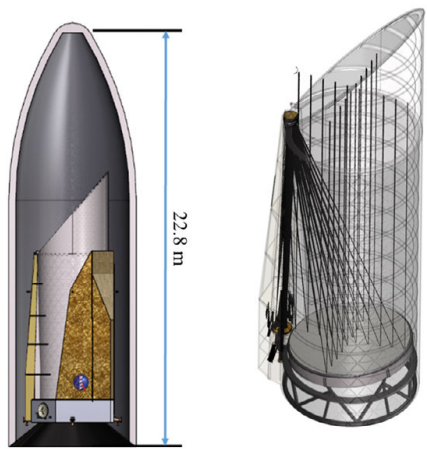

**Figure 5.5-2.** *Left:* HabEx in the SLS. *Right:* Transparent view of the HabEx telescope.

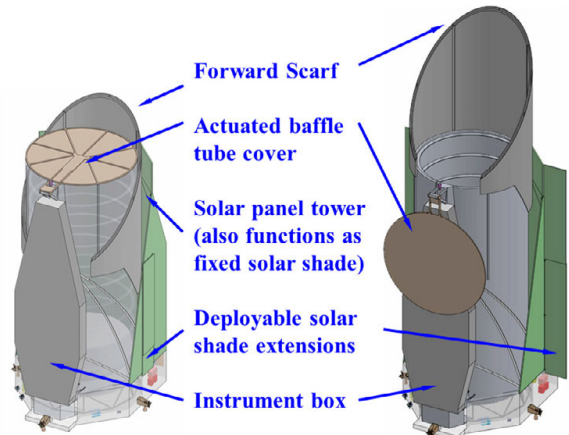

**Figure 5.5-4.** The HabEx telescope spacecraft concept.

A 4 m primary mirror (M1) directs light to the secondary (M2) then onto the tertiary (M3). Instruments are arranged near the tertiary. There are basically four ways to lay out the TMA (Lampton and Sholl 2007): two annular field configurations and two full field configurations. In the annular field design, the Cassegrain focus is spatially larger than the exit pupil and this widely separates the telescope fields on the tertiary mirror. In a three-mirror design, the output beam is directed towards the secondary mirror, so fold mirrors are normally used to bring the beam back behind the tertiary mirror. By placing a mirror between the tertiary and the exit pupil, the individual field can be extracted and passed to an instrument. This design also allows for different optical coatings at different locations on the tertiary, or separate tertiary mirrors with instrument-specific coatings, to aid transmission efficiency with some instruments. Since the telescope has to work into the UV, a protected aluminum coating is required on at least the first two mirrors.

The instruments have been arranged on the side of the telescope near M3 (see **Figure 5.5-5**). An alternative often employed (particularly with on-axis telescope designs) is to place the instruments behind the primary mirror, above the bus. The side mounting allows easier extraction of the instrument modules for servicing and takes advantage of existing volume created by the presence of the tertiary mirror beside the primary.

**Table 5.5-1.** HabEx telescope optical design parameters.

| Optic | M1 | M2 | M3 |
|---|---|---|---|
| Diameter | 4,000 mm | 450 mm | 680 mm |
| Radius of curvature | 19,800 mm | -1,953 mm | -2,168 mm |
| Thickness | 420 mm | 100 mm | 100 mm |
| Spacing to next optic | 9,030 mm | 9,080 mm | |
| Coating | Protected Al | Protected Al | Varies |

Furthermore, it creates easy access to radiators needed for cooling the detectors. Rear mounting would make use of space behind the primary, between its supporting structure and the bus, but may increase the overall telescope spacecraft height. Cooling paths would generally be longer and extraction of individual modules might be more complex. However, both design concepts are viable and would occupy similar volumes.

### 5.5.2.2 Telescope Instruments

**Figure 5.5-5** shows the instruments arranged near the tertiary mirror. Rays from the secondary come from the left of the figure. **Figure 5.5-6** shows the fields of view of the instruments. As can be seen in the figure, the UVS occupies the center of the field and the coronagraph views a smaller field to the left. The starshade instrument views a region to the right and slightly upwards while the HWC views a wide area to the right and slightly downwards. Arranged around the rest of the tertiary mirror are the four fine guidance sensor areas.

After striking the tertiary, the rays are collimated and converge towards a common on-axis pupil plane. Before reaching that plane,

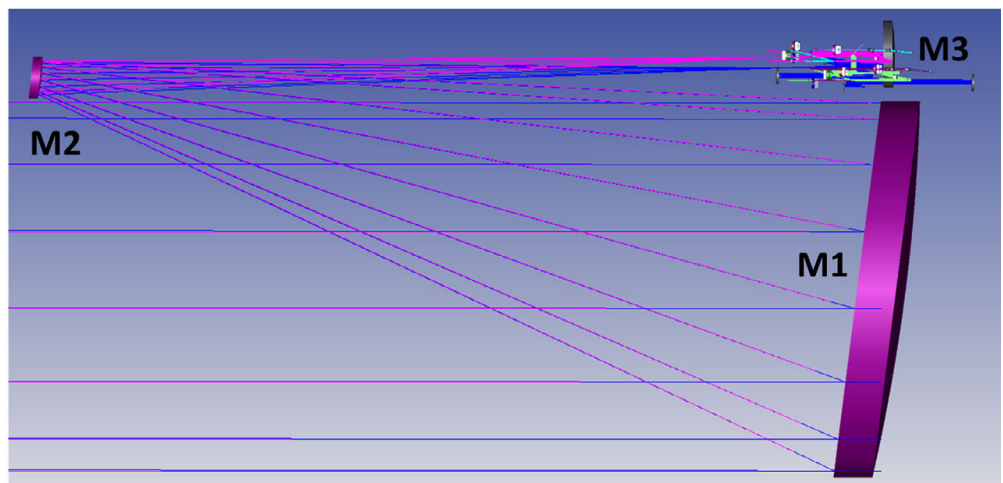

**Figure 5.5-5.** The principal HabEx telescope optics: the primary mirror M1, the secondary M2, and tertiary M3. The instruments are deployed near to M3.

however, the beams are extracted by fold mirrors and directed to steering mirrors at the pupil plane. These steering mirrors direct the light into the instruments. An exception is the UVS, which has its own tertiary and the beams are extracted with the minimum number of folds in order to preserve optical throughput. How these steering mirrors are used is discussed below and in Section 5.7.

### 5.5.2.2.1 Starshade Instrument Overview

Starshade operation requires both science and formation flying cameras, viewing the starshade simultaneously. A high suppression (dark) shadow region exists behind the starshade and the telescope is placed as far back as possible within this shadow (~124,000 km) while maintaining high starlight suppression. The telescope can move laterally ±1 m within the shadow and the tips of the starshade form an angle of 60 mas to the line of sight when operating in the 0.3 to 1.0 μm spectral band.

The HabEx starshade is of the numerically optimized type, rather than hyper-Gaussian, producing a designed high-suppression wavelength band. Light of both shorter and longer wavelengths is attenuated but leaks into the shadow region and is used for starshade positioning. **Figure 5.5-7** shows the starshade transmission functions for the three planned science bands, 0.2 to 0.67 μm, 0.3 to 1.0 μm, and 0.54 to 1.8 μm. When performing science at longer "red" wavelengths, shorter wavelength "blue" light is used for guiding and vice versa.

In normal operation, the starshade has a suppression band from 0.3 to 1.0 μm. To obtain suppression down to 0.2 μm in the UV, the starshade can be moved further away from the telescope, potentially achieving an IWA of 28 mas. For infrared science, the starshade moves closer to the telescope and the IWA will increase in proportion to the wavelength. **Table 5.5-2** shows the science bands, starshade/telescope separation, and IWAs.

The starshade instrument, shown schematically in **Figure 5.5-1**, contains six beam paths to accommodate three optical channels: UV, visible, and infrared (**Figure 5.5-8**). Light entering the starshade camera is split by dichroic optics into UV, then visible, and IR beam paths, so all of these channels can be operated simultaneously. Camera and spectrograph properties are shown in **Table 5.5-3**. The UV channel carries a simple slit spectrograph employing a grism with R = 7. The visible channel carries a broadband integral field spectrograph (IFS) capable of covering the wavelength range from 0.45 to 1.0 µm, plus an imaging camera for more rapid and wide-field system imaging. The infrared channel carries an IFS with R = 40 to enable disc and object spectroscopy.

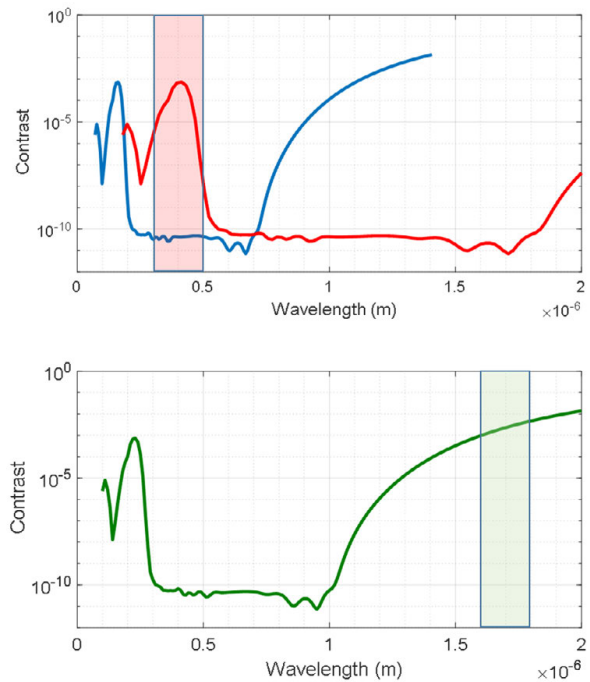

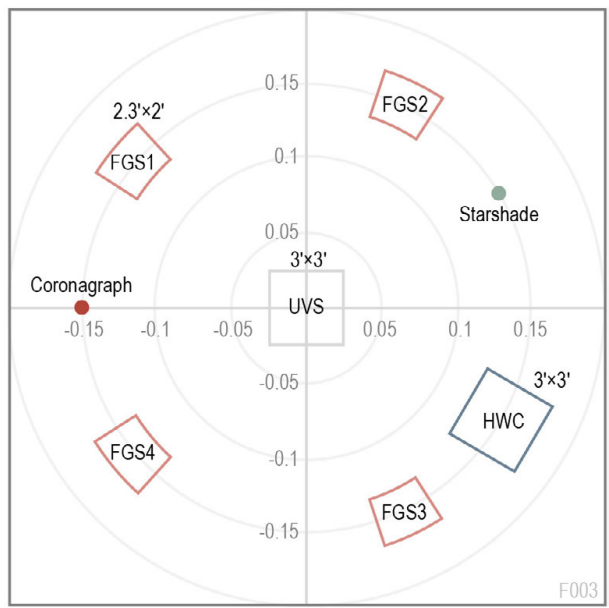

**Figure 5.5-6.** The instrument fields of view on the sky. The axes are scaled in degrees.

**Figure 5.5-7.** Starshade transmission functions. *Top:* the IR science band extends from 0.54 to 1.8 μm and its guide band is from 0.3 to 0.45 μm, which is detected on the UV guide channel. Also shown is the UV science band. Its guide band will be in the IR, as shown in the bottom chart. *Bottom:* the visible band showing deep contrast between 0.3 and 1.0 μm. The guide band is shown on the right: the infrared sensor detects the leakage light between 1.6 and 1.8 μm.

**Table 5.5-2.** HabEx starshade science bands and working distances with corresponding guide bands.

| Science Band | UV | Visible | IR |
|---|---|---|---|
| Wavelength band (µm) | 0.2–0.67 | 0.3–1.0 | 0.54–1.8 |
| Starshade separation (km) | 186,000 | 124,000 | 69,000 |
| IWA (mas) | 40 | 60 | 108 |
| Guide band (µm) | 1.6–1.8 | 1.6–1.8 | 0.3–0.45 |

*Starshade Guiding*

The ultraviolet and infrared channels have guide camera modes, which project a pupil image onto the focal plane. With a selected channel in science mode, an optic is introduced into the corresponding guide channel to place an image of the pupil on the guide CCD. The starshade's lateral position is sensed from an image of the light distribution at the telescope entrance pupil (see **Figure 5.5-9**) so that the pixel resolution is given in centimeters in **Table 5.5-3**. At the entrance pupil, the starshade shadow has some structure, typically with a much diminished "spot of Arago" at the center. The lateral position of the telescope is sensed by imaging this structure onto the focal plane and comparing with a library of expected images. **Figure 5.5-9** shows an image of the starshade shadow structure in infrared light when the starshade is set up for visible science. The central dot appears directly on the line-of-sight between the center of the starshade and the star and thus forms the target for the guide system. Outside this core, there are two faint rings and then the flux increases with a monotonic slope towards the edge. Outside this region, a reflective pattern of the starshade geometry appears with, in this case, 24 peaks around a circumference, the HabEx starshade design having 24 petals. The central peak has a diameter

$$= d \frac{\lambda}{D_{SS}}$$

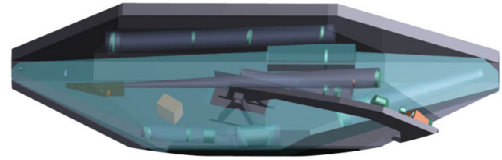

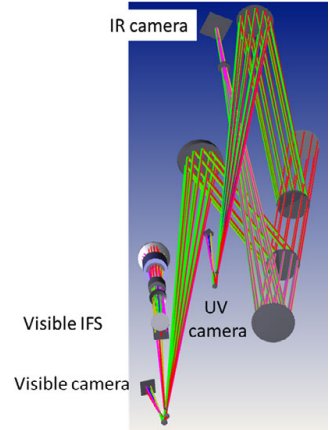

**Figure 5.5-8.** *Top:* The starshade instrument. *Bottom:* The principal components of the starshade instrument.

where d is the separation distance between the starshade and the telescope, λ is the wavelength, and $D_{SS}$ is the starshade diameter. In the case modeled, the central peak is about 3 m in diameter. The diameter of the smooth, sloping region is approximately 40 m and beyond this a pronounced pattern exists for about another 40 m.

**Table 5.5-3.** Starshade camera specifications.

| Cameras | UV Channel | Visible Channel | IR Guide Channel |
|---|---|---|---|
| FOV | 10.2" | 11.9" | - |
| Wavelength bands | 0.2–0.45 µm | 0.45–1.0 µm | 0.975–1.8 µm |
| Pixel resolution | 14.2 mas | 14.2 mas | 12 cm |
| Telescope resolution | 21 mas | 21 mas | - |
| IWA (at longest λ) | 40 mas | 60 mas | - |
| Detector | 1×1 CCD201 | 1×1 CCD201 | 1×1 LMAPD |
| Array width (pixels) | 1024 | 1024 | 256 |

| Spectrometers | UV Channel | Visible Channel | IR Channel |
|---|---|---|---|
| FOV | 10.2" | 1.9" | 3.8" |
| Wavelength bands | 0.2–0.45 µm | 0.45–1.0 µm | 0.975–1.8 µm |
| Spectrometer resolution | 7 | 140 | 40 |
| Spectrometer type | Slit/grism | IFS | IFS |
| Detector | 1×1 CCD201 | 1×1 CCD282 | 2×2 LMAPD |
| Array width (pixels) | 1024 | 4,096 | 2,048 |

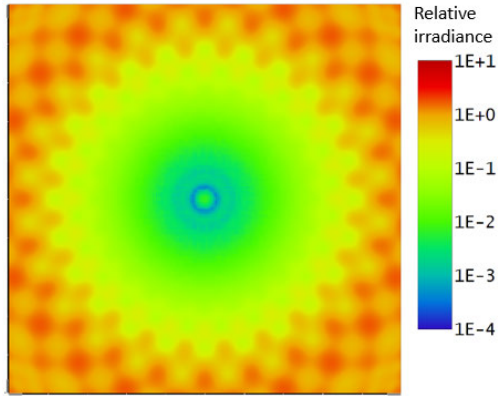

**Figure 5.5-9.** Starshade shadow at 1.7 μm wavelength. The central spot, which is used for guiding when in science mode, has an intensity ~7e-3 of the starlight. Away from the center, with the starshade misaligned, the gradient of the shadow is used to locate the direction of the center of the shadow.

Starshade navigation, covered in detail in Section 5.8, utilizes this pattern to bring the starshade into line with the star in the acquisition and science modes. Within the patterned region, the starshade follows the gradient down to the center. Once centered, the system maintains the central spot in the telescope pupil by periodically (every few 100 s) firing thrusters on the starshade. This alignment is precise because the target stars are bright and the attenuation by the starshade in the guide bands is poor. When a thruster firing occurs, science data taking is briefly suspended (~1 s). This is because the thruster plumes are illuminated by the Sun and would contaminate the data. However, they rapidly dissipate and the flux at the detectors is small; the detectors are electronically cleared and resume data acquisition.

### Visible Channel

The visible channel is the principal science channel and carries a camera and an IFS. The layout is shown schematically in **Figure 5.5-10**. Light from the telescope M3 strikes the fold mirror and then the FSM at the entrance to the starshade instrument. It then passes through a dichroic optic, which reflects UV light. The remaining visible and infrared light passes to a second dichroic where the visible light is reflected to an off-axis paraboloidal (OAP) mirror and thence to a focus where field stops are inserted to limit the field of view, one for the imaging mode (11.9" diameter) and a second for spectroscopy (1.9" diameter). This focus is reimaged by an ellipsoidal mirror to the focal plane. A filter wheel is inserted after the ellipse with filters to select wavebands appropriate for different starshade-to-telescope distances. For example, with the starshade at the nominal distance for visible work, the filter would pass 0.45 to 1.0 μm light. With the starshade more distant as set up for UV science, the spectral range would be 0.45 to 0.67 μm (see **Table 5.5-4**). Further filters and polarizing optics for polarization studies could be inserted here such as the science filters provided for coronagraphy (**Tables 5.5-7, 5.5-8, 5.5-9, and 5.5-10**).

The imaging focal plane consists of a single EMCCD operated at 153K. The chosen type is a modified CCD201 with delta doping and a thickened substrate together with a broadband "astro" coating giving response out to 1.0 μm.(Nikzad et al. 2017) The pixel scale is as shown in **Table 5.5-3**. During a thruster firing, the sensor is read out at 1 kHz to keep the accumulated photon count appreciably below full well.

For spectroscopy, an additional ellipsoidal mirror is inserted into the beam following the first ellipse, producing a large increase in the f/# from 47 to 1330. Via a fold mirror, this beam is focused onto a microlens array (MLA), which forms the entrance to the IFS. The IFS consists of the MLA, a matching multiple-aperture mask to restrict stray light, a set of lenses to collimate the beam, prisms to disperse the wavelengths and a second set of lenses to focus onto the focal plane. This type of IFS is described in McElwain et al. (2016) (**Figure 5.5-11**). The IFS operation can be

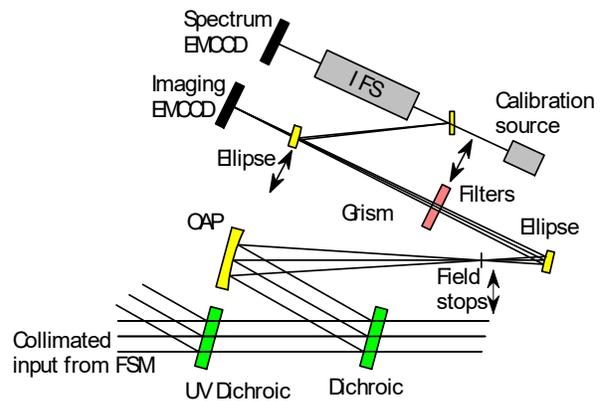

**Figure 5.5-10.** Schematic layout of the HabEx starshade visible channel.

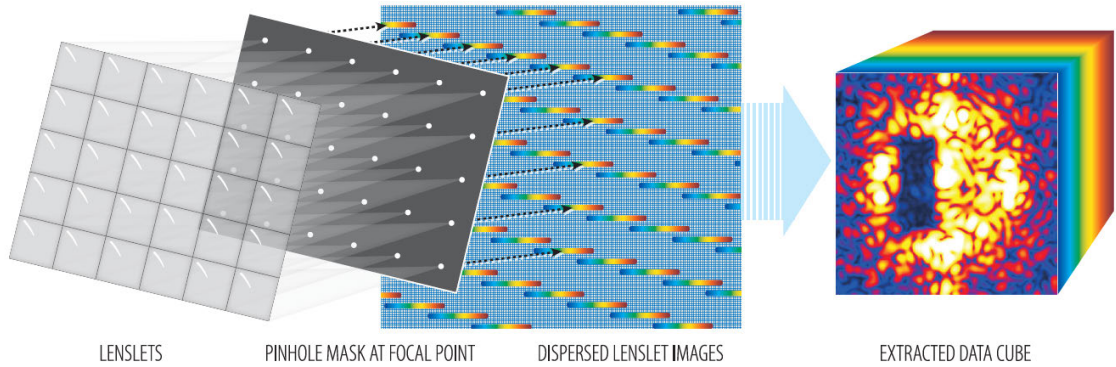

LENSLETS      PINHOLE MASK AT FOCAL POINT     DISPERSED LENSLET IMAGES     EXTRACTED DATA CUBE

**Figure 5.5-11.** Schematic form of the PISCES IFS from McElwain et al. (2016) showing the lenslet array, arrayed spectra on the focal plane and the resulting "data cube" illustrating a coronagraph image.

visualized thus: for each microlens array element, a spectrum is produced on the focal plane. The optical geometry involving the MLA ensures that the spectra do not overlap or interfere with each other. An image can be formed of the scene at one wavelength by using all the pixels on the focal plane that correspond to the same wavelength. A series of images known as "slices" can be assembled into a "data cube" with sides corresponding to the directions of the field of view and height corresponding to wavelength. Thus, the scene is reproduced in a stack of images representing narrow wavelength bands. To calibrate the images, it is necessary to provide a calibration source with at least one known wavelength in the band, and that illuminates the entire MLA. The calibration source light is injected when required through the fold mirror, which has a small leakage ~2%.

The HabEx IFS's focal plane consists of a large single electron multiplying CCD (Teledyne/e2v CCD282) operated at 163K. The chosen type is a modified version of the off-the-shelf item with delta doping and a thickened substrate together with a broadband "astro" coating giving response out to 1.0 μm. The pixel scale is as shown in **Table 5.5-3**. The format is an 8k×4k array with frame store areas at both sides of the 8k length, and a 4k×4k center imaging area. The spectral images are produced on the 4k square center area and moved into the frame stores before readout at high EMCCD gain. During a thruster firing, the sensor is read out as fast as possible to keep the accumulated photon count appreciably below full well.

*Ultraviolet Channel*

This channel (**Figure 5.5-12**) carries a low-resolution spectrometer and is also used as the guide channel for IR science. Light from the telescope M3 strikes the fold mirror and then the FSM at the entrance to the starshade instrument. It then reaches a dichroic optic that reflects the UV light to an off-axis paraboloidal mirror and thence through a field stop to an ellipsoidal mirror. Following the ellipse is a filter wheel to allow filter selection as shown in **Table 5.5-4**. Two field stops are provided, one to allow a field of view up to 10.2" diameter, and another with 0.02" diameter to select individual objects. The field stops are mounted on a piezoelectric stage to allow selection and positioning. The beam is then refocused to the focal plane, passing through a filter placed at the intermediate exit pupil which removes light of wavelengths longer than 450 nm. Also at this exit pupil, a grism can be introduced for low-resolution spectroscopy. With the grism removed, the camera forms an undispersed image. With the introduction of a mirror further downstream, the exit pupil is relayed to the focal plane, forming a pupil image suitable for starshade guiding. The pupil scale need not be large; a 32×32 pixel image is formed with each pixel covering a 12 cm square section of the entrance pupil.

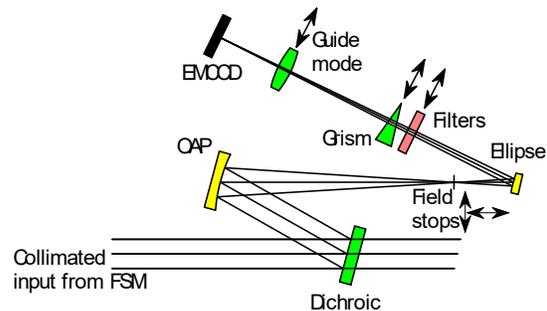

**Figure 5.5-12.** Starshade instrument schematic layout of UV channel.

The focal plane consists of a single EMCCD (CCD201) operated at 153K. The chosen type is a modified version of the off-the-shelf item optimized for high UV sensitivity by deep depletion and delta-doping processes (Nikzad et al. 2012), together with a broad band coating to improve response down to 0.2 μm. Pixel scale is as shown in **Table 5.5-3**. The format is a 1k×1k array with adjacent frame store. Again, the CCD will be read out with high gain to minimize read noise and during a thruster firing, the sensor is read out at 1 kHz to keep the accumulated photon count down.

### Infrared Channel

The infrared channel is the primary guide channel used for both visible and UV science, and also carries an infrared IFS. When the IFS is being used, guiding is on the UV guide channel. Infrared light entering the instrument passes through both dichroics and is reflected off a paraboloidal mirror (**Figure 5.5-13**). Between the second dichroic and the paraboloid, a filter wheel operates to allow band selection as shown in **Table 5.5-4**. The subsequent layout follows a similar scheme to the UV channel with a focus, ellipsoid and conditioning optics to reach the desired F#s. At the focus, a fixed field stop limits the field of view to 4", slightly larger than the IFS MLA FOV.

The guide channel consists of a lens to relay the exit pupil following the ellipsoid to the focal plane with the magnification providing 32 pixels across the telescope aperture. The focal plane consists of a single linear mode avalanche photodiode (LMAPD) array detector based on a HgCdTe sensor (Saphira array by Selex, formerly Leonardo). The avalanche gain-mode allows the effective read noise to be reduced (but not yet to the extent possible in EMCCDs). The detector is cooled to 60K to minimize dark current. Note that an IR imaging mode is easily provided in this layout, though not called for in the STM (see Section 4).

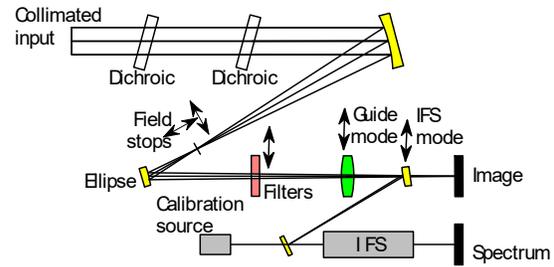

**Figure 5.5-13.** HabEx starshade instrument IR channel schematic layout.

The science channel consists of a powered relay mirror inserted near the guide channel relay lens. This provides the necessary larger focal length to the MLA. Before reaching the MLA, the beam is folded at a plane mirror. A calibration source is provided behind this mirror, injecting through it, so that the position of the spectrum can be identified on the IFS focal plane. The IFS utilizes a planned variant of the Saphira detector with a 1k×1k format and smaller pixels (12 micron) operated at 60K. Four of these detectors are arrayed in a 2×2 format to provide a full FOV of 3.8"×3.8" at R= 40. The IFS optics follow the same general design as for the visible IFS, with appropriate optical prescription changes. The field of view is Nyquist sampled by the lenslets and likewise, the spectrum is Nyquist sampled at the detector. During a thruster firing, due to its parallel output format and windowing capability, the sensor can be read out at very high frame rates to keep the accumulated charge down and thus avoid contamination of the science data caused by charge persistence.

#### 5.5.2.2.2 Coronagraphs

With 20% bandwidth being the current state of the art for a high-contrast coronagraph, covering the wavelength range 0.45 to 1.0 μm requires four observations. To reduce observation time, two coronagraphs are specified as shown schematically in **Figure 5.5-1**, covering the spectral ranges shown in **Table 5.5-5**. Light entering the coronagraphs is split into two bands, wavelengths shorter than 0.67 µm being passed to the "blue" channel and longer wavelengths to the "red" channel. Within the channels, dichroic filters set the optical bandwidth to 20%, so that to cover the range requires two observations as seen in the table. Each channel carries a camera and an IFS, selected by inserting a mirror. An infrared channel also resides on the red side and is selected similarly. It carries a slit spectrograph with R = 40 and covers the band 0.95–1.8 µm. An infrared avalanche diode detector is used on this channel.

**Table 5.5-4.** Science and guide mode filters.

| Instrument Channels | UV Science | Visible Science | IR Science |
|---|---|---|---|
| UV bandpass (μm) | 0.2–0.45 (science) | 0.3–0.45 (science) | 0.3–0.45 (guide) |
| Visible bandpass (μm) | 0.45–0.67 (science) | 0.45–1.0 (science) | 0.54–1.0 (science) |
| IR bandpass (μm) | 1.6–1.8 (guide) | 1.6–1.8 (guide) | 0.975–1.8 (science) |

Two deformable mirrors (DMs) are used to correct the wavefront phase and amplitude. To obtain a compact layout, increasing stability and lowering the mass, it is desirable to use a small DM actuator spacing. For this reason, 64×64 actuator DMs of a commercially available type were specified with 0.4 mm actuator pitch yielding outer working angles (OWAs) shown in **Table 5.5-5**.

The coronagraphs (**Figure 5.5-14a** and depicted schematically in **Figure 5.5-14b**) follow a similar layout to the WFIRST coronagraph design, while attempting to minimize the number of mirrors needed so as to maintain optical throughput. A more detailed geometric layout is shown in **Figure 5.5-15**. Following the common fine-steering mirror, the red and blue channels are separated in collimated space by a dichroic beam splitter. An initial relay of two off-axis parabolas (OAPs) sets the magnification to place the pupil on the DM. Telecentricity is not preserved at the entrance but is restored after the relay so that DM1 is positioned at a pupil plane. Following a fold, the beam strikes DM2 and is then focused onto the coronagraphic mask. Light reflected by the mask is directed to a low order wavefront sensor.

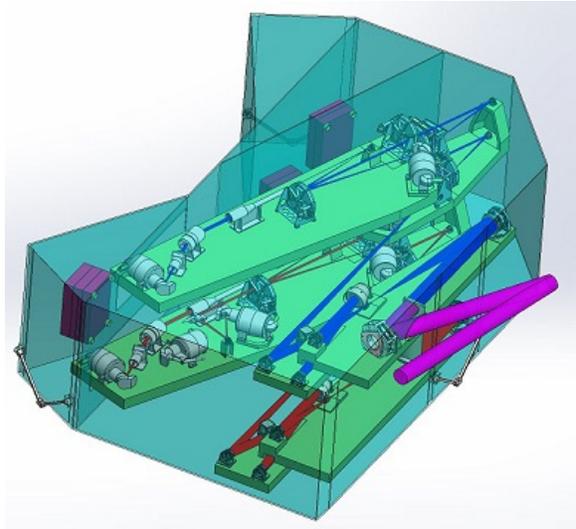

**Figure 5.5-14a.** Coronagraph instrument (DM drive electronics not shown).

Following recollimation of the science beam, it is apertured at the Lyot stop. After the Lyot stop the light can be directed via selector mirrors to the IFS, to the camera, or in the case of the red channel, to the IR camera/spectrograph.

**Table 5.5-5.** Coronagraph channel specifications.

|  | "Blue" Channel | "Red" Channel | IR Channel |
|---|---|---|---|
| **Cameras** | | | |
| FOV | 1.5″ | 2.2″ | 3.1″ |
| Wavelength bands | 0.45–0.55 µm<br>0.55–0.67 µm | 0.67–0.82 µm<br>0.82–1.0 µm | 0.95–1.8 µm |
| Pixel resolution | 11.6 mas | 17.3 mas | 29.9 mas |
| Telescope resolution | 23 mas (at 0.45 µm) | 35 mas (at 0.67 µm) | 49 mas (at 0.95 µm) |
| IWA (2.4 λ/D) | 56 mas (at 0.45 µm) | 83 mas (at 0.67 µm) | 118 mas (at 0.95 µm) |
| OWA (as) | 0.74 | 1.11 | 1.57 |
| Detector | 1×1 CCD201 | 1×1 CCD201 | 1×1 LMAPD |
| Array width | 1024 | 1024 | 256×320 |
| **Spectrometers** | | | |
| FOV | 1.5″ | 2.2″ | 3.1″ |
| Spectrometer resolution λ/Δλ | 140 | 140 | 40 |
| Spectrometer type | IFS | IFS | Slit |
| Detector | 1/4 CCD282 (EMCCD) | 1/4 CCD282 (EMCCD) | 1×1 LMAPD |
| Array width (pixels) | 2048 | 2048 | 256×320 |
| Deformable mirror | 64×64 0.4 mm pitch | 64 × 64 0.4 mm pitch | 64×64 0.4 mm pitch |

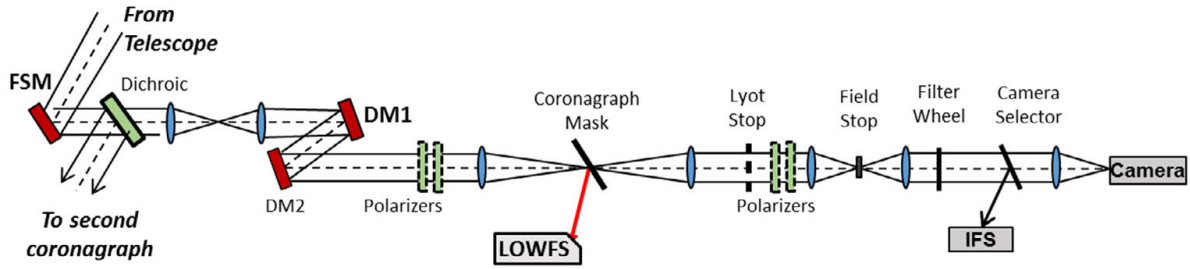

**Figure 5.5-14b.** Simplified schematic layout of coronagraphs.

Polarizers are included in the beam train to allow operation with existing vector vortex masks and also to allow selection of polarized light from the science targets, for example during disk imaging.

### *Vortex and Hybrid Lyot Coronagraph Architectures*

The baseline telescope architecture for HabEx is a 4 m, off-axis monolith. Four main coronagraph families were initially considered: the shaped pupil (SP) and apodized pupil Lyot coronagraph (APLC), the phase-induced amplitude apodization complex mask coronagraph (PIAACMC), the hybrid Lyot coronagraph (Trauger et al. 2016), and the VVC (Mawet et al. 2005, Foo, Palacios, and Swartzlander 2005). The VVC family was found to present the most favorable trade-off between IWA and immunity to low-order aberrations.

The VVC is a phase-based coronagraph that imprints a phase screw dislocation of the form $e^{il\theta}$ on the Airy diffraction pattern at the instrument focus, where $\theta$ is the azimuthal coordinate in the focal plane. When the star is centered on the phase ramp, the screw dislocation forced upon the electric field generates a singularity or optical vortex. While the phase undergoes rapid changes around the singularity, the field amplitude is zeroed out locally creating a dark hole. Upon propagation to the downstream Lyot stop (**Figure 5.5-16**), the dark hole grows to fill the entire pupil geometric area. Parameter $l$ is called the topological charge and quantifies the number of times the vortex phase ramp goes through a full $2\pi$ radian cycle. Mawet et al. (2005) demonstrated that perfect starlight rejection within the downstream geometric area can be achieved with an unobscured circular aperture and VVC of even topological charges. Moreover, the topological charge can be seen as a knob allowing a trade-off of IWA for immunity to low-order aberrations (Mawet, Pueyo, et al. 2010, Ruane et al. 2017). Indeed, the higher the charge, the lower the sensitivity to low-order aberrations, but the larger the IWA.

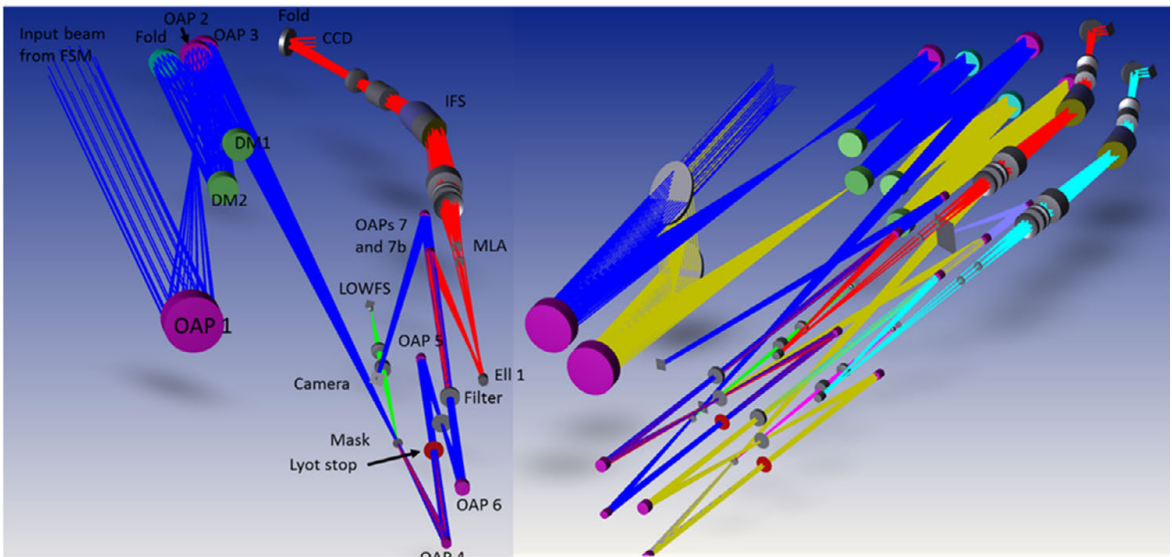

**Figure 5.5-15.** Coronagraphs: *Left:* The blue coronagraph channel. *Right:* Red channel beneath blue channel.

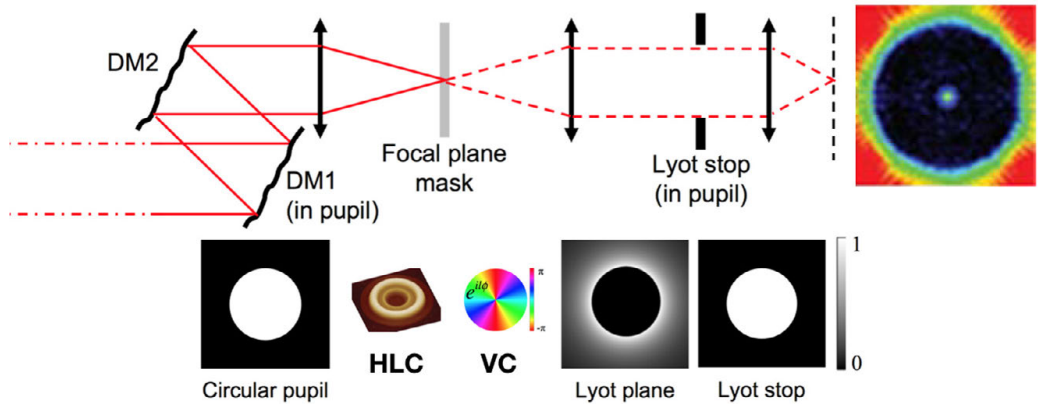

**Figure 5.5-16.** Classical Lyot-type two-plane coronagraph. The focal plane mask can either be an HLC or a VVC.

The VVC has been particularly popular on land-based adaptively corrected telescopes. Its small IWA, layout simplicity, intrinsic achromaticity, and high throughput makes it an attractive solution for high contrast imagers. The vortex coronagraph is currently in operations at Palomar (Serabyn, Mawet, and Burruss 2010, Mawet, Serabyn, et al. 2010, Mawet et al. 2011, Bottom et al. 2015, Bottom et al. 2016), VLT (Mawet et al. 2013, Absil et al. 2013), Subaru (Kühn et al. 2017), Keck (Absil et al. 2016, Serabyn et al. 2017, Mawet et al. 2017, Reggiani and TEAM 2017), and the Large Binocular Telescopes (Defrere et al. 2014).

The optimal coronagraph for the HabEx 4 m, off-axis architecture was identified to be the vortex coronagraph of topological charge 6. Topological charge 6 presents the optimal trade-off between inner working angle (2.4 λ/D at 50% total off-axis throughput), and immunity to low-order aberrations (tip-tilt, defocus, astigmatism, coma, spherical, see **Table 5.5-6**). No other coronagraph architecture for an off-axis telescope that was considered in this report has matched the charge 6 vortex's optimal trade-off. However, the HLC is also considered as a backup due to its high technology readiness (TRL 5) and traceability to the WFIRST CGI.

The notional hybrid Lyot/vortex coronagraph design can be described as follows. Collimated light from the telescope's M3 mirror enters the coronagraph instrument, creating a pupil image at the FSM. The FSM is used for pointing control and jitter suppression within the instrument. The collimated beam is then passed to a pair of DMs needed for wavefront correction. Following the DMs, the beam is converged to a focal point using a parabolic mirror. Focal plane masks are introduced into the beam at the focal point. Since the primary difference between the HLC and the VVC is the mask, both coronagraphs can be implemented in the same optical train by installing a mask wheel at the focal point, allowing either mask to be introduced into the coronagraph configuration. Both the HLC and VVC masks have a small reflective dot at the center of their masks that will reflect the mask-rejected light to the Low-Order Wavefront Sensor (LOWFS). LOWFS is used in conjunction with the DMs and FSM to correct wavefront error and drift. Light not rejected by the mask is recollimated with a second parabolic mirror creating a pupil image, then passed through a Lyot stop to block light at the perimeter of the image. Finally, the light hits another pair of focusing and collimating mirrors to resize the collimated beam, and then enters either a camera or an IFS where imaging and spectral measurements are completed.

**Table 5.5-6.** Aberration sensitivity of the vortex coronagraph charge 4 to 10, at 2e-11 raw starlight suppression (at 2.5–3.5 λ/D). The table on the right shows how the aberration sensitivity is affected by stellar size of 1 mas.

| Aberration | Indices | | Allowable RMS Wavefront Error (nm) per Mode | | | | Allowable RMS Error (nm) per Mode | | |
|---|---|---|---|---|---|---|---|---|---|
| | $n$ | $m$ | Charge 4 | Charge 6 | Charge 8 | Charge 10 | Charge 6 | Charge 8 | Charge 10 |
| Tip-tilt | 1 | ±1 | 1.1 | 6.1 | 16 | 29 | 5.5 | 18 | 31 |
| Defocus | 2 | 0 | 0.8 | 4.6 | 13 | 32 | 4.6 | 15 | 36 |
| Astigmatism | 2 | ±2 | 0.0068 | 1.1 | 0.92 | 4.8 | 0.36 | 1.0 | 4.6 |
| Coma | 3 | ±1 | 0.0064 | 0.69 | 0.84 | 5.4 | 0.44 | 0.95 | 5.5 |
| Spherical | 4 | 0 | 0.0049 | 0.53 | 0.75 | 7 | 0.32 | 0.81 | 6.7 |
| Trefoil | 3 | ±3 | 0.0073 | 0.0064 | 0.59 | 0.68 | 0.0065 | 0.35 | 0.71 |

Specific to the HabEx coronagraph design, a dichroic beam-splitter is introduced into the beam following the FSM. The beam-splitter divides the beam into a visible channel and an infrared channel; both having all the elements of the preceding notional coronagraph description, with the visible channel on the instrument's upper deck and the IR channel on the lower deck.

Descriptions of HabEx specific coronagraph optical components are included in the next sections.

*Fine-Steering Mirror*

The fine-steering mirror is used to stabilize the optical system line-of-sight by keeping the target star image centered on the coronagraph mask as the spacecraft attitude wanders within the limits of its control capability. The FSM is located at the pupil image formed by the telescope's tertiary mirror. Placing the FSM at the pupil minimizes the beam walk downstream as the FSM steers the beam. Beam walk places the beam onto different parts of the downstream optics, and slight changes in the surface imperfections subtly change the wavefront error in the beam and therefore adversely affect the contrast.

The maximum tip and tilt of the FSM is sufficient to handle small spacecraft pointing biases and has an angular resolution that is small compared to pointing error corrections that would be made. The FSM function is implemented as a two-axis tip/tilt stage carrying a plane fold mirror.

*Deformable Mirrors*

Deformable mirrors are at the core of the HabEx coronagraph instrument. Using two DMs enable phase and amplitude control over both sides of the high contrast image, an important requirement for HabEx.

There are two DM technologies available to the HabEx concept. The first utilizes lead-magnesium-niobate (PMN) electrostrictive ceramic actuators on a 1 mm pitch to drive a continuous fused-silica mirror face sheet. This technology is currently baselined on the WFIRST CGI. The second is based on microelectromechanical systems (MEMS) deformable mirror technology. Each actuator in the MEMS DM can be individually deflected by electrostatic actuation to achieve the desired pattern of deformation without hysteresis. The MEMS technology is a backup to the PMN on WFIRST.

The number of DM actuators drives the coronagraph OWA, while the pitch and number of actuators contribute to overall instrument size. HabEx has baselined a 64×64 MEMS DM with a 0.4 mm actuator pitch; the actuator count allows the coronagraph to reach a 32 $\lambda/D$ OWA, while the small actuator pitch helps minimize overall instrument size. Simulations show that the HabEx DM configurations is sufficient to provide a capability in wavefront control in both amplitude and phase domains, correcting minute wavefront errors due to fabrication and alignment inaccuracies in the system and facilitating the ability to achieve the required deep ($10^{-10}$) starlight suppression for this instrument.

*Coronagraphic Masks*

The collimated beam reflecting off DM2 is brought to a focus by an off-axis parabolic mirror with a focal ratio of f/30. The focused star image has a PSF core Airy disc diameter of 40 microns (at a 0.55 µm wavelength). The coronagraphic mask element is placed at this focal plane. To cover the entire HabEx bandwidths within both the visible and infrared channels, multiple masks are needed to provide the best starlight suppression over the full wavelength range. These masks are carried by a wheel mechanism, with the appropriate mask rotated into position depending on the science waveband selected for observation.

The vortex coronagraph is a phase-mask coronagraph, requiring only a focal-plane mask and a standard circular Lyot stop. The vortex phase mask comes in various "topological charges," which parametrize the height of the phase ramp. The topological charge allows us to trade IWA for insensitivity to stellar size and low-order aberrations, with higher coronagraph charges bringing larger IWAs and less sensitivity to aberrations. The state of the art vortex coronagraph masks have all been based on the vector vortex coronagraph concept, using one of the following technologies: liquid crystal polymers (LCP), photonics crystals (PC), or subwavelength gratings. The best lab results have been obtained on the High Contrast Imaging Testbed (HCIT) with the LCP approach (Serabyn and Trauger 2014).

The HLC mask uses a partially opaque spot to block the majority of the target star light and an overlaid phase modulation pattern provided by an optimized dielectric layer. The HLC design includes optimized DM shapes that help make the coronagraph achromatic and mitigate sensitivity to low-order aberrations. The HLC has demonstrated the deepest starlight suppression on the HCIT ($6\times10^{-10}$ over 10% BW from 3 to 16 $\lambda$/D), and is one of the two baseline coronagraphs of the WFIRST CGI (Moody and Trauger 2012).

In the HabEx implementation, both the HLC and the VVC masks are slightly tilted to the beam path and each type of mask has a reflective spot at the mask center, sending incident starlight into the FGS/LOWFS, the elements of which are discussed later in this section.

### Lyot Stops

The Lyot stop design for the vortex coronagraph does not depend on wavelength, and so in principle, a single Lyot stop is needed. The HLC requires a series of Lyot stops optimized for each waveband, so a Lyot stop wheel would be required.

### Low-Order Wavefront Sensor (LOWFS)

For the HabEx coronagraph, the LOWFS uses the rejected starlight from coronagraph to sense the low-order wavefront error, which includes LOS pointing error and thermal-induced low-order wavefront drift. Once the spacecraft has been slewed to a target star and stabilized, an acquisition process results in the star being centered on the coronagraph occulting mask, and the starlight reflecting off the mask. This light is re-imaged by the optical elements onto the LOWFS detector, creating a pseudo-interferogram. Motion of the telescope creates a change in the interferogram and the resulting error signal is fed back to the FSM to correct.

**Table 5.5-7.** Visible channel filter set.

| Band # | Wavelength Start (µm) | Wavelength End (µm) | Bandwidth (%) |
|---|---|---|---|
| 1 | 0.45 | 0.55 | 20 |
| 2 | 0.495 | 0.605 | 20 |
| 3 | 0.585 | 0.715 | 20 |
| 4 | 0.7 | 0.86 | 20 |
| 5 | 0.82 | 1.0 | 20 |

**Table 5.5-8.** Optimal photometric bands for identifying Earth-like exoplanets (Krissansen-Totton et al. 2016)

| Band # | Wavelength Start (µm) | Wavelength End (µm) | Bandwidth (%) |
|---|---|---|---|
| 6 | 0.431 | 0.531 | 20 |
| 7 | 0.569 | 0.693 | 20 |
| 8 | 0.77 | 0.894 | 20 |

**Table 5.5-9.** Narrowband filters for giant planet color characterization.

| Band # | Wavelength Start (µm) | Wavelength End (µm) | Bandwidth (%) | Comments |
|---|---|---|---|---|
| 9 | 0.45 | 0.5 | 10 | Rayleigh + weak $CH_4$ |
| 10 | 0.51 | 0.57 | 10 | Weak $CH_4$ |
| 11 | 0.6 | 0.66 | 10 | Weak/medium $CH_4$ & $NH_3$ |
| 12 | 0.695 | 0.765 | 10 | Intermediate $CH_4$ & $H_2O$ |
| 13 | 0.85 | 0.94 | 10 | Strong $CH_4$ & $H_2O$ |

**Table 5.5-10.** Infrared channel filter set.

| Band # | Wavelength Start (µm) | Wavelength End (µm) | Bandwidth (%) |
|---|---|---|---|
| 14 | 1.0 | 1.2 | 20 |
| 15 | 1.19 | 1.46 | 20 |
| 16 | 1.45 | 1.8 | 20 |

The LOWFS sensor is a Zernike wavefront sensor (ZWFS) similar to the WFIRST CGI's LOWFS. (Shi et al. 2017) The ZWFS is based on the Zernike phase contrasting principle where a small ($\sim$1 $\lambda$/D) phase dimple with phase difference of $\sim\pi/2$ is placed at center of the rejected starlight PSF. The modulated PSF light is then collimated and forms a pupil image at the LOWFS camera. The interferences between the light that passes inside and outside the phase dimple convert the wavefront phase error into the measurable intensity variations in the pupil image on the LOWFS camera. The spatial sampling of the pupil image on the LOWFS camera depends on the spatial frequency of WFE to be sensed. There is a design trade between number of sensed modes, photons per pixel, and the LOWFS camera frame rate. To improve the signal-to-noise ratio the LOWFS uses broadband (>20%) light. On WFIRST, the LOWFS camera is running at high temporal frequency (~1 kHz frame rate) in order to sense the fast LOS jitter from the vibration sources such as telescope's reaction wheels. For HabEx, this rate would be reduced to match the disturbance profile of microthrusters.

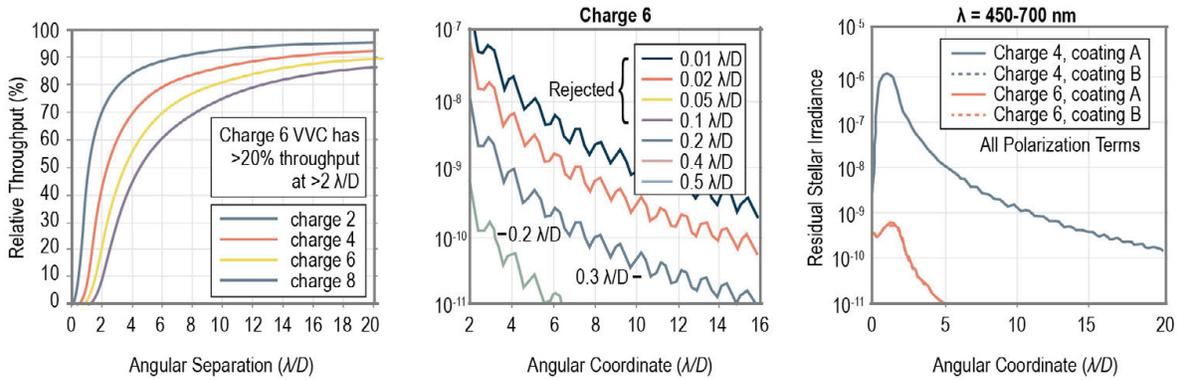

**Figure 5.5-17.** *Left:* Off-axis throughput of the vortex charge 2 to 8. *Middle:* Leakage due to finite stellar size for a charge 6 vortex coronagraph by stellar diameter. *Right:* Sensitivity of the charge 4 and 6 vortex coronagraph to the nominal HabEx polarization aberrations.

The LOWFS sensed tip-tilt errors are used to control the FSM for LOS disturbances correction. Similar to WFIRST CGI, the FSM LOS control loops contain a feedback loop to correct the telescope's LOS drift (Shi et al. 2017). The LOWFS sensed low-order wavefront errors beyond tip-tilt will be corrected using one of the DMs.

*Focal Planes*

Detector arrays for the visible channels are EMCCD types, selected because of their exceptionally low effective read noise. The imaging focal planes (blue and red channels) consist of a single EMCCD per channel operated at 153K. The chosen type is a modified CCD201 with delta doping and a thickened substrate together with a broadband "astro" coating giving response out to 1.0 μm. The pixel scale is as shown in **Table 5.5-5**. The corresponding IFS focal planes consist of a modified, cut down version of CCD282, which is a 4k×4k device with a 2k×4k frame store at each end. The device format allows it to be cut in half along both axes forming a 2k×2k sensor area with a 2k×2k frame store. The device would be operated at 163K to minimize dark current.

For the infrared, an HgCdTe avalanche photodiode array (Saphira LMAPD) is baselined. The detector is cooled to 60K to minimize dark current. It has a low read noise with avalanche gains of 50 or more available.

### 5.5.2.2.3  High-Resolution UVS

The UVS instrument is designed to enable high-resolution spectroscopy down to 0.115 μm in the UV. The driving science requirements for the instrument are discussed in Section 5.2.4. The UVS will access a large number of diagnostic emission and absorption lines available at wavelengths shorter than 0.3 μm. The driving science cases include understanding the life cycle of baryonic material as it is moved into and out of galaxies and into stars and planets, determining the escape fraction of hydrogen-ionizing photons from star-forming galaxies and whether this can explain the reionization of the universe at early ages, and the life cycle and impact of massive stars on their environments and the subsequent generations of stars and planets that follow the first generation of massive stars. The needed capabilities include a wide field of view and the ability to perform multi-object spectroscopy (MOS) within that field. The science also calls for access at the shortest wavelengths possible. The baseline design specifies reflectivity down to 0.115 μm using Al coated mirrors protected with MgF₂, and there is a stretch goal of reaching down to 0.1 μm using Al mirrors protected using LiF and either AlF₃ or MgF₂ deposited using atomic layer deposition (ALD). The science also calls for a range of spectral resolutions to enable measurement of both line shape and separation of specific lines in both emission and absorption.

**Table 5.5-12.** Spectral bands for UVS instrument.

| Resolution R | λ min | λ max | Δλ | Resolution R | λ min | λ max | Δλ |
|---|---|---|---|---|---|---|---|
| λ/Δλ | μm | μm | pm | λ/Δλ | μm | μm | pm |
| 60,000 | 0.115 | 0.127 | 2.01 | 25,000 | 0.115 | 0.146 | 5.41 |
| 60,000 | 0.127 | 0.139 | 2.21 | 25,000 | 0.146 | 0.186 | 6.88 |
| 60,000 | 0.139 | 0.153 | 2.44 | 25,000 | 0.186 | 0.236 | 8.74 |
| 60,000 | 0.153 | 0.169 | 2.68 | 25,000 | 0.236 | 0.300 | 11.11 |
| 60,000 | 0.169 | 0.186 | 2.95 | 12,000 | 0.115 | 0.186 | 12.29 |
| 60,000 | 0.186 | 0.204 | 3.25 | 12,000 | 0.186 | 0.300 | 19.86 |
| 60,000 | 0.204 | 0.225 | 3.58 | 6,000 | 0.115 | 0.300 | 32.15 |
| 60,000 | 0.225 | 0.248 | 3.94 | 3,000 | 0.120 | 0.300 | 64.29 |
| 60,000 | 0.248 | 0.273 | 4.33 | 1,000 | 0.120 | 0.300 | 185.00 |
| 60,000 | 0.273 | 0.300 | 4.77 | 500 | 0.120 | 0.300 | 185.00 |

The UVS utilizes a microshutter array (MSA) situated at the two-mirror Cassegrain focus to enable selection of objects of interest from a 3'×3' FOV. **Table 5.5-11** shows key design parameters for the UVS. With a maximum resolution of 60,000, the UVS needs a large set of gratings to cover the wavelength band (**Table 5.5-12**). The detector area is large, requiring about 30,000×17,000 pixels (or "pores" in the case of microchannel plate detectors) to cover the FOV. This area will be covered using a 3×5 array of approximately 100×100 mm microchannel plate (MCP) detectors, or alternatively, a larger array of delta-doped, UV-optimized CCDs. Both types of detector have similar performance and technology readiness levels.

With a Nyquist sampling criterion for the field of view at 0.4 μm, the pixel width is equal to $\lambda/2d$. In the spectral domain, the criterion for spectral elements to be resolved is the same so that a spectral resolution element $\Delta\lambda$ covers two pixels. For example, with R = 60,000 at 0.12 μm, $\Delta\lambda = 2$ pm (**Table 5.5-12**) and the number of spectral elements needed to cover the first band 0.012 μm wide is 6,000. Thus, a single spectrum on the detector will cover 12,000 pixels, resulting in the rectangular shape of the focal plane.

**Table 5.5-11.** High-resolution UV spectrograph parameters

| | UVS |
|---|---|
| FOV | 3'×3' |
| Wavelength bands | 20 bands covering 0.115 to 0.3 μm |
| Spectral resolutions | 60,000; 25,000; 12,000; 6,000; 3,000; 1,000; 500 |
| Telescope resolution | Diffraction limited at 0.4 μm |
| Detector | 3×5 MCP array, 100 mm sq each |
| Array width | 17,000 × 30,000 pixels (pores) |
| Microshutter aperture array | 2×2 array of 171×365 200×100 μm apertures |

Because of the relatively low reflectance of available mirror coatings in the UV (the reflectivity of aluminum is ~81% @ 0.12 μm), it is important to minimize the number of reflections from the primary mirror to the focal plane. Thus, a conventional TMA approach is precluded. After extensive design trials, a 4-reflection design was arrived at. In the design shown in **Figure 5.5-18**, the beam from an off-axis tertiary mirror strikes a powered grating before falling on the focal plane. This system resides in front of the main tertiary mirror. The 4-reflection design yielded 50% throughput at 0.125 μm. However, it requires larger gratings, about 100 mm diameter compared with 50 mm for a 5-reflection design, but this size is still considered to be practical. In addition, the 4-reflection design does not require an optical bench structure extending behind the primary mirror. Therefore, the 4-reflection design was selected for the HabEx UVS baseline design. The set of gratings will be mounted in two rows set into a wheel of ~0.5 m diameter.

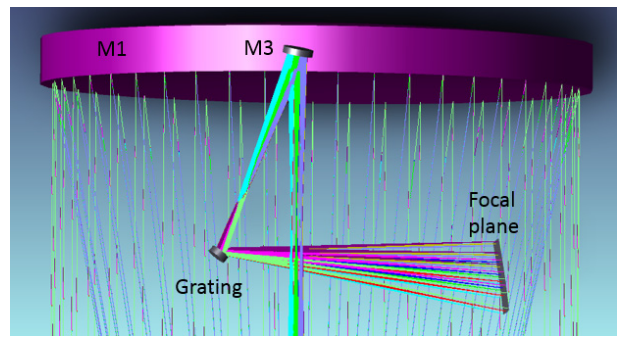

**Figure 5.5-18.** High-resolution UVS. Light from the telescope secondary mirror strikes a dedicated M3 and then the grating before arriving at the focal plane. For clarity, the other optical systems around this area are not shown.

The optical design is constrained by the telescope's first two mirrors and so a wide design space was explored to reach a solution. The best-corrected focus available is on-axis at the Cassegrain focus formed by the primary and secondary mirrors, so this is where the MSA is located. The tertiary mirror is an off-axis aspheric surface modified by polynomial and Zernike terms, producing a beam focused near the detector. The grating, located a little beyond the exit pupil is weakly aspheric and carries an unevenly spaced groove pattern with approximately 2 μm spacing for the shortest wavelength and highest dispersion. This surface can be fabricated using conventional optical polishing plus electron beam lithographic techniques. The resultant design yields the 3'×3' field, corrected at the telescope diffraction limit of 0.4 μm. Gratings are individually optimized for each waveband and the design includes one optic without grating lines, so that an undispersed UV image is formed.

The detector is a photon-counting device consisting of a MCP array utilizing large-format plates (~100 mm width). The MCPs are glass capillary arrays (GCAs) consisting of thin-walled hexagonal tube assemblies, and are fabricated in very low-Pb glass for a low x-ray cross section. The tubes (micropores) are arranged with a small angle typically ~15 degrees off normal. Each plate consists of a micropore array of two layers with opposing pore angles for efficiency. Furthermore, the materials used are very pure and contain few radioactive isotopes, leading to a very low dark count. Atomic layer deposition is used to create the resistive and emissive layers (GaN and multialkali) of the cathode, producing improved performance specifications over conventional MCPs. The assembled plate is enclosed in a frame and covered with an $MgF_2$ window, also coated, so that a high vacuum can be pulled to assure long-term performance of the emissive layers during integration and testing through launch.

$100 \times 100$ mm$^2$ MgF2 windows are considered possible with the large crystal boules now being made. Beneath the plate a lattice of wires forms the anode. Incoming UV photons produce a cascade of electrons with gain of $10^6$ or more, and the charge cloud emerges at the base of the MCP assembly to impact the anode wires. High-speed analog-to-digital converters (ADCs), digitizing 8 bits at 10 MHz, collect charge from the wires. An application-specific integrated circuit (ASIC) postprocessor outputs a stream of data consisting of charge cloud centroid position (x and y), peak height, coincidence flag, and time stamp. In the case of two or more photons arriving at the same time, the postprocessor rejects the event based on peak height. Behind the anode is a plastic scintillator viewed by a miniature avalanche photodiode or photomultiplier tube that acts as a cosmic ray detector. In the case of a detection here coincident with an event on the wire grid, the coincidence flag is set and the event rejected. Thus, a clean signal can be generated in the presence of a cosmic ray background. The spatial digitization is at the micropore spacing. While the charge cloud spreads upon emission from the base of the plates, the large electron count allows localization of the event at the micropore level. The ADC rate (10 MHz), a determinant of photon flux, can handle up to approximately $10^7$ photons/sec with an efficiency loss due to the coincidence of a portion of events. Most targets will be much weaker, so that typically hundreds of objects may be observed simultaneously.

Forming a 3×5 array of plates produces a 3'×3' field of view with the long axis accommodating the spectral dispersion. Since the plates are surrounded by the frame of the vacuum assembly there will be small gaps of coverage as is often the case with detector arrays.

### 5.5.2.2.4  Workhorse Camera

The HabEx workhorse camera (HWC) is a general purpose instrument providing UV through near-IR imaging and spectroscopy, with objectives ranging from solar system science to detailed studies of galaxies and quasars at the epoch of reionization to cosmology. The HWC would enable detailed follow-up of interesting targets, such as those identified from the wide-field surveys of the 2020s, such as Euclid, LSST, and WFIRST. Specifically, the instrument is designed to provide unique scientific capabilities compared to the facilities expected in the 2030s. For example, nearly all of the first-generation instruments on the new 30 m class telescopes (e.g., TMT, GMT, and ELT) are near-IR instruments because ground-based adaptive optics (AO) are not expected to be effective for wavelengths much shorter than about 1 µm. The HWC would provide both unique capabilities, including: (1) UV science, (2) high-spatial resolution imaging, (3) a stable platform for both photometry and morphology, and (4) access to spectral regions inaccessible on the ground due to telluric absorption.

The design, like the Wide-Field Camera 3 (WFC3) on the HST, has two channels that can simultaneously observe the same field of view: a UV/optical channel using delta-doped CCD detectors providing good throughput from 0.15 µm to 0.95 µm, and a near-IR channel using Hawaii-4RG HgCdTe arrays providing good throughput from 0.95 µm to 1.8 µm, at which point thermal backgrounds dominate over most celestial targets.

Both channels will have imaging and spectroscopic modes, and a MSA assembly provides for slit spectroscopy of targeted sources, significantly reducing the backgrounds and source confusion compared to the slit-less spectroscopic modes available on HST. The two modes of operation share the same optical path and cameras. In the spectrographic mode, the MSA and grism sets are introduced into the beam paths. It is intended that the MSA be attached to a mechanism and thereby removable for the best imaging function. **Table 5.5-13** shows the design parameters for the HWC's two channels. For good imaging, the pixel magnification is chosen to Nyquist sample the PSF. To obtain sufficient field of view, the visible channel has a 3×3 array of 4k square CCD detectors, and the IR channel utilizes 2×2 H4RG10.

**Figure 5.5-19** shows the layout of the HWC instrument. After reflecting off M3 and the fold

**Table 5.5-13.** HWC design specifications.

| | UV/VIS Channel | IR Channel |
|---|---|---|
| **FOV** | 3'×3' | 3'×3' |
| **Wavelength bands** | 0.15–0.95 µm | 0.95–1.8 µm Stretch goal >=2.5 µm |
| **Pixel resolution** | 15.5 mas | 24.5 mas |
| **Telescope resolution** | 30.9 mas | 49 mas |
| **Design wavelength** | 0.6 µm | 0.95 µm |
| **Detector** | 3×3 CCD203 | 2×2 H4RG10 |
| **Detector array width** | 12,288 pixels | 8,192 pixels |
| **Spectrometer** | R = 2,000 | R = 2,000 |
| **Microshutter array** | 2×2 arrays; 200×100 µm aperture size; 171×365 apertures | |

mirror, the input beam strikes a fine-steering mirror used for image dithering and small pointing adjustments and is normally fixed during an observation. The beam then passes through a relay formed by a pair of biconic paraboloidal mirrors, then on to a dichroic where the UV/visible light is separated from the IR light. Note that this system, while efficient in terms of throughput, does limit optical performance at the shorter wavelengths. Improved imaging below 0.6 µm would be achieved using additional optics to form the relay, at the expense of throughput in the UV. This is an area for a trade. In spectroscopy mode, a microshutter aperture array is inserted into the focal plane of the relay, enabling selection of particular targets. This array is identical to the set of arrays installed in JWST's Near-Infrared Spectrograph (NIRSPEC) (STScI).

*UV/Visible Channel*

At the dichroic, UV and visible light is reflected and passes through a filter wheel to a camera. The filter wheel is mounted at a pupil plane and enables selection of different wavelengths of interest in the image. A grism is also placed in the wheel to allow spectroscopy in conjunction with the MSA at R = 2,000. The camera consists of a three-mirror relay and the focal plane itself. The performance is diffraction limited at 0.6 μm. The focal plane is designed for Nyquist sampling of the 3'×3' field at the same wavelength. The selected array is CCD203, a conventional low-noise CCD with 12 μm pixel size and 4k×4k format. A set of nine of these CCDs, cooled to 153K, forms the focal plane. For UV performance these arrays would be deep depletion, delta-doped devices.

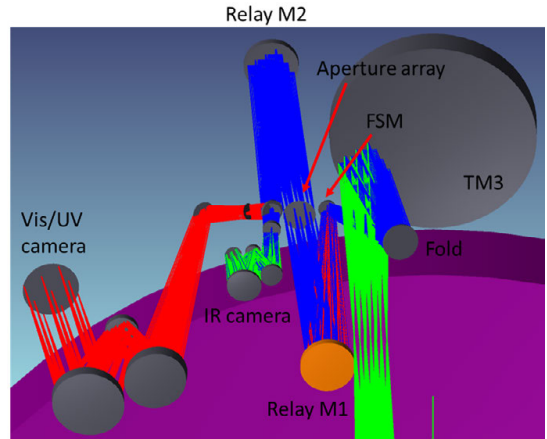

**Figure 5.5-19.** The HabEx Workhorse Camera (HWC).

*Infrared Channel*

At the dichroic, infrared light from 0.95 to 1.8 μm is transmitted and passes through a filter wheel to a camera. As in the UV/visible channel, the filter wheel is mounted at a pupil plane and enables selection of different wavelengths of interest in the image. Again, a grism is placed in the wheel to allow spectroscopy in conjunction with the MSA at R = 2,000. The camera consists of a three-mirror relay leading to the focal plane. Performance is diffraction limited at 0.95 μm. The focal plane is designed for Nyquist sampling of the 3'×3' field at the same wavelength. The selected array is the Teledyne H4RG10, a low-noise hybrid HgCdTe/CMOS bump-bonded array with 10 μm pixel size and 4k×4k format. These focal plane arrays (FPAs) are currently being developed for WFIRST. A set of four FPAs cooled to 100K forms the focal plane.

### 5.5.2.3 Fine Guidance Sensor

The coronagraph is very sensitive to wavefront changes. Initially, WFE introduced by the optical system is corrected on two DMs included in the coronagraph beam train. During observations, which may take many hours, the wavefront slowly evolves under small thermal changes. Changes to tip/tilt and focus will be detected at the LOWFS. This sensor allows detection of the wavefront error at sub-milliarcsecond levels, so that tip/tilt can be corrected by the FSM at the entrance to the coronagraph.

Assuming the telescope rolls slightly around its optical axis, there is an induced motion of the coronagraph FOV on the sky, which appears as a tilt of the wavefront of the observed object. This "tilt" will also be detected by the coronagraph's LOWFS and corrected by the FSM. However, a residual wavefront error remains, caused principally by the beam footprint moving a small amount across the tertiary mirror. Different parts of the mirror have different residual surface irregularities, and the effect is to introduce an uncorrected wavefront error at the coronagraph input. The roll stability requirements depend particularly on: the surface quality on the mirrors that experience most beam walk, in this case M3 and the following fold mirror; the accuracy of the estimate of the PSF center; and the availability of bright guide stars. Current technology allows the production of extremely well-figured optics for UV lithography with 1 nm RMS surface figure error, so special optics can be made for the sensitive locations in the beam train (M3 and the fold mirror). The challenge then is to accurately measure the positions of a sufficient number of guide stars by which the roll of the telescope can be measured. **Table 5.5-14** shows a roll sensitivity comparison between two sensors, one with a small field of view (about 3'×3', corresponding to the HWC), and one with a larger FOV across the well-corrected annular field of the telescope (0.3° across).

The HWC would be an adequate detector for pointing the telescope 98% of the time. However, it is an inadequate detector for roll because there are too few well-separated stars and the instrument has insufficient angular resolution. For these reasons, a dedicated FGS is included in the HabEx design, utilizing some of the unused annular field of the TMA. With the FGS, roll estimation accuracy will be below ~1 mas for stars 17th magnitude or brighter. To estimate the sky coverage of the FGS, a numerical model was used (GSFC 2018) that generates an average across the whole sky of the number of stars in a given field of view. This model shows that the FGS will see sufficiently bright stars ~100% of the time. Naturally, there are some parts of the sky where the density of stars is low, but a gradual degradation in performance would be expected since the FGS can guide on just two stars and obtain both pointing and roll.

**Table 5.5-14.** Fine guidance sensor roll sensitivity comparison.

|  | Small FGS | Large FGS |  |
|---|---|---|---|
| **Angular roll resolution** | 5.8 | 0.65 | arcsec |
| **Angle between FGS and coronagraph** | 0.29 | 0.15 | degrees |
| **On-sky tilt angle** | 29.6 | 1.70 | mas |

### 5.5.2.4    Telescope Mechanical Design

The HabEx telescope structure is designed to achieve the WFE and LOS ultra-stability required for coronagraphy. **Figure 5.5-20** shows the major components of the telescope baseline design. The primary mirror truss is connected to the secondary mirror by a rigid secondary mirror support tower (**Figure 5.5-20**), which also contains the science instruments. To maximize stiffness, the tower is integral with the stray light baffle tube. The tube and its internal stray light baffles provide lateral and bending stiffness support. However, since the telescope is off-axis, the internal baffles are discontinuous and external gussets complete the support. The composite material for the tube and truss structure is M46J with quasi-isotropic laminate properties.

The baseline optomechanical design meets the required optical component rigid body alignment stability necessary to achieve the LOS and WFE stability specifications of **Table 5.2-6**. The telescope's stability performance has been analyzed for three different mechanical disturbance environments: (1) JWST reaction wheel assembly (RWA) specification with JWST two-stage passive isolation; (2) JWST RWA specification with active isolation; and (3) microthrusters replacing reaction wheels with no additional isolation. The baseline telescope structure meets the stability requirement for Case 1, has a 5× margin for Case 2 and over 10× margin for Case 3. The methodology and analysis details of this analysis are summarized in Section 5.5.2.5.

### 5.5.2.5    Telescope Optomechanical Stability

An ultra-stable optomechanical telescope structure is required to achieve WFE and LOS stability specifications identified in **Table 5.2-6**. The function of this structure is to align the primary, secondary, and tertiary mirrors to each other and maintain that alignment. Rigid body motions of the primary, secondary, and tertiary mirrors introduce WFE and LOS errors. This section discusses the method for defining the telescope WFE and LOS rigid body motion requirements, as well as modeled performance.

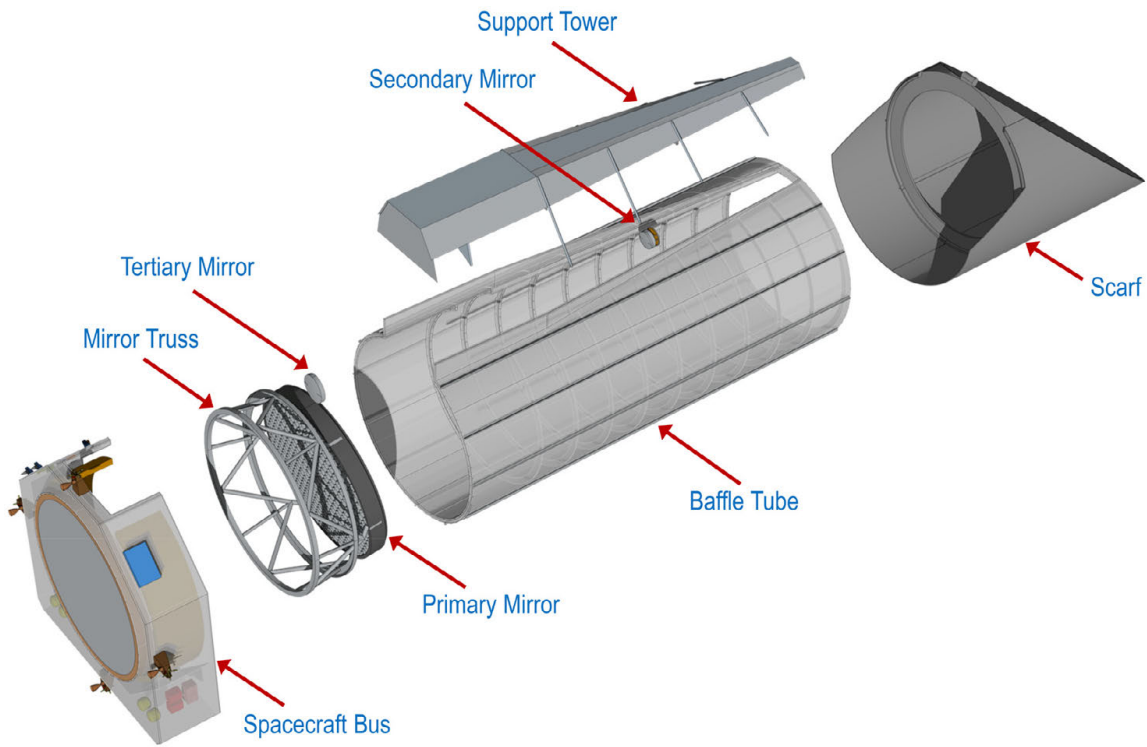

**Figure 5.5-20.** HabEx structural design for the baseline 4 m off-axis telescope.

### 5.5.2.5.1 Telescope Mirror Rigid Body Tolerance for LOS and WFE Stability

A tolerance analysis was performed to derive the LOS stability specification for the baseline optical design. The analysis evaluated the LOS sensitivity to rigid body motions of the primary, secondary and tertiary mirrors at the FSM. These sensitivities were then converted to rigid body allocations for the primary, secondary, and tertiary mirrors to achieve the LOS stability specification. Please note: since the optical design has a magnification of 80, the 0.7 mas LOS stability specification is 56 mas at the FSM. The allocation for each mirror derived is reported in **Table 5.5-15**. Since the modal responses with the highest impact on LOS are primary mirror decenter, allocations for these motions are large.

WFE instability arises from mechanical and thermal sources. Thermally driven WFE occurs when a telescope is slewed relative to the Sun. Thermal load changes cause the structure holding the mirrors to expand/contract (resulting in alignment drift) and the mirrors themselves to change shape. Fortunately, thermal effects are typically low spatial frequency, occur slowly, and can easily be corrected by the coronagraph's LOWFS and control system.

Mechanical forces (e.g., reaction wheels, cryocoolers) can excite inertial motion and vibrational modes in the mirrors and their supporting structure. For these mechanical disturbances, temporal frequency is important. A LOWFS can only sense and correct low-order errors up to about 10 Hz. Preliminary analysis of the baseline HabEx optomechanical structure indicates that all rigid body modes C stability occur at frequencies above 20 Hz and are thus uncorrectable by the coronagraph. To mitigate these errors HabEx is designing the optomechanical structure to be as stiff as possible, including the use of a laser truss system, and has eliminated the use of reaction wheels which have been the primary source for mechanical disturbance on past telescope missions that included them.

**Table 5.5-15.** HabEx optical component rigid body stability tolerance specification.

| Alignment | LOS (0.5 mas) | WFE (VVC 6) | Units |
|---|---|---|---|
| PM X-Decenter | 15 | 400 | nm |
| PM Y-Decenter | 15 | 400 | nm |
| PM Z-Despace | 8 | 500 | nm |
| PM X-Tilt (Y-Rotation) | 0.25 | 5 | nrad |
| PM Y-Tilt (X-Rotation) | 0.25 | 5 | nrad |
| PM Z-Rotation | 0.5 | 5 | nrad |
| SM X-Decenter | 4 | 400 | nm |
| SM Y-Decenter | 4 | 400 | nm |
| SM Z-Despace | 8 | 500 | nm |
| SM X-Tilt (Y-Rotation) | 0.5 | 5 | nrad |
| SM Y-Tilt (X-Rotation) | 0.5 | 5 | nrad |
| SM Z-Rotation | 0.5 | 5 | nrad |
| TM X-Decenter | 10 | 1000 | nm |
| TM Y-Decenter | 10 | 1000 | nm |
| TM Z-Despace | 1000 | 1000 | nm |
| TM X-Tilt (Y-Rotation) | 10 | 1000 | nrad |
| TM Y-Tilt (X-Rotation) | 10 | 1000 | nrad |
| TM Z-Rotation | 1000 | 1000 | nrad |

Any temporal or dynamic change in WFE can result in dark-hole speckles that produce a false exoplanet measurement or mask a true signal. The key issue is how large of a WFE can any given coronagraph tolerate. The leading candidate for HabEx is the vector vortex coronagraph (VVC $N$) where $N$ indicates the 'charge' or azimuthal shear. The higher the 'charge' the more low order error it can tolerate, but the larger its IWA and lower its throughput. This being so, a VVC 4 is more desirable for IWA and throughput reasons but is less able to tolerate WFE compared to higher charge VVCs. **Table 5.5-16** summarizes specifications for several VVCs capturing the maximum amount of WFE as a function of spatial frequency, expressed using a Zernike polynomial expansion series.

WFE stability specification was handled in a similar manner as the LOS: an optical tolerance analysis identifying WFE sensitivities to telescope mirror misalignments, followed by alignment allocations for each mirror. The VVC 6 maximum WFE values set the constraint, and misalignment allocations can be compared against the

**Table 5.5-16.** Wavefront stability required by VVC.

| Aberration | Indices | | Allowable RMS Wavefront Error (nm) per Mode | | | |
|---|---|---|---|---|---|---|
| | n | m | Charge 4 | Charge 6 | Charge 8 | Charge 10 |
| Tip-tilt | 1 | ±1 | 1.1 | 5.9 | 14 | 26 |
| Defocus | 2 | 0 | 0.8 | 4.6 | 12 | 26 |
| Astigmatism | 2 | ±2 | 0.0067 | 1.1 | 0.90 | 5 |
| Coma | 3 | ±1 | 0.0062 | 0.66 | 0.82 | 5 |
| Spherical | 4 | 0 | 0.0048 | 0.51 | 0.73 | 6 |
| Trefoil | 3 | ±3 | 0.0072 | 0.0063 | 0.57 | 0.67 |
| 2nd Astig. | 4 | ±2 | 0.0080 | 0.0068 | 0.67 | 0.73 |
| 2nd Coma | 5 | ±1 | 0.0036 | 0.0048 | 0.69 | 0.85 |
| 2nd Spher. | 6 | 0 | 0.0025 | 0.0027 | 0.84 | 1 |
| Quadrafoil | 4 | ±4 | 0.0078 | 0.0080 | 0.0061 | 0.53 |
| 2nd Trefoil | 5 | ±3 | 0.0051 | 0.0056 | 0.0043 | 0.72 |
| 3rd Astig. | 6 | ±2 | 0.0023 | 0.0035 | 0.0034 | 0.81 |
| 3rd Coma | 7 | ±1 | 0.0018 | 0.0022 | 0.0036 | 1.18 |
| 3rd Spher | 8 | 0 | 0.0018 | 0.0018 | 0.0033 | 1.49 |

Garreth Ruane, June 2017

■ not rejected
■ first-order rejection
■ >first-order rejection

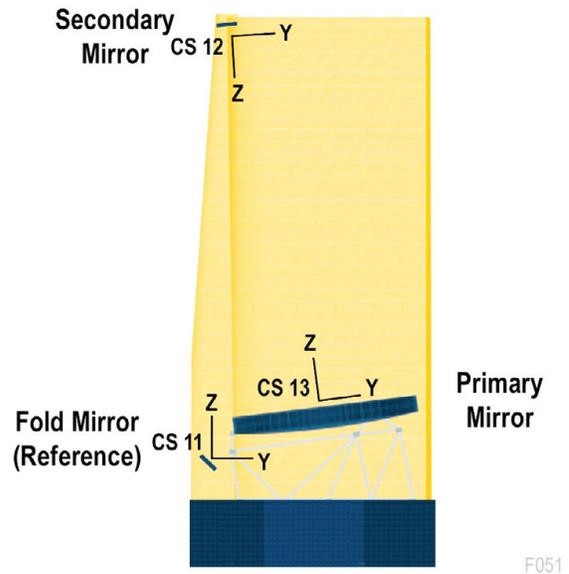

F051

**Figure 5.5-21.** Mirror displacements are relative to fold mirror in coordinate system with Z-axis normal to optical surface.

constraint using the alignment sensitivities. The WFE telescope mirror alignment allocations are included in **Table 5.5-15** as well. Telescope LOS mirror alignments are the more demanding of the two.

### 5.5.2.5.2 Telescope Structure Dynamic Optomechanical Performance

To determine if the baseline optomechanical design can meet the required telescope mirror rigid body alignment stability needed to achieve the LOS and WFE stability specifications, a finite element model of the telescope and spacecraft structure was constructed with critical damping set to 0.05%. The NASTRAN Multi-Point Constraint (MPC) function was used to determine the rigid body displacements of the primary mirror and secondary mirror relative to the fold mirror (**Figure 5.5-21**) for frequencies up to 500 Hz. The model was exposed to three different mechanical disturbance spectrums.

**Figures 5.5-22, 5.5-23,** and **5.5-24** show the amplitude of the primary mirror's rigid body degrees of freedom as a function of temporal frequency.

**Figure 5.5-22** shows the response for Case 1: JWST RWA specification with JWST two-stage passive isolation. JWST's first stage is an 8 Hz isolator between the reaction wheels and spacecraft. The second stage is a 2 Hz isolator between the spacecraft and telescope. **Figure 5.5-23** shows the response of Case 2: JWST RWA specification with 40 dB active isolation. The active isolation system senses and corrects low frequency vibrations. The Case 2 analysis assumes a single-stage 1-Hz active system that can attenuate low frequency vibrations by 100× (40 dB) with 15% damping. Finally, **Figure 5.5-24** shows the response of Case 3: microthrusters replacing reaction wheels with no additional isolation. The Case 3 analysis assumes a white noise spectrum of 0.1 microNewton. In all figures, the amplitudes were multiplied by a 2× MUF for frequencies below 20 Hz and a 4× MUF for frequencies above 20 Hz. The constraining red lines are the tolerances summarized in **Table 5.5-15**.

The baseline telescope structure meets the stability requirements for Case 1 for all frequencies except the first modes of the secondary tower and primary mirror assembly. And, the baseline structure meets all rigid body motion stability specifications for both Case 2 and Case 3.

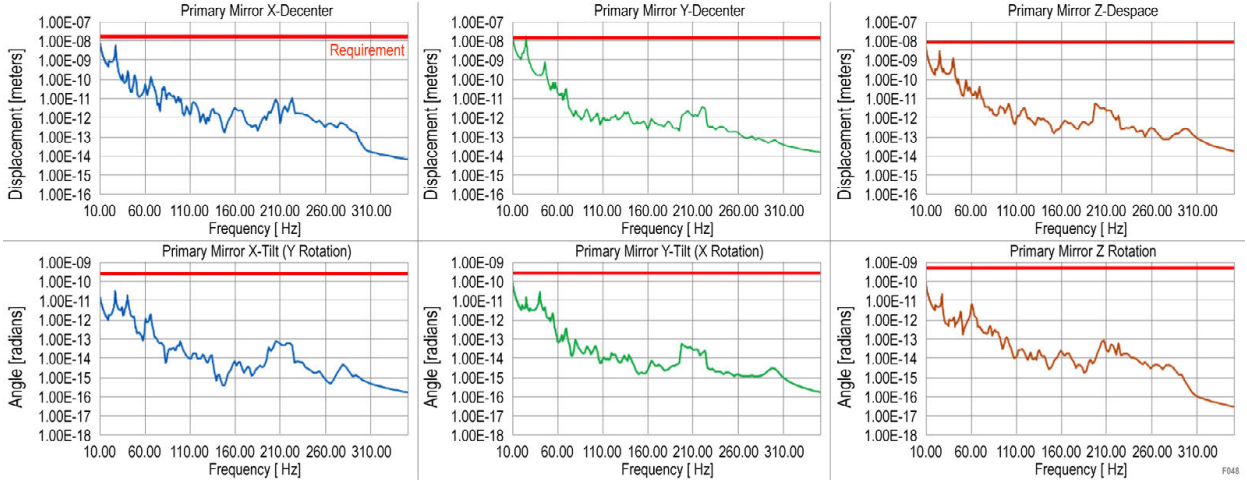

**Figure 5.5-22.** Primary mirror rigid body amplitudes for JWST reaction wheels and JWST passive 2-stage isolation.

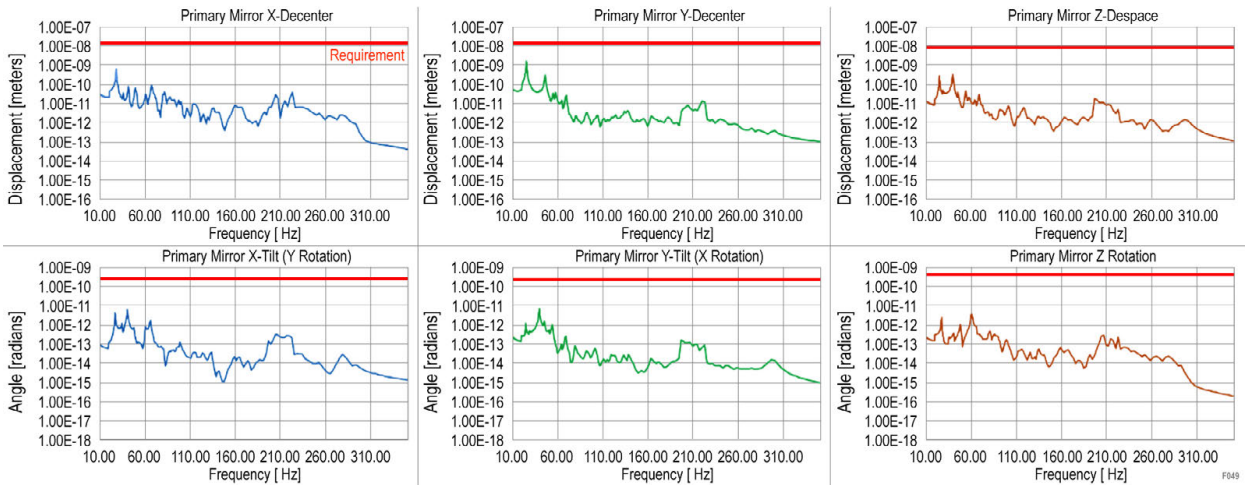

**Figure 5.5-23.** Primary mirror rigid body amplitudes for JWST reaction wheels and 40 dB active isolation.

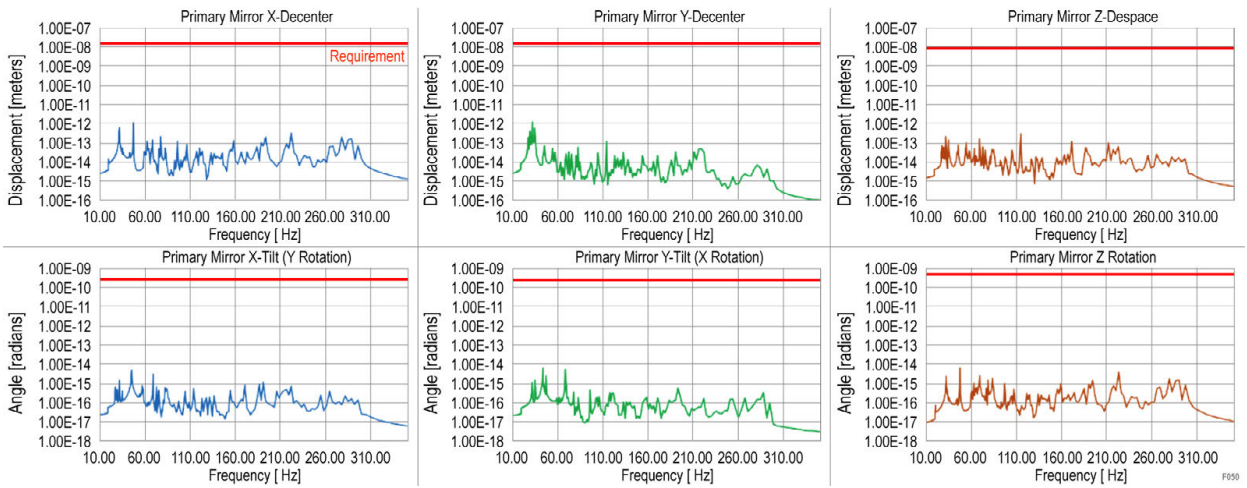

**Figure 5.5-24.** Primary mirror rigid body amplitudes for microthrusters and no additional isolation.

The first mode of the primary mirror is a lateral translation (**Figure 5.5-25a**). The X-translation has a slightly different frequency of 27 Hz from the Y-translation at 25 Hz. The second mode of the primary mirror is a rocking or tilt mode at 40 Hz (**Figure 5.5-25b**). And, the first mode of the secondary mirror at 28 Hz is a lateral translation produced by a bending mode of the stray light baffle tube (**Figure 5.5-26**). Without the tube, the first mode of a free-standing secondary mirror would be less than 5 Hz.

It is important to note that, for the VVC 4, astigmatism sensitivity is driving the optical component rigid body alignment stability specification. Because the optical design is off-axis, Z-despace between the primary and secondary mirrors introduces both defocus and astigmatism. But, while a given despace produces 22× more defocus than astigmatism, the VVC 4 is 120× more sensitive to astigmatism than to defocus. Similarly, while the VVC 4 can accept nearly identical amounts of astigmatism, coma and trefoil, the primary mirror decenter and tilt tolerance is driven only by astigmatism, because a given PM decenter or tilt introduces 4× more astigmatism than coma and 10× more astigmatism than trefoil. Rigid body degree-of-freedoms (DOF) simply cannot introduce enough coma or trefoil for their sensitivities to be important. The only significant source for these WFEs are from inertial or modal bending of the primary mirror.

### 5.5.2.6 Laser Metrology Subsystem

The Laser Metrology Subsystem provides sensing and control of the rigid body alignment of the telescope. In a closed loop with actuators, MET actively maintains alignment of the telescope front-end optics, thereby eliminating the dominant source of wavefront drift. With an internal laser source, MET is not photon-starved and can operate at high bandwidth. Furthermore, laser metrology maintains wavefront control even during attitude maneuvers such as slews between target stars. The result is an almost infinitely stiff truss supporting the telescope optics.

#### 5.5.2.6.1 Laser Metrology

Laser metrology for large coronagraph-equipped space-born observatories was first proposed for the Terrestrial Planet Finder Coronagraph (Shaklan et al. 2004). The early versions consisted of large optical benches populated with discreet optical beam splitters, retroreflectors and lenses. Recently, the optical bench has been replaced by planar lightwave circuit (PLC) technology, resulting in a compact lightweight beam launcher (**Figure 5.5-27**). PLC technology was developed for the optical communications industry so circuits in the communications band are easily mass produced.

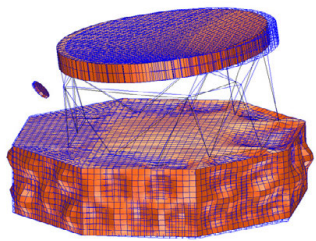

**Figure 5.5-25a.** Primary mirror first mode: 25 Hz lateral.

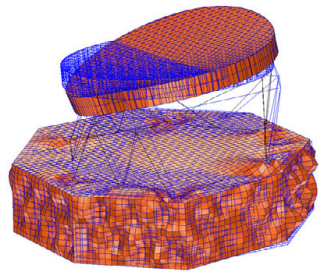

**Figure 5.5-25b.** Primary mirror second mode: 40 Hz tilt.

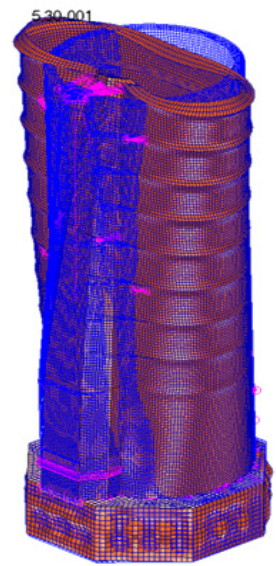

**Figure 5.5-26.** Secondary mirror first mode: 28 Hz tube bend.

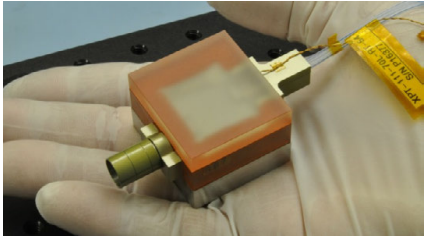

**Figure 5.5-27.** PLC beam launcher.

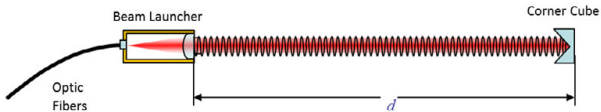

**Figure 5.5-28.** Cartoon of a laser gauge.

The principle of operation remains much the same as the original laser heterodyne interferometer system proposed for the Terrestrial Planet Finder Coronagraph. The beam launcher transmits a collimated beam through free-space to a corner cube retroreflector (**Figures 5.5-28** and **5.5-29**). The reflected beam couples back into the beam launcher where it mixes with a reference beam.

The heterodyne metrology beam and an internal heterodyne reference beam are sensed and the signals passed through a phase meter. The change in the phase differences between the two signals indicate the change in the distance between the beam launcher and the target retroreflector.

The MET system senses with a 1 kHz bandwidth, while the full control loop operates with a 10 Hz bandwidth. The MET control bandwidth is sufficient to counteract any thermal related disturbances in the structure.

Furthermore, a greater control bandwidth is unnecessary given that the use of microthrusters during observations produces almost negligible disturbances outside this bandwidth.

### 5.5.2.6.2 HabEx Laser Metrology Truss

The HabEx laser metrology truss connects the telescope secondary mirror and tertiary mirror assembly to the primary mirror. As shown in **Figure 5.5-30**, three points on the circumference of the primary mirror are linked to three points on the secondary mirror. Similarly, three points on the tertiary mirror assembly are also linked to the secondary mirror.

By monitoring each leg of this truss with a beam launcher gauge and retroreflector pair, and measuring relative changes in the distances, rigid body motions of M2 and M3 in six degrees of freedom are solved from the geometric truss equation. Commands are then sent to the rigid body actuators on the secondary and tertiary mirrors to counteract these changes and maintain the truss in its original state.

With an uncorrelated gauge error of 0.1 nm per gauge, and a laser truss based on the 4 m HabEx off-axis telescope configuration, MET is capable of maintaining the position of the M2 to less than 1 nm and 1 nrad, and M3 to less than 3 nm and 1 nrad (with the exception of M3 clocking at <5 nrad; see **Table 5.5-17**).

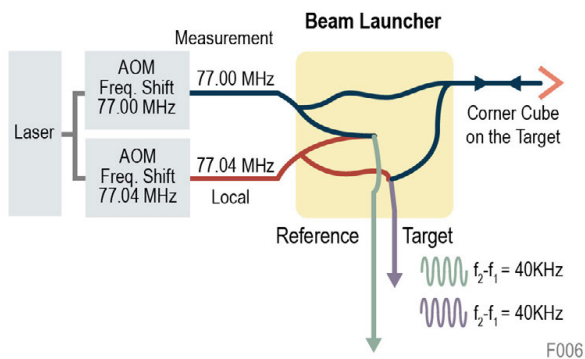

**Figure 5.5-29.** The heterodyne technique eliminates common mode phase shifts between the target phase and the reference phase measured at the phasemeter.

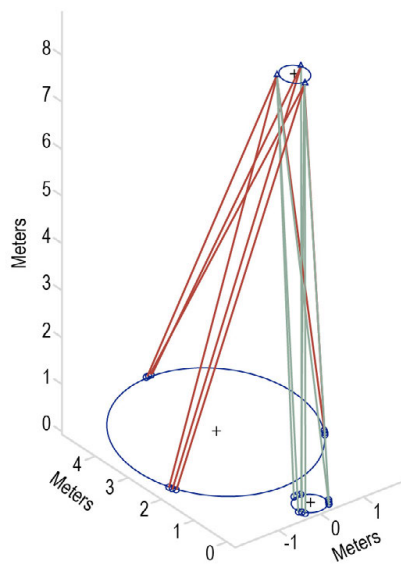

**Figure 5.5-30.** Model of the HabEx metrology truss. Primary, secondary, and tertiary mirrors shown.

**Table 5.5-17.** MET rigid body motion residuals for a 0.1 nm uncorrelated gauge uncertainty per gauge for the HabEx MET truss.

| DOF | Secondary Mirror | Tertiary Mirror |
|---|---|---|
| Θx (nrad) | 0.1781 | 0.2705 |
| Θy (nrad) | 0.2032 | 0.3962 |
| Θz (nrad) | 0.817 | 4.6612 |
| Δx (nm) | 0.21833 | 2.9186 |
| Δy (nm) | 0.2217 | 2.266 |
| Δz (nm) | 0.0436 | 0.1391 |

## 5.6    Telescope Bus

This section describes the key design features of each of the telescope subsystems and discusses some of the driving parameters for the telescope design. **Table 5.6-1** offers a mass breakdown of the telescope concept; the total mass (current best estimate, CBE) of the telescope spacecraft flight system is estimated to be 12,389 kg, with 29% average contingency and an additional 14% system margin, leading to a MPV dry mass and wet mass of 17,717 kg and 19,425 kg, respectively. The HabEx telescope is a Class A system with redundant subsystems.

**Table 5.6-1.** HabEx telescope flight system mass breakdown per subsystem. CBE: current best estimate.  MEV: maximum expected value.

| | CBE (kg) | Cont. % | MEV (kg) |
|---|---|---|---|
| **Payload** | | | |
| Telescope and Instruments | 7,846.6 | 30% | 10,200.6 |
| **Spacecraft Bus** | | | |
| ACS | 17.7 | 3% | 18.2 |
| C&DH | 20.6 | 11% | 22.8 |
| Power | 237.4 | 28% | 303.3 |
| Propulsion: Monoprop | 204.6 | 5% | 215.4 |
| Propulsion: Microthruster | 163.4 | 33% | 218.0 |
| Structures & Mechanisms | 3,102.9 | 30% | 4,033.8 |
|   Spacecraft side adaptor | 0.0 | 0% | 0.0 |
| Cabling | 337.9 | 30% | 439.2 |
| Telecom | 36.1 | 28% | 46.3 |
| Thermal | 422.1 | 30% | 548.8 |
| **Bus Total** | 4,542.8 | 29% | 5,845.8 |
| **Spacecraft Total (dry) CBE & MEV** | **12,389.4** | | |
|   Subsystem heritage contingency | 3,657.0 | | |
|   System contingency | 1,670.4 | | |
| **Spacecraft with Contingency (dry)** | **17,717** | | |
|   Monoprop and pressurant | 1,596 | | |
|   Colloidal propellant | 112.6 | | |
| **Total Spacecraft Wet Mass** | **19,425** | | |
|   Launch vehicle side adaptor | 3,709.2 | | |
| **Total Launch Mass** | **23,135** | | |

## 5.6.1    Structures and Mechanisms

The telescope flight system consists of a hexagonal bus with a body-fixed solar array, which also acts as a thermal shield for the telescope and instruments. Mechanisms have been minimized to eliminate disturbance sources within the flight system. There is no articulated high-gain antenna; a phased array is used instead. Similarly, the solar array is body-mounted rather than articulated. The only mechanism outside of the instruments is the telescope door (see Section 5.5.2), which is only moved when the telescope is not observing.

To keep the structure as stiff (and simple) as possible, the number of deployments has been minimized. The telescope possesses two deployable solar shield wings. These deployed shields reduce the thermal impact of attitude changes on the telescope system.

The telescope tube possesses a deployable scarf. This choice was made in order to ensure that the combined stack height of the starshade, launch vehicle adapter, and telescope would fit in the fairing of an SLS Block-1B. Without a co-launched starshade, the deployed scarf is unnecessary and the telescope would be designed with a fixed scarf. The full configuration of the deployed telescope is presented in **Figure 5.6-1**.

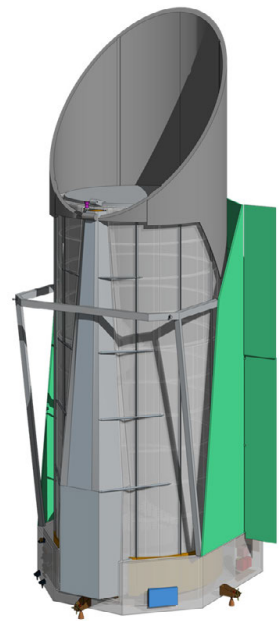

**Figure 5.6-1.** HabEx baselined configuration showing the telescope with its scarf deployed.

### 5.6.2 Thermal

The driver for the thermal design is to limit induced wavefront error on the telescope's optics. The key thermal requirements for the thermal design of the telescope are to maintain the primary mirror at 270K, with <50 mK temporal stability over the integration period, and <50 mK change in spatial gradient. The approach taken in this thermal design follows the state of the practice: the flight system will be cold-biased and heat will be added back into the system to maintain desired temperatures. Keeping a 4 m primary mirror at 270K while it is facing 3K space takes considerable power. HabEx is helped somewhat by having a telescope barrel. Maintaining the barrel walls at a much higher temperature than 3K space reduces the radiative heat loss from the primary mirror. The HabEx design holds the telescope barrel at 230K. The primary mirror is contained in its own thermal insulation "can" and is held at 270K. Consequently, the thermal power requirement for the telescope was found to be 3.4 kW for the telescope's inner barrel, and a further 50 W for the primary mirror.

### 5.6.3 Power

The baseline power system uses a dual-string "cold spare" approach, and small-cell technology Li-Ion batteries. Like WFIRST, the solar array is assumed to not be serviceable and is designed for a lifetime of 20 years. The solar arrays are sized for the science mode, which requires 6.4 kW of power, leading to a 39 m² array. The batteries are sized to survive a 3-hour launch scenario while maintaining the depth-of-discharge above 70%. Two 66 Ah lithium ion batteries are needed.

### 5.6.4 Propulsion

The telescope's propulsion system consists of a monopropellant propulsion system and a separate microthruster propulsion system. The monopropellant system consists of one 445 N main engine, four 22 N thrust vector control engines, and sixteen 4.45 N attitude control engines; the monopropellant system is capable of 3-axis control and has been sized to perform trajectory maneuvers, station-keeping, slewing, attitude control, and disposal. A monopropellant system was selected over a bipropellant system since monopropellant is easier to refuel and is less complicated than biprop. The mass penalty for the lower specific impulse that comes with monopropellant was not seen as an issue since the concept's baseline launch vehicle is the SLS Block 1B. The second system—the microthruster system—is based on 10 Busek BET-100 Microspray engines (Busek 2016) and is responsible for maintaining fine-pointing by counteracting the effects of solar pressure.

The propellant sizing was designed assuming a 5-year nominal mission plus a 5-year extended mission. The propellant system was also made to be serviceable by the use of a Vacco Type II interface resupply valve. 1,596 kg of hydrazine and 112 kg ionic liquid are needed for the monopropellant and Busek thrusters, respectively, assuming the worst-case spacecraft dry mass.

### 5.6.5 Attitude Determination and Control

As telescope apertures increase, pointing requirements grow ever tighter. Traditionally, pointing has been handled with reaction wheel-based pointing control systems that introduce jitter to the observatory and require increasingly sophisticated vibration isolation systems to defeat this unwanted side effect. Current passive vibration isolation technology is insufficient to meet the HabEx jitter suppression needs (see the telescope description in Section 5.5.2.5). Active isolation is a new technology that may be able to reach the HabEx requirements but only after additional development and with no certainty of success.

A new approach has arrived that could reduce jitter at least two orders of magnitude over current capabilities. By eliminating the reaction wheels and adopting microthrusters to carry the telescope fine pointing, HabEx can easily meet the telescope's jitter requirement; this key technology exists today. Several astrophysics missions with demanding pointing requirements have already blazed this trail. NASA's Gravity Probe B (GP-B), and the European Space Agency's (ESA) Gaia, Laser Interferometer Space Antenna (LISA) Pathfinder, and Microscope missions have all dropped reaction wheels in favor of microthruster pointing control. ESA's upcoming Euclid and LISA missions are also baselining microthruster systems. This technology is flight-proven and capable of supporting HabEx.

The HabEx attitude determination and control functions are handled in the following manner. Slewing, attitude control, orbit maintenance, and trajectory corrections are all handled by a conventional monopropellant propulsion system. The microthruster system exists solely to maintain pointing on target. Since the primary disturbance force at Earth-Sun L2 is solar pressure, the microthruster system operates continuously to counteract the effects of solar pressure. This is the same arrangement as used on the successful Gaia mission (Milligan 2017).

The HabEx fine-pointing system incorporates feedback control loops using conventional star trackers, a fine-guidance system, and sensors and fine-steering mirrors within the coronagraph and starshade instruments. Details on the pointing control system are given in Section 5.7.

The HabEx telescope was designed to be able to slew through 10 degrees in less than 60 minutes. Propellant has been sized to support two slews per day and maintain orbit over 10 years.

### 5.6.6    Telecommunication

The telescope telecommunication system was designed to support a crosslink between the telescope and the starshade, NASA's Deep Space Network (DSN) tracking for navigation, and downlinking of science data without disrupting on-going observations.

An S-band patch antenna is used for crosslinking between the telescope and starshade, while Ka- and X-band are used for telescope to DSN communications. The Ka-band phase array antennas allow for high-rate science downlink and can operate while conducting observations. The two X-band low-gain antennas (LGAs) offer nearly 4π steradian coverage. Command, engineering data, and navigation will be handled over the X-band link.

The S-band crosslink would allow for 100 bps communication with the starshade with 6.0 dB margin. The Ka-band would permit a downlink rate of 6.5 Mbps with 3.0 dB margin, while the X-band would permit downlink at 100 kbps with 9.0 dB margin.

### 5.6.7    Command & Data Handling and Flight Software

The telescope command & data handling (CDH) subsystem would be mostly built-to-print based on the JPL reference bus CDH design. The JPL reference bus provides standard CDH capabilities including spacecraft operations, communication, and data storage. Of particular note, the CDH subsystem was designed to provide 1 Tbit of storage, allowing the ability to minimize data downlinks to the DSN to 1 hour twice per week while maintaining ample memory margin. Furthermore, the CDH enables telescope-to-starshade S-band communications by adding a low-voltage differential signaling (LVDS) interface from the built-to-print design to the telescope transponders.

The flight software would be designed based on the JPL core product line, which was designed to work with the JPL reference bus CDH subsystem. Some mission-specific changes would be made to accommodate science requirements, telescope-to-starshade communication, and attitude control integration.

### 5.7    Attitude Control System (ACS) and Coronagraph Fine-Pointing

Direct imaging of exoplanets in the habitable zone of nearby sunlike stars with a coronagraph levies some of the most challenging pointing requirements ever met by a space telescope. With LOS error on the HabEx telescope set at 2 mas, HabEx would need to meet HST's best pointing performance on a routine basis. Fortunately, HabEx has three advantages. First, HabEx's diffraction limited angular resolution is two-thirds that of HST, allowing for tighter angular sensing. Second, the environment at Earth-Sun L2 has significantly less thermal and gravitational gradient disturbances than those experienced by HST. Third, without reaction wheels, HabEx's self-induced jitter is essentially nonexistent.

This section describes how HabEx would achieve the necessary LOS pointing for its telescope and instruments. The discussion covers the pointing requirements, pointing control architecture, operational modes and the expected pointing performance of the telescope and of the most demanding instrument: the coronagraph.

### 5.7.1  Requirements

The telescope pointing requirement, levied by the instruments, is 2 mas RMS per axis at the FGS—about $1/10^{th}$ of the 21 mas full width at half maximum (FWHM) of its diffraction-limited PSF at 0.4 µm wavelength. This amount of error reduces the Strehl ratio from the nominal 80% (diffraction limited) to 77.5%. Thus, the peak of the PSF for a chosen target will be reduced by only 3%: a small effect on observing efficiency. For the starshade instrument, the workhorse camera, and the UV spectrograph, this level of pointing is sufficient. For the coronagraph instrument, additional internal pointing refinement is required.

High-precision pointing is key to attaining the required levels of contrast in the HabEx coronagraph, and this drives the optical, mechanical, and ACS designs. The pointing requirements arise from the coronagraph error budget (see Section 5.3) and are summarized in **Table 5.7-1**. In turn, these requirements derive fundamentally from the contrast degradation caused by small wavefront fluctuations as the telescope LOS drifts away from the target star. Internally, while the coronagraph FSM corrects the telescope pointing error, there is a residual error caused by the input beam "walking" across the optics. This error appears as a variation in the speckle pattern in the coronagraph dark field and drives the requirements on the telescope LOS error.

Table 5.7-1. Attitude Determination and Control Subsystem (ADCS) pointing requirements.

| ADCS Requirements | |
| --- | --- |
| ACS pointing stability: | 2 mas rms/axis |
| Coronagraph pointing stability: | 0.7mas rms/axis |

There are three key disturbance sources on the ACS pointing system ahead of back end compensation by the coronagraph instrument. These minimize beam walk upstream of the coronagraph:

- Quasi-static observatory drift (drift between telescope instrument boresight and FGS field star sensors), corrected by the FSM,

- Low-frequency observatory jitter, corrected by the FSM, and

- High-frequency observatory jitter (i.e., ACS residual control error near the ACS bandwidth frequency), not corrected by the FSM.

Both the corrected and uncorrected residual jitter must be very small, and the telescope must be designed to mitigate it. After the FSM, there remains WFE arising from residual control error from the ACS and from high frequency jitter. Ultimately, these requirements are driven by the need to maintain a stable speckle pattern and set the maximum coronagraph internal pointing error of 0.7 mas RMS per axis at 1-sigma.

### 5.7.2  Control Architecture

The HabEx optical telescope assembly is designed to minimize LOS deviations due to internal misalignments. The design includes a very stiff precision metering structure that consists of very low CTE carbon fiber composite materials to provide passive control. In addition, active control of the telescope internal geometry is also provided.

Pointing control is handled by a multistage control loop architecture (see **Figure 5.7-1**). A conventional stage uses star trackers and gyros as sensors, and monopropellant thrusters as the actuators. This loop is responsible for slewing and other typical ACS functions. Once the target is acquired and within the FOV of the telescope, the next stage takes control. This stage uses the telescope's fine-guidance system to sense position against field stars in the FGS, and the microthrusters to hold telescope position. The loop is responsible for counteracting environmental disturbances such as solar pressure and the L2 gravity gradient. The third stage is internal to the coronagraph. Pointing alignment is monitored using the coronagraph's LOWFS camera, which directly detects the tilt of the incoming wavefront. The tip/tilt is then corrected by the FSM. Furthermore, focus error can be detected in the LOWFS, which can be corrected using the DMs.

In addition to the loops for sensing and correcting target position, HabEx also includes telescope thermal control and a laser truss (MET) to fix the relative positions of the first three mirrors.

The thermal control loop consists of precision thermistors on the telescope optics, structure, and barrel-sensing temperature changes to 50 mK. Strip heaters compensate for changes in environmental heat input based on the thermistor measurements. The effect of the loop is to control drift in the relative instrument boresights and changes in surface figure in the optics.

The relative positions of M1, M2, and M3 within the telescope are maintained using the laser metrology truss, which detects nanometer scale changes in spacing and alignment of the primary, secondary, and tertiary mirrors. Displacements are corrected using rigid body actuators on the secondary and tertiary mirrors. Details of the laser truss are given in Section 5.5.2.6.

### 5.7.3 Pointing Modes

The primary pointing modes for HabEx are *slew mode*, *target acquisition mode*, and the *science modes* for each of the four instruments. Since the four instruments have separate fields of view, it would be possible to operate all of the instruments simultaneously. Typically, the HWC and UVS instruments could undertake deep field observations while the starshade instrument or coronagraph is observing an exoplanetary system.

*Slew mode* begins by firing the monopropellant thrusters to initiate rotation of the telescope toward the next target for observation. The thrusters are fired again to stop rotation once the target is reached and within the FOV of the telescope. The star trackers are used to determine telescope orientation during this phase.

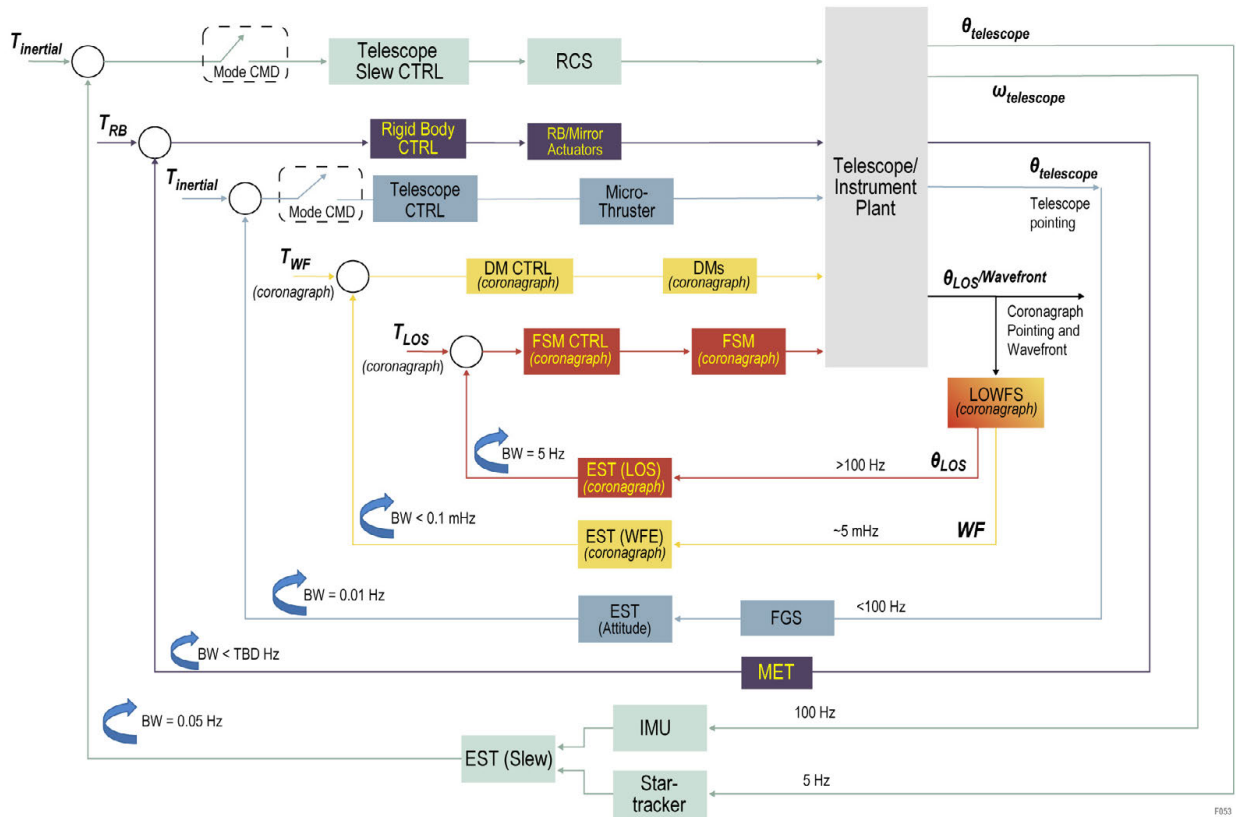

**Figure 5.7-1.** Pointing control loops.

*Target acquisition mode* begins with the position sensing handoff from the star trackers to the fine-guidance system for refinement of the telescope's line-of-sight error. Initial actuation is handled by the monopropellant thrusters until they reach near their minimum impulse bit limit, at which point the microthrusters take over actuation to bring the LOS error below the required 2 mas. The microthrusters continue to operate during instrument observations to counteract environmental disturbances, primarily solar-pressure-induced torque, on the telescope. Once the telescope has acquired the target and achieved 2 mas or better LOS error, *science modes* begin. Instruments remain in *science mode* until the science data has been collected, at which time, the telescope is ready to slew to the next target and the cycle repeats.

### 5.7.4    Slew Mode

Navigational needs for the HabEx spacecraft, slewing and repositioning, are handled by a combination of a multi-head star-tracker sensor system, an inertial measurement unit (IMU) with four fiber optic gyros for spacecraft position, and attitude sensing, and a hydrazine monopropellant system for rotational and translational movement of the spacecraft. The star tracker and IMU are internally redundant, and the thrusters are numerically redundant. All components are flight-proven and commercially available.

The sensing system provides a telescope pointing accuracy of 1 arcsec RMS, which would bring observational targets within the telescope's FGS FOV (greater than 2 arcmin). Sensing is then handed off to the telescope's fine-guidance system.

For the purpose of slew propellant estimation, the telescope was assumed to slew twice a day with a slew rate of 10 degrees per hour, for 10 years. These assumptions result in a slew propellant mass estimate of 210 kg. The overall hydrazine mass for the mission is about 1,600 kg so most of the hydrazine (1,400 kg) is needed for routine ACS functions such as the trajectory correction maneuvers required to establish the spacecraft in L2 orbit, and L2 orbit maintenance. This hydrazine need is essentially the same with or without reaction wheels, so the decision to go without reaction wheels has little impact on the overall system mass. It should also be noted that faster and more frequent slews are possible, but at the cost of additional hydrazine mass. For the final report, this slew allocation will be replaced with a science-based slew budget and related propellant estimate.

### 5.7.5    Acquisition Mode

Once in acquisition mode, sensing for telescope pointing is handed over to the telescope's FGS. The FGS is part of the telescope's payload and is described in Section 5.5.2.3. The system looks through the aperture of the telescope at bright, known stars in the FOV. Using actuated mirrors to position chosen field stars on the system's CCD detectors, the system can measure any deviation of the telescope's LOS. That information is supplied to a control loop, actuated with the microthrusters, to acquire the desired observational target and maintain the telescope's LOS on the target during science observations. As noted earlier, the FGS is composed of four CCD detector arrays looking through the aperture at four widely separated fields of view. This configuration allows the FGS to sense down to LOS errors of less than 1 mas over most of the sky.

Precise telescope LOS actuation control is accomplished with microthrusters. The conventional monopropellant thrusters are used with the star trackers and IMU to reduce LOS error down to 1 arcsec in the slew mode. At this point, actuation is transferred to the microthrusters, which reduce telescope LOS error to within the required performance level of 2 mas, and counteract environmental disturbances while the observing instruments collect data.

### 5.7.6    Science Mode

In science mode, telescope pointing is maintained using the FGS and microthrusters in a control loop. Typically, with the starshade instrument operating, the HWC and UVS would also be operating. Over time, the relative boresights of the FGS and the SSI would evolve within the constraints imposed by the telescope thermal control system. Assuming a CTE of the structure of $10^{-6}$, a linear separation of 1 m between focal planes of the FGS and starshade instrument, and 50 mK thermal control, the net result would be a maximum relative shift between the guide star and the science target of 0.004 pixels (12 µm pixel assumed) corresponding to 88 mas. This shift is extremely small relative to the angular pixel resolutions of the HWC, UVS, and starshade instrument (~12 to 30 mas) and is negligible. Even for the coronagraph this shift is small but would be corrected by the LOWFS control loop. Therefore, no special capability needs to be included to monitor relative boresights. Between the separate FGS optical paths, the relative boresights will also evolve, providing a verification of overall pointing stability over the time of an observation.

In the case of the coronagraph, LOS error needs to be reduced to achieve the contrast levels needed for science observations. As shown in the error budget in Section 5.3, the internal LOS error must be reduced below 0.7 mas. To achieve this, a fine-pointing control loop internal to the coronagraph is engaged (**Figure 5.7-1**). Tip/tilt sensing is done with the coronagraph's LOWFS and corrected by the FSM. A small, low-noise, high-resolution focal-plane camera forms the sensor and supports high readout speeds (≥100 Hz). The FSM is a precision piezo-electric actuated steering mirror. The loop brings LOS error within the requirement and reduces any disturbance components (currently not expected) up to a ~5 Hz controller bandwidth, with a measurement error of 0.2 mas per tip/tilt axis. Higher LOWFS sampling rates are possible up to 1 KHz, which would allow for faster control rates as well as feedforward approaches. Current simulations show the settling time to be short—much less than a minute—with a steady state inertial pointing performance of less than 0.1 mas per tip/tilt axis (the requirement is at 0.7 mas; **Figure 5.7-3**).

### 5.7.7 Microthruster Attitude Control

Currently, HabEx is considering two types of microthrusters as part of this study. The first is a nitrogen cold-gas microthruster used on ESA's LISA-Pathfinder and Microscope (Lienart, Doulsier, and Cipolla 2017), and currently in use on Gaia (Chapman et al. 2011). The second is a colloidal microthruster, which was used on NASA's ST7, which was onboard ESA's LISA-Pathfinder. Both are being evaluated for use on ESA's upcoming LISA mission. The cold-gas system has had more time on orbit and will have reached the HabEx baseline mission duration of five years by the time the HabEx study final report is released (2019). The colloidal microthrusters will undergo lifetime testing to verify lifetime models by the end of FY2022 as part of TRL 6 qualification for the LISA mission. The specific impulse is higher for the colloidal microthrusters (200–250 sec vs. 50–60 sec for cold-gas) so less fuel is required with that choice.

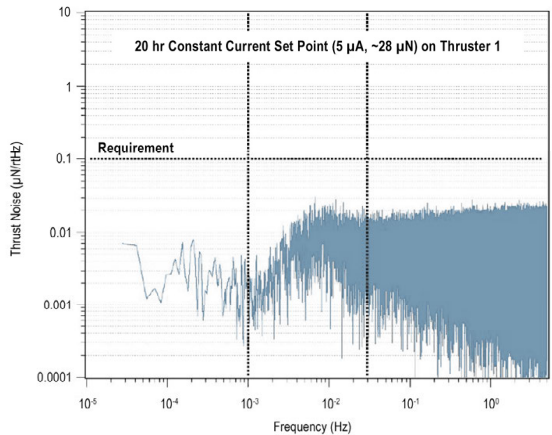

**Figure 5.7-2.** Noise profile for colloidal microthrusters (Ziemer et al. 2010). The noise is well below the requirement of 0.1uN/rtHz for LISA Pathfinder.

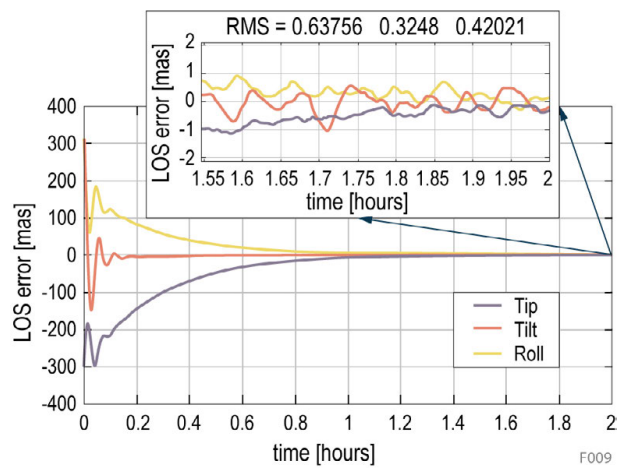

**Figure 5.7-3.** A sample LOS pointing simulation using the FGS and cold-gas microthrusters, with steady state performance <0.7 mas RMS per axis (see zoomed in portion). The colloidal microthruster performance is comparable.

Early performance estimates are encouraging for both options. For the cold-gas system, Aerospace Corp. has done a feasibility study (not yet published, but results and methodology presented to JPL). In their study, Aerospace reported a total noise level of 3.6 μN RMS, with a flat distribution at about 0.1 μN/√Hz level from 0.1 Hz to 4 Hz. This noise profile is approximately 10 times higher than the measured colloidal thruster noise (**Figure 5.7-2**). Thrust ranges from 1 μN to 1 mN per thruster. The Aerospace study has shown that the pointing performance for sample missions can be better than 0.7 mas RMS per axis, which is consistent with the results in the HabEx simulation (**Figure 5.7-3**). The settling time to reach the required <2 mas RMS per axis LOS error level is under 2 hours, primarily due to low thrust levels.

In addition to the higher specific impulse, colloidal microthrusters bring a lower noise profile than the cold-gas thrusters. However, the performance of the colloidal microthrusters in simulations is similar to the cold-gas microthrusters, indicating that at these low microthruster noise levels, the LOS performance is mainly sensor driven. The LISA-Pathfinder colloidal microthrusters nominally contained a cluster of nine emitters, with additional emitters added as needed to satisfy the required torque capability.

How the two systems size for the HabEx mission is also a trade consideration. The torque capability is a driver for fuel consumption and overall system size. Fuel consumption for these two microthruster options depends, in part, on microthruster location and on the spacecraft's center of pressure/center of mass (CPCM) offset value. The CPCM offset creates the solar-pressure-induced torque that the microthrusters must mitigate. Given these parameters, computing a torque capability space for a microthruster geometry is possible. Based on the required torque capability, the geometry and microthruster numbers are determined.

Fuel estimation was done for both microthruster systems and for two microthruster configurations: "middle" and "bus" (see **Table 5.7-2**). The "middle" configuration refers to placing four microthruster clusters on four sides of the spacecraft at the mid-height point (8.64 meters from the base of the spacecraft bus) for tip/tilt control. The "bus" configuration refers to placing 4 microthrusters clusters on 4 sides of the spacecraft at the bus height (1.5 meters). In addition, the fuel trade also looked at adding a "sail" to the spacecraft to reduce the CPCM offset and, consequently, the necessary fuel masses. This "sail" was envisioned as a 10 m×5 m panel that would deploy like a single panel solar array and would create more area below the center of mass to lower the center of pressure closer to the center of mass.

From **Table 5.7-2**, the fuel required for the colloidal microthrusters attached in the "middle" configuration and without a sail for the spacecraft would be about 120 kg for a 10-year mission. Cold-gas microthrusters also fulfill requirements, but a "sail" should be considered since the fuel saved by reducing the CPCM offset will exceed the mass of the added structure.

For the purpose of this study, HabEx is currently adopting the colloidal microthrusters due to superior fuel mass and microthruster noise performance, but the trade is still being evaluated and the microthrusters could be changed for the final report.

**Table 5.7-2.** Cold-gas and colloidal microthruster fuel consumption based on different microthruster locations.

| HabEx Case | Fuel: ISP = 200 s μthruster | Fuel: ISP = 50 s cold gas | Additional Mass for Sail |
|---|---|---|---|
| No sail THR "bus" \|CMCP\| = 3.31 m | 1 hour = 7.2e-3 kg 5 years = 321 kg | 1 hour = 29e-3 kg 5 years = 1284 kg | No |
| No sail THR "middle" \|CMCP\| = 3.31 m | 1 hour = 1.3e-3 kg 5 years = 56.3 kg | 1 hour = 5.1e-3 kg 5 years = 225 kg | No |
| With sail THR "bus" \|CMCP\| = 0.7 m | 1 hour = 2.1e-3 kg 5 years = 91.2 kg | 1 hour = 8.3e-3 kg 5 years = 365 kg | Yes; 92 kg |
| With sail THR "middle" \|CMCP\| = 0.7 m | 1 hour = 0.5e-3 kg 5 years = 20.1 kg | 1 hour = 1.8e-3 kg 5 years = 80.4 kg | Yes; 92 kg |

[1] Ref: Ziemer/ST7-DRS
[2] Ref: Ziemer/Gaia

## 5.8 The Starshade Occulter

The starshade, flying in formation with the telescope, creates a deep shadow suppressing the light from the parent star and thereby revealing the reflected light from the exoplanets in the system. The optical design and position of the starshade occulter, along with the resolution and performance of the telescope and starshade instrument, determines the depth of the contrast in the dark field.

For clarity, the occulter is the part of the starshade responsible for blocking the starlight, as opposed to the whole starshade system which includes the occulter, formation flying hardware, and typical spacecraft subsystems.

The optical performance of the starshade occulter is almost entirely an optomechanical problem. Its size necessitates a deployable architecture that is passively shape controlled, both mechanically and thermally. The function of the starshade mechanical system is to reliably deploy on orbit, and meet the specified shape accuracy, shape stability, and solar glint requirements. A 0.33 rpm rotation of the starshade reduces temperature gradients and improves shape stability.

Currently, there exist two proposed architecture solutions to the mechanical deployment to achieve the on-orbit requirements for HabEx. The two solutions will be hereafter referred to as the "furled petal" and "folded petal" architectures. Each of the architectures are described in this section, including the architecture approach and heritage, mechanical design and deployment, and relevant structural and thermal analysis for meeting the driving requirements.

The starshade occulter must also be highly mobile, since it must be placed along the LOS of the telescope to the target star. With a nominal separation of 124,000 km, the starshade spends up to 80% of the mission slewing from one target to the next. Accordingly propulsion and formation coordination are key capabilities required of the spacecraft bus.

The Formation Flying Control System, which positions starshade relative to the LOS of the telescope is also presented in this section.

### 5.8.1 Starshade Optical Designs

This section evaluates the starshade performance for a 72 m starshade concept. In particular, a discussion considering observations of exoplanets appearing between the petals, where throughput is reduced and scatter is increased, but observational completeness is improved, is included.

#### 5.8.1.1 Key Allocations

The building and testing of starshade disks and petals has been underway for several years through the NASA Technology Development for Exoplanet Missions (TDEM) program and, consequently, has evolved a list of the most likely and significant mechanical perturbations affecting the starshade contrast performance. **Table 5.8-1** lists the allocations of key starshade parameters and the resulting image plane contrast at the nominal IWA (60 mas), and longest wavelength (1.0 μm) for the starshade. The values given are the tolerances for global (bias) terms and the 3-sigma allowances for random terms. The corresponding contrast contributions for each term are also shown. The terms fall into two categories: petal position and petal shape. Formation flying tolerances are also considered. The starshade is designed to allow a 1 m lateral shift and 250 km of longitudinal shift (along the line of sight to the star) from the nominal position.

**Table 5.8-1.** Allocation of key parameters.

|  | HabEx-72 m | Contrast * 10⁻¹¹ |
| --- | --- | --- |
| **Manufacture** | | |
| Petal segment shape (bias) | 28 μm | 0.3 |
| Petal segment shape (random) | 170 μm | 0.4 |
| Petal segment placement (bias) | 17 μm | 0.3 |
| Petal segment placement (random) | 156 μm | 0.7 |
| **Pre-launch deployment** | | |
| Petal radial position (bias) | 500 μm | 1.2 |
| Petal radial position (random) | 1,500 μm | 0.2 |
| **Post-launch deployment** | | |
| Petal radial position (bias) | 500 μm | 1.2 |
| Petal radial position (random) | 1,500 μm | 0.2 |
| **Thermal** | | |
| Disk-petal differential strain (bias) | 30 ppm | 1.7 |
| 1–5 cycle/petal width (bias) | 20 ppm | 0.5 |
| **Formation flying** | | |
| Lateral displacement | 1 m | 0.5 |
| Longitudinal displacement | 250 km | 0.3 |
| Reserve | | 2.5 |
| **Total:** | | **10** |

The largest manufacturing/deployment level contributor to image plane contrast is the bias on petal positions, to which 500 µm is allocated both pre- and post-launch. This term describes the average radial position of the petals relative to their ideal position. The disk-to-petal thermal strain (allocated 30 ppm) is the largest single contrast contributor. This term affects the petal position relative to the size of the disk like the manufacturing bias. The uniform strain term arises from differences in the overall CTE of the petals relative to the truss, their average temperatures, and their temperature differences.

In the modeled design, the starshade petal shape is assumed to be formed from precision edge segments 3 m long and positioned on the petal mechanical structure. The segment positioning tolerances are ±156 µm (random) while the bias terms are substantially tighter at 17 µm. Petal thermal deformations have also been considered and are expressed in terms of spatial frequencies of 1–5 cycles along the petal edge. The tolerance on these terms is 20 ppm of width change integrated over the spatial frequencies considered.

### 5.8.1.2 Performance vs. Working Angle and Wavelength

Starlight leaking around the starshade appears to come from along the starshade petal edges. The telescope PSF convolves the light, resulting in some energy appearing both beyond and within the starshade tip. However, unlike a high-contrast coronagraph, which is subject to large-angle scatter originating on the telescope optics, the scattered energy continues to decrease with larger angular radii. **Figure 5.8-1** shows how the starshade contrast changes with working angle. The chart includes a curve showing geometric throughput, which will differ slightly from the actual throughput when diffraction is included. The starshade is tolerated for $10^{-10}$ contrast at 60 mas and a wavelength of 1.0 µm. At the 50% throughput point, 51 mas off-axis, the contrast decreases to $1.6\times10^{-10}$. In the middle of the band, the performance degrades from $5\times10^{-11}$ at the petal tips to $9\times10^{-11}$, a factor of 1.8, at the 50% throughput point. At the short end of the band, the performance degrades from $1.2\times10^{-11}$ at 60 mas to $2.8\times10^{-11}$ at 51 mas, a factor of 2.8 degradation.

Design experience on HabEx and other starshades indicates that the scattered light background from starshade shape errors increases by about a factor of 1.6 as the working angle moves inward to the 50% transmission point. A corollary is that to maintain the contrast observed at the full radius of the starshade, the allocations must decrease by about $\sqrt{1.6} = 1.3$. For the broadband HabEx starshade, even though the performance at the end of the bandpass degrades by almost a factor of 3, the contrast remains better than $10^{-10}$ at the 50% throughput point at 51 mas.

A feature of starshades is that they maintain their suppression characteristics for a fixed value of their Fresnel number, $F = r^2/(\lambda z)$, where $r$ is the starshade radius, $\lambda$ is the wavelength, and $z$ is the separation from the telescope. Note that for IWA$= r/z$ **Figure 5.8-2** shows the contrast for the HabEx starshade used at two different distances, with two different wavelengths, maintaining constant $F$. With the exception of the point at 27 mas, which is shifted due to a modeling resolution issue, the points in the 0.35 µm curve closely match the corresponding points in the 0.7 µm curve, for the same tolerances. This verifies that tolerancing in one band equally describes tolerancing at a different band, for a corresponding change in IWA.

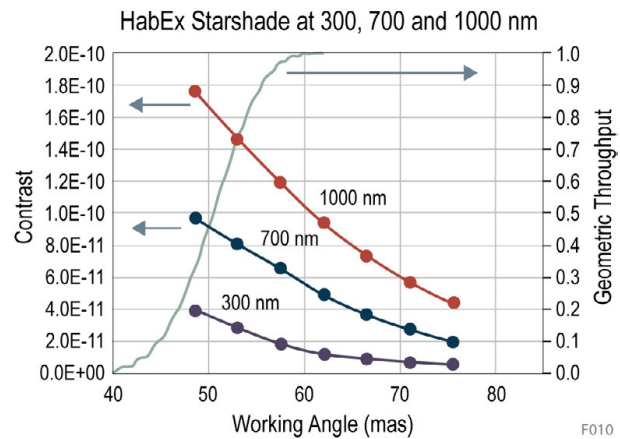

**Figure 5.8-1.** HabEx modeled performance vs. working angle and wavelength.

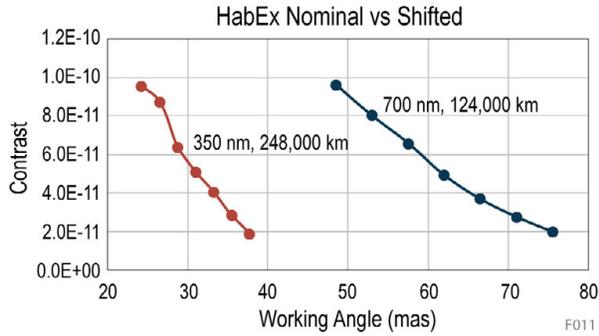

**Figure 5.8-2.** Performance when the starshade is shifted for observation in a different band with a corresponding shift in working angle.

### 5.8.1.3 Final Comments

The most important lessons from this work are that tolerancing scales roughly linearly with starshade dimension and that contrast performance at working angles within the projected starshade image degrades by about 60% where the throughput is 50%. The scaling law helps to evaluate the challenges of building starshades for large telescopes. The performance parameters will be used in future studies to more accurately compute starshade performance and to optimize the starshade diameter while improving science yield.

### 5.8.2 Furled Petal Design (Baseline)

The furled petal starshade design has been in development at JPL since the 2010 Astrophysics Decadal Survey identified the need to advance starshade technology. Great progress has been made in design since then. This section explains the architecture and heritage of the HabEx design, and describes its mechanical features. Work done on structural and thermal evaluation of the design is also included.

#### 5.8.2.1 Mechanical Architecture Approach and Heritage

The approach of the furled petal architecture was to leverage existing heritage deployable structure technology to formulate a concept that would minimize uncertainty in technology development. The approach allows the starshade mechanical system to be functionally separated into two distinct subsystems that have separable requirements, can be developed in parallel, and validated with separate technology demonstrations.

**Figure 5.8-3** illustrates the furled starshade mechanical design. It also shows the two major subsystems—the inner disk and the petal. By design, each subsystem deploys independently in

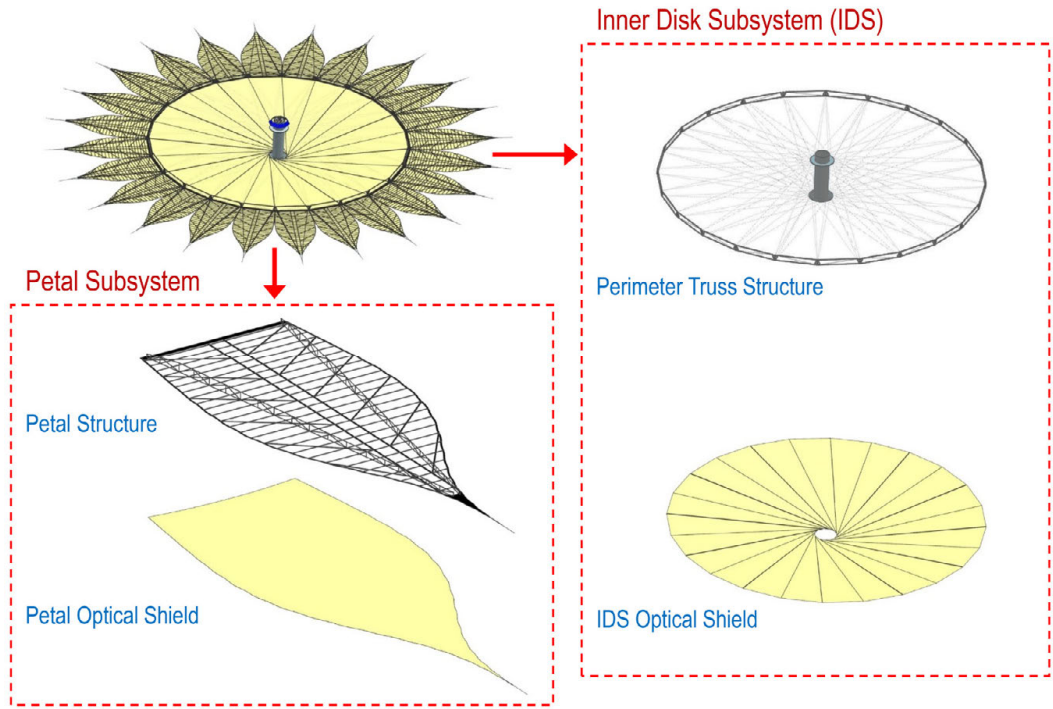

**Figure 5.8-3.** Starshade furled petal architecture subsystem breakdown and principle features.

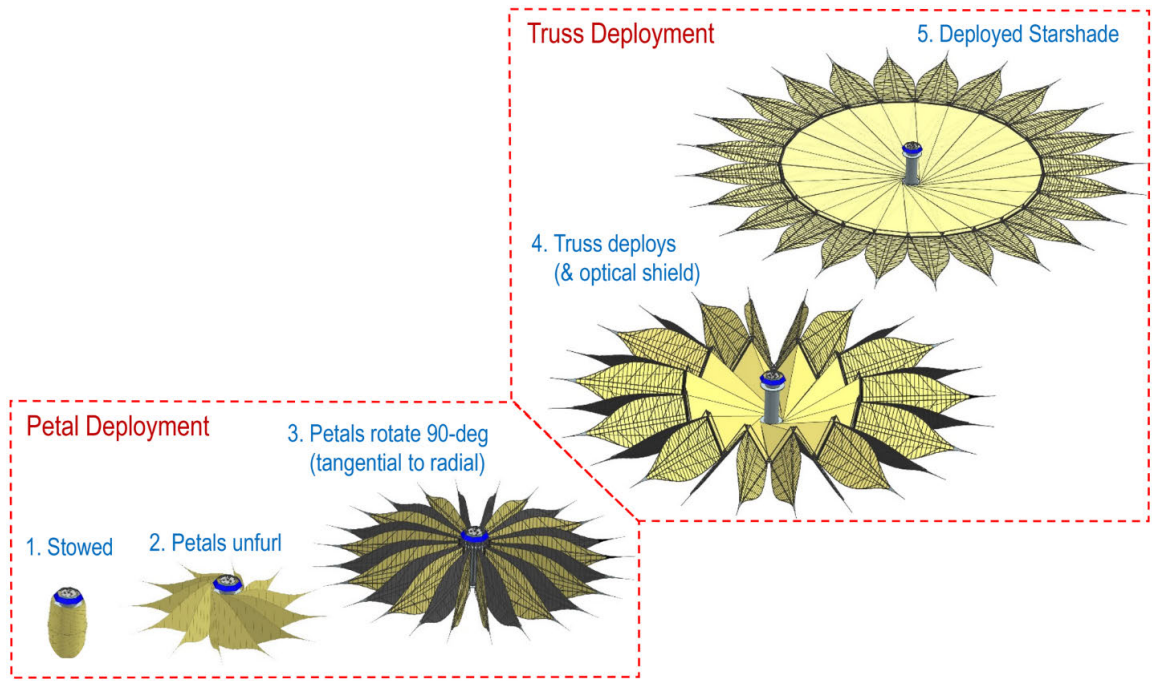

**Figure 5.8-4.** Deployment sequence for the starshade furled petal design is separated into two stages, petal unfurling, and the inner disk deployment, which is driven by the perimeter truss. Note the petals are passively translated and rotated to their final position on-orbit.

sequential stages. **Figure 5.8-4** illustrates the deployment sequence and highlights the two-stage deployment—unfurling of the petals followed by the inner disk deployment.

The furled petal architecture draws on heritage from two flight-proven deployable technologies—the Astromesh antenna and the Lockheed Martin Wrap-rib antenna (NRC 2010).

The starshade inner disk is an adaptation of the Astromesh antenna, and is the core of the structure to which the petals attach. The Astromesh antenna is lightweight, precise, and has a high deployed-diameter-to-stowed-diameter ratio enabling very large deployed diameters to fit within a small launch vehicle fairing volume. Importantly, the Astromesh antenna has successfully deployed at least nine times on orbit, providing credibility to this deployment technology. The application of this technology to the starshade is illustrated in **Figure 5.8-5**, highlighting the replacement of the precision, gold-coated geodesic mesh that forms the antenna surface with the tensioned, linear spokes, resulting in a tensegrity hoop that is rigid in plane to ensure starshade in-plane shape accuracy. The adaptation results in a ring that is less deep and better suited for attaching petals but retains the same deployment kinematics and mechanism upon which the antenna's perimeter truss architecture is based. The starshade perimeter truss is centered on a rigid, structural hub, which houses the starshade spacecraft and propellant, as well as providing a stiff interface to the deployed starshade. The use of a central hub requires that the petals furl, or wrap, in order to fit within the launch fairing.

The addition of the petals to the inner disk perimeter truss can be seen in **Figure 5.8-5**. To wrap the petals, the Lockheed Martin wrap-rib antenna approach was applied to the petals, resulting in the petals spirally wrapping around the stowed perimeter truss and central spacecraft for launch. The wrap-rib approach has been successfully deployed hundreds of times on orbit, again adding credibility to the use of this approach. **Figure 5.8-6a** illustrates the similarity between thin, radially oriented petals before wrapping for launch, and the thin, radial ribs of the wrap-rib rib antenna in **Figure 5.8-6b**. **Figure 5.8-6c** illustrates the wrapping of the ribs

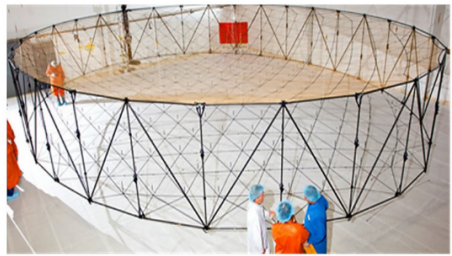

Replace precision geodesic antenna mesh
with linear spokes

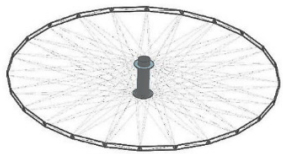

Add petals and loose fitting
kapton blanket

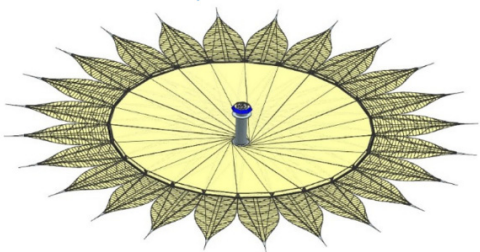

**Figure 5.8-5.** Traceability between Astromesh antenna technology and starshade design. The Astromesh antenna technology serves as the core to which the starshade petals are attached.

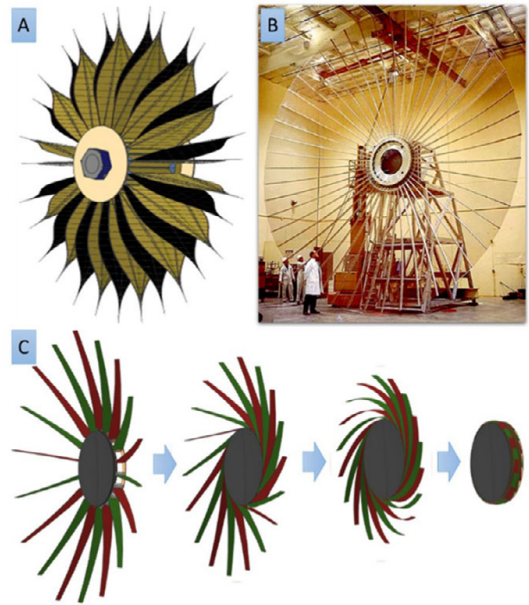

**Figure 5.8-6.** Comparison of Lockheed Martin wrap-rib antenna technology to starshade petal wrapping architecture.

of the wrap-rib antenna around a large hub, which is the approach the petals follow. It is important to note that wrapping of the petals is in the out-of-plane direction, so as not to disturb the in-plane shape of the petal, the critical dimension for petal performance. Unfurling the petals is accomplished quasi-statically with a separate "unfurler," unlike the dynamically deployed wrap-rib antenna. The unfurler subsystem is not considered a technology gap, but rather an engineering development.

The unique challenge of the HabEx configuration compared to that of the Exo-S, is fitting a starshade that is roughly twice the diameter and four times the area, within the same 5 m launch fairing as that of the Exo-S. The 5 m fairing was adopted as a design constraint for two reasons. First it ensures that the starshade can fit in a shared launch configuration with the telescope in the SLS Block 1B. Second, it allows the starshade to be launched separately—if programmatic reasons require it or if a replacement starshade is needed in the future—without requiring a second SLS launch vehicle. A series of early configuration studies determined that for a given starshade size and length of petal, the driving mechanical parameter to meet the stowed diameter requirement was the number of petals, which directly translates into the wrap radius of the stowed structure. The much larger diameter starshade is able to fit within the 5 m fairing because the increased diameter of the truss only increases the height of the stowed starshade.

As described above, the deployed starshade comprises two mechanical subsystems, the petals and the inner disk. The inner disk serves as the core to which the petals attach, and as stowed, forms the barrel-like structure around which the petals are wrapped for launch. Encaging the petals for launch restraint is the Petal Launch Restraint & Unfurler Subsystem (PLUS) shown in **Figure 5.8-7**. After launch, the PLUS quasi-statically unfurls the petals in a controlled fashion, ensuring the petals edges are not damaged.

### 5.8.2.2    Furling Starshade Mechanical Deployment

Deployment of the furling starshade involves multiple steps to transition from the compact, stowed system that fits within the launch vehicle

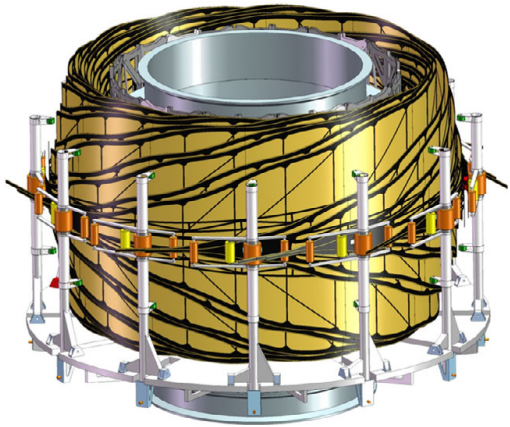

**Figure 5.8-7.** Petal Launch Restraint and Unfurl Subsystem (PLUS) slowly unfurls the petals on orbit to reduce dynamic excitation of the petals, and is then jettisoned to reduce starshade retargeting mass.

fairing to the fully deployed operational system. Each step is described below.

### 5.8.2.2.1 Step One: Unfurling the Petals

The PLUS is a large carousel assembly that rotates about the starshade spacecraft hub's long axis. For launch, the PLUS is locked in rotation, and the vertical cage posts around its perimeter serve as an external boundary condition that preloads radially aligned launch-restraint interfaces on the spirally wrapped stack of petals. The petal tips protrude beyond the stack of furled petals, and thus dynamic excitation during launch must be controlled with a separate mechanism. This is accomplished via two roller assemblies that extend tangentially from the vertical cage posts. Once on orbit, the petal preload mechanism on the cage posts is released, allowing the furled petal strain energy in the petals to lightly press the petals against a roller assembly on the post. The roller assembly is centered vertically on the petal, aligning with the petal central spine. The carousel rotational constraint is then released, and a single, redundant motor system slowly and deterministically rotates the carousel with respect to the wrapped petals, allowing for controlled release of the petal furled strain energy and ensuring no damage to the petal edges.

### 5.8.2.2.2 Step Two: Rotating the Petals

Once the petals have fully unfurled, they are passively rotated to a radial orientation via torsion springs in the hinges that attach the petals to the perimeter truss. Once the petals are radial, and radially out of the way of the vertical cage posts, the cage posts are then rotated radially down and out of the way of the petals/truss system, allowing for the entire PLUS subsystem to be jettisoned before truss deployment.

### 5.8.2.2.3 Step Three: Truss and Petal Deployment

Once unfurled the petals deploy passively from vertical to horizontal along with the active deployment of the perimeter truss. The perimeter truss design and deployment are fundamentally the same as those used on the Astromesh antenna, with deployment controlled via a braided steel cable that serpentines the diagonals of the truss, and is reeled in with a motor onto a spool, expanding the perimeter truss. This deployment technology has been utilized successfully more than nine times on orbit.

The truss is composed of thermally stable carbon fiber composite tubes, called longerons, which form a perimeter ring. This ring is placed in compression upon final deployment by the radial, thermally stable carbon fiber composite spokes that connect the ring to the central spacecraft hub. The tension and compression in the carbon fiber structure creates a strain stiffened, and thus precise, thermally stable structure to which the petals are attached. The petals, attached to the perimeter ring, are rotated 90 degrees into position as the truss deploys.

The entire disk and petals are covered with multiple layers of carbon impregnated black kapton—a material that intrinsically meets the HabEx opacity requirements—that unfolds as the truss deploys. Separation between the kapton layers mitigates the effect of micrometeoroid impacts by reducing the percentage of micrometeoroid puncture holes that will provide a direct path for starlight to pass through the starshade and enter the telescope. The deployment of the truss pulls out the spirally wrapped opaque optical shield.

### 5.8.2.3 Petal Structure

The starshade petal, unlike the inner disk, does not require the articulation of any joints or tensioned members to create its structure. Pictured in **Figure 5.8-8**, the petals are a planar gossamer carbon fiber composite structure that, as manufactured, meets the in-plane shape requirements of HabEx. The shape critical width of the petal is provided by thermally stable carbon fiber composite tubes, called "battens", that hold the petal structural edge at the periphery of the petal. The optical shape profile is able to meet shape requirements because it is produced in discreet 1 m segments that are precisely bonded to the structural edge in the correct location. The edge profile is formed by a thin, amorphous metal alloy, which is chemically etched to produce a sharp beveled edge that limits solar glint from the edge into the telescope. The entire petal is then loosely covered with the same opaque optical shield as the inner disk. Because the petal structure is thin to allow for the petals to wrap for launch, out-of-plane stiffness of the petals is provided via two piano-hinged ribs that passively deployed via a reliable and redundant over-center sprung hinge strut. These ribs are attached near the base of the petal to the perimeter truss, which provides a stiff out of plane connection from the petal to the perimeter truss ring.

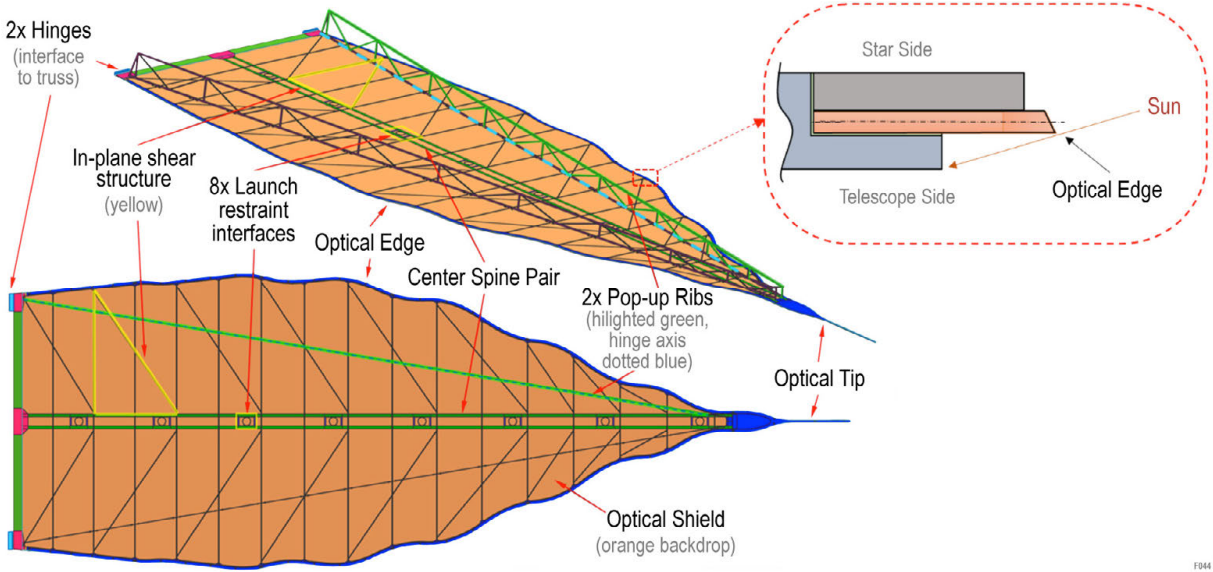

**Figure 5.8-8.** Details of a starshade petal including a cross section of the optical edge. Petal shape is largely dominated by the width controlling elements, the battens. Solar glint is minimized by reducing the edge radius of the optical edge as well as its reflectivity.

### 5.8.2.4  Structural Analysis

In order to meet deployed shape performance requirements, the starshade structure must be sufficiently stiff and damped to ensure the structure maintains on-orbit shape during observations. To this end, the structure is first analyzed to determine the fundamental frequency, which is desired to be above 0.5 Hz, for position control of the spacecraft. Additionally, the lower-order mode shapes are assessed against critical performance error budget terms. Finally, the structure is assessed for the duration of its damping response to thruster firings during observations. In addition to meeting on-orbit shape, the structure must be suitable for ground handling during integration and test activities, as well as survive launch loads.

#### 5.8.2.4.1  Launch Structural Analysis

The starshade stowed structure must be assessed to be suitably strong and stiff to survive launch loads. An initial assessment of the starshade hub structure carrying the mass of the starshade payload will be compared to launch vehicle guidelines for the final report.

#### 5.8.2.4.2  On-Orbit Structural Analysis

The starshade structure comprises two types of structures, the inner disk, which is a tensegrity structure, and the petals, which are cantilevered structures that extend from the inner disk perimeter truss. Construction of a finite element model of the HabEx petal allowed for assessment of the lower mode shapes against the key error budget deformations, all of which are in-plane petal deformations. Assessment of the lower-order mode shapes showed out-of-plane deformations that would not excite key error budget terms (**Figure 5.8-9**). The first mode of 0.92 Hz, which is dominated by the petal tip and primary structure bending out of plane about the base of the petal, with significant corresponding strain energy in the out-of-plane ribs. The second mode at 1.50 Hz shows petal tip bending once again, but also introduces a mild "taco" shape, in which the lateral edges of the petal curve downward. This is the first mode in which the petal battens participate. The battens are responsible for the petal in-plane shape, however, an out-of-plane deformation only indirectly affects in-plane shape. The third mode at 1.54 Hz, introduces greater participation of the battens in out-of-plane bending. The fourth mode at 1.67 Hz introduces a petal twist mode about the long axis. This twist only indirectly affects the petal in-plane shape.

Next, a full system structural finite element model was constructed to analyze fundamental frequency and mode shapes of the overall structure. Importantly, the system fundamental response is above 0.5 Hz, and there is ample separation between the petal and overall starshade structure fundamental frequencies. This is needed to avoid coupling between petal and overall structure modes. It is also desired that system response not excite key error budget terms for shape deformation, therefore the mode shapes were assessed for in-plane deformations and were found to not excite in-plane deformations. The first deformed (non-rigid body) mode of the starshade structure is 0.72 Hz and exhibits petal "flapping," or bending about the base of the petal (**Figure 5.8-10**). Variations, or clusters, of this mode shape (with petals 1–24 in varying up and down configurations) dominate the first four deformed mode shapes. The second cluster of modes starting at 0.73 Hz (deformed mode shapes 5–8) is similar to the first mode shape in that it exhibits petal bending. However, the overall deformed shape is anticlastic, or potato chip. The 20th mode at 0.81 Hz is the third unique mode shape, which exhibits petal bending at a higher frequency with the petals alternating up and down at a period of four petals. The fourth unique deformed shape is mode 31 at 1.36 Hz, which is another clustered mode shape. This mode is the first mode to exhibit petal twisting about the petal long-axis and includes some petal bending out of plane.

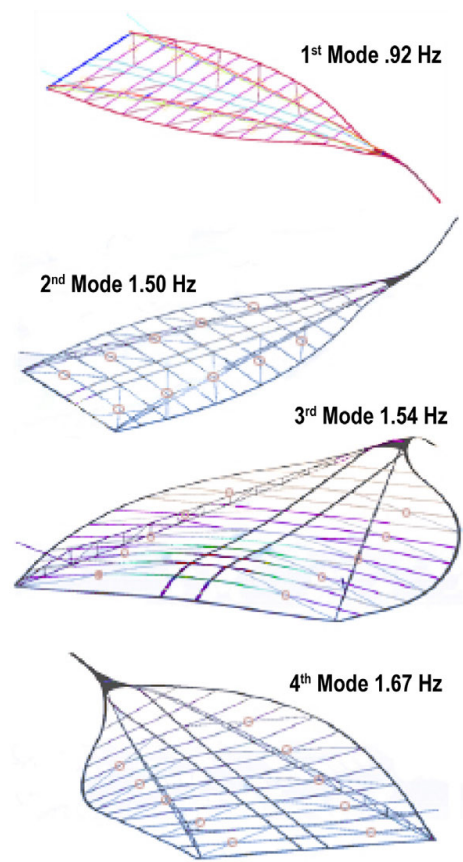

**Figure 5.8-9.** Finite element model modal results for the HabEx starshade petal assuming connection to a rigid interface at the petal base.

The starshade structure will be assessed against the HabEx requirements for duration and amplitude of response to thruster firings before the HabEx study final report. Based on previous studies, this is not expected to be an issue.

### 5.8.2.5 Thermal Analysis

The starshade instrument is sensitive to perturbations of in-plane shape during observations. The starshade structure is therefore designed to limit distortion of the shape due to the on-orbit environment. As a passively shape controlled instrument, the starshade allows the temperature of the structure to vary based on the

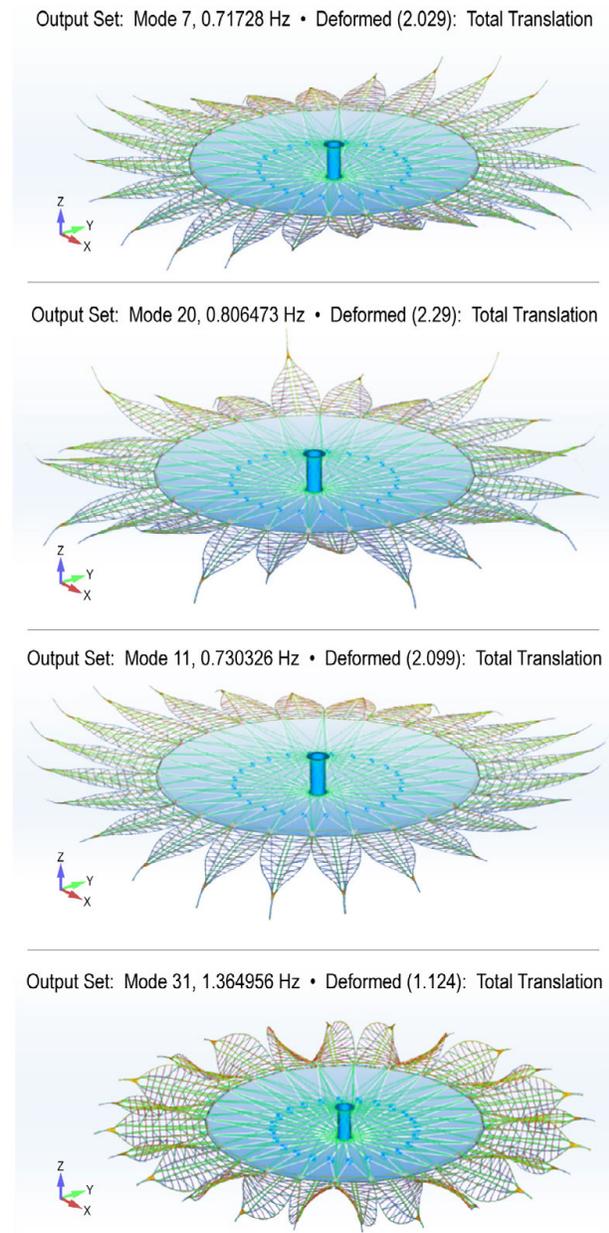

**Figure 5.8-10.** Structural modal analysis of the starshade showing representative mode shapes, all of which are out-of-plane shape deformations, which only indirectly contribute to degradation of starshade performance.

Sun angle with respect to the starshade during observations. The thermal design of the structure is such that it limits deformations to within the error budget allocation over the known temperature variation. The key term and challenge for shape error is the uniform relative expansion and contraction of the petals with respect to the inner disk, which in flight will be caused by a relative temperature difference between the inner disk and the petals, as well as a difference in the coefficient of thermal expansion between the inner disk and petals. For the HabEx starshade, the uniform expansion or contraction of the petals with respect to the disk must not exceed 30 ppm.

The starshade is designed to be thermally stable to required levels by utilizing a carbon fiber composite for the structure, throughout the petal, and inner disk. In particular, the battens of the petals, which directly control the width of the petal, and the longerons, the circumferential member of the truss to which the petals attach, have been identified as the key elements for controlling uniform expansion and contraction of the petals relative to the truss. For this reason, the petal battens are constructed of a commercially available carbon fiber pultruded composite that, by virtue of manufacturing process, has a very consistent CTE of 0.2 ppm/K. The longerons, which will need to have greater cross axis stiffness and strength than the battens, will therefore be constructed of a quasi-isotropic layup of carbon fiber reinforced polymer. The layup will be designed to optimize for the desired longeron CTE to balance the combination of the CTE of the battens and the difference in temperature between the petal battens and the truss longerons, limiting relative uniform expansion and contraction between the petals and inner disk.

The thermal analysis using finite modeling methods performed during this study will answer the question of the expected temperatures for the shape-critical starshade components as well as the desired CTE to limit shape deformations to within error budget requirements. This work is underway and will be completed for the final report.

### 5.8.3 Folded Petal Design (Alternative)

The Starshade Folding Petal design is an alternative architecture developed to meet the requirements necessary to suppress starlight and allow direct imaging of exoplanets around that star. The Folding Petal shape design differs from the Furled Petal design in that there are fewer petals (16 vs. 24), the petals are longer, and the shape of the petals is slightly different. Furthermore, the Folding Petal design is taller in the stowed configuration, making it incompatible with a telescope co-launch on an SLS Block 1B. While this alternative design cannot co-launch with the telescope, this mechanical deployment architecture provides strong heritage for a potentially larger starshade launching in a 5-meter fairing, should the need arise. Regardless of these differences, starshade occulter design solutions vary widely enough to allow taking advantage of a variety of deployment mechanisms, all while achieving sufficient starlight suppression contrasts.

### 5.8.3.1    Mechanical Architecture Approach and Heritage

The current state of the art for a deployable, multilayer membrane structure is the JWST sunshield membrane architecture. That structure is approximately 15×20 meters in size with five membrane layers made of Kapton E material and specialized coatings to meet the solar rejection requirement needed to thermally isolate JWST from the Sun. These membranes have been fully flight-qualified for that mission. The JWST membrane qualification provides a wealth of data, along with a roadmap of the work required to develop a membrane architecture and plans for handling, stowing, and membrane management during launch and deployment of the large structure. All of this information is documented in detail in the JWST TRL 6 report.

Development of the JWST sunshield is particularly relevant to the HabEx Folding Petal starshade occulter. The JWST membrane design for stowing and controlled deployment, as well as venting the stowed structure during launch and ascent, is similar to what is required of the starshade occulter. The Folding Petal design described here leverages the technologies developed for the JWST sunshield.

In the process of this study, super heavy-lift launch vehicle (SHLLV) 5 m fairing packaging limitations and their effect on starshade sizing and design, were examined. This alternative starshade design uses a deployment approach with commonality to all other Northrop Grumman starshade designs and deployment approaches.

### 5.8.3.2    Folding Starshade Mechanical Deployment

The starshade deployment involves multiple actuations/deployments to transition from a stowed system that fits within launch vehicle volume requirements to the fully deployed operational system (**Figure 5.8-11**). These deployments require a high level of accuracy in order to meet the specified shape requirements of the starshade. The full deployment procedure is broken down into chronological steps in the subsections below.

#### 5.8.3.2.1    Step One: Release and Deployment of Launch Lock Support Structure

The launch lock support structure provides the necessary stiffness for the stowed petal structure to survive launch loads. Launch lock support arms provide this support and are constrained via a perimeter hoop restraint located approximately mid-height of the stowed system (**Figure 5.8-12**, left). Additional restraints at the support structure tips may be added to the launch lock system if required by stowed stiffness or structural frequency.

Deployment begins once the release of the launch lock structure is triggered. The perimeter hoop restraint system releases all support structure arms in a near-simultaneous fashion. The deployment of the arms is accomplished by preloaded springs within a simple hinge at the root of the arms. The deployment spring preload is expected to be sufficient to hold the arms in the deployed position after release. **Figure 5.8-12** shows the launch lock system in both its stowed (left) and its deployed (right) configuration. The concept presented uses a four-segment perimeter hoop and 16 support arms.

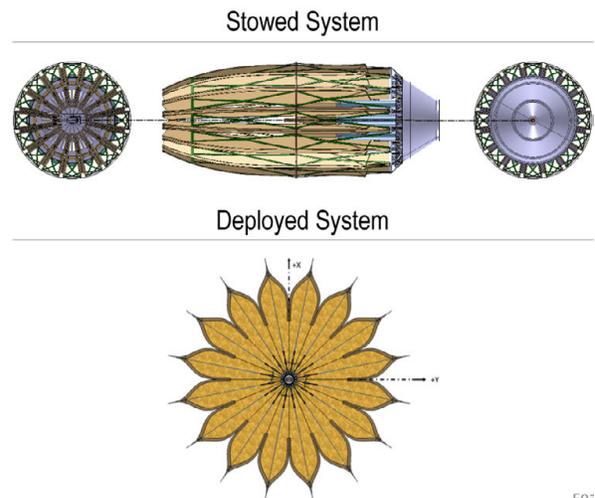

**Figure 5.8-11.** The folded petal starshade fully stowed (top) and fully deployed (bottom).

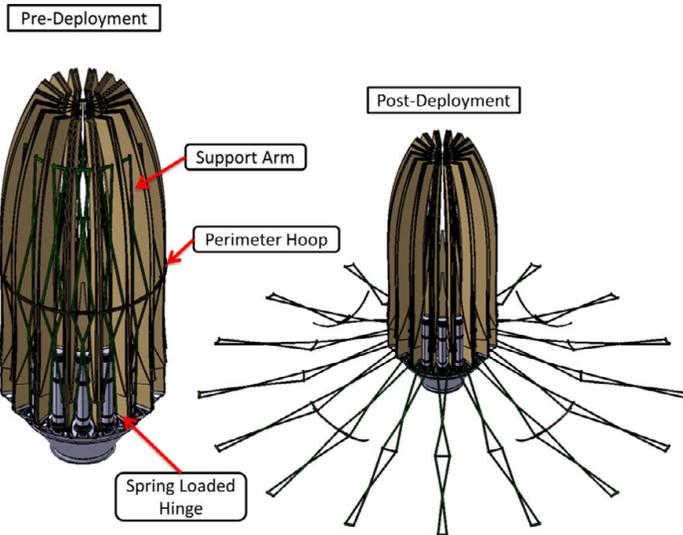

**Figure 5.8-12.** Launch lock mechanisms for the stowed starshade system. The perimeter hoop applies compressive force into the launch lock support arms. When the hoop is released, spring loaded hinges cause the support arms to deploy downward. This image shows a design with 16 support arms, but the hoop/support arm design could change in a future trade.

### 5.8.3.2.2  Step Two: Vertical Driving of Petals to Clear Base Restraints

When stowed, the lower ends of the folded petals are seated in base restraints located on the main structure of the system (**Figure 5.8-13**, left).

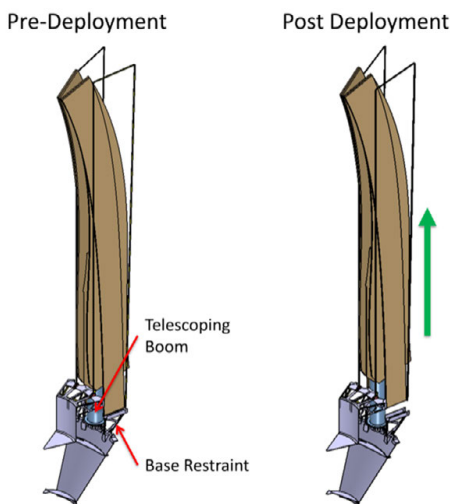

**Figure 5.8-13.** Images of a single boom before (left) and after (right) the STEM deployment required for the folded petal edges to clear the base restraints.

To clear the base restraints, the petals are driven upward to a sufficient deployment height so that the petals will clear the base restraints during the subsequent rotation of the boom systems (step 3 below). This deployment is shown in **Figure 5.8-13**.

This vertical motion is achieved through the use of the Storable Tubular Extendible Member (STEM) systems. These STEM systems are drivers that are used in step four (see below) to fully extend the 16 telescoping booms.

Because each individual petal is tied to the next at the valley locations, all STEM systems must drive vertically simultaneously. Each STEM drive system has start and stop capabilities and contains rotary potentiometers tied to their output feeds, allowing a high degree of telemetry and control throughout the deployment cycle.

### 5.8.3.2.3  Step Three: Telescoping Boom Angular Deployment (Unfolding)

As in step two, near simultaneous deployment of all petals is a requirement to prevent excessive stress on the valley joints, and to prevent possible snags or tears of the membranes near the valley joints.

The booms are driven to their deployment angles via a central deployment tower, tied to each boom via linkages (**Figure 5.8-14**). The linkages

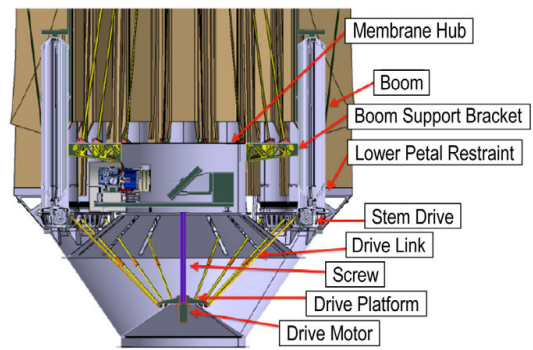

**Figure 5.8-14.** Details of the boom deployment mechanism. The 16 booms are driven simultaneously by a single drive motor, which drives a platform vertically up a worm gear. Linkages between the platform and the booms generate the desired angular deployment.

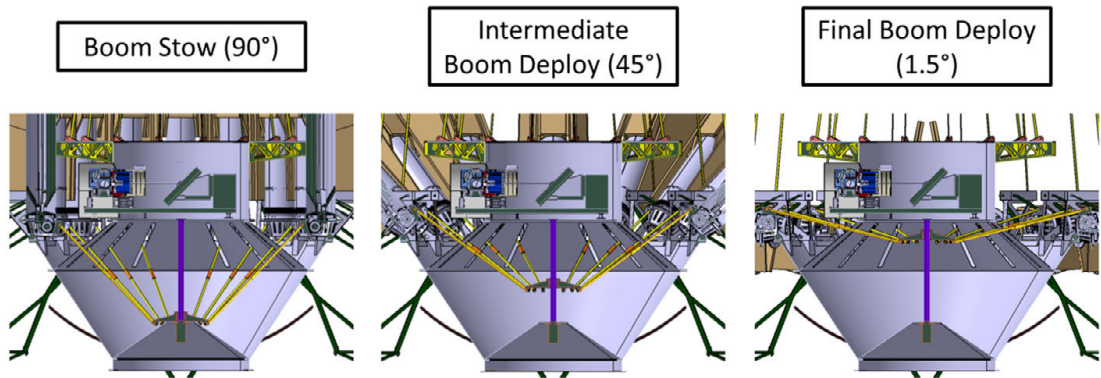

**Figure 5.8-15.** Snapshots of the angular boom deployment. Initially (left) the booms are in a vertical stow position. As the platform is driven up the worm gear, the booms move angularly downward until they reach the final deployed position, 1.5 degrees above horizontal (center, right)

are connected to a drive platform, which is driven up a worm gear by the drive motor. As the platform moves upward, the booms are deployed angularly downward. This single motor option guarantees the synchronization of the unfolding step, controlling the inter-petal stresses. It also has a mechanical advantage, as the centralized tower system with deployment linkages allows back driving of the deployment if necessary, and holds steady at its final position, precluding the need for any positive latching of the deploying boom or STEM elements. Snapshots of the angular deployment are provided in **Figure 5.8-15**.

### 5.8.3.2.4    Step Four: STEM Deployment (Telescoping Boom Extension)

After the telescoping booms are driven to their final angle, the petals are fully extended by driving the STEM motors. **Figure 5.8-16** illustrates this deployment step for a single petal. As the telescoping boom extends, the petal edge unfolds naturally, creating the required petal shape that makes up the starshade outline. Once fully extended, the petal systems latch into place via latching hinges. This latching makes a ridged structural petal edge, helping achieve the desired petal shape within the required accuracy.

Tension-links that run from the petal valley location to the central hub of the starshade are pulled out by the deploying booms in order to generate a positive tension within the starshade petal frame, as well as to control the MLI during deployment. Future trade studies will determine if more degrees-of-freedom at the petal valley (i.e., the petal-to-petal joint) will need to be controlled.

After this motion is complete, the starshade forms a tensegrity structure, with the telescoping booms in compressions and the petal edges and 3 bar links in tension. Each petal is also in tension circumferentially, transferring from one to the next at the mechanical valley joint.

### 5.8.4    Formation Flying

Formation flying—defined as two or more spacecraft autonomously controlling relative position or attitude based on inter-spacecraft

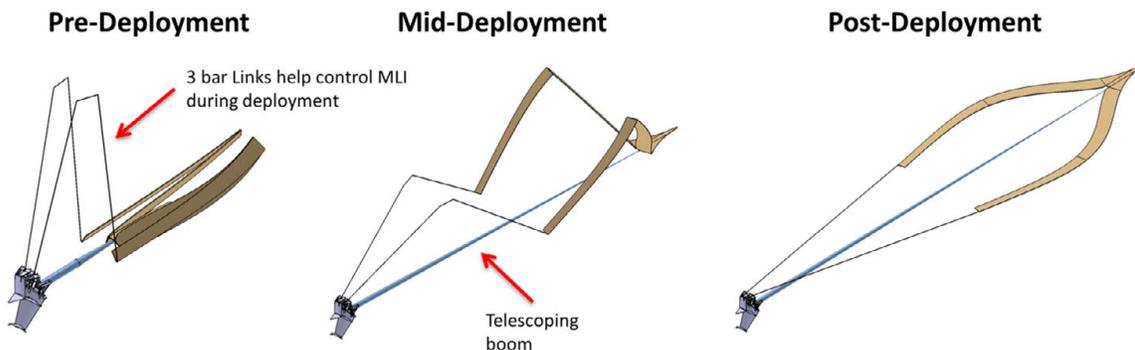

**Figure 5.8-16.** Snapshots of the STEM deployment for a single petal. As the telescoping boom extends, the petal outline unfolds. Once full extension is achieved, petal hinges lock in place in order to hold the desired shape.

measurements—is needed for HabEx to align the starshade and telescope for science observations, and to repoint this synthesized observatory at a new target star. Repointing is done by translating the starshade relative to the telescope, which is referred to as "retargeting."

The overall concept of operations (CONOPS) for formation flying, and the accompanying translational control requirements, are shown in **Figure 5.8-17**. This CONOPS heavily leverages the extensive studies and engineering analyses that were performed for Exo-S (Worlds 2010), and that are being performed for a potential starshade mission that would rendezvous with the WFIRST telescope. Each of the operational modes—initialization, acquisition, science, and retargeting—are discussed subsequently.

The formation flying architecture for HabEx shown in **Figure 5.8-17** has the starshade maneuvering relative to the telescope. This arrangement allows the telescope to perform independent science during the days to weeks required for starshade retargeting. Additionally, this architectural choice results in the so-called Leader/Follower formation control architecture that is commonly used in rendezvous and docking in low Earth orbit (LEO) and which makes control design, and stability and performance analyses straightforward (Murphey 2009).

The driving requirement for formation flying is to align the starshade to within 1 m radially, that is, laterally, of the telescope-star line at spacecraft separations of 69,000–186,000 km for science mode. While aligning two spacecraft to 1 m at a separation of up to 186,000 km appears daunting, this formation flying problem is more tractable than might be expected for the following two reasons. First, even though the inter-spacecraft distances are immense, the relative dynamics remain benign: the gravity gradient at Sun-Earth L2 and the maximum planned separation is less than approximately 1e-4 m/s² (Seager et al. 2015) and differential solar radiation pressure is orders of magnitude smaller. The HabEx gravity gradient is equivalent to that experienced at tens-of-meter separations in low-Earth orbit, which is similar to the berthing distance used at the International Space Station (ISS). Although the telescope and starshade are far apart, they are not "flying apart."

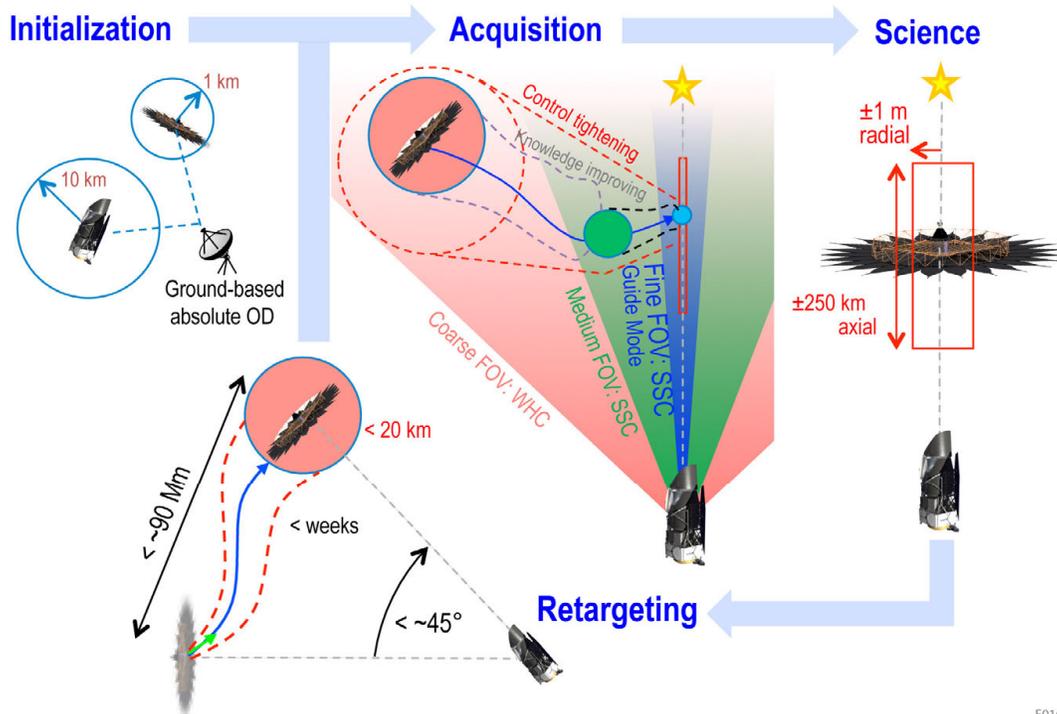

**Figure 5.8-17.** Starshade concept of operations (CONOPS) for formation flying.

This observation regarding thruster-firing intervals leads to the second reason that formation flying is tractable for HabEx, namely, controlling formation flying spacecraft to the sub-meter level is commonly done for rendezvous and docking with the ISS. A typical radial control requirement for the terminal docking phase is 10 cm (Scharf, Hadaegh, and Ploen 2004). For example, ESA's Automated Transfer Vehicle, which has a mass around 20,000 kg, controls to 10 cm. Formation control to 1 m is not only tractable but also already commonly demonstrated in flight.

The principal challenge for HabEx formation flying is *sensing* the lateral offset of the starshade from the telescope-star line to a fraction of 1 m at hundreds of megameters, while the starshade is obstructing the star. A HabEx technology gap (see **Table 6-1**) formalizes this challenge as sensing the lateral offset of the starshade to 0.2 m.

Note that inter-spacecraft range measurements with ~1 km precision will be made by a radio frequency (RF) inter-spacecraft link (ISL) that also provides low-bandwidth communication for coordination. This RF link is not considered a technology challenge.

### 5.8.4.1 The Principal Formation Challenge: Fine Lateral Sensing

There has been extensive work on solving this lateral sensing challenge, which is referred to as fine lateral sensing. A short survey and further references can be found in Sirbu, Karsten, and Kasdin (2010). The general approach is to utilize the telescope's primary mirror and the light of the target star that "leaks" around the starshade outside the wavelength bands for science, where the attenuation of the target star is only on the order of 1e-3.

The fine lateral sensor uses the starshade instrument in what is referred to as "guide" mode, producing pupil-plane images. The sensor's pupil-plane images are matched using least squares to a library of pre-generated images of the "shadow structure" of the starshade. Image matching is done on the telescope and the resulting lateral offset is sent to the starshade over the ISL. NASA's Starshade Technology Project (S5), managed by the Exoplanet Exploration Program Office, plans to demonstrate this sensing approach to TRL 5 by the end of FY2020.

Current analyses show that performance of 15 cm 3σ is possible with ~5 s exposures of an 8th magnitude star when sensing in UV and operating within 1 m of alignment (Kelly and Cryan 2016). Sensing in the UV is done when science is being done in the IR. Conversely, when doing science in the visible or UV, sensing is done in the IR. Expected target stars have lower flux at UV wavelengths and instrument losses are greater. Hence, a 5 s exposure is considered worst case; tenths of a second is more typical. Even so, since thruster firings are needed only on the order of every hundreds of seconds, many formation measurements can be made, thereby improving relative velocity knowledge for efficient formation control.

Even in areas of low shadow structure (e.g., only smooth gradients), image matching can still be performed to ~25 cm 3σ. As a result, the pupil-plane image-matching fine lateral sensor can be used out to lateral offsets equal to the radius of the starshade (~36 m).

Since the fine lateral sensor functions only when the starshade is within ~36 m of alignment, the acquisition mode is needed to move the starshade from the end of retargeting and initialization modes to within 1 m of alignment for science mode.

### 5.8.4.2 Initialization Mode

When the starshade first rendezvouses with the telescope after launch, it would be operated from the ground. Similarly, if during regular operations, either the telescope or the starshade enters safe mode and relative position knowledge is lost, the ground would recover the spacecraft. In both cases, the ground tracks both spacecraft and determines a trajectory for the starshade to align between the telescope and the target star at the desired distance. The trajectory is uploaded to, and executed by, the starshade.

The position of the telescope is determined via standard tracking methods to better than 10 km 3σ (Scharf et al. 2016), which can require ~30 min of tracking a day for 1–2 weeks; if Delta Differential One-way Ranging (DDOR) is used, telescope position can be established in several days.

The position of the starshade can also be determined from the ground with much greater accuracy by utilizing the laser beacon carried by the starshade for use in acquisition mode. A visible, 1 W laser beacon with a 2.5 deg angular spread ensures the starshade appears as at least a 15th magnitude star at Earth while the 2.5 deg FOV subtends 5 Earth radii from Earth-Sun L2. Using the DSN radar to range to the starshade and astrometry with even just meter-class ground-based telescopes, the position of the starshade can be determined to 1 km 3σ in under an hour (Bottom et al. 2017). The relative position uncertainty is then the root sum-square of 10 km and 1 km, which is effectively 10 km. On the HabEx telescope, this level of uncertainty is sufficient to guarantee that the HWC will see the laser beacon of the starshade when it is pointed at the target star. The HWC FOV is 3'×3' (±30 km) at the minimum range of 69,000 km.

### 5.8.4.3 Retargeting Mode

At the end of the science mode, the lateral position and velocity of the starshade is known to better than 5 cm and 1 mm/s, and the axial position and velocity to better than 15 m and 4 cm/s, respectively (Truong, Cuevas, and Slojkowski 2003). This relative state knowledge is the initial condition for the retargeting trajectory.

The retargeting trajectory is planned on the ground and uploaded to the starshade. Using SEP, the starshade executes the planned trajectory. When necessary, intermittent ground tracking and update of the retargeting trajectory will be executed as the starshade traverses to the next target. However, mid-course tracking will require pointing the laser beacon at Earth, which may briefly interrupt thrusting.

Retargeting mode concludes with the starshade decelerating into its next observing position. After completing its deceleration, the starshade and its laser beacon should be within tens-of-kilometers of lateral alignment, ready for acquisition by the FOV of the HWC.

### 5.8.4.4 Acquisition Mode

At the end of initialization or retargeting, the starshade laser beacon is turned down to tens of milliwatts to match the expected flux of the target star, and the starshade points the laser beacon at the telescope. The HWC images the laser beacon and the unobstructed target star on the same detector, producing bearing measurements with a resolution of ~1e-7 radians (a 7.3 m offset at 69,000 km).

The lateral position and velocity of the starshade will be known to better than 5 m and 1 m/s, with less than an hour of measurements. This level of knowledge is enough to adjust the starshade velocity with chemical thrusters to achieve target final alignment. Since the bipropellant thrusters are placed at various angles and can generate a thrust vector in any directions, the starshade does not need to reorient for this adjustment. A change in velocity of 0.5 m/s is enough to move the starshade to alignment in 5 hours or less.

Any errors that accumulate (e.g., inexact knowledge of the gravity gradient or solar pressure) will be corrected using model predictive control throughout the acquisition mode. This control approach is illustrated in **Figure 5.8-18** (e.g., Morari, Garcia, and Prett 1988). After the thruster firing, HWC measurements continue, and relative position and velocity knowledge improve. If the starshade drifts sufficiently far off the planned trajectory, additional, much smaller adjustments are applied. The subsequent velocity changes are more efficient because both the relative state knowledge and the estimated differential acceleration between the starshade and telescope have improved.

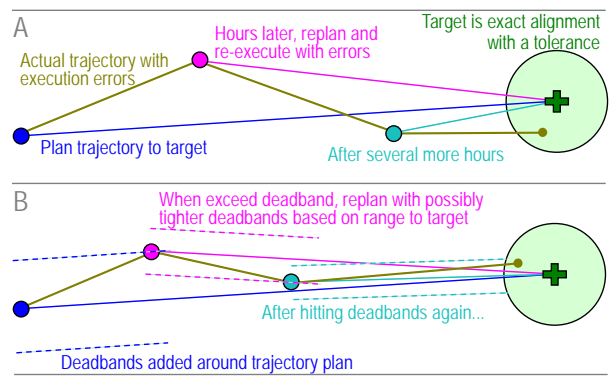

**Figure 5.8-18.** Example model predictive control approaches for controlling the starshade during acquisition.

When the starshade passes to within 2 km of alignment, the telescope slews to place the starshade instrument on the target star. The starshade instrument half-FOV is 2 km at 69,000 km. The HWC's resolution is sufficient to steer the starshade to within 1 km of alignment—easily within the starshade instrument's FOV. While in visible imaging mode, the starshade instrument, like the HWC, measures the bearing between the starshade laser beacon and the target star on the same detector. The resolution of the starshade instrument is 5.5 m at 69,000 km and 13.9 m at 186,000 km.

The starshade instrument measurements and model predictive control continue as the starshade approaches to within 36 m of alignment. At this point, the laser beacon is deactivated and the starshade instrument switches to guide mode, becoming the fine lateral sensor. While the starshade instrument resolution prior to switching to guide mode is ~14 m worst-case, previous analyses for Exo-S indicate that the position knowledge is generally 5 times better with an estimator. Estimator knowledge of ~3 m is sufficient to steer to within 36 m of alignment.

Once the fine lateral sensor acquires the starshade (indicated by a pupil-plane image match with low residual), model predictive control continues to steer the starshade until, finally, it is within 1 m of alignment, at which point science mode begins.

Autonomous logic is needed on the starshade to transition between sensors, switching estimators as sensors hand-off and coordinating with the telescope. Previous missions have demonstrated complex, sensor-based mode logic, such as Mars Science Laboratory and Orbital Express.

### 5.8.4.5   Science Mode

With the starshade within 1 meter of alignment, science observations can be performed. During science mode, the starshade instrument provides measurements to 15 cm $3\sigma$. These measurements are used to estimate the relative position and velocity of the starshade and the differential acceleration between the starshade and telescope. When the starshade approaches the 1 m alignment limit, thrusters fire to correct the alignment. The starshade moves back into alignment until the gravity gradient and other environmental factors eventually move the starshade back to the 1 m limit, causing the cycle to repeat. This control method creates a "one-sided" dead-band where the starshade moves in a ballistic trajectory within the 1 m alignment limit. Once within the 1 m alignment circle, the starshade fires thrusters, imparting just enough velocity to reach the other edge while opposing gravitational and other environmental forces. These environmental forces return the starshade to its approximate starting position where the cycle repeats. Each time the starshade fires its chemical thrusters to formation-keep the observation must be suspended to avoid corruption by the brightness of the thruster plumes. Therefore, thruster firing is coordinated between the telescope and starshade so that the observation data can be protected. As a result, maximizing the drift time within the 1 m alignment zone reduces the overall time spent in science mode.

**Figure 5.8-19** shows an example of optimal dead-banding within the formation control requirement circle for the science mode. As an example, consider the starshade coasting into the final 1 m of alignment at point 1. As the starshade coasts across the circle, the estimate of the differential acceleration is continually updated. At point 2, the formation control algorithm uses its estimate at that time, $\mathbf{a}_1$, to fire the thrusters,

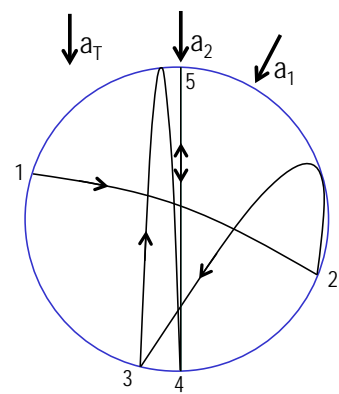

**Figure 5.8-19.** Example of optimal deadbanding for the starshade Science mode.

targeting a drift to the "bottom" of the circle as indicated by the vector $\mathbf{a}_1$. The true differential acceleration in this example is the vector $\mathbf{a}_T$. During the ensuing coast, the differential acceleration estimate improves to $\mathbf{a}_2$.

When the boundary of the control requirement is encountered again at point 3, the thrusters are fired to coast to the bottom of the circle as indicated by $\mathbf{a}_2$. For the purposes of this example, assume $\mathbf{a}_2$ is close to $\mathbf{a}_T$. Thereafter, thrusters are fired to traverse the diameter of the circle aligned with $\mathbf{a}_T$ from point 4 to point 5 and back to point 4. The departing velocity at the bottom of the circle is sized to bring the starshade to zero relative velocity at the "top" of the circle and then "fall" back down again.

To execute the thruster firings, a thrust allocator uses the configuration of thrusters and the current estimate of the spacecraft attitude to compute thruster firing durations that give the desired force impulse. Thrust allocators function with a spinning starshade as well. An example thruster configuration is shown in **Figure 5.8-20**. Thrust allocators also handle motion of the starshade center of mass as propellant is expended.

### 5.8.4.6 Summary

While formation flying to 1 meter at separations of up to ~14 Earth diameters initially appears daunting, the relative dynamics are similar to tens-of-meters separation in LEO, and the control performance has been previously demonstrated by much larger spacecraft in Earth orbit docking with the ISS. The principal challenge is sensing the lateral offset of a starshade from a target star to tens-of-centimeters while the target star itself is obscured by the starshade.

Several lateral sensing approaches exist, and NASA's Starshade Technology Project is maturing one approach to TRL 5 by the end of FY2020. The pupil-plane image-matching approach being matured does not require a laser beacon for fine guiding, has performance better than required with just seconds of exposure, functions even with secondary obscuration, and provides measurements from the edge of the starshade to the center.

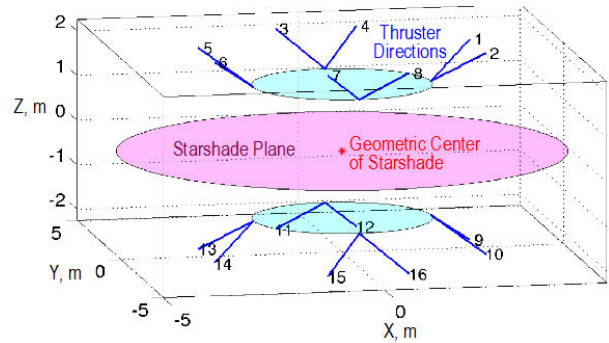

**Figure 5.8-20.** Example thruster configuration. Thruster identification numbers and thrust vectors are shown.

## 5.9    Starshade Bus

This section describes the key design features of each of the starshade bus subsystems. **Table 5.9-1** presents the starshade mass breakdown; the total dry mass is estimated to be 6394 kg, including 27% average contingency and an additional 9% system margin. Wet mass with contingency is 13,401 kg.

### 5.9.1    Starshade Mechanical Design Overview

The HabEx starshade consists of a 40 m diameter disk surrounded by 24 petals with a tip-to-tip diameter of 72 m. As illustrated in **Figure 5.9-1**, each petal is 16 m long with a 5.25 m wide base and a maximum width of 5.8 m. The starshade has a 4.6 m diameter while stowed. This allows it to be launched separately from the telescope using a smaller faring. The starshade spacecraft bus is in the center of the starshade, fully within the hub. The total CBE mass of the petal and disk system (excluding the hub) is 2,520 kg, with an additional 500 kg allocated to the PLUS deployment system, which will be jettisoned after deployment. See Section 5.8.2 for details on the starshade mechanical design.

**Table 5.9-1.** Starshade flight system mass estimate. CBE: current best estimate. MEV: maximum expected value.

| | CBE (kg) | Cont. % | MEV (kg) |
|---|---|---|---|
| **Payload** | | | |
| Starshade Petals and Disk (no hub) | 2,520 | 30% | 3,276 |
| **Spacecraft Bus** | | | |
| ACS | 15.8 | 8% | 17.1 |
| Communications & Data Handling | 16.1 | 19% | 19.3 |
| Power | 248.4 | 28% | 317.6 |
| Prop Biprop | 140 | 3% | 144.4 |
| Prop SEP | 848.3 | 21% | 1,026.8 |
| Structures & Mechanisms | 636.2 | 30% | 827.1 |
|   Spacecraft side adaptor | 26.5 | 30% | 34.5 |
| Cabling | 87.8 | 30% | 114.1 |
| Telecom | 27.5 | 18% | 32.4 |
| Thermal | 130.2 | 30% | 169.3 |
| **Bus Total** | **2,180.4** | **24%** | **2,707.1** |
| **Spacecraft Total (dry)** | **4,700.4** | | |
|   Subsystem heritage contingency | 1,282.7 | | |
|   System contingency | 410.9 | | |
| **Spacecraft with contingency (dry)** | **6,394** | | |
|   Biprop and pressurant | 1,406.8 | | |
|   Xenon and pressurant | 5,600.0 | | |
| **Total Spacecraft Wet Mass** | **13,401** | | |
|   Launch vehicle side adaptor | 80 | | |
|   PLUS system | 500 | | |
| **Total Launch Mass** | **13,981** | | |

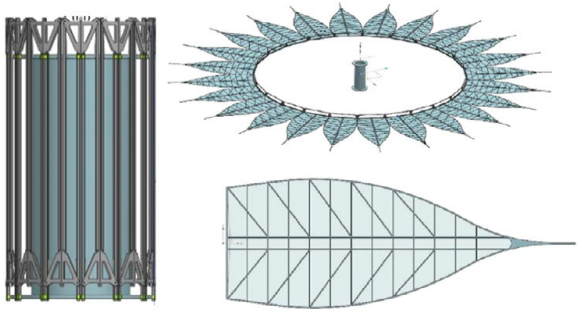

**Figure 5.9-1.** The HabEx starshade baseline configuration.

### 5.9.2 Bus Structures and Mechanisms

The starshade hub structure was designed to fully contain the spacecraft bus subsystems as well as the hydrazine and Xenon propellants. The hub is a 4.6 m diameter honeycomb cylindrical structure with reinforced aluminum rings to prevent buckling and joints for reinforcements at attachment points. The inner core of the hub is 1.88 m in diameter. The starshade's hub structure is designed to attach directly to the launch vehicle's adapter ring, providing the best possible load path during launch.

Aside from the starshade payload deployment mechanisms, no additional mechanisms are needed on the flight system.

### 5.9.3 Thermal

The goal of the thermal subsystem is to maintain the starshade's bus subsystems' temperatures within their allowable flight temperatures (AFTs). The starshade petals were designed to be passively thermally stable and do not need any active heating. See Section 5.8.2.5 for details on the starshade payload thermal design and analysis.

The driving design issue for the bus thermal system is maintaining the temperature of the propellant tanks within operational limits. Approximately 150 W of heater power would be required to maintain the propellant temperatures during normal operations. However, only 30 W of make-up heater power would be required during the launch, downlink, and safe modes. The thermal design is an actively controlled system using thermistors to sense tank and subsystem temperatures, and strip heaters to add heat when needed. The subsystem also includes multi-layer insulation (MLI) blankets, a set of variable conductance heat pipes to redistributed waste heat, and a 3.8 m$^2$ radiator to dissipate excess heat.

### 5.9.4 Power

The starshade baseline power system contains two different power strings: a 28 V string to power the bus and a 600 V string for the SEP system. It also contains 2 different types of arrays: a 28 V rigid array on top of the hub, which is used to power the bus before the starshade is deployed, and a flexible array mounted directly onto the starshade inner disk. The majority of the flexible cells are strung together to form a 33.3 kW high-voltage array to power the SEP system (a single thruster requires 12.5 kW). The pointing range for the starshade is 40 to 83 degrees from the spacecraft-Sun line and the bus power requirement is approximately 1,000 W. Consequently, a set of 28 V strings are also needed on the starshade to provide additional bus power during worst-case science operations.

Battery sizing is set by the launch-phase power requirements, where it is assumed that the bus will be powered by batteries for up to 3 hours. The starshade requires two 66 Ah Lithium ion batteries to avoid the battery depth-of-discharge dipping below 70% during that period of time.

### 5.9.5 Propulsion

The starshade bus would possess a hybrid propulsion system: a bipropellant chemical system for TCM-1 and -2, station-keeping within the formation flying box, and pointing, and a SEP system for retargeting. The chemical propulsion system carries twelve 22 N thrusters that are responsible for TCMs and station-keeping.

The SEP thrusters provide 0.52 N of thrust each and have an ISP of 3,000 seconds. Because the flexible array must be illuminated for the SEP engines to operate and the starshade must be able to translate in any direction in order to meet the observation needs, there must be functional SEP engines on each side of the starshade. Two engines are required to achieve sufficient thrust for efficient retargeting. Best practices require that an additional engine be flown on each side in case of failure for a total of six SEP engines needed to support the HabEx mission.

Assuming a 6,394 kg dry mass, the 1,407 kg of bipropellant and 5,600 kg of xenon will yield at least 100 targeted starshade observations. Additionally, the propulsion system was sized to meet the mission's 5-year requirement and was also designed to be refuelable.

### 5.9.6    Attitude Determination and Control

The attitude determination and control subsystem (ADCS) requirements for the starshade are presented in **Table 5.9-2**. In addition to these requirements, the starshade must also carry a laser beacon to support formation flying (see Section 5.8.4 for details).

The baseline starshade bus is currently designed to provide spin-stabilization; the alternative starshade bus is three-axis stabilized. Trade studies are ongoing to assess which approach will work better with a starshade of this size and will be presented in the HabEx final report.

Attitude determination is achieved with star trackers and gyros, including additional gyros and Sun sensors as backup. Once the starshade is within sensor range of the telescope, formation flying control takes over to maintain position relative to the telescope.

**Table 5.9-2.** Starshade ADCS requirements.

| Parameter | Value |
|---|---|
| Control | ±1° (3σ) zero-to-peak |
| Knowledge | 0.5° (3σ) zero-to-peak |
| Stability | ±1 arcsec/sec (3σ) per axis |

### 5.9.7    Telecommunication

The starshade does not directly generate any science products. Its telecommunication requirements are driven by its needs to communicate to the ground for commanding and ranging in X-band (1 kbps downlink requirement) and to communicate with the telescope in S-band for data transfer (100 bps) and ranging. The starshade telecommunication system would therefore be an exact replica of the telescope system, but without the Ka-band capability. It would be fully redundant and carriy two universal space transponders (UST), two X-band low-gain antennas, and an S-band patch antenna. This system would easily meet the HabEx starshade downlink and cross-link requirements, with at least 6 dB of margin in all operational cases.

### 5.9.8    Command & Data Handling and Flight Software

To help reduce cost, the starshade CDH subsystem would be an exact replica of the telescope CDH, which is described in Section 5.6.7. The starshade flight software would be somewhat simpler than that of the telescope since it only has a single deployment and lacks science instruments. However, the formation flying requirements, the spacecraft cross-link communication, and the ADCS approach will all require some customization from JPL core software products. Nonetheless, the development risk associated with this type of software is expected to be low.

### 5.10    System Integration

Many aspects of the HabEx integration and test (I&T) activities are like those found on other astrophysics missions. One or two commercial bus vendors are responsible for the bus integrations, spacecraft integrations and launch support. The telescope provider handles not only telescope testing but also end-to-end optical testing with the imaging instruments, leveraging established procedures, and existing test facilities and support equipment.

There are, however, several unique characteristics of the HabEx concept that need to be addressed in the integration and test planning. The first consideration is that not all requirements will be verifiable by test (such as end-to-end system functional performance with a starshade); some will be handled by analyses, modeling, or simulations. While not ideal, this condition has been addressed successfully in a number of past situations (e.g., Mars atmospheric deceleration, primary mirror 0 g relaxation, spacecraft docking, etc.). For HabEx, formation flying cannot be tested in an 'as-you-fly' configuration on the ground since the spacecraft are separated by distances ranging from 69,000 km to 186,000 km. Formation flight components (e.g., ranging radios, beacons and thrusters) can be tested in routine spacecraft subsystem tests. But the primary formation flight challenge lies with sensing lateral performance errors which cannot be tested in a flight configuration and must be verified through modeling and simulation.

The formation flight algorithms and software, leveraging proto-flight software developed and ground-demonstrated for StarLight, Terrestrial Planet Finder Interferometer (TPF-I), and PROBA-3 (PRoject for OnBoard Autonomy-3) (Scharf et al. 2016), will be fully exercised in a Control Analysis Simulation Testbed (CAST) and a flight software (FSW) testbed. All functionality and performance requirements are first verified in CAST, then re-verified in the FSW testbed under flight-like operating conditions.

Another HabEx I&T challenge lies with the integration and deployment testing of the starshade. At 72 m in diameter, most flight integration facilities cannot support the fully deployed starshade. The vendor performing the starshade payload I&T must have the requisite test facility for deployment. Testing is analogous to large deployable antenna systems, requiring similar gravity compensation fixtures.

HabEx has just begun developing the integration and test material for the baseline concept. A more detailed discussion, covering the telescope, coronagraph, UVS, HWC, starshade, and formation flying, will be given in the final report.

## 5.11 Serviceability

All future large space-based observatories are congressionally mandated to be serviceable. Given the cost and time required to construct these facilities, and their importance to science advancement and the nation, having the capability to extend their useful life should be part of the baseline design. Both spacecraft are serviceable on HabEx. The telescope spacecraft would follow the WFIRST model; it would be refuelable, and instruments and avionics would be replaceable. The starshade spacecraft would be refuelable and have accessible avionics to facilitate replacement. Details of the HabEx servicing design are currently in development and will be presented in the HabEx study final report.

# 6   Technology Maturation

The HabEx design evolved from a strategy of choosing mature technologies and minimizing risk to meet the requirements set by the top-level science goals (see Section 4). The design team sought to make extensive use of the state of the art and as high a Technology Readiness Level (TRL) as possible to minimize risk. **Table 6-1** summarizes the HabEx technology challenges, including the TRL expected by the end of 2019. Many of the enabling technologies are at, or expected to be at, TRL 5 by 2019 and the remaining technologies at TRL 4. One starshade-related technology, currently at TRL 3, is expected to mature to TRL 5 by 2022 as part of the Starshade to TRL 5 (S5) project. Technology roadmap flow-plans and completion dates to TRL 5 are included in Appendix E.

Two mission-enhancing technologies are also considered. These technologies are not required but can enhance the science performance. HabEx observatory science could be enhanced by extending the ultraviolet (UV) wavelength down to 0.1 μm if a compatible mirror coating is available. In addition, delta-doped charge-coupled devices (CCDs) capable of operating in the UV could simplify UV instrument designs. If these technologies mature in time to be included in HabEx instruments, they would be worthwhile additions to the mission performance.

## 6.1   Notes on the Starshade

Unlike the coronagraph instrument and other HabEx technologies, system-level performance verification (e.g., contrast, inner working angle [IWA], observatory alignment) is simply not possible for the starshade due to the starshade/telescope separation distance needed for flight-like testing. HabEx will qualify the starshade system based on performance model validation and key subsystem ground testing. These starshade technology gaps and maturation plans were first identified in the NASA-commissioned, community-led, Exo-S study report (Seager et al. 2015). They have since been adopted by the NASA Exoplanet Exploration Program (ExEP) and are now tracked in the ExEP Technology Plan Appendix. These five gaps are:

1. Optical performance verification and model validation (Section 6.3.1)
2. Solar edge scatter (Section 6.4.1)
3. Formation sensing and control (Section 6.4.2)
4. Petal shape control and stability (Section 6.2.2)
5. Petal deployment accuracy (Section 6.2.1)

Since the Exo-S study, ExEP has established the S5 technology development task to mature these technologies for a notional 26 m starshade mission. In doing so, S5 matures the majority of the starshade gaps for the HabEx mission as well. Each of the technology development areas—and their connections to S5—will be discussed in the identified subsections.

### 6.1.1   Background on S5

In November 2016, the Starshade Readiness Working Group recommended to the NASA Astrophysics Director a plan to validate starshade technologies "that is both necessary and sufficient prior to building and flying" a starshade science mission. With the full concurrence of an independent Technical Advisory Committee, it was determined that "a ground-only development strategy exists to enable a starshade science flight mission" and "a prior flight technology demonstration is not required" (Blackwood 2016).

At the core of the S5 activity, "starshade shape accuracy and stability requirements are derived from a comprehensive error budget that will be verified by mechanical and optical performance models anchored to subscale ground tests." These performance models address the five technologies listed above (Ziemer 2018).

The S5 reference baseline design of a future flight starshade is 26 meters in diameter with a 10-meter diameter disk and 8-meter-long petals.

**Table 6-1.** HabEx 4 m Baseline Architecture Technology Gap List.

| Title | Description | Section | State of the Art | Capability Needed | ExEP Assessed 2017 TRL | Expected 2019 TRL |
|-------|-------------|---------|------------------|-------------------|------------------------|-------------------|
| **Technology Challenges for the 4-Meter Architecture** | | | | | | |
| Starshade Petal Deployment Position Accuracy | Deploy and maintain petal position accuracy in L2 environment | 6.2.1 | • Petal deployment tolerance (≤0.15 mm) verified with multiple deployments of 12 m flight-like perimeter truss and no optical shield<br>• No environmental testing | • Petal radial deployment accuracy on 40 m perimeter truss: ±500 µm (3σ) bias<br>• Position stability in operational environment: ±1.5 mm (3σ) random | 3 | 3 |
| **Technologies Approaching TRL 5** | | | | | | |
| Starshade Petal Shape and Stability | Starshade petal shape maintained after deployment, thermal at L2 | 6.3.1 | • Manufacturing tolerance (≤100 µm) verified with low fidelity 6 m prototype<br>• No environmental tests<br>• Petal deployment tests conducted on prototype petals to demonstrate rib actuation<br>• No deployed petal shape measurements | • Petal shape manufacture: ±170 µm (3σ)<br>• Postdeployment 16 m petal shape demonstrated to ≤ ±156 µm (3σ)<br>• Stability (thermal): disk to petal strain ≤30 ppm, 1–5 cycle/petal width ≤20 ppm | 3 | 4 |
| Large Mirror Fabrication | Large monolith mirror that meets tight surface figure error and thermal control requirements at visible wavelengths | 6.3.2 | • Schott demonstrated computer-controlled-machining lightweighting to pocket depth of 340 mm, 4 mm rib thickness (E-ELT M5)<br>• State-of-the-practice (SOP) lightweighting has yielded large mirrors of aerial density 70 kg/m²<br>• Zerodur® can achieve 5 parts per billion/K CTE homogeneity<br>• Wavefront stability: ~10 nm rms | • Current state-of-the-art lightweighting is sufficient to meet SLS launch capabilities<br>• Wavefront thermal stability of ~1 nm rms between consecutive low-order wavefront updates which are 100 s of seconds apart<br>• First mode ≥ 60 Hz | 4 | 4 |
| Large Mirror Coating | Mirror coating with high spatial uniformity over the visible spectrum | 6.3.3 | • IUE, HST, and GALEX used MgF₂ on Al to obtain >70% reflectivity from 0.115 µm to 2.5 µm<br>• Reflectance non-uniformity <0.5% of protected Ag on 2.5 m TPF Technology Demonstration Mirror<br>• Operational life: >28 years on HST | • Reflectivity from 0.115–1.8 µm<br>• Reflectance uniformity <1% over 0.45–1.0 µm<br>• Operational life >10 years | 4 | 4 |
| Coronagraph Architecture | Suppress starlight by a factor of ≤1E-10 at visible and near-IR wavelengths | 6.3.4 | • Hybrid Lyot: $6 \times 10^{-10}$ raw contrast at 10% bandwidth across angles of 3–16 λ/D demonstrated with a linear mask and an unobscured pupil in a static vacuum lab environment<br>• Vector vortex charge 4: $5 \times 10^{-10}$ raw contrast monochromatic across angles of 2–7 λ/D demonstrated with an unobscured pupil in a static vacuum lab environment | • Raw contrast ≤$1 \times 10^{-10}$<br>• Raw contrast stability of ≤$2 \times 10^{-11}$<br>• IWA ≤ 2.4 λ/D<br>• Coronagraph throughput ≥10%<br>• Bandwidth ≥20% | 4 | 4 |

| Title | Description | Section | State of the Art | Capability Needed | ExEP Assessed 2017 TRL | Expected 2019 TRL |
|---|---|---|---|---|---|---|
| **Technologies Approaching TRL 5, continued** | | | | | | |
| LOWFS | Sensing and control of low-order wavefront drift | 6.3.5 | • <0.5 mas rms per axis LOS residual error demonstrated in lab with a fast-steering mirror attenuating a 14 mas LOS jitter and reaction wheel inputs; ~100 pm rms sensitivity of focus (WFIRST Coronagraph Instrument Testbed)<br>• WFE stability of 25 nm/orbit in low Earth orbit. Higher low-order modes sensed to 10–100 nm WFE rms on ground-based telescopes | • LOS error < 0.5 mas rms<br>• Wavefront stability:≤~100 pm rms over 1 second for vector vortex<br>• WFE <0.76 nm rms | 3 | 4 |
| Deformable Mirrors | Flight-qualified large-format deformable mirror | 6.3.6 | • Micro-electromechanical DMs available up to 64 × 64 actuators with 6 nm RMS flattened WFE<br>• Smaller DMs supported coronagraph demonstrations of 2×10⁻⁷ raw contrast at 10% bandwidth in a static test | • 64×64 actuators.<br>• Enable coronagraph raw contrasts of ≤1×10⁻¹⁰ at ~20% bandwidth and raw contrast stability ≤2×10⁻¹¹ | Not assessed | 4 |
| Starshade Edge Scattering | Limit edge-scattered sunlight and diffracted starlight with optical petal edges | 6.3.7 | • Machined graphite edges solar glint flux: 25 visual magnitudes in two main lobes<br>• Metal edges meet all specs but in-plane shape tolerance | • Petal edge in-plane shape: 40 µm<br>• Solar glint: 28 visual magnitudes in two main lobes | 3 | 4 |
| **Technologies at TRL 5 or Higher** | | | | | | |
| Starshade Modeling and Performance | Validate at flight-like Fresnel numbers the equations that predict the contrasts | 6.4.1 | • 6×10⁻⁶ suppression in pupil at F1.0 ~15<br>• 6×10⁻⁸ suppression in pupil, 2.5×10⁻¹⁰ contrast demonstrated at F1.0 ~27 (monochromatic) | • Experimentally validated models with total starlight suppression ≤ 1E–8 in scaled flight-like geometry, with F1.0 between 5 and 40 across a broadband optical bandpass. Validated models are traceable to 1E-10 contrast system performance in space | 3 | 5 |
| Starshade Lateral Formation Sensing | Lateral formation flying sensing to keep telescope in starshade's dark shadow | 6.4.2 | • Simulations have shown centroid star positions to ≤1/100th pixel with ample flux to support control loop<br>• Sensing demonstration of lateral control has not yet been performed | • Demonstrate sensing lateral errors ≤0.20 m accuracy at scaled flight separations (≤1 mas bearing angle)<br>• Control algorithms demonstrated with scaled lateral control errors corresponding to ≤1 m | 4 | 5 |
| Microthrusters | Jitter is mitigated by using microthrusters instead of reaction wheels during exoplanet observations | 6.4.3 | • Colloidal microthrusters 5–30 µN thrust with a resolution of ≤ 0.1 µN , 100 days on-orbit on LISA-Pathfinder<br>• Cold-gas micronewton thrusters flown on Gaia (TRL 9), 0.1 µN resolution, 1 mN max thrust, 4 years of on-orbit operation | • Thrust capability: 0.35 mN<br>• Operating life: 5 years | 3 | 5 |

| Title | Description | Section | State of the Art | Capability Needed | ExEP Assessed 2017 TRL | Expected 2019 TRL |
|---|---|---|---|---|---|---|
| **Technologies at TRL 5 or Higher, continued** | | | | | | |
| Laser Metrology | Sensing for control of rigid body alignment of telescope front-end optics | 6.4.4 | • Thermally stabilized Planar Lightwave Circuit fully tested<br>• Nd:YAG ring laser and modulator flown on LISA-Pathfinder<br>• Phase meters flown on LISA-Pathfinder | • Sense at 1 kHz bandwidth<br>• Uncorrelated per gauge error of 0.1 nm | Not assessed | 5 |
| Delta-Doped Visible Electron Multiplying CCDs | Low-noise visible detectors for exoplanet characterization via integral field spectrograph | 6.4.5.1 | • 1k×1k EMCCD detectors (WFIRST)<br>  ○ dark current of $7 \times 10^{-4}$ e-/px/s<br>  ○ CIC of $2.3 \times 10^{-3}$ e-/px/fram<br>  ○ read noise ~0 e- rms (in EM mode)<br>  ○ Irradiated to equivalent of 6 year flux at L2<br>• 4k × 4k EMCCD fabricated with reduced performance | • 0.45–1.0 µm response;<br>• Dark current < $10^{-4}$ e-/px/s<br>• CIC < $3×10^{-3}$ e-/px/s;<br>• Effective read noise <0.1e- rms<br>• Tolerant to a space radiation environment over mission lifetime at L2<br>• 4k × 4k format | 5 | 5 |
| Linear Mode Avalanche Photodiode Sensors | Near infrared wavelength (0.9 µm to 2.5 µm), extremely low noise detectors for exo-Earth IFS | 6.4.5.2 | • HgCdTe photodiode arrays have read noise <~2 e- rms with multiple non-destructive reads; dark current <0.001 e-/pix; very radiation tolerant (JWST)<br>• HgCdTe APDs have dark current ~ 10–20 e-/s/pix, read noise <<1 e- rms, and < 1k × 1k format<br>• eAPD have 0.0015 e-/pix/s dark current, <1 to 0.1 e rms readout noise (SAPHIRA) | • Read noise <<1 e- rms<br>• Dark current <0.002 e-/pix/s<br>• In a space radiation environment over mission lifetime | 5 | 5 |
| UV Microchannel Plate (MCP) Detectors | Low-noise detectors for general astrophysics as low as 0.115 µm | 6.4.5.3 | • MCPs: QE 44% 0.115-0.18 µm with alkalai photocathode, 20% with GaN; dark current ≤0.1–1 counts/cm²/s with ALD borosilicate plates | • Dark current <0.001 e-/pix/s (173.6 counts/cm²/s), in a space radiation environment over mission lifetime,<br>• high QE for 0.115–0.3 µm wavelengths | 5 | 5 |
| Delta-Doped UV Electron Multiplying CCDs | Low-noise detectors for general astrophysics as low as 0.1 µm | 6.4.5.4 | • Delta-doped EMCCDs: Same noise performance as visible with addition of high UV QE ~ 60–80% in 0.1–0.3 µm, dark current of $3×10^{-5}$ e-/pix/s beginning of life. 4k × 4k EMCCD fabricated with reduced performance. Dark current <0.001 e-/pix/s, in a space radiation environment over mission lifetime, ≥4k × 4k format for spectrograph, high QE for 100–350 nm wavelengths | • Dark current <0.001 e-/pix/s, in a space radiation environment over mission lifetime,<br>• ≥4k × 4k format for spectrograph,<br>• high QE for 0.1–0.3 µm wavelengths | 5 | 5 |
| **Enhancing** | | | | | | |
| Far-UV Enhanced Coatings | General astrophysics imaging as low as 0.1 µm | 6.3.3.2 | • For a 0.1 µm cutoff, Al + LiF + AlF3 has been demonstrated at the lab proof-of-concept level with test coupons achieving reflectivities of 80%+ for >0.2 µm and 60% at 0.1 µm and 3-year lab environment stability | • Reflectivity from 0.3–1.8 µm: >90%<br>• Reflectivity from 0.115–0.3 µm: >80%<br>• Reflectivity below 0.115 µm: >50%<br>• Operational life: >10 years | 3 | 3 |

## 6.2 Technology Challenges for the 4-Meter Architecture

Most of the S5 TRL5 milestones will result in TRL 5 for HabEx as well. The starshade petal deployment position accuracy involves deployment of a subscale truss and petals. The current HabEx 72 m baseline design may be too great a scaling from the S5 test article to be considered TRL5. If so, an additional larger test article may be needed beyond S5. Following this interim report, HabEx will perform a trade study of smaller starshades and science performance. It is possible that a smaller starshade may be within reasonable scaling from the S5 test article and then this one remaining technology gap may be completed to TRL 5 through the S5 effort.

### 6.2.1 Starshade Petal Deployment Position Accuracy

The starshade must have the ability to stow, launch, and deploy the petals and inner disk to within the deployment tolerances budgeted to meet the shape, and ultimately, the contrast requirements. The optical shields within both the petals and the inner disk must fully deploy intact with no damage.

For the S5 activity a sub-scale flight-like structure will be built to demonstrate deployment tolerances. In April 2018, the Starshade Mechanical Architecture Trade Study presented evaluation of the NGAS deployment architecture and the wrapped (JPL) deployment architecture and recommended the wrapped architecture for S5 technology development.

The HabEx design uses the JPL wrapped petal design as the baseline, which enables the co-launch of both the starshade and the telescope flight system in a single SLS Block 1B fairing. Following the interim report, a trade study will consider co-launch and separate launch.

As concluded by Shaklan et al. (2017), "for a given level of performance, the physical tolerances on a starshade scale roughly linearly with starshade size." The S5 baseline reference starshade is 26 m in diameter with a 10 m diameter central disk and 8 m long petals, as compared to the 72 m diameter HabEx starshade with a 40 m diameter central disk and 16 m long petals a factor of 4 larger in scale than the S5 sub-scale demonstrator, which would not be sufficient to advance the HabEx petal position and accuracy technology readiness. HabEx would need to plan beyond S5 for a demonstrator with 20-meter diameter perimeter truss.

Following this interim report, the HabEx study will assess alternative starshade designs that will potentially reduce the starshade diameter such that a modest extension of the S5 development plan would bring petal deployment position accuracy for HabEx to TRL 5 sooner than if HabEx were to continue with a 72 m starshade.

## 6.3 Technologies Approaching TRL 5

A number of key technologies needed for HabEx are currently at TRL 4 (large optics, mirror coatings, and coronagraph architecture) or are expected to be at TRL 4 by the release of the HabEx final report in 2019 (low order wavefront sensing and control [LOWFS], starshade edge scatter and formation flying lateral sensing). For most of these technologies, advancement to TRL 5 is expected by the end of FY2022. The exceptions are large mirror and mirror coating technology gaps, which are blocked from advancing to TRL 5 by the need for a prototype mirror. Development of a 4 m mirror and construction of a suitable coating chamber is a significant commitment toward the large monolithic telescope architecture and unlikely to occur without prioritization of a monolithic flagship concept by the Decadal Survey.

This section discusses these approaching technologies, their current performance and their paths forward to TRL 5.

### 6.3.1 Starshade Petal Shape and Stability

The starshade must be able to deploy the petals to the proper shape, as well as maintain the shape during any given observation. The shape stability is dependent on the orientation of the starshade with respect to the Sun given the variation of temperatures across the passively temperature-controlled structure.

S5 is planning to build a 4 m petal with the width of an 8 m petal as part of its TRL 5 demonstration. This demonstration provides a test article that is half-scale in width, the critical stability dimension, for a HabEx petal. With this factor of two scaling, HabEx petal technology will advance to TRL4 when the S5 technology task achieves TRL5 on their test petal in 2022 (Willems 2018).

### 6.3.2 Large Aperture Monolithic Primary Mirror

HabEx has selected a 4 m monolithic mirror design that is TRL 4. Two critical choices enable this design: 1) the Space Launch System (SLS) allows for a generous mirror mass so that state-of-the-art mirror lightweighting is sufficient and 2) replacing reaction wheels with microthrusters for pointing control during observations reduces jitter disturbances sufficiently to allow the use of a low first frequency monolith mirror designs. In assessing the TRL of a large aperture monolithic mirror, the material, thermal stability, mechanical stability, and manufacturability were considered.

As a material, Zerodur® is TRL 9. Over 30 Zerodur mirror systems have flown in space (Döhring et al. 2009); the largest in the visible band are the 0.8 m, 73% lightweighted mirrors on the meteosats. NASA's Chandra space telescope has cylindrical mirrors that are also Zerodur, with the largest having a surface area of 1 m × 3 m. The Zerodur coefficient of thermal expansion (CTE) is an excellent match to carbon-fiber metering structure material and it can be 'tuned' to provide zero-CTE over a range of operational temperatures. Introduced in 2012, Zerodur Special achieves CTE homogeneity of 10 ppb/K and Zerodur Extreme achieves CTE homogeneity of 7 ppb/K, and CTE homogeneity of 1–5 ppb/K has been shown through the thickness of the boule which is import for thermal-induced focus error (Jedamzik and Westerhoff 2017). The Zerodur CTE homogeneity was verified using Schott's extremely lightweight Zerodur mirror

(**Figure 6.3-3**) via thermal cycle testing at Marshall Space Flight Center (Brooks et al. 2017).

The ability to manufacture mirror blanks as large as 4 m has also been demonstrated by Schott. Schott routinely makes 4.2 m diameter × 42 cm thick mirror blanks. A recent example is the Daniel K. Inouye Advanced Solar Telescope (DKIST) mirror (**Figure 6.3-1**). In addition, Schott regularly manufactures 2 m × 40 cm lightweight ultra-stiff structures from Zerodur with ultra-CTE homogeneity for its lithography bench product line (Westerhoff and Werner 2017). Scott uses computer-controlled-machining to produce ribs as thin as 2 mm (**Figure 6.3-1**)

The mechanical stability required by using microthrusters is a ~60 Hz first mode. Multiple

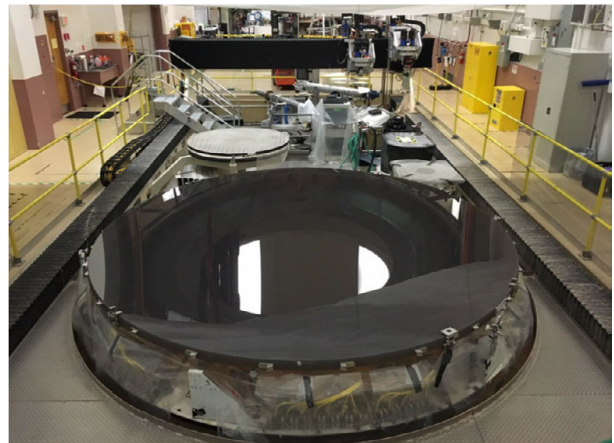

**Figure 6.3-1.** 4-meter Daniel K. Inouye Solar Telescope primary mirror (Oh et al. 2016).

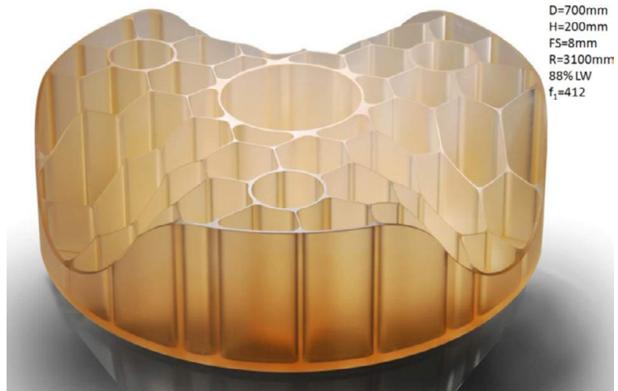

D=700mm
H=200mm
FS=8mm
R=3100mm
88% LW
$f_1$=412

**Figure 6.3-2.** SCHOTT 700 mm diameter and 200 mm high Zerodur® demonstration piece showing advanced lightweighting, cells with 2 mm machined walls, and contouring of the back. The back of the facesheet within each pocket is conformal to the facesheet. Credit:Schott

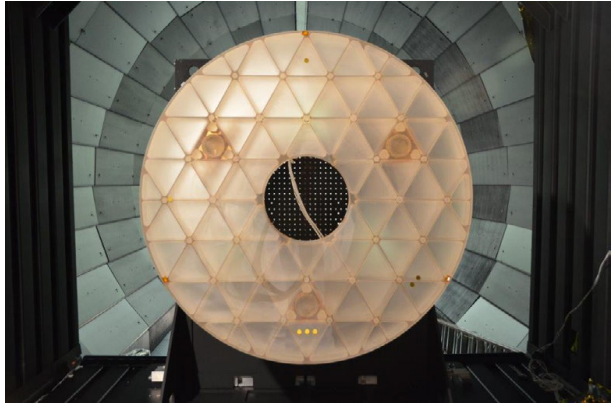

**Figure 6.3-3.** SCHOTT 1.2 m diameter and 125 mm thick Zerodur ELZM mirror in MSFC XRCF thermal/vacuum test chamber (Brooks et al. 2017).

designs are low mass and have first mode >60 Hz. Schott has published a 4 m open-back point design that has a ~80 Hz first mode and a 718 kg mass (Hull et al. 2013). The best performing 4 m design came from United Technology Aerospace Systems (UTAS); it has a first mode of 120 Hz while mounted on bipod supports, and only weighs 1,200 kg. For the HabEx interim report, a 64 Hz first-mode, 1,356 kg Marshall Space Flight Center (MSFC) design was adopted (**Figure 6.3-5**), though the UTAS design will be evaluated going forward.

For manufacturability of the surface error, UTAS has demonstrated the ability to fabricate mirrors with power spectral density appropriate for coronagraphy: total surface figure errors of <6 nm rms; mid-spatial-frequency error of

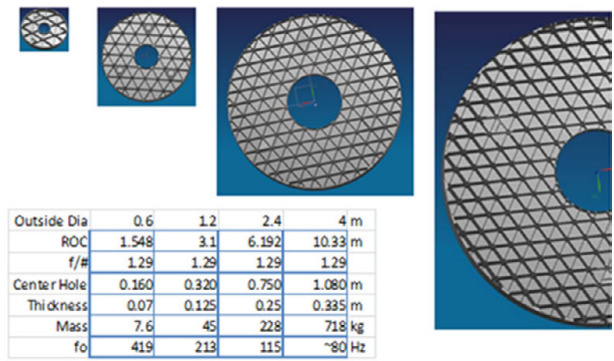

**Figure 6.3-4.** Results of analysis of 0.6, 1.2, 2.4, and 4 m lightweight Zerodur mirror substrates by the SCHOTT process. Masses represented are consistent with most present and anticipated OTAs for spaceborne missions. Each case was constrained to satisfy launch load with strength margin, although launch locks are assumed for the 4 m case. Credit: Schott

| Outside Dia | 0.6 | 1.2 | 2.4 | 4 m |
|---|---|---|---|---|
| ROC | 1.548 | 3.1 | 6.192 | 10.33 m |
| f/# | 1.29 | 1.29 | 1.29 | 1.29 |
| Center Hole | 0.160 | 0.320 | 0.750 | 1.080 m |
| Thickness | 0.07 | 0.15 | 0.25 | 0.335 m |
| Mass | 7.6 | 45 | 228 | 718 kg |
| fo | 419 | 213 | 115 | ~80 Hz |

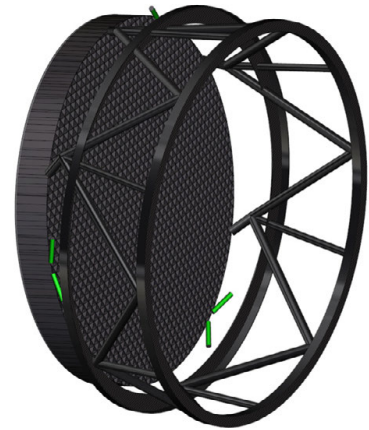

**Figure 6.3-5.** Baseline 4 m × 40 cm thick flat-back open-back isogrid core Zerodur mirror.

<2 nm rms; and surface roughness of <1 nm rms. On Chandra, UTAS produced Zerodur mirror surfaces with a roughness of 0.2 nm rms. Other mirror fabricators capable of meeting this level precision include L3 Brashear, University of Arizona Optical Sciences Center, and REOSC.

A mirror system's self-weight deflection (i.e., gravity sag) and the accuracy to which it can be removed from a one gravity (1 G) measurement is a critical limitation for producing a zero gravity (0 G) mirror. The challenge is that the magnitude of gravity sag grows larger as mirror stiffness decreases. As gravity sag grows, so does the associated back-out error.

Any design of a 4 m mirror assembly must consider metrology capabilities to guide the optical surface finishing and demonstrate zero gravity surface figure error. Specific items of attention must include surface figure and slope errors anticipated at different stages of manufacturing and under gravity load. Optical metrology accuracy, dynamic range, and spatial resolution are critical.

UTAS has TRL 9 experience designing and manufacturing 0 G mirrors as large as 2.5 m and ground-based mirrors as large as 4 m. The gravity flip metrology method allows empirical determination of gravity deformation in order to meet 0 G surface figure requirements without FEA-derived gravity compensation. Their method is self-verifiable by comparing results from different gravity flip orientations. UTAS has demonstrated TRL 9 ability to back-out gravity sag errors in mirrors as large as 2.5 m to an accuracy of less than 3 nm rms (Yoder and Vukobratovich 2015). Typically, this is limited by metrology uncertainty. Actuators on the back of the primary mirror will be considered for the final report to reduce risk with gravity sag back out.

### 6.3.3   Large Mirror Coating and Uniformity

All HabEx instruments are affected by the telescope mirror coating performance so all instruments must be considered when defining the mirror reflective properties. Telescope mirror coatings for the HabEx mission require the following fundamental properties:

- Spectral coverage with high throughput from 0.115 to 1.8 µm

- Uniformity of reflectivity—both amplitude and phase—of ≥99% over the full aperture are critical to achieve coronagraph contrast in the $10^{-10}$ level

- Consistent reflective properties for at least 10 years. Since HabEx is serviceable but cannot replace its mirrors, a coating that can last 20 or 30 years is highly desirable

This section describes both the high-heritage baseline coating, and better-UV-performing alternatives that could be considered if they are technologically mature at the time of the future mission.

#### 6.3.3.1   Baseline Al+MgF₂ Coating

The current state-of-the-art for space telescope mirror coatings is summarized in Table 6.3-1. HabEx selected a Hubble Space Telescope (HST)–like coating: an aluminum reflecting surface with a magnesium-fluoride protective overcoat. The materials and processes have been flight-proven by HST over the last 27 years and are at TRL 9. Silver and gold coatings do not meet the spectral range needed by HabEx, and though the lithium-fluoride overcoat used on the Far Ultraviolet Spectroscopic Explorer (FUSE) went below 0.1 µm in spectral coverage, the coating had degradation issues during the FUSE mission (Fleming et al. 2017). Work to develop a stable LiF protected aluminum coating for spectral coverage below 0.115 µm continues. Should one be developed in time for a future HabEx mission, the improved coating would offer a significant enhancement to the current ultraviolet science case.

Aluminum mirrors overcoated with MgF₂ have been used on space telescopes since the 1970s. Most notable is the mirror coating for the long-operating HST observatory. **Figure 6.3-6** shows a model of reflectance performance of a HST-like mirror coating in comparison with ideal Al with no overcoat. The coating on HST provides high reflectivity at wavelengths greater than ~0.12 µm. Below 0.115 µm, the reflectivity drops sharply to less than 20% due to the absorption edge of MgF₂. This level of performance is sufficient to meet HabEx baseline requirements.

Uniformity of the 4 m mirror coating is the primary coating issue for HabEx. Coating uniformity—specifically reflectivity phase and amplitude—is mainly a result of the coating process controls relevant to the specific chamber geometry. As such, engineering development is needed to build a sufficiently large chamber for the 4 m primary, and to optimize manufacturing processes to ensure a coating with less than 1% variability, as desired for HabEx.

**Table 6.3-1.** State-of-the-art coatings for large aperture space telescope primary mirrors.

| | HST | Kepler | JWST | FUSE |
|---|---|---|---|---|
| PM Size | 2.4 m monolith | 1.4 m monolith; 950 mm entrance aperture | 18 hexagonal Be mirror segments (~1.52-m wide) with total collecting area of 25 sq m | Four mirrors of 38.7×35.2 cm each |
| Spectral Range | 0.115–2.5 µm | 0.3–1.2 µm | 0.7–20 µm | 0.0905–0.1187 µm |
| Operational Lifetime | >27 years | > 9 years | 10 years max (Design Life) | 8 years (had significant Al coating degradation) |
| Coating | Protected Al (MgF₂ on Al) | Protected Ag (multilayer on Ag) | Protected Au | LiF on Al on 2 mirrors and SiC on other 2 |
| Uniformity | | <30 nm PV; Reflectivity variation <2% | <1% thickness variation among the 18 segments. <10 nm pv; Reflectance variation <0.5% in the IR | |

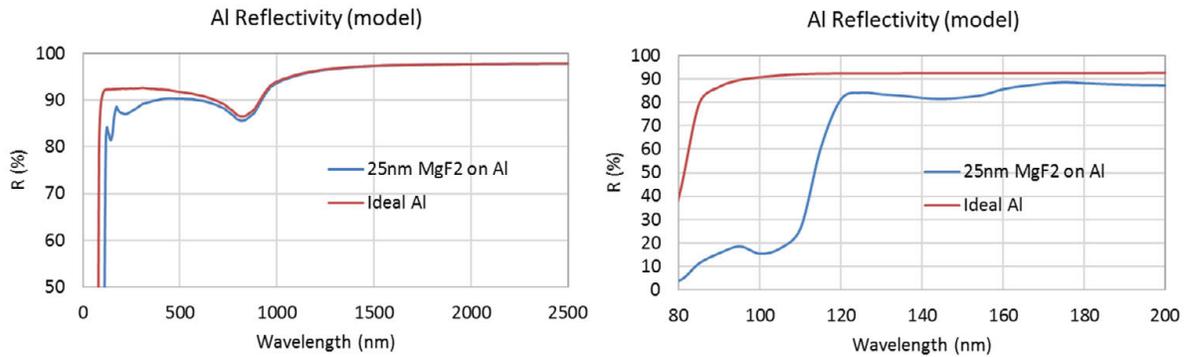

**Figure 6.3-6.** Aluminum reflectivity with and without a protective layer of MgF₂ (**HST-like model prediction**): the spikes and dips between 0.09 and 0.2 µm are a consequence of interference effects and absorption due to the protective layer and depends critically on the optical constants of the material, which depend on the coating process. The dip at ~0.83 µm is due to the native absorption property of Al.

The Kepler 1.4 m primary mirror has a protected silver coating generated using ion assisted deposition with a moving source, resulting in a thickness uniformity of about 30 nm peak-to-valley with about 2.5% reflectivity variation (Sheikh, Connell, and Dummer 2008). Better uniformity has been achieved on JWST. The JWST gold mirrors showed <10 nm peak-to-valley thickness non-uniformity with <0.5% reflectance non-uniformity in the infrared among its 18 hexagonal segments (Lightsey et al. 2012).

In 2004, Kodak (now Harris Corp, Rochester) demonstrated reflectivity variability of less than 0.5% for a high reflectivity protected silver coating over a 2.5 m diameter optic as part of the Terrestrial Planet Finder Technology Demonstration Mirror project (Cohen and Hull 2004).

These historical examples of large space mirrors with highly uniform protected metal coatings are subscale manufacturing demonstrations for a future 4 m HabEx mirror with an HST-like Al+MgF₂ coating.

### 6.3.3.2 Far-UV Enhanced Coatings

Extending the coating performance down to ~0.1 µm in the far-UV (FUV) requires technology development. At present, some of the coatings capable of this spectral range are at ~TRL 3.

Significant research and development work is underway at JPL and Goddard Space Flight Center (GSFC) to accomplish FUV spectral coverage combined with long-term stability.

Some of the more promising candidates are new lithium-fluoride evaporation techniques and adding a second thin coating layer of aluminum-fluoride to protect the lithium-fluoride layer. Work at GSFC has explored evaporation of LiF at elevated substrate temperatures, which has been shown to improve performance and environmental stability over legacy LiF coatings such as those used on the FUSE mission. Although improved, these coatings still exhibit degradation of reflectance in moderate humidity storage conditions. This has motivated JPL research into a stacked approach where the GSFC LiF coating is itself protected by a second layer of AlF₃.

Early lifetime stability tests of the Al+LiF+AlF₃ are encouraging. Samples have been tested for reflectivity changes over a 3-year period. The samples were stored in normal laboratory conditions with relative humidity ranging from 20 to 50% at nominally 68°F. Measured performance is shown in **Figure 6.3-7** (Pham and Neff 2016), and (Balasubramanian et al. 2015).

Extending the spectral range of the HabEx telescope optics down to 0.1 µm is not in the current baseline design due to the technological maturation needed. Should continued investment in this technology result in a demonstrably stable coating able to meet uniformity requirements, then a future HabEx

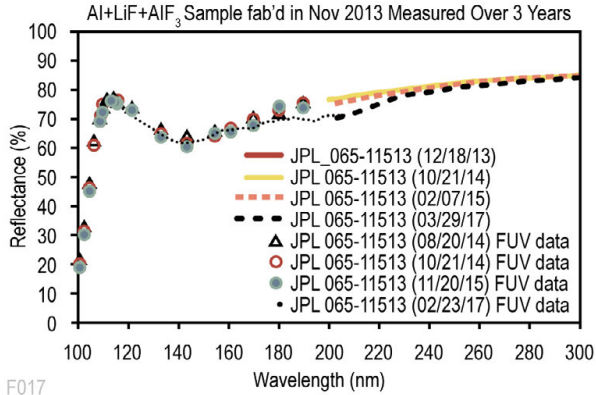

**Figure 6.3-7.** AlF$_3$ overcoat prevents LiF moisture degradation in lab environment (Balasubramanian et al. 2017).

mission could elect to use such a coating in place of the current baseline HST-like coating.

### 6.3.4    Coronagraph Architecture

Tremendous progress in coronagraph performance has been made over the last decade. Through the efforts on the WFIRST technology demonstration coronagraph and several strategic technology investments by the NASA Exoplanet Exploration Program, exoplanet direct imaging contrast performance is nearing the levels required to detect Earth-sized planets in the habitable zone of nearby stars.

This section covers the state of the art for the two coronagraph architectures under consideration by HabEx—the vortex coronagraph (VVC) and the hybrid Lyot coronagraph (HLC)—and the work needed to advance these technologies to TRL 5 with respect to HabEx's requirements.

#### 6.3.4.1    VVC and HLC Architectures

As noted earlier, the current HabEx design uses the VVC as the baseline design and the HLC as a backup option. Details of the coronagraph design and decision rationale are discussed in Section 5.5.2.2.2.

The block diagram in **Figure 6.3-8** identifies the major coronagraph elements common to both the VVC and the HLC: a fine-steering mirror (FSM) to control pointing and mitigate jitter; two 64×64 deformable mirrors (DMs) to correct wavefront error (WFE); and a LOWFS to detect WFE. These architectures have nearly the same optical layout so they are of similar size and footprint, and could be exchanged even in a fairly advanced design with minimal impact, adding flexibility to any future mission development.

The vortex coronagraph uses a focal-plane phase mask (**Figure 6.3-9**) to redirect the on-axis starlight to the outside of a subsequent pupil image, where it is blocked. The vortex phase pattern consists of an azimuthal phase ramp that reaches an even multiple of $2\pi$ radians in one circuit about the center of the mask. The very center of the vortex mask is usually covered by a small opaque spot, in order to mask defects near the phase pattern's central singularity, where the desired spatial orientation gradient is too large.

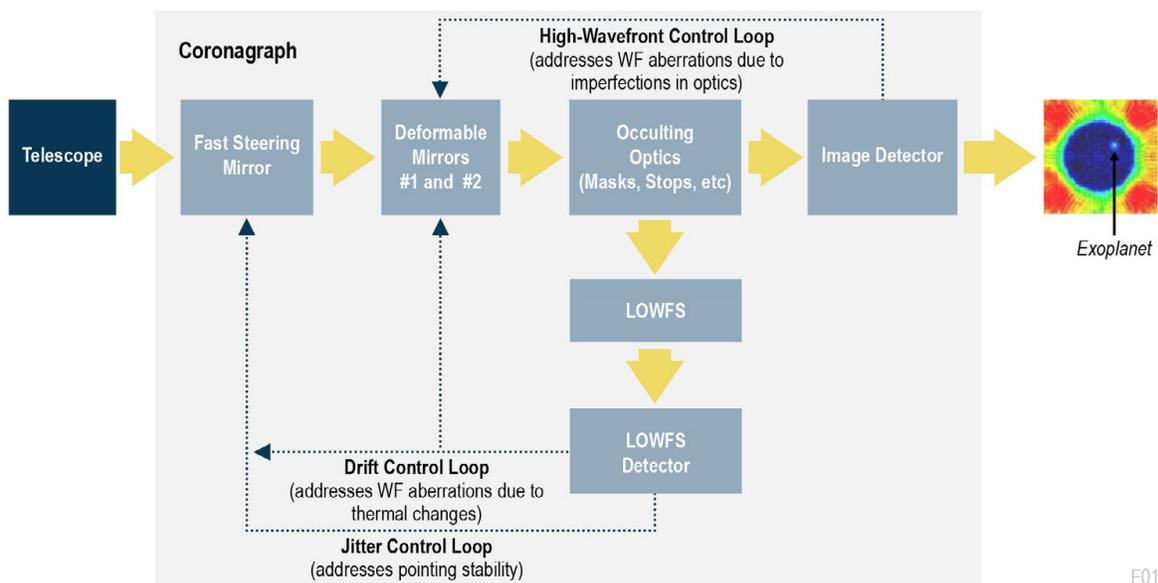

**Figure 6.3-8.** Coronagraph control loop block diagram

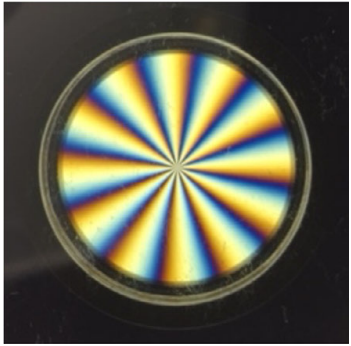

**Figure 6.3-9.** A charge 6 liquid crystal polymer vector vortex mask as seen through crossed polarizers. Credit: E. Serabyn

The HLC mask uses a partially opaque spot to block the majority of the target starlight and an overlaid phase modulation pattern provided by an optimized dielectric layer. The HLC design includes optimized DM shapes that help make the coronagraph achromatic and mitigate sensitivity to low-order aberrations. In the HabEx evaluation, both the VVC and the HLC masks are slightly tilted and their central obscurations are reflective, sending incident starlight into the coronagraph's fine-guidance sensor (FGS) and LOWFS.

### 6.3.4.2    State of the Art

The HLC has demonstrated the deepest starlight suppression to date—$6\times10^{-10}$ over 10% bandwidth from 3 to 16 $\lambda$/D—and is one of the two baselined coronagraphs on the WFIRST coronagraph instrument (Trauger et al. 2015). While this is close to the HabEx requirement ($1\times10^{-10}$ contrast over a 20% bandwidth with an inner working angle at 2.4 $\lambda$/D), there is still work needed.

In the first vortex-related Technology Demonstration for Exoplanet Mission (TDEM), carried out in the original high-contrast imaging testbed (HCIT) chamber, monochromatic contrasts of $5\times10^{-10}$ were demonstrated (Serabyn et al. 2013) for dark holes extending both from 3 to 8 $\lambda$/D, and from 2 to 7 $\lambda$/D (**Figure 6.3-10**), demonstrating very good performance all the way into 2 $\lambda$/D (**Figure 6.3-11**). Since then, the goal has shifted to broadband performance. For broadband testing under a second TDEM, a new HCIT at JPL is being used. As a result, both the chamber and the vortices are being improved in tandem. Even so, contrasts for 10% bandwidths quickly reached the level of a few $10^{-8}$ of total leakage, which currently is dominated by a fairly uniform incoherent background, with the coherent contribution being a little above the $10^{-9}$ level. These tests will resume in May 2018.

### 6.3.4.3    Maturing the Technology

The HabEx coronagraph architecture follows the same development path as the one successfully completed by the WFIRST coronagraph instrument (the WFIRST coronagraph instrument testbed is shown in **Figure 6.3-12**). First, the coronagraph will undergo a narrow-band static contrast test, then a broadband static test, and finally a broadband dynamic test. All testing will be carried out in the NASA HCIT. Both the VVC and the HLC will undergo the full development testing process. This approach will provide alternatives should one of the coronagraphs not clear a development test. The technology development roadmap is captured in Appendix E.

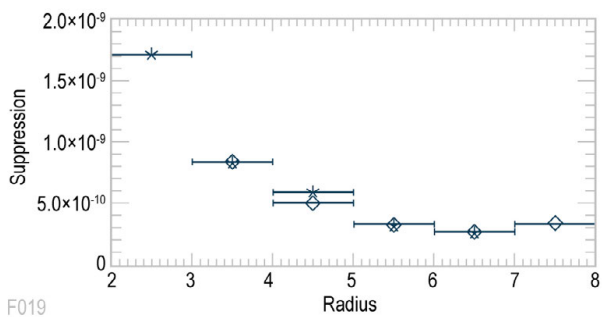

**Figure 6.3-10.** Cross-cuts through vortex dark holes of 2 to 7 $\lambda$/D (asterisks) and 3 to 8 $\lambda$/D (diamonds) show $5\times10^{-10}$ contrast.

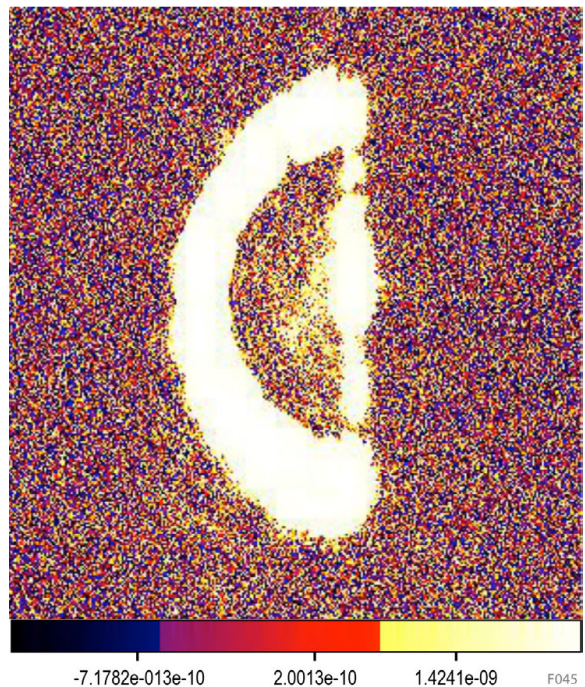

**Figure 6.3-11.** A vortex coronagraph monochromatic dark hole at the $5\times10^{-10}$ level covering 2 to 7 $\lambda$/D (Serabyn et al. 2013).

The initial narrow-band test is necessary to demonstrate that each coronagraph is capable of reaching the required $10^{-10}$ raw contrast level. While the contrast level for HabEx is more demanding than for WFIRST, HabEx benefits from an aperture without obscurations. The coronagraph optics will be in a flight-like layout and include two 64×64 DMs. The testing will be carried out at 0.5 µm, under vacuum, in a static environment. The test will be repeated with DM resets between each test to demonstrate consistency of performance.

The broadband test will demonstrate that the coronagraphs can operate without significant chromatic problems. The test will be conducted with a 20% band centered at 0.5 µm, in a static environment. To address chromatic issues, wavefront control is required. A control loop between the imaging detector and the DMs will be added at this point in the development. Data from the test will be used for post-processing algorithm development and optimizing wavefront control algorithms.

Finally, the HabEx coronagraphs will demonstrate broadband contrast under flight-like environmental conditions. To achieve this, a FSM and a LOWFS system will be added to the testbed. The LOWFS control loops correct for drift and jitter using the FSM and the DMs. Again, the test will be conducted in a 20% band centered at 0.5 µm and will be repeated several times to demonstrate the consistency of the results. Once successfully completed, the coronagraphs will be at TRL 5.

One last note: the HabEx coronagraph is significantly larger than the WFIRST coronagraph instrument, so an evaluation of the HCIT facilities must be conducted ahead of this technology development effort. Some facility upgrades may be necessary to complete the HabEx coronagraph testing program.

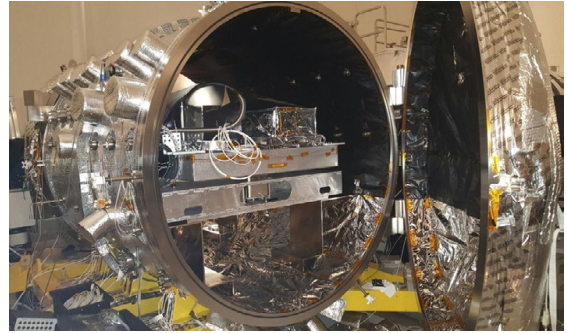

**Figure 6.3-12.** WFIRST coronagraph instrument testbed.

### 6.3.5    LOWFS and Control

The coronagraph LOWFS uses the rejected starlight from the coronagraph to sense the low order WFE, which includes line-of-sight (LOS) pointing error and thermal-induced low-order wavefront drift. The LOWFS module consists of four elements fed by light reflecting from the coronagraph mask for both the HLC and the VVC.

The LOWFS sensor is a Zernike wavefront sensor (ZWFS) similar to the WFIRST coronagraph instrument's LOWFS. The ZWFS is based on the Zernike phase contrasting principle where a small (~1–2 $\lambda$/D) phase dimple with phase difference of ~$\lambda$/2 is placed at center of the rejected starlight point spread function (PSF). The modulated PSF light is then collimated and forms a pupil image at the LOWFS camera. The interferences between the light passing inside and outside the phase dimple convert the wavefront phase error into the measurable intensity variations in the pupil image on the LOWFS camera. The spatial sampling of the pupil image on the LOWFS camera depends on the spatial frequency of WFE to be sensed. There is a design trade between number of sensed modes, photons per pixel, and the LOWFS camera frame rate. To improve the signal-to-noise ratio, the LOWFS uses broadband (>20%) light. The LOWFS camera is running at high temporal frequency (~1 KHz frame rate) in order to sense the fast LOS jitter.

The LOWFS-sensed tip-tilt errors are used to control the FSM for LOS disturbances correction. Similar to the WFIRST coronagraph instrument, the FSM LOS control loops contain a feedback loop to correct the telescope's LOS drift and a feedforward loop to suppress LOS jitter. The LOWFS sensed low-order wavefront errors beyond tip-tilt will be corrected using one of the DMs.

ZWFS-based LOWFS has been developed, designed, and testbed demonstrated for WFIRST coronagraph instrument at JPL's LOWFS testbed and occulting mask coronagraph (OMC) dynamic testbed, which has two coronagraph modes: HLC and shaped pupil coronagraph (SPC). Testbed results have shown that ZWFS is very sensitive, capable of sensing LOS less than 0.2 milliarcsec and low-order WFE as small as 12 picometers (rms). Recent OMC dynamic test results have demonstrated that with the LOWFS FSM and DM control loops closed, both the HLC and SPC are able to maintain their contrasts to better that $10^{-8}$ with the presence of WFIRST-like LOS variations (~14 mas drift and ~2 mas jitter) and slow-varying low-order WFE disturbances (~1 nm rms at ~1 mHz) (Shi et al. 2017). Although HabEx's LOWFS architecture is traceable to that of the WFIRST coronagraph instrument, this LOWFS technology will have to be demonstrated in conjunction with the HabEx nominal VVC.

### 6.3.6 Deformable Mirrors

The baseline DMs are 64-actuator by 64-actuator by Boston Micromachines Corporation (BMC) shown in **Figure 6.3-13**. The DMs are microelectromechanical systems (MEMS) made using semiconductor device fabrication technologies. The DM has a continuous facesheet for the surface of the mirror; the actuators pull on the back of the mirror using capacitance with an electrode in the back plane. The 4,096-actuator DM has been used in ground based coronagraphy on the Gemini Planet Imager (Macintosh et al. 2014). The 4,096 DM has a 3.5 µm stroke and 400 µm pitch.

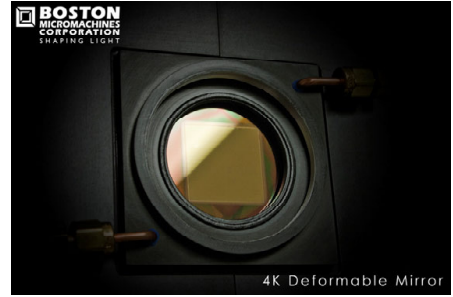

**Figure 6.3-13.** BMC 64×64 deformable mirror. Credit: BMC

The current BMC DMs are undergoing modifications to improve performance. Recent advances have reduced facesheet quilting to 3.3 nm rms. Facesheet scalloping is also being reduced by using a lower operating voltage. The best flattened WFE achieved in air to date is measured at about 6 nm rms. Additionally, the fabrication process of the DM has difficulty achieving 100% actuator yield. These challenges are currently being improved through a NASA Small Business Innovation Research (SBIR) program.

BMC DMs have proven useful for ground-based coronagraph instruments and have been demonstrated in a suborbital sounding rocket (Douglas et al. 2018). Additional use of the BMC DMs is underway in more ground-based instruments, a high precision testbed, and in space on a CubeSat (**Table 6.3-2**).

**Table 6.3-2.** Boston Micromachines Corporation DMs in current and planned astronomical use.

| | Actuators | Instrument | Location |
|---|---|---|---|
| Ground | 140 | ROBO-AO | Palomar 2012, Kitt Peak 2015 |
| Ground | 1,024 | Shane-AO | Lick Observatory 2013 |
| Ground | 2,040 | SCExAO | Subaru 2013 |
| Ground | 4,092 | GPI | Gemini South 2013 |
| Space | 1,024 | PICTURE-B | Sounding Rocket 2015 |
| Ground | 2,040 | MagAO-X | U of Az, *In work* |
| Ground | 492 | Rapid Transit Surveyor | U of H, *In work* |
| Ground | 952 | Keck Planet Imager and Characterizer | Keck, *In work* |
| Testbed | 1,000 segments | Caltech HCST | Testbed, *In work* |
| Space | 140 | DEMI | CubeSat, *In work* |

The BMC DM has not yet been environmentally tested, although a TDEM is currently underway to put 16 DMs through environmental testing with pre-test and post-test performance characterization (Bierden 2013). Environmental testing of the BMC DMs as part of the TDEM is scheduled for 2018. The deepest raw contrast achieved by a coronagraph using the BMC DM was $2\times10^{-7}$ over 2–11 λ/D at 0.65 µm central wavelength and 10% bandwidth in the Exoplanetary Circumstellar Environments and Disk Explorer (EXCEDE) proposal testbed (Sirbu et al. 2016). Further coronagraph testing with BMC DMs will be carried out in the Exoplanet Exploration Program Office Decadal Survey Testbed starting this year. Should the new coronagraph tests achieve a $1\times10^{-10}$ raw contrast under simulated environmental disturbances then the coronagraph and the BMC DMs will be assessed at TRL 5.

### 6.3.7   Starshade Edge Scatter Suppression

The primary goal of the starshade optical edges is to provide the correct apodization function needed to suppress starlight to levels suitable for exoplanet direct imaging. To do this, light emanating from other sources—principally petal edge-scattered sunlight—must be addressed since this scattered light can significantly degrade image contrast. The intensity of this scattered light must be limited to below the exozodiacal background.

Light scatter is driven by both the area and reflectivity of the scattering surface. As such, to mitigate edge-scatter, the starshade optical edges must have a sharp, beveled edge, and a surface with low reflectivity. Achieving these two edge characteristics has been the focus of edge-scatter technology work since 2015.

To resolve this technology gap, two parallel development efforts were initiated: a JPL effort focused heavily on creating sharp edges, and a Northrop Grumman Aerospace Systems (NGAS) effort centered on low reflectance. In 2015, JPL funded an effort to produce prototype optical edges. These edges were constructed using thin strips of amorphous metal as the absence of material grain structure allows for extremely sharp edges. Chemical etching techniques were used to manufacture the edges as it provides a means to produce the necessary beveled edge and can be implemented on meter-scale edge segments with micron-level in-plane tolerances. Multiple coupon samples were constructed and their geometry characterized using scanning electron microscope (SEM) images (**Figure 6.3-14**). A terminal radius of <0.5 µm was achieved with low levels of variability across each coupon (Steeves et al. 2018). The solar glint performance of these coupons was also established using a custom scattered-light testbed; measurements indicate that the scattered flux is dimmer than the predicted intensity of the background zodiacal light over a broad range of sun angles. Further improvements to this performance can be achieved through the addition of low-reflectivity coatings on the optical edge—an area currently in development. While suitable solar glint performance was demonstrated at the coupon level, meter-scale development is still underway. Demonstration at the edge segment level—and accordingly, achievement of TRL-5—is expected by FY2020 through the S5 task (Willems, in progress).

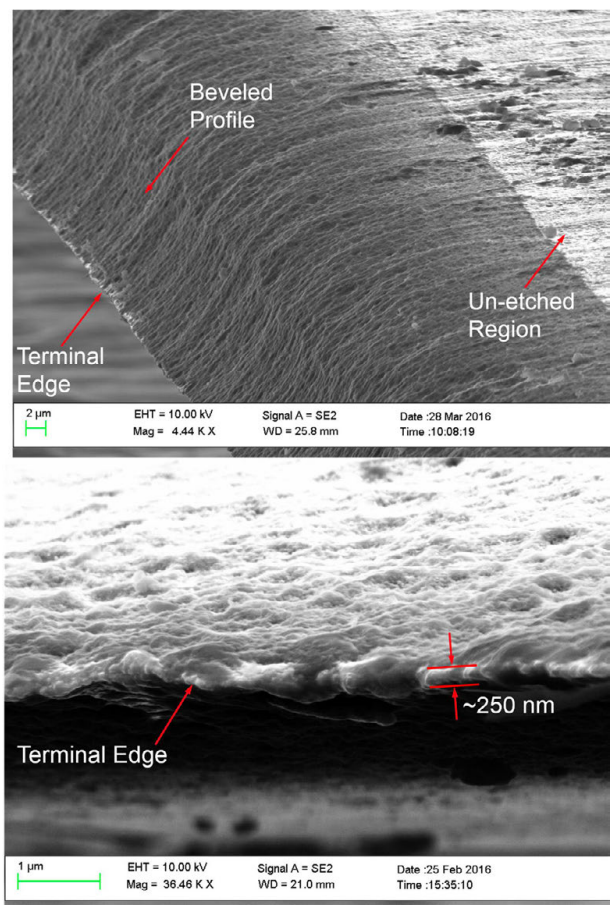

**Figure 6.3-14.** SEM images of the starshade beveled edge and terminal edge

## 6.4    Technologies at TRL 5 or Higher

Some of the technologies in the HabEx design will be at TRL 5 or higher by the release of the HabEx final report (in 2019), including: microthrusters, deformable mirrors, low-noise visible and UV detectors, starshade formation flying, and laser metrology. This section describes the state of the art of these technologies and how the state of the art is applied to the HabEx design.

### 6.4.1    Starshade Starlight Suppression and Model Validation

As noted earlier, system-level performance testing of a flight-like starshade (contrast and IWA validation) is not possible on the ground due to the separation distance (nominally 124,000 km) required between the telescope and the starshade occulter. Validation must be done through models and reduced-scale testing. To achieve the required image contrast, the two key model parameters are the Fresnel number and the ratio of the radius of the starshade's shadow at the telescope aperture, to the starshade's radius. The Fresnel number is defined as

$$F = \frac{r^2}{\lambda L}$$

Where $r$ is the starshade radius, $L$ is the starshade/telescope separation distance, and $\lambda$ is the wavelength of the light being measured.

For HabEx, the Fresnel number is 10.45 (at $\lambda = 1$ μm) and the aperture-shadow-to-starshade ratio is 0.083. Any systems sharing these values will have the same contrast performance regardless of the starshade occulter size and system separation distance. This model behavior allows ground testing of scaled-down versions of the starshade direct imaging system, to assess the optical performance of the flight system.

Such scaled testing is already underway through the S5 project. A testbed has been built at Princeton University (**Figure 6.4-1**) for starshade optical performance testing and model verification. The testbed illuminates the starshade mask with a 0.633 μm laser. Contrast levels are detected with an electron multiplying charge coupled device (EMCCD) observing on the 50 mm starshade through a 4-mm aperture. The starshade and aperture are separated by 50 m. The deepest starshade model suppression achieved so far has been $1\times10^{-7.5}$ suppression at a flight like Fresnel number of 14.5 with a corresponding image plane contrast of $1\times10^{-9}$ (Kim et al. 2017). A deeper suppression of $6.0\times10^{-8}$ with contrast $2.5\times10^{-10}$ has been achieved at Fresnel umber of 27 (Harness et al. 2018). The expectation is to achieve $<1\times10^{-10}$ contrast with model validation and TRL 5 in FY2020 (Willems 2018).

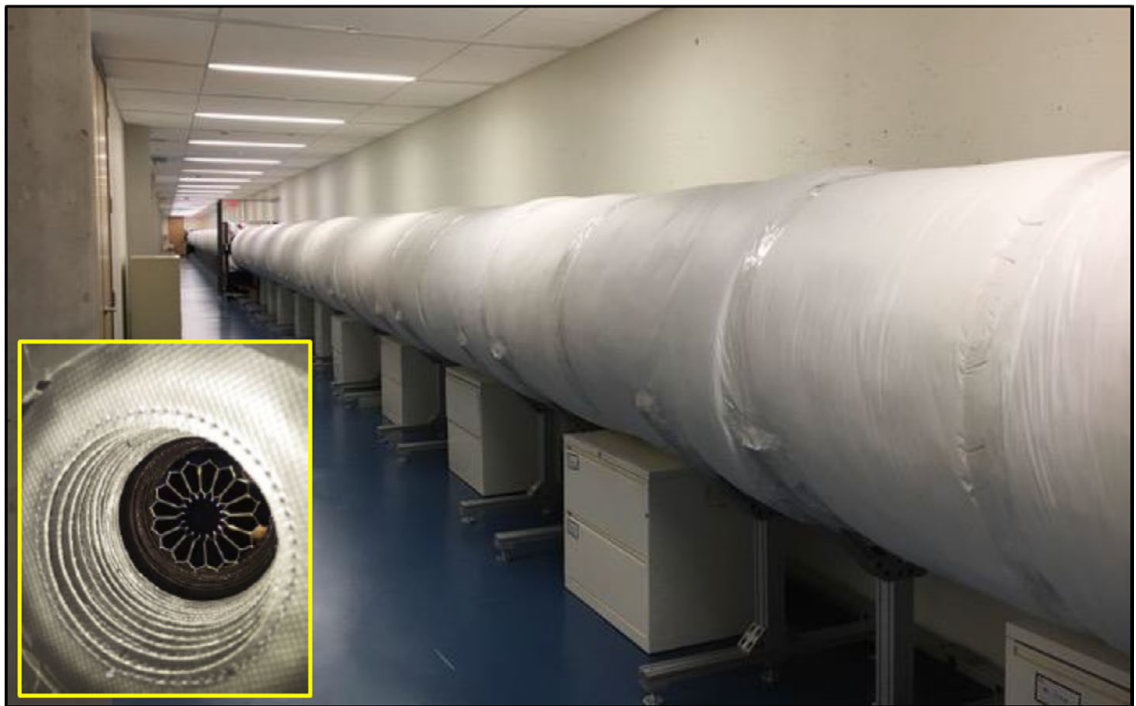

**Figure 6.4-1.** S5's starshade model validation testbed at Princeton. Model starshade shown in inset.

### 6.4.2   Starshade Lateral Formation Sensing

Technology work on formation flying is also addressed through the S5 technology development task. S5 will fully address the HabEx formation flying gap and bring the technology to TRL 5 in FY19. Specifically, S5 will demonstrate lateral sensing of the starshade to less than 0.2 m, and control to within 1 m radial.

The sensing approach uses pupil-plane images at a wavelength outside the science band, where the starshade's attenuation of the starlight is only $\sim10^{-4}$ (see **Figure 6.4-2**). The shadow has sufficient structure at the center, that matching de-trended pupil-plane images to a library of pre-generated images can determine lateral position of the starshade to 15 cm (3σ) with short exposures. See Section 5.8.4 for more details on the design and operation of the HabEx formation flying system.

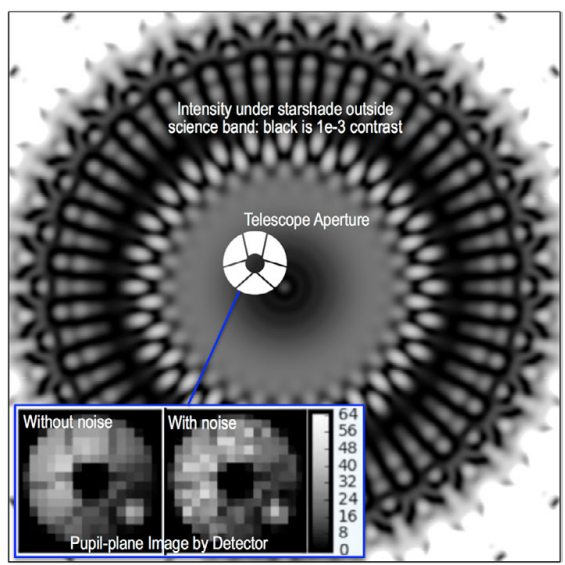

**Figure 6.4-2.** Illustration of approach for starshade precision lateral sensing using pupil-plane image matching.

By the end of FY19, S5 will have demonstrated this sensing approach in a scaled 2 m low-contrast testbed operating at a realistic Fresnel number with a 6 mm starshade, and with the detector mounted on a motion stage. The detector will be placed at a number of positions including corner cases (position extrema), and the output of the sensing algorithm will be recorded. By comparing the sensor output and the truth positions from the motion stage, the predicted accuracies and precisions of the sensing approach will be verified to TRL 5. An example comparison of a predicted image and an initial image from the low-contrast testbed is shown in **Figure 6.4-3**.

Following closure of the sensing gap, formation control will be demonstrated in simulation using a noise model for the lateral sensor. The noise model will have been verified during the earlier formation sensing work. The control simulation will include realistic thruster dynamics that require thrust allocation, thruster minimum impulse, and errors in attitude knowledge of the starshade. The dynamics will use a representative maximum gravity gradient and the control will be done with an optimal circular deadbanding algorithm, including representative drift times between thruster firings. In addition, an estimator combining the lateral sensor and radio frequency (RF) ranging measurements will be developed. Realistic actuator misalignments and mass property uncertainties will also be included. Work will be completed, and the HabEx formation flying technology will be at TRL 5 by the end of FY19 (Ziemer 2018).

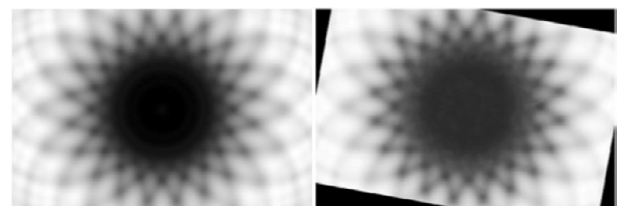

**Figure 6.4-3.** Preliminary results from the Low-Contrast Testbed. *Left:* Simulated image. *Right:* Testbed image.

### 6.4.3   Microthrusters

Colloidal microthrusters provide low-noise, precise thrust and drag-free operation of spacecraft against disturbances—mainly solar pressure—for ultra-fine pointing and telescope stability control. Busek Co., Inc. has worked with JPL to provide two clusters of 4 colloidal microthrusters (**Figure 6.4-4**) for the NASA Space Technology 7 Disturbance Reduction System (ST7-DRS) mission in 2008. ST7-DRS was launched on board the European Space Agency's (ESA's) LISA-Pathfinder Spacecraft in December 2016, and accumulated over 100 days of operation on orbit. The LISA-Pathfinder colloidal microthrusters were single string designs, intended for only 90 days of operation. Each thruster emits a finely controlled electrospray (electrostatically accelerated charged droplets) using an ionic liquid propellant, producing between 5–30 $\mu$N of thrust with 100 nN resolution All eight thrusters demonstrated full thrust range and controllability after 8 years of ground storage. As a system, thrust noise has been measured using ESA's inertial sensor on the LISA Technology Package at levels $\leq 0.1\ \mu$N/$\sqrt{\text{Hz}}$ (average per thruster).

NASA's Physics of the Cosmos (PCOS) Program is currently developing the colloidal microthruster technology as a potential contribution to the ESA-led LISA mission. During the next three years, the colloidal microthrusters will be redesigned to be fully redundant with sufficient capacity to support a 12-year mission. Minor changes over the ST7's TRL 7 design are expected to be needed to meet lifetime and redundancy requirements; the redesigned thrusters are expected to reach TRL 6 at the end of the PCOS technology program. For HabEx, the reliability and lifetime technology development activities for LISA would provide a strong basis for colloidal microthruster use. Studies are currently ongoing to determine thrust range requirements and propellant-minimizing thruster configurations for HabEx.

HabEx is also continuing to consider the Gaia cold-gas microthrusters as an alternative. The Gaia thrusters will have been operational at L2 for 5 years when the HabEx final report is delivered in 2019 so they will have reached the HabEx baseline lifetime requirement. However, their specific impulse is lower than the colloidal microthrusters so more fuel would be required. While HabEx is baselining the colloidal microthrusters for this interim report, the final decision on microthrusters will come after the report release.

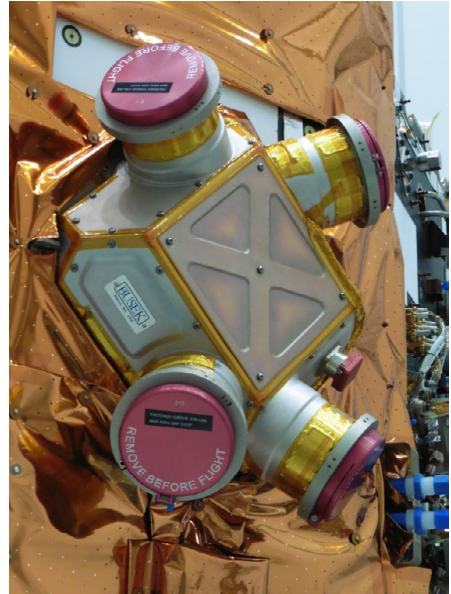

**Figure 6.4-4.** A single cluster of four Busek Co. colloidal microthrusters integrated on the LISA-Pathfinder Spacecraft just prior to launch. Image courtesy ESA / Airbus.

### 6.4.4   Laser Metrology

As noted in Section 5.5.2.6, a laser metrology truss provides the sensing end of a Laser Metrology Subsystem (MET) rigid body control loop for the telescope optics. Using rigid body actuators on the secondary and tertiary mirrors, MET actively maintains alignment of the front-end optics, removing the primary source of telescope wavefront drift. This breakthrough technology operates at high bandwidth and can maintain control throughout all phases of the mission, effectively creating a near perfect, infinitely stiff, telescope truss.

The backbone of MET is the laser metrology gauge, which monitors any changes in the distance to a retroreflector. Each planar lightwave circuit (PLC) beam launcher, or gauge, requires a stable laser source and a phase meter to operate. Each of these components are at TRL 6 or higher.

The laser source for MET at JPL has historically been a Nd:YAG non-planar ring oscillator (NPRO). A similar, TRL 9, Nd:YAG ring laser has flown on LISA-Pathfinder for the laser metrology system monitoring the test masses (Voland et al. 2017). Since laser metrology operates as a heterodyne system, a thermally stabilized PLC beam launcher (**Figure 6.4-5**) is sufficient for the purposes of HabEx.

The phase meter monitors the heterodyne measurement signal and compares it to the reference signal. Changes in the phase between the two signals are directly related to the change in the distance between the PLC beam launcher and the retroreflector. The LISA-Pathfinder phase meter (Hechenblaikner et al. 2010) is an example of a suitable phase meter at TRL 9.

The final component of the MET system is the beam launcher. For HabEx, the beam launcher is the beam splitting/combining system that must be mounted to the telescope optics and therefore must be small and of similar construction. The PLC beam launcher evolved from the large, external metrology beam launchers developed for the Space Interferometry Mission (SIM) and has been refined into a very compact, stable component (Zhao et al. 2003). Continued development at JPL beyond the cancellation of SIM in 2010 has brought these beam launchers up to TRL 6 (Nissen et al. 2017).

### 6.4.5    Detectors

HabEx can achieve its primary exoplanet scientific objectives with detectors that operate within the 0.3–1.0 μm spectral range. High performance in this range can be achieved using existing silicon-based detectors (e.g., CCDs and CMOS, arrays) with high TRL. Extending the spectral range at both ends enables a greater return for the exoplanet science and is required to meet the observatory science requirements. This extended spectral coverage necessitates a closer look at existing detector capabilities in the UV down to 0.115 μm and in the near-IR out to 1.8 μm.

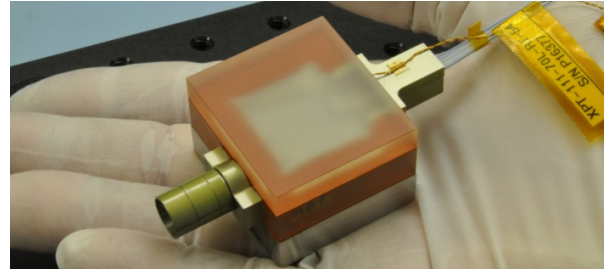

**Figure 6.4-5.** Planar lightwave circuit (PLC) beam launcher.

This section introduces detector candidates that have been selected by careful examination of the performance and latest status of the available technologies. The principles of operation for these detectors are briefly described and information is provided on the performance of their major relevant parameters, TRL status, and the path of further development.

#### 6.4.5.1    Delta-Doped Visible EMCCDs

Electron multiplying charge-coupled devices (EMCCDs) are otherwise conventional CCDs that possess high S/N by the virtue of having an additional serial register. This so-called 'gain' register produces gain via avalanche multiplication in a stochastic process. Gains of greater than 1,000 can be achieved and photon counting can be performed. The Teledyne e2v's CCD 201, which has been baselined for the WFIRST coronagraph, has also been optimized for high efficiency and high stability in the 0.4–1.0 μm range using a delta doping process.

The delta-doping utilizes JPL's low temperature (<450°C) molecular beam epitaxy (MBE) growth process to inject dopant atoms in a highly localized layer. "Delta-doping creates very high electric fields near the surface that drive photogenerated charge away from the back surface and suppress the generation of excess dark current from the exposed silicon surface" (Hoenk et al. 2009).

Extensive radiation testing for WFIRST has been carried out as part of the WFIRST coronagraph instrument technology development program (Harding et al. 2015, 2018). The CCD201 is currently at TRL 5 for WFIRST, which has a nearly identical environment as HabEx. A larger format of the EMCCD (4k × 8k, CCD 282) has been demonstrated (Daigle et al. 2018) and is baselined for HabEx.

### 6.4.5.2 Linear Mode Avalanche Photodiode Near-IR Detectors

Leonardo-ES Ltd in Southampton, UK, has been developing HgCdTe avalanche photo diode (APD) sensors for astronomy in collaboration with the European Southern Observatory and the University of Hawaii since 2008. The devices use metalorganic vapor phase epitaxy (MOVPE) grown on gallium arsenide (GaAs) substrates. This, in combination with a mesa device structure, produces a detector that achieves a noiseless avalanche gain, very low dark current (due to band gap engineering) and a near-ideal spatial frequency response. A device identified as "Saphira"—a 320×256, 24 µm pixel detector—has been developed for wavefront sensors, interferometry, and transient event imaging and is currently in use in a number of ground-based telescopes including Subaru and NASA's Infrared Telescope Facility.

Saphira has demonstrated read noise as low as 0.26 electrons rms and single photon imaging with avalanche gains of up to 500. An avalanche gain of 5 can be achieved with dark current of less than 0.002 electrons per second per pixel. This dark current translates into nearly a factor of five improvement in S/N for signals of the order of 100 photons, or a factor of 25 improvement in observation time. The Saphira detectors have been assessed at TRL 5, which means they have been tested in a relevant environment. There is a current ESA program to assess radiation (gamma and proton) resilience and, to date, there has been no change in the detector performance after exposures of 50 krads of gamma radiation.

Currently, there is a NASA program funding development of a custom MOVPE design for low-background/high-gain imaging aimed at extending the gain and reducing dark current of the Saphira detectors even further.

### 6.4.5.3 UV Microchannel Plate Detectors

Microchannel plates (MCPs) have been the workhorse of ultraviolet instruments for several decades. They are image tube–based detector technologies in which a photocathode material is used to absorb photons in the desired spectral range, creating electrons that are accelerated in vacuum and multiplied in the tube. The gain achieved in this fashion and the low background noise renders MCPs highly applicable in photon-counting applications. The charge packet exiting the MCP tubes is detected through various readout schemes depending on the application.

Microchannel plate detectors have flown on UV astronomical missions (FUSE, GALEX, HST-COS). More recent MCP detector developments include atomic layer deposition (ALD) on borosilicate microcapillary arrays. An ALD MCP detector has flown on the LITES ISS instrument (Siegmund et al. 2017).

Continued development of UV MCP detectors will improve performance and packaging in the coming years. In 2012, a NASA Strategic Astrophysics Technology (SAT) grant was awarded to raise the TRL of a 50 mm square cross-strip MCP detector from 4 to 6. The team was also funded in 2016 with a follow-on SAT to scale this detector to a flight qualified 100×100 mm format (Vallerga et al. 2016). Even larger formats (200×200 mm) are also being developed. MCP detectors currently baselined for HabEx are TRL 5.

### 6.4.5.4 Delta-Doped UV EMCCDs

Delta-doped UV EMCCDs offer an alternative to current MCPs in the 100–300 nm wavelength range. High efficiency (>60% quantum efficiency) in the 100–400 nm range has also been demonstrated on EMCCDs. JPL has been working closely with e2v to develop the end-to-end processing for CCD 201 and has focused on the high efficiency and photon-counting performance of the detector. A delta-doped EMCCD with coatings to optimize the performance at 205 nm has been delivered to FIREBall, a balloon-based UV experiment, which is expected to launch this year. Another EMCCD that is optimized for 120–150 nm range is baselined on the sounding rocket SHIELDS, which is expected to fly in early 2019 and would advance to TRL 6.

# 7  CONCLUSION AND PATH FORWARD

In this report, the science motivation for the Habitable Exoplanet Observatory, or HabEx, has been described. As envisioned, HabEx would be a flagship-class mission that would launch mid-2030s. The three primary science applications of HabEx are as follows. (1) Direct detection and characterization of exoplanets orbiting nearby sunlike stars, and in particular the detection and characterization of potentially habitable planets. (2) A broad range of observatory science inquiry areas, ranging from precision measurements of the local expansion rate of the universe, detailed study of resolved stellar populations in nearby galaxies, tracing the complete life cycle of baryons, the matter that makes up galaxies, stars, planets, and life. (3) A variety of studies of solar system objects and phenomena, from planetary aurora, to cryovolcanism in some of the most fascinating bodies in the solar system, such as Europa and Enceladus, to an improved understanding of the nature of atmospheric escape, and finally to uncovering the origin of the Earth's water.

A fortunate confluence of several circumstances has made the HabEx mission not only technologically viable within the next few decades, but also particularly timely. As described in the Introduction, revolutions in the understanding of planets orbiting other stars, the contents, geometry, and indeed entire history of the universe, and the contents of our own solar system and broad array of phenomenology that has been observed using remote and in-situ methods, have made the time ripe for a mission with the capabilities of HabEx.

Perhaps more importantly, it is now known that potentially habitable planets—rocky worlds that may have thin atmospheres like the Earth, located at the right distances from their parent stars such that they could support liquid water on their surface—are likely relatively common, with roughly 20% of sunlike stars hosting such planets. Since the minimum aperture required to detect and characterize a given number of Earth-like planets decreases with the occurrence rate of such planets, this implies that relatively small apertures, such as the 4 m aperture architecture detailed in this report, are capable of achieving the lofty goals of searching for habitable conditions and even evidence for life outside the solar system.

Simultaneously, rapid advances of starlight suppression technologies have made the contrast and resolution requirements to detect and characterize Earth-like planets around the nearest sunlike stars achievable within the immediate horizon. Other technological developments, in particular in the areas of very sensitive detectors, large mirror fabrication, and spacecraft pointing and vibration control, have also enabled many of the science applications of HabEx described in this report.

The HabEx architecture detailed in this report is based on a monolithic, off-axis, 4 m aperture telescope equipped with a suite of four instruments. Two of these instruments are dedicated to high-contrast direct imaging of exoplanets. The other two are designed to maximize the unique strengths of HabEx for astrophysics and solar system studies: a large aperture diffraction-limited at 0.4 μm, and thus high resolution and large photon collection area, combined with exceptional ultraviolet sensitivity.

The HabEx architecture relies on the two most mature starlight suppression technologies: coronagraphs (which allows starlight to enter the telescope aperture but suppress it to the requisite contrast levels using sophisticated optics and wavefront control) and starshades (which use a large, opaque structure located tens to hundreds of thousands of kilometers away from the telescope to block the starlight before it enters the telescope aperture). Both technologies have different intrinsic strengths and weaknesses.

Coronagraphs are more technologically advanced, and much more nimble: the starlight suppression happens interior to the telescope, allowing for rapid pointing and revisits of multiple targets. However, they operate over narrow bandwidths and so require multiple channels or serial observations for spectra covering a wide wavelength range. Coronagraphs also require highly stable telescopes with active wavefront control systems, resulting in more complex optical trains and relatively lower throughput than starshade-based instruments.

Starshades are less technologically mature. However, they have the advantage that the starlight never enters the telescope aperture. As a result, the instrument and telescope design are significantly simpler than for a coronagraph. For a starshade, the telescope stability requirements are dramatically relaxed. The primary difficulty with a starshade is that the number of targets that can be monitored is limited by the time and fuel required to move between targets, and thus the total number of slews. For HabEx, roughly 100 starshade slews are currently being baselined.

Fortunately, the two methods of starlight suppression are very complementary. The nimbleness of the coronagraph complements the ability to achieve deep and wide-wavelength spectra of planets with relatively low integration times with the starshade. The strength of the starshade is the ability to acquire precise spectra over a wide wavelength range in a relatively short integration time. The strength of the coronagraph is its ability to rapidly detect planets, including those potentially in the habitable zone; characterize their phase-resolved colors; and measure their orbits. Together these two technologies allow for both the detection, orbit characterization, and spectral characterization of a vast variety of planets (potentially habitable or not), for a large number of systems. The HabEx architecture takes advantage of the strengths and complementarity of both methods to maximize its exoplanet science capabilities.

Following the delivery of the HabEx interim report, the HabEx team will continue work on the baseline design in a number of areas. Two key trades will need to be addressed. Determining which coronagraph (vortex charge-6, vortex charge-8, or hybrid-Lyot coronagraph [HLC]) is the best choice for the final design and assessing the science impact of a reduced size starshade (~50 m) will be resolved before the final report. In addition, more detailed design work is needed in certain areas. Servicing will be developed in more detail; telescope spacecraft mass will be reduced; system-level jitter modeling will be completed; and a system-level structural/thermal/optical (STOP) model verification of performance will be performed.

### Telescope Spacecraft Mass Reduction

The current load-path from the telescope bipods into the spacecraft structure may not be the most efficient way to move launch loads from the telescope to the launch vehicle. HabEx will look at a different bus structure configuration that could significantly reduce the mass of the bus structure.

### Pointing Control

An initial pointing system sizing study and a rough pointing simulation have been completed for this report. The simulation will be updated with the final structural design, thruster locations, and the most recent thruster noise profiles available. The simulation will characterize the telescope's line-of-sight (LOS) jitter as input to a STOP analysis of the overall telescope/coronagraph system.

### STOP Modeling

The STOP modeling integrates the structural, thermal and optical models of a design to estimate the time-varying effect of thermal and mechanical disturbances on optical performance. For HabEx, the optical performance being assessed is the coronagraph's dark field. The thermal behavior of the telescope when moved from star to star or when rolled about the boresight, will be modeled to bound the size of thermal transients likely to be encountered during observations. The associated mechanical distortions to the telescope optics will be simulated to assess the effectiveness of telescope thermal control and laser metrology to mitigate these transients. Structural dynamics will be modeled to verify that, when supported by the LOS and jitter control loops, the mechanical design sufficiently suppresses likely disturbances. Simulated thermal and mechanical distortions are then combined with the coronagraph's wavefront sensitivity to assess their impact of the coronagraph's dark field.

### Coronagraph Trade

The baseline design for this report assumes a vortex charge-6 coronagraph due to its insensitivity to disturbances in the low-order Zernike modes, which greatly simplifies the telescope design for jitter and thermal distortion mitigation. The vortex charge-8 design has even greater insensitivity than the vortex charge-6 but at a cost of increased inner working angle (IWA) and degraded throughput. Also under consideration, the HLC has the best measured contrast performance to date—nearly reaching the required $10^{-10}$ contrast in static tests—and has significantly better throughput than either of the two vortex coronagraphs. A STOP analysis of the whole HabEx telescope system will determine which coronagraph offers the best performance with the baseline design.

### Starshade Sizing Trade

Starshade mass is highly sensitive to starshade diameter. Not only does dry mass grow approximately with the square of the diameter, but larger starshades must operate at greater distances from the telescope, which results in greater slewing distances when moving from between targets. More fuel is needed to move not only the heavier mass, but also to cover greater distances in roughly the same time. Gravity gradients and solar pressure for the larger starshades are greater as well, so bipropellant requirements also grow quickly with size. Reducing the starshade size will greatly improve the number of slews possible for a fixed total system mass. In addition, a smaller starshade will have a stronger heritage story to the deployment accuracy and shape stability work done by the Starshade to Technology 5 (S5) project, reducing its development risk and speeding its maturation to Technology Readiness Level (TRL) 5. However, a smaller starshade will increase the IWA for part of the baseline bandwidth. Starshade mass estimates must be scaled down to the appropriate diameter and the science yield must be recalculated. The STDT must then decide on which size offers the best balance of science, cost, and technical risk.

### Observatory Science and Design Reference Mission Trades

The science potential of the high-resolution UV spectrograph (UVS) and HabEx workhorse camera (HWC) will be further investigated and developed by the final report (in particular through a broad "Science with HabEx" community meeting scheduled on October 15 and 16, 2018, to be held at the Flatiron Institute Center for Computational Astrophysics. Further Design Reference Mission (DRM) studies will also be conducted by the final report to optimize the fraction of observatory time that should be dedicated to a GO program, i.e., separate from the two notional exoplanet surveys described in Section 2 and Appendix B. The latest HabEx DRM simulations indicate that a 50% time fraction could be dedicated to GO observations during the primary 5-year HabEx mission—without counting parallel deep field observations—and preserve most of the exoplanet surveys yield presented in this report.

### Slew Budget

While HabEx can accommodate a significant number of slews, propellant will be a limited resource. A science-based assessment of the likely number of required slews is needed to accurately size required propellant.

### System I&T

The final report will include an initial assessment of the integration and test (I&T) flow for both spacecraft. In particular, the report will assess the engineering units, facilities, and specialized test equipment needed for the starshade and telescope I&T.

### Servicing

The current HabEx spacecraft designs will be updated to facilitate future servicing missions. Avionics will be attached to replaceable panels and located for maximum ease of access; an instrument replacement strategy will be formulated.

### Ground Segment and Science Operations

The expected functions and processes of the ground science operations center, as well as data volume issues and the overall mission operations concept will be described in detail in the final report.

### Architecture Trades

This report describes one HabEx architecture, with its hybrid starlight suppression system and two Observatory science instruments (the UVS and HWC). At least one additional point design will be presented in detail in the final report, with full science yield and cost estimates. In addition to this second architecture, tabular comparisons with other possible options will be provided, as recently recommended by the Large Concept Studies Report Team. The explicit goal of the HabEx final report will be to demonstrate the sensitivity of science return, mission cost, and technical risk over a plausible architecture tradespace.

### Section 4

### Section 5

## Section 6

### Appendix B

### Appendix C
none

### Appendix D

### Appendix E
None

# 9  ACRONYMS

| | | | |
|---|---|---|---|
| ACS | Advanced Camera for Surveys | COS | Cosmic Origins Spectrograph |
| ACS | attitude control system | CPCM | center of pressure/center of mass |
| ADC | analog to digital converter | | |
| ADCS | attitude determination and control subsystem | CRIRES | CRyogenic high-resolution InfraRed Echelle Spectrograph |
| AFT | allowable flight temperature | | |
| AFTA | Astrophysics Focused Telescope Asset | CSA | Canadian Space Agency |
| | | CTE | coefficient of thermal expansion |
| ALD | atomic layer deposition | DDOR | Delta Differential One-way Ranging |
| ALMA | Atacama Large Millimeter/ submillimeter Array | | |
| | | DI | directly imaged |
| AO | adaptive optics | DLR | Deutschen Zentrums für Luft- und Raumfahrt |
| AOM | acousto-optic modulator | | |
| APD | (NASA) Astrophysics Division | DM | deformable mirror |
| APD | avalanche photodiode | DOF | degree of freedom |
| APLC | apodized pupil Lyot coronagraph | DRM | design reference mission |
| ASMCS | Astrophysics Strategic Mission Concept Study | DSN | Deep Space Network |
| | | E-ELT | European Extremely Large Telescope |
| ATLAST | Advanced Technology Large Aperture Space Telescope | | |
| | | ee | encircled energy |
| AU | astronomical unit | EEC | exo-Earth candidate |
| AUI | Associated Universities, Inc. | ELT | Extremely Large Telescope |
| AYO | altruistic yield optimization | ELZM | Extreme Lightweight Zerdodur Mirrors |
| BMC | Boston Micromachines Corporation | | |
| | | EMCCD | electron multiplying CCD |
| BOSS | Big Occulting Steerable Satellite | EPRV | Extremely Precise Radial Velocities |
| CAST | Control Analysis Simulation Testbed | | |
| | | eROSITA | extended Roentgen Survey with an Imaging Telescope Array |
| CATE | Cost Appraisal and Technical Evaluation | | |
| | | E-S | Earth-Sun |
| CBE | current best estimate | ESA | European Space Agency |
| CCD | charge coupled device | ESO | European Southern Observatory |
| CDH | command and data handling | ESPRESSO | Echelle SPectrograph for Rocky Exoplanet and Stable Spectroscopic Observations |
| CDM | cold dark matter | | |
| CGI | (WFIRST) coronagraph instrument | | |
| | | ExEP | Exoplanet Exploration Program |
| CGM | circumgalactic matter | Exo-C | Exo-Coronagraph |
| CMB | cosmic microwave background | EXOPAG | Exoplanet Program Advisory Group |
| CMOS | complementary metal-oxide semiconductor | | |
| | | Exo-S | Exo-Starshade |
| CMT | colloidal microthruster | FEA | finite element analysis |
| CNES | Centre National d'Etudes Spatiales | FGC | formation guidance channel |
| | | FGS | fine guidance sensor |
| CONOPS | concept of operations | | |
| COR | cosmic origins | | |
| FINESSE | Fast INfrared Exoplanet Spectroscopic Survey Explorer | ISL | interspacecraft link |
| | | ISM | interstellar medium |
| | | ISS | International Space Station |
| FIREBALL | Faint Intergalactic-medium Redshifted Emission Balloon | IWA | inner working angle |
| | | JAXA | Japan Aerospace Exploration Agency |
| FOV | field of view | | |
| FPA | focal plane array | JPL | Jet Propulsion Laboratory |
| FSM | fast-steering mirror | JWST | James Webb Space Telescope |
| FSW | flight software | KT | Kepner-Tregoe |
| FUV | far ultraviolet | LBTI | Large Binocular Telescope Interferometer |
| FY | fiscal year | | |
| G-CLEF | GMT-Consortium Large Earth Finder | LCP | liquid crystal polymer |
| | | LEO | low Earth orbit |
| GaAs | gallium arsenide | LGA | low-gain antenna |
| GALEX | Galaxy Evolution Explorer | LMAPD | linear mode avalanche photodiode |
| GMT | Giant Magellan Telescope | | |
| GN&C | guidance, navigation, and control | LOS | line of sight |
| GO | Guest Observer | LSST | Large Synoptic Survey Telescope |
| GP-B | Gravity Probe B | LUVOIR | Large UV Optical Infrared Surveyor |
| GPI | Gemini Planet Imager | | |
| GSFC | Goddard Space Flight Center | LV | launch vehicle |
| Gyr | gigayear | LVDS | low-voltage differential signaling |
| HabEx | Habitable Exoplanet Imaging Mission | mas | milliarcsecond |
| | | MBE | molecular beam epitaxy |
| HCST | High Contrast High-Resolution Spectroscopy for Segmented Telescopes Testbed | MCP | microchannel plate |
| | | MEMS | microelectromechanical systems |
| | | MET | laser metrology and control |
| HD | Henry Draper (Catalogue) | MEV | maximum estimated value |
| HgCdTe | mercury cadmium telluride | MIDEX | Mid-sized Explorer |
| HIP | Hipparcos Catalog | MIT | Massachusetts Institute of Technology |
| HIRES | (E-ELT) High Resolution Spectrograph | | |
| | | MLA | microlens array |
| HITRAN | high-resolution transmission molecular absorption (database) | MLI | multilayer insulator |
| | | MOS | multi-object spectroscopic |
| | | MOVPE | metalorganic vapor phase epitaxy |
| HOI | halo orbit insertion | MPC | multi-point constraint |
| HST | Hubble Space Telescope | MPV | maximum possible value |
| HWC | HabEx workhorse camera | MSA | microshutter array |
| HZ | habitable zone | MSFC | Marshall Space Flight Center |
| I&T | integration and test | MUF | model uncertainty factor |
| IFS | integral field spectrograph | Myr | million years |
| IGM | intergalactic medium | NAOJ | National Astronomical Observatory of Japan |
| IMF | initial mass function | | |
| IMU | inertial measurement unit | NASA | National Aeronautics and Space Administration |
| IPAG | Institut de Planétologie et d'Astrophysique de Grenoble | | |
| | | NGAS | Northrop Grumman Aerospace Systems |
| IR | infrared | | |
| IRAS | Infrared Astronomical Satellite | NICM | NASA Instrument Cost Model |
| NICMOS | Near Infrared Camera and Multi-Object Spectrometer | S/C | Space craft |
| NIR | near-infrared | S5 | Starshade to Technology 5 project |
| NIRCam | Near Infrared Camera | SAG | Study Analysis Group |
| NPRO | Nd:YAG non-planar ring oscillator | SAT | Strategic Astrophysics Technology |
| NRAO | National Radio Astronomy Observatory | SBIR | Small Business Innovation Research |
| NRC | National Research Council | SCExAO | Subaru Coronagraphic Extreme Adaptive Optics |
| NSF | National Science Foundation | | |
| NUV | near ultraviolet | SDSS | Sloan Digital Sky Survey |
| OAP | off-axis parabola | SEM | scanning electron microscope |
| OFTI | Orbits for the Impatient | SEP | solar electric propulsion |
| OTA | optical telescope assembly | SHHLLV | super heavy-lift launch vehicle |
| OWA | outer working angle | SIDM | self-interacting dark matter |
| pc | parsec | SIMBAD | Set of Identifications, Measurements, and Bibliography for Astronomical Data |
| PC | photonics crystal | | |
| PCOS | Physics of the Cosmos | | |
| PDR | Preliminary Design Review | | |
| PIAACMC | phase-induced amplitude apodization complex mask coronagraph | SLS | Space Launch System |
| | | SM | secondary mirror |
| | | SN | supernova |
| PICTURE-B | Planet Imaging Coronagraphic Technology Using a Reconfigurable Experimental Base | SNe | supernovae |
| | | SNR | signal-to-noise ratio |
| | | SOA | state of the art |
| | | SOFIA | Stratospheric Observatory for Infrared Astronomy |
| PLATO | PLAnetary Transits and Oscillations of stars | SP | shaped pupil |
| | | SPECULOOS | Search for habitable Planets EClipsing ULtra-cOOl Stars |
| PLC | planar lightwave vircuit | | |
| PLF | payload fairing | | |
| PLUS | Petal Launch Restraint & Unfurler Subsystem | SPHERE | (VLT) Spectro-Polarimetric High-contrast Exoplanet Research (instrument) |
| PM | primary mirror | | |
| PMN | lead-magnesium-niobate | SRON | Netherlands Institute for Space Research |
| PROBA | PRoject for OnBoard Autonomy | | |
| PSF | point spread function | SS | starshade |
| QE | quantum efficiency | SSI | starshade imager |
| QSO | quasi-stellar object | STDT | Science and Technology Definition Team |
| R&D | research and development | | |
| RCS | reaction control system | STEM | Storable Tubular Extendible Member |
| RECONS | REsearch Consortium On Nearby Stars | | |
| | | STIS | Space Telescope Imaging Spectrograph |
| RF | radio frequency | | |
| RMS | radio-millimeter-submillimeter | STM | Science Traceability Matrix |
| ROSES | Research Opportunities in Space and Earth Sciences | TCM | trajectory correction maneuver |
| | | TDEM | Technology Development for Exoplanet Missions |
| RV | radial velocity | | |
| RWA | reaction wheel assembly | | |
| TESS | Transiting Exoplanet Survey Satellite | UTC | United Technologies Corp. |
| TGAS | Tycho-Gaia astrometric solution | UV | ultraviolet |
| THEIA | Telescope for Habitable Exoplanets and Interstellar/ Intergalactic Astronomy | UVOIR | near-UV to far-infrared |
| | | UVS | ultraviolet spectrograph |
| | | Vis | visible |
| TMA | tertiary mirror assembly | VLT | Very Large Telescope |
| TMA | three-mirror anastigmat | VVC | vector vortex coronagraph |
| TMT | Thirty Meter Telescope | WBS | Work Breakdown Structure |
| TPF | Terrestrial Planet Finder | WFE | wavefront error |
| TPF-I | Terrestrial Planet Finder Interferometer | WFIRST | Wide-Field Infrared Survey Telescope |
| TRL | Technology Readiness Level | WFPC2 | Wide-Field Planetary Camera 2 |
| TSI | transit stellar intensity | WFPC3 | Wide Field Camera 3 |
| UMBRAS | Umbral Missions Blocking Radiating Astronomical Sources | WG | Working Group |
| | | WISE | Wide-field Infrared Survey Explorer |
| UST | universal space transponder | XRCF | X-ray and Cryogenic Facility |
| UTAS | UTC Aersopace Systems | Z | redshift (value) |
| | | ZWFS | Zernike wavefront sensor |

# A  EXOPLANET SCIENCE IN THE 2030s

Between now and the expected launch of HabEx in the mid-2030s, many new observatories will come online, both space-based and ground-based, both large and small, some that are currently under construction or being planned, and some that have yet to be envisioned. Notable examples include the James Webb Space Telescope (JWST), the Wide Field Infrared Survey Telescope (WFIRST), and the next generation of giant segmented telescopes with apertures of 24, 30, and 39 m. All of these missions and observatories will have a variety of capabilities and have impressive science goals. This appendix reviews the expectations for the landscape of exoplanet studies leading up to and concurrent with the HabEx Observatory, focusing on how HabEx fits into this menagerie with the new and complementary capabilities it will provide.

## A.1  Exoplanet Science in the 2030s

Planetary systems consist of an enormous diversity of planets: gas giant planets, ice giants, sub-Neptunes, super-Earths, rocky terrestrial planets, and belts of small bodies that generate debris particles. Ongoing research, upcoming developments in ground-based facilities, and the launch of new space missions will continue to advance our knowledge of the variety and nature of these exoplanetary system components over the next decade and a half. Even so, a flagship exoplanet direct imaging and spectroscopy mission like HabEx will provide unique capabilities. The following subsections set the likely context for exoplanet science at the time of the launch of HabEx.

### A.1.1  Exoplanets from Stellar Reflex Motion

Radial velocity (RV) surveys have detected 658 exoplanets as of late 2017,[i] with a median orbital period of roughly 1 year. The median RV semi-amplitude of these detections is 37 m/s. To date, only a dozen planets have been reported with RV semi-amplitudes below 1 m/s. The planets with the lowest claimed RV amplitudes to date are tau Ceti h and f (Feng et al. 2017), both at roughly 0.4 m/s. The current best measurement precision of 50 cm/s is expected to improve toward 10 cm/sec through the development of a new generation of RV spectrographs on large telescopes, observations at higher cadence, and improved calibration methods (Fischer et al. 2016). Stellar RV jitter arising from spots and activity sets a natural noise floor near 2 m/sec (Bastien 2014). Through careful averaging, filtering, and detrending of the data, as well as detailed studies of change in the spectral line shape, the noise from stellar activity may be mitigated, allowing for RV detections of planets with semi-amplitudes below 1 m/sec. Control of systematics at levels considerably better than 10 cm/s level will be required to both identify Earth analogs in the habitable zones (HZs) of sunlike HabEx targets and measure their masses. It is unclear at this writing whether future RV performance improvements will extend to these levels by 2035 for a significant fraction of the stars in the HabEx target sample.

RV surveys to date have detected most of the Jupiter-mass planets within a few AU of late-type stars, but generally lack sensitivity to Neptune-mass planets outside a few tenths of an AU (Fulton et al. 2016). By 2035, a dedicated observing program with instruments and precisions available today could achieve sensitivity to planets of Saturn-mass and greater with orbital periods up to 20 years, and to super-Earths ($\sim$8 M$_\oplus$) with periods of several years. Complementary measurements of stellar astrometric wobbles by the European Space Agency (ESA) Gaia all-sky survey will be available by 2022. Gaia should detect and measure the full orbits for planets of Jupiter mass or larger with periods <5 years across most of the HabEx sample. Altogether, a complete census along with an accurate measurement of the orbital elements of inner giant planets of nearby stars should be well in-hand by 2035.

---

[i] http://exoplanetarchive.ipac.caltech.edu

### A.1.2 A Nearly Complete Statistical Census of Exoplanets: Kepler and WFIRST

Transit surveys have detected 2,789 planets as of late 2017,[ii] largely thanks to NASA's Kepler mission. By 2035, this number will have increased by at least a factor of 10, taking into account the expected results from the transit survey missions Transiting Exoplanet Survey Satellite (TESS) (Ricker et al. 2015) and PLAnetary Transits and Oscillations of stars (PLATO). The launches of these missions are expected in 2018 and 2025, respectively. Nearly all of the detections via transits will be for short orbital periods <1 year. The TESS and PLATO detections will enable mass, radius, and density constraints on the detected planets, which will also be suitable for follow-up transit spectroscopy. In particular, TESS should complete the survey of bright (V < 9) field stars distributed around the sky. With the RV follow-up that will be possible for such targets, the frequency of planets as a function of their mass, and the planetary mass-radius relationship, should be well established for short orbital periods by the time HabEx launches in 2035. Note that both of these missions will have difficulty detecting rocky planets in the habitable zones of sunlike stars due to their long orbits and small transit signals.

The WFIRST observatory, due to launch in 2025, will include a microlensing survey for exoplanets. With its dramatically higher data quality over ground-based microlensing surveys, WFIRST microlensing will discover thousands of exoplanets orbiting beyond the snow line with masses as small as Mars. The results will robustly define the mass and separation distribution of planets orbiting beyond a few AU in a representative sample of the galactic stellar population. It will also enable the determination of the compact object mass function (including free-floating planets) over nearly eight orders of magnitude from objects with the mass of Mars to ~30 solar mass black holes. In combination with the results from Kepler, WFIRST will enable the completion of the statistical census of planets with mass greater than the Earth and separations from zero to infinity (**Figure A.1-1**). These empirical data will be the ultimate check to validate planet formation models that span the full scale of planetary systems, and set expectations for the range of planet sizes and orbital locations that HabEx will study.

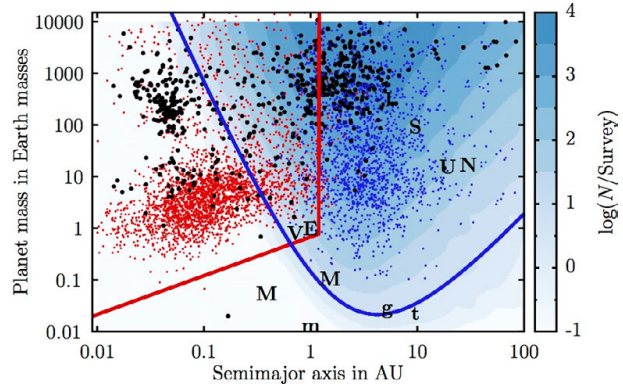

**Figure A.1-1. Together, Kepler and WFIRST will complete the statistical census of exoplanets.** The red points show a subset of the detections from Kepler, whereas the black points shows planets detected by other techniques. The blue points show a prediction of the planets that will be detected by WFIRST. The red and blue curves show the sensitivity limits of Kepler and WFIRST, where three planets would be detected if every star hosted a planet at that mass and semi-major axis. The letters are planets in our solar system, as well as several giant moons. (WFIRST Microlensing Penny Plot)

### A.2 Characterizing Exoplanet Atmospheres: Transiting Planets

There are three primary methods of studying the atmospheres of transiting planets. First, one can measure a thermal emission or reflection spectrum by measuring the drop in the total of the planet plus star flux as the planet passes behind the star. Second, one can measure the absorption spectrum of the planetary atmosphere as the starlight is filtered through the atmosphere and the starlight at wavelengths corresponding to atomic and molecular transitions is absorbed. Finally, one can measure the variation in brightness (the phase curve) as the planet orbits its star. To date, spectral observations of giant exoplanets in transit have confirmed identifications of Na I, $H_2O$, and Ly α, but planets smaller than Neptune lack any detectable molecular absorptions (Deming et al. 2009). It also clear that high-altitude hazes are

---

[ii] http://exoplanetarchive.ipac. caltech.edu

likely extant in a significant fraction of irradiated planets, although the physical mechanisms that dictate whether or not a planet will exhibit hazes is unclear.

By focusing on red dwarf stars, the K2 (Howell et al. 2014) and TESS (Ricker et al. 2015) missions, along with specially-designed ground-based surveys such as MEarth (Charbonneau et al. 2009) and Search for habitable Planets EClipsing ULtra-cOOl Stars (SPECULOOS),[iii] can detect transiting rocky planets with transit depths of 0.1–1.0%, JWST will obtain spectra of a small sample of super-Earth atmospheres and a large sample of mini-Neptune atmospheres. Mid-IR wavelengths should penetrate haze layers that have hampered the detection of absorption features in near-IR transit spectra to date. These observations should be able to establish definitive trends in atmospheric composition and cloud properties as a function of planet or host star properties. Overall, the JWST mission should provide spectra for dozens of warm to hot (mildly to highly irradiated) exoplanets, measuring their temperatures, albedos, and composition with greater sensitivity than ever before (Cowan et al. 2015).

Potentially more intriguing is that JWST could—with an optimal target, a large amount of observing time, and some luck—detect habitable conditions on a rocky exoplanet in the HZ of a nearby mid-to-late-type red dwarf. Detections of biosignatures (such as $O_2/O_3$ in disequilibrium with $CH_4$) will be difficult, and may only be possible for bright red dwarfs hosting large (but not too massive) exoplanets in their HZ. This will also require excellent control of systematics to definitively measure the exceptionally small (tens of parts per million) signal, as well as a concurrent major investment in observing time to reach the statistical precision needed to detect such signals.

Excitingly, the exoplanet community is in the enviable position of having 'two paths' to search for life outside the solar system—the transit spectroscopy method of TESS and JWST to search around red dwarfs for "small black shadows" and the HabEx direct-imaging method to search around sunlike stars for "pale blue dots."

## A.2.1 Characterizing Exoplanet Atmospheres: Direct Imaging

Forty-four potential planetary-mass companions have been imaged around young stars in the near-IR, although less than a dozen that have masses securely below 13 Jupiter masses. For a few of these, detections of $CH_4$, CO, and $H_2O$ have been achieved. The Gemini Planet Imager (GPI) and the Very Large Telescope (VLT) Spectro-Polarimetric High-contrast Exoplanet REsearch (SPHERE) ground-based coronagraphs have now been operating since 2014, although they have detected fewer self-luminous exoplanets than anticipated. Their best planet/star contrast achieved to date on a typical science target is $\sim 10^{-6}$ at a separation of 0.4 arcseconds (Macintosh et al. 2015).

Construction has commenced on two Extremely Large Telescopes (ELTs) in Chile: the European ELT (E-ELT) set for completion in 2024, and the Giant Magellan Telescope (GMT) scheduled for first-light in 2024. Presumably, these telescopes will eventually be equipped with the extreme adaptive optics (AO) systems needed to enable high-contrast coronagraphic imaging, and such systems will likely be built and operating by 2035. For broadband direct imaging, they will be limited to contrasts no better than $10^{-8}$ (ultimately limited by the Earth's atmosphere) in the near-IR, which should be sufficient to spectrally characterize a modest sample of giant planets detected by RV surveys.

The use of high-dispersion spectroscopy to isolate exoplanet signals from starlight has recently shown significant advances (Snellen et al. 2014, Birkby et al. 2017). This method cross-correlates an observed spectrum with a model template spectrum of the exoplanet. It relies on there being a large number of molecular absorption features in the exoplanet atmosphere spectrum to provide a measurable correlation

---

[iii] http://www.amaurytriaud.net/Main/SPECULOOS/index.html

signal. Performance models suggest Earth-like planets located in the habitable zones of a small sample of nearby red dwarf stars can be found when this technique is combined with coronagraphy data from ELTs (Snellen et al. 2015, Wang, Zhou, and Meng 2017). The newly discovered low-mass planets in the habitable zones of Proxima Centauri (Anglada-Escudé et al. 2016) and Ross 128 (Bonfils et al. 2017) will be prime targets. Many years of ELT work with spectral template cross-correlation will have taken place by the time HabEx launches. The contrast and IWA capabilities for these detections ($\sim 3\times10^{-8}$, $\sim30$ mas) will limit the results to the HZs of perhaps a dozen of the nearest red dwarf stars, likely at wavelengths greater than 1 μm.

In conclusion, while there are many promising avenues to explore potentially habitable planets around low-mass, red dwarf stars in the next few decades, spectral characterization of the reflected light of rocky planets in the HZs of sunlike stars will require a space mission, as exemplified by HabEx.

**Figure A.2-1** clearly illustrates that point by placing HabEx in the context of existing and future facilities. Only a mission like HabEx may provide the huge performance improvement required to extend current characterizations of bright self-luminous giant exoplanets to mature rocky Earth-like planets orbiting sunlike stars.

### A.2.2 Circumstellar Disks and Dust

Almost 300 resolved disks around nearby stars are known today. Their internal structures are of great interest, as they can be driven by perturbations from unseen planets. Over the last two years, there has been a surge in the number of resolved disks in continuum and line emission, due primarily to the Atacama Large Millimeter Array (ALMA). Rings and gaps have been observed in the disks of HL Tauri (Ricci et al. 2015) and TW Hydrae (Andrews et al. 2016), and perhaps suggest the presence of forming planets. Protoplanetary disks in the nearest star-forming regions (distances of ~150 pc) are ideal ALMA targets, as their optical depths give them high surface brightness in the submillimeter continuum. For these targets, ALMA's ultimate spatial resolution of 0.01 arcsec will be achievable. An exciting prospect for the 2030s is the follow-up of ALMA disk images with AO coronagraphic imaging on ground-based ELTs: imaging the protoplanets within the disk gaps and directly observing the planet/disk interaction. The ~20 mas inner working angles (IWAs) provided by the ELTs will be enabling for such studies. By the time HabEx launches, ALMA will have thoroughly explored the nearby populations of protoplanetary disks and defined the key targets for HabEx follow-up—mapping the distribution of the smaller particles relative to larger ones.

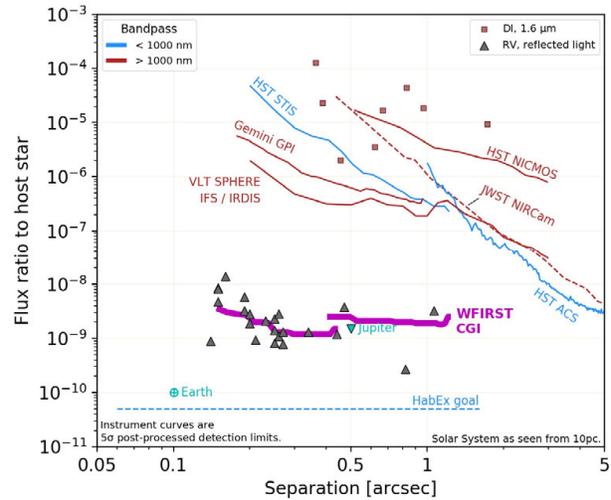

**Figure A.2-1. The HabEx planet-to-star flux ratio performance goal in the context of known planets and existing and planned high-contrast direct imaging instruments.** Shown is the flux ratio between a planet and its star (points for individual planets) or between the dimmest source detectable (solid and dashed lines, assuming a 5σ detection after post-processing) and its star (for instrument performance curves) versus the projected separation in arcseconds. The black triangular points are estimated reflected light flux ratios for known gas giant radial velocity-detected (RV) planets at quadrature, with assumed geometric albedo of 0.5. Red squares are 1.6 μm flux ratios of known self-luminous directly imaged (DI) planets. Cyan points represent the Earth and Jupiter at 10 pc. Figure courtesy of K. Stapelfeldt, T. Meshkat, V. Bailey.

Debris disks are distinct from protoplanetary disk as they are found around older main-sequence stars that have almost certainly ceased giant planet formation. As in our own solar system, they are likely signposts of extant planets and their ongoing dynamical sculpting of the reservoirs of small bodies by collisions. These collisions supply the small dust particles that reflect the starlight, ultimately revealing the existence of these belts and perhaps planets. Many are located relatively nearby with distances of only ~25 pc. They are optically thin with a much lower dust content and much fainter submillimeter continuum emission. It is therefore challenging even for ALMA to resolve their detailed structure. ALMA will map a limited number of the brightest debris disks (i.e., those with total fractional luminosity greater than $10^{-4}$ of their host star) at 0.1 arcsec resolution (e.g., **Figure A.2-2** and MacGregor et al. 2017). Scattered light observations with large diffraction-limited telescopes provide comparable resolution, but not comparable sensitivity, and show a strong detection bias towards debris disks inclined close to edge-on. Nevertheless, interesting systems have been discovered coming out of the Gemini Planet Finder (GPI) and SPHERE (e.g., Currie et al. 2015, Bonnefoy et al. 2016). There are hundreds of nearby (unresolved) Kuiper debris disks with a fractional luminosity of less than $10^{-4}$ observed by Herschel, Spitzer, and Wide-field Infrared Survey Explorer (WISE) studies that neither ALMA nor AO coronagraphy are able to directly detect. JWST will image some of these in thermal emission around a small sample of nearby luminous stars, but only with 0.3" resolution at $\lambda= 20$ µm. In 2035, most of the nearby debris disks detected by Spitzer, Herschel, and WISE will still be too faint for the available detection methods, and thus will be ripe for a space observatory like HabEx—with the sensitivity, contrast floor, and resolution—to make the first high-resolution images.

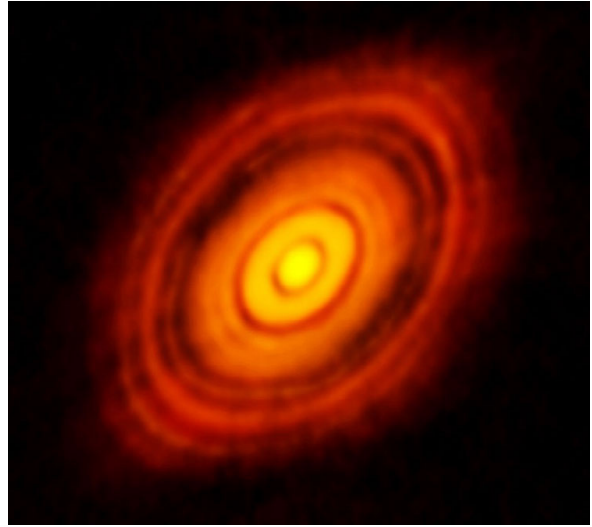

**Figure A.2-2.** ALMA image of the young star HL Tau and its protoplanetary disk. This best image ever of planet formation reveals multiple rings and gaps that herald the presence of emerging planets as they sweep their orbits clear of dust and gas. Credit: ALMA (NRAO/ESO/NAOJ); C. Brogan, B. Saxton (NRAO/AUI/NSF).

As of this writing, the Large Binocular Telescope Interferometer (LBTI) has observed a dozen solar-type stars for dust in their habitable zones. The median exozodi level is no greater than 29 zodis (Ertel et al. 2018), and could be substantially less. When complete, the full survey results will provide an improved estimate. The design of the WFIRST technology demonstration coronagraph should achieve comparable sensitivity to dust in the HZ of nearby stars. If a WFIRST exozodi science program takes place, it could observe HabEx targets inaccessible to LBTI and provide constraints on the dust albedo. Together these two datasets will improve HabEx observational efficiency by prioritizing targets and estimate needed integration times (Howell et al. 2014).

# B SCIENCE YIELD ASSUMPTIONS AND COMPUTATIONS

This appendix details the methodology, instrumental, and astrophysical assumptions used to derive the planet yield estimates summarized in Section 2.3. It also provides further information about the exoplanet surveys' operation concept, and presents the full yield results obtained for different planet types under a broad range of planet occurrence rate assumptions, ranging from pessimistic to optimistic.

## B.1 Methodology for Estimating the Yield of the HabEx Direct Imaging Surveys for Exoplanets

The estimate of the yield of directly imaged planets assumed that HabEx must conduct a blind survey to search for and characterize potentially Earth-like exoplanets. While the efficiency of the HabEx exoplanet survey and the quality of its data products would benefit from a precursor survey identifying potentially Earth-like planets, this study conservatively assume such a survey does not exist at the time of launch. Thus, the yield of such a blind survey is a probabilistic quantity, which depends on HabEx's capabilities using the coronagraph or starshade, the occurrence rate of planets of various types, their detectability, and the unknown distribution of planets around individual nearby stars.

To calculate expected exoplanet yields, the Altruistic Yield Optimization (AYO) yield code of Stark et al. (2014) was used, which employs the completeness techniques introduced by Brown (2005). Briefly, for each star in the HabEx master target list, a random distribution of a large number of synthetic planets of a given type was made, forming a "cloud" of synthetic planets around each star, as shown in **Figure B.1-1**. Planet types are defined by a range of radii, albedo, and orbital elements. The reflected light flux was calculated for each synthetic planet, giving its properties, orbit, and phase, and then determining the exposure time required to detect it at SNR = 7. Based on these detection times and the exposure time of a given observation, the fraction of the synthetic planets that are detectable (i.e., the completeness, as a function of exposure time) was calculated. The completeness simply expresses the probability of detecting that planet type, if such a planet exists. The average yield of an observation is the product of the completeness and the occurrence rate of a given planet type. This process is repeated for every observation until the total mission lifetime is exceeded, arriving at an average total mission yield. In reality, yields may vary from this average due to the random distribution of planets around individual stars; this source of uncertainty was incorporated in the study's yield calculations by accounting for the Poisson probability distribution of planets for each star.

The techniques of Stark et al. (2015) and Stark et al. (2016), which optimize the observation plan to maximize the yield of potentially Earth-like planets, were employed. For a coronagraph-based search, this involves optimizing the targets selected for observation, the exposure time of each observation, the delay time between each observation of a given star, and the number of observations of each star (Stark et al. 2015). For a starshade-based search, a similar optimization was made, but the time between observations was not allowed to be optimized due to expected scheduling constraints; instead the balance between fuel use and exposure time was optimized (Stark et al. 2016).

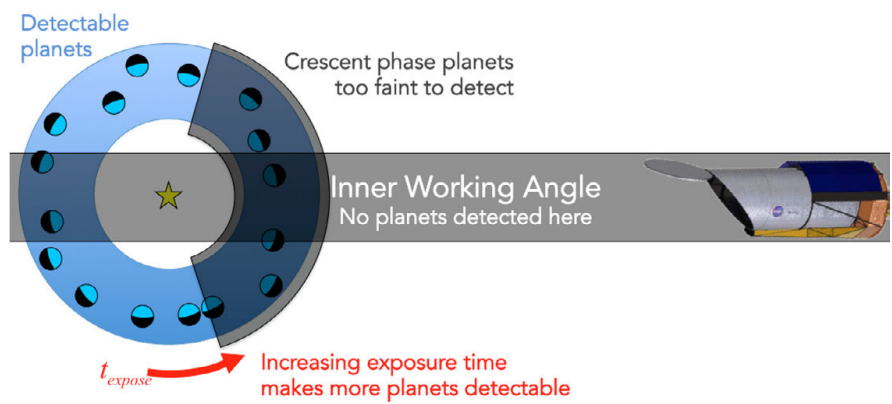

**Figure B.1-1.** The completeness of an observation is the fraction of detectable planets to total planets and is a function of exposure time. The yield of an observation is the product of completeness and the probability that such a planet actually exists (the occurrence rate).

Observations are not directly scheduled. As discussed in Section B.3.5, the baseline 4 m HabEx architecture would detect planets and measure orbits with a coronagraph, then measure spectra with a starshade. The ability to schedule the observations is expected to have a negligible impact on the coronagraph-based search given HabEx's large field of regard. However, the ability to schedule the observations would be more of an issue for the starshade, which has a smaller field of regard and requires direct scheduling with realistic mission dynamic elements, such as solar angle constraints, to optimally schedule starshade observations and more realistically determine the quality and/or quantity of spectra obtained with the starshade, as shown in Section B.3.8.

## B.2    Inputs and Assumptions

Yield estimates require simulating the execution of a mission at a high level. They are therefore dependent on a large number of assumptions about the target stars, planetary systems they host, and the capabilities of the mission. Given the inherent uncertainties in many of these assumptions, consistency between yield analyses is of primary importance. This study adopts inputs and assumptions that are consistent with the choices made by the Exoplanets Standard Definitions and Analysis Team and those made by the LUVOIR STDT. Fiducial assumptions about the parameters that affect the yield of both the coronagraph and the starshade are now reviewed and justified.

### B.2.1    Astrophysical Assumptions

Astrophysical assumptions include planet types and associated occurrence rates, the brightness and extent of exozodiacal and zodiacal dust that will affect observational performance, and the quality of the data in the target catalog.

### B.2.1.1    Planet Types & Occurrence Rates

HabEx followed the planet categorization scheme of Kopparapu et al. (submitted), which consists of a 3 by 5 grid of planets (**Figure B.2-1**) binned by temperature (hot, warm, and cold) and planet radius: small rocky planets (0.5–1 Re), rocky super-Earths (1–1.75 Re), sub-Neptunes (1.75–3.5 Re), Neptune-size planets (3.5–6 Re) and giant planets (6–14.3 Re). Each planet type was assigned a single albedo (listed in **Figure B.2-1**), a Lambertian scattering phase function, and all planets were assumed to be on circular orbits. The semi-major axis boundaries that define the temperature bins of each planet type are assumed to scale with the bolometric stellar insolation, such that they scale with the square root of the bolometric stellar luminosity.

For exo-Earth candidates, this study adopted the region within the green outline in **Figure B.2-1**. By this definition, exo-Earth candidates are on circular orbits and reside within the conservative HZ, spanning 0.95–1.67 AU for a solar twin (Kopparapu et al. 2013). Only planets with radii smaller than 1.4 Earth radii and radii larger than or equal to $0.8a^{-0.5}$, where $a$ is the semi-major axis for a solar twin, were included. The lower limit on this definition of the radius of exo-Earth candidate is derived from an empirical atmospheric loss relationship derived from solar system bodies (Zahnle and Catling 2017). The upper limit on planet radius is a *conservative* interpretation of an empirically measured transition between rocky and gaseous planets at smaller semi-major axes (Rogers 2015). All exo-Earth candidates were assigned Earth's geometric albedo of 0.2, assumed to be valid at all wavelengths of interest.

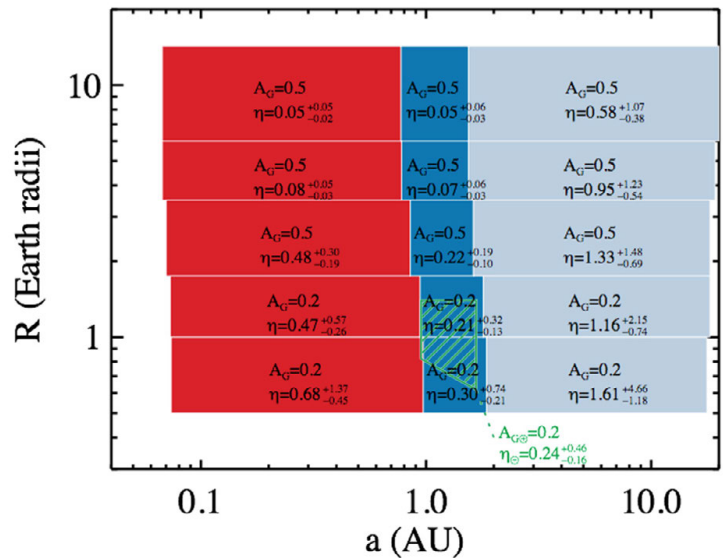

**Figure B.2-1.** Planet classifications for a solar twin used for yield modeling, including bin-integrated occurrence rates (η) and geometric albedos (AG). Planets are binned into hot (red), warm (blue), and cold (ice blue) temperature bins and 5 size bins ranging from small rocky planets to giant planets. The green outline indicates the boundaries of exo-Earth candidates. The semi-major axis boundaries shown are for a solar twin; semi-major axis boundaries are scaled to maintain a constant bolometric insolation.

The occurrence rate values were adopted from the SAG13 meta-analysis of Kepler data (Kopparapu et al., submitted), given by

$$\frac{d^2N(R,P)}{d\ln R\, d\ln P} = \Gamma R^{\alpha} P^{\beta},$$

where $N(R,P)$ is the number of planets per star in a bin centered on radius $R$ and period $P$, $R$ is in Earth radii and $P$ is in years, and $[\Gamma, \alpha, \beta] = [0.38, -0.19, 0.26]$ for $R < 3.4\ R_\delta$ and $[\Gamma, \alpha, \beta] = [0.73, -1.18, 0.59]$ for $R \geq 3.4\ R_\delta$. **Figure B.2-1** lists the occurrence rates when integrating over the boundaries of each planet type. Within each planet type, the SAG13 radius and period distribution given above were used. With this distribution, within a given planet temperature bin, small planets outnumber large planets.

The SAG13 meta-analysis is a crowd-sourced average of published and unpublished occurrence rates, averaged over FGK spectral types. Uncertainties on the SAG13 occurrence rates are not well understood, and are simply set to the standard deviation of the crowd-sourced values. Because of the large uncertainties in the SAG13 occurrence rates, this analysis has weak constraints on how occurrence rates change with spectral type. Thus, the analysis simply assumes that the occurrence rates for each planet type bin are independent of spectral type.

In particular, for exo-Earths in the HZ of sunlike stars, the resulting SAG13 occurrence rate estimate is $\eta_{Earth} = 0.24^{+0.46}_{-0.16}$. This value is consistent with what is arguably the most careful estimate of $\eta_{Earth}$ (and its statistical and systematic uncertainties) by the Kepler team itself (Burke et al. 2015). Burke et al. (2015) notes, however, that different but equally plausible methods of treating various systematic errors can change this value by factors of several in either direction. Partly this is due to the fact that any estimate of $\eta_{Earth}$ from the Kepler survey is necessarily an extrapolation. Nevertheless, pending a more robust estimate of $\eta_{Earth}$ accounting for all Kepler data, this study adopts the SAG 13 value and uncertainty.

**Table B.2-1** summarizes the key astrophysical assumptions underlying the exo-Earth candidate yield calculations.

**Table B.2-1.** Summary of astrophysical assumptions.

| Parameter | Value | Description |
|---|---|---|
| $\eta_\oplus$ | 0.24 | Fraction of sunlike stars with an exo-Earth candidate |
| $R_p$ | [0.6, 1.4] $R_\oplus$ | Planet radius[a] |
| $\alpha$ | [0.95, 1.67] AU | Semi-major axis[b] |
| $e$ | 0 | Eccentricity (circular orbits) |
| $Cos\ i$ | [−1, 1] | Cosine of inclination (uniform distribution) |
| $\omega$ | [0, 2π] | Argument of pericenter (uniform distribution) |
| $M$ | [0, 2π] | Mean anomaly (uniform distribution) |
| $\Phi$ | Lambertian | Phase function |
| $A_G$ | 0.2 | Geometric albedo of planet from 0.55–1 μm |
| $z_c$ | 23 mag arcsec$^{-2}$ | Average V band surface brightness of zodiacal light for coronagraph observations[c] |
| $z_s$ | 22 mag arcsec$^{-2}$ | Average V band surface brightness of zodiacal light for starshade observations[c] |
| $x$ | 22 mag arcsec$^{-2}$ | V band surface brightness of 1 zodi of exozodiacal dust[d] |
| $n$ | 3 | Number of zodis for all stars |

[a] Distribution is a function of $\alpha$ according to the SAG13 occurrence.
[b] $\alpha$ given for a solar twin. The habitable zone is scaled by $\sqrt{L_*/L_\odot}$
[c] Local zodi calculated based on ecliptic pointing of telescope. On average, starshade observes into brighter zodiacal light.
[d] For solar twin. Varies with spectral type, as zodi definition fixes optical length.

### B.2.1.2 Exozodiacal & Zodiacal Dust

Exozodiacal dust adds background noise, thereby reducing the SNR of a planet detection relative to the case of no exozodiacal dust. Unfortunately, current constraints on the amount of zodiacal dust around the HabEx target stars are weak. Therefore, this study simply adopted a baseline exozodi level of 3 zodis, a value consistent with the LBTI exozodi survey intermediary measurements of the median exozodi level of sunlike stars (Ertel et al. 2018). The definition of 1 zodi is a uniform (optically-thin) optical depth producing a V band surface brightness of 22 mag arcsec$^{-2}$ at a projected separation of 1 AU around a solar twin. Thus, the exozodi surface brightness drops off as the inverse square of the projected separation (Stark et al. 2014). Because the HZ boundaries scale for the *bolometric* stellar insolation, the V band surface brightness of 1 zodi of exozodi varies with spectral type (Stark et al. 2014).

The solar system's zodiacal brightness varies with ecliptic latitude and longitude; the closer one observes toward the Sun, the brighter the zodiacal cloud will appear. The zodiacal brightness for each target star is calculated by making simple assumptions about typical telescope pointing (Leinert et al. 1998). For the coronagraph, HabEx assumed that observations could be made near where the local zodi is minimized and adopted a solar longitude of 135 degrees for all targets. For the starshade, the field of regard is limited to solar elongations between 40 and 83 degrees; a constant solar elongation of ~60 degrees is adopted. As a result, the starshade's line of sight through the zodiacal cloud is ~2.5 times brighter than that of the coronagraph.

### B.2.1.3 Target Catalog

The input star catalog was formed using the methods of Stark et al. (in prep). Briefly, the target list is equivalent to the union of the original Hipparcos catalog and the Gaia TGAS catalog. For each star, HabEx adopted the most recent measured parallax value from the original Hipparcos, updated Hipparcos, and Gaia TGAS catalogs, then down-selected to stars within 50 pc. BVI photometry and spectral types were obtained from the Hipparcos catalog. Additional bands and missing spectral types were supplemented using SIMBAD. All stars identified as luminosity class I-III were filtered out, leaving only main sequence stars, sub-giants, and few unclassified luminosity classes.

While the accuracy of any individual star's parameters may be important when planning actual observations, yield estimates can be very robust to these inaccuracies, as their effects average out when considering a large target sample. Accordingly, the blind search portion of HabEx's broad exoplanet survey is expected to be fairly robust to these uncertainties. The exo-Earth yield for the deep survey portion of HabEx's exoplanet search, on the other hand, would be much more sensitive to the uncertainties of the nine individual stars observed.

### B.2.2 Binary Stars

Detecting exoplanets in binary star systems presents additional challenges. Light from companion stars outside of the coronagraph's field of view, but within the field of view of the telescope, will reflect off the primary and secondary mirrors. Due to high-frequency surface figure errors and contamination, some of this light is scattered into the coronagraph's field of view. For some binary systems, this stray light can become brighter than an exo-Earth.

The stray light from binary stars in the final image plane was directly calculated. The numerical stray light models of Sirbu et al. (in prep) were utilized. These models predict the power in the wings of the point spread function (PSF) at large separations assuming a $\lambda/20$ RMS surface roughness and an $f^3$ envelope, where $f$ is the spatial frequency of optical aberrations. Stray light was assumed to be measureable, or able to be modeled; it was included simply as an additional source of background noise. This study made no artificial cuts to the target list based on binarity, and allowed the benefit-to-cost optimization in the AYO yield code to determine whether or not stray light noise makes a target unobservable. In practice, the AYO prioritization does reject a number of binary systems with contrast ratios close to unity and/or close separations. It should be noted that including the full amount of light scattered by the companion is actually conservative, as the companion scattered starlight could be actively reduced with specialized observation methods. For example, HabEx could use the starshade to block the companion starlight while observing with the coronagraph (Sirbu, Belikov, et al. 2017), or use multi-star or super-Nyquist wavefront control coronagraphic techniques (Thomas, Belikov, and Bendek 2015, Sirbu, Thomas, et al. 2017).

### B.2.3   Mission Parameters

For all mission concepts investigated, a total lifetime of 5 years was assumed. For the baseline hybrid mission architecture, 3.75 years of total exoplanet science time (including overheads) was allocated, leaving 1.25 years for dedicated observatory science (not counting parallel observations). Of these 3.75 years, 3.5 years are devoted to a broad exoplanet survey optimized for potentially Earth-like planets, and 3 months are devoted to "deep survey" observations of 9 nearby stars using the starshade. Coronagraph observations were assigned a constant 5 h overhead on all exposures for wavefront control (dark hole generation) while starshade observations were assigned a 30 min alignment overhead after slewing to the target. Total exposure time and overheads were required to fit within the exoplanet science time budget. Slew time of the starshade did not count against the exoplanet science time; it was assumed that during the slew the either the coronagraph or general astrophysics instruments would be observing targets.

For planet detections, this analysis required an SNR = 7 evaluated over the full bandpass of the detection instrument, where both signal and noise are evaluated in a simple photometric aperture of 0.7 $\lambda/D$ in radius. The SNR was evaluated according to Eq. 7 in Stark et al. (2014), which includes a conservative factor of 2 on the background Poisson noise to account for a simple background subtraction. A background term for detector noise was also included and is discussed in Section B.2.4.3. For spectral characterizations, HabEx required a spectrum with R = 140 and

SNR = 10 per spectral channel, which was evaluated at a wavelength of 1 micron.

### B.2.4   Instrument Performance Assumptions

#### B.2.4.1   Coronagraph Assumptions

Coronagraph performance was estimated via a wave propagation model, assuming an idealized optical system and perfect wavefront control. The baseline for this report used the vector vortex coronagraph (VVC 6) described in detail in Section 5. Leaked starlight was simulated as a function of stellar diameter and the off-axis PSFs as a function of angular separation, providing inputs to the yield calculations according to the standards of Stark and Krist (2017).

The wave propagation model does not include some known systematic noise sources, such as residual spatial speckle noise caused by dynamic wavefront errors. Realistic estimates would require full end-to-end simulations of a well-defined telescope, instrument, and observing procedure. Although these effects are mostly independent of the coronagraph design, they will impact the coronagraph noise floor, the properties and frequency of false positives, and the final yield. Because the raw contrast of the vortex coronagraph is expected to be limited by these unmodeled effects, this analysis included a raw contrast floor of $10^{-10}$, which sets the level of shot noise coming from leaked starlight in the SNR calculations. A constant $10^{-10}$ floor is assumed all over the coronagraphic dark hole up to the OWA, meaning that the local level of raw contrast is always set to the worse of $10^{-10}$ and any value predicted by detailed simulations.

**Table B.2-2** summarizes the coronagraph performance that HabEx adopted. Note that although these metrics may provide a useful high-level understanding of coronagraph performance, some metrics should be interpreted with caution. For example, the inner working angle (IWA) estimates where the planet's throughput reaches 50% of the maximum value, but this does not mean that there is no planet signal interior to the IWA. On the contrary, the VVC does provide useful (albeit lower) throughput down to ~2 $\lambda/D$, such

**Table B.2-2.** Summary of adopted vortex coronagraph performance. Listed contrast is for a theoretical point source; contrasts used in simulations included the effects of finite stellar diameter. While only the spatially averaged raw contrast and coronagraph throughput are indicated, AYO simulations used their actual values at the planet angular separation.

| Parameter | Value | Description |
|---|---|---|
| $\zeta$ | $10^{-10}$ | Raw contrast[a] |
| $\Delta mag_{floor}$ | 26 | Systematic noise floor (faintest detectable point source) |
| $T_{core}$ | 0.48 | Coronagraphic core throughput[a] |
| $T$ | 0.18 | End-to-end facility throughput, excluding core throughput |
| $IWA$ | $2.4\lambda/D$ | Inner working angle[b] |
| $OWA$ | $32\lambda/D$ | Outer working angle |
| $\Delta\lambda$ | 20% | Bandwidth |

[a] Average value between the IWA and OWA.
[b] Separation at which core throughput reaches half the maximum value.

that bright, short-period planets may be detectable interior to the quoted 2.4 $\lambda/D$ IWA (= 62 mas at 0.5 µm). *However, only the planets detected by the coronagraph outside of 60 mas can be spectrally characterized by the starshade between 0.3 and 1.0 µm, and only those planets count towards the yields computed here and summarized in Section 2.3.*

The outer working angle (OWA) is the maximum angular separation where planets can be detected, which is set by the size of the dark hole generated by the deformable mirrors (DMs) in the coronagraph. The angular radius of the dark hole is limited to $(N_{act}/2) \times (\lambda/D)$, where $N_{act}$ is the number of DM actuators across the beam diameter; the assumed 32 $\lambda/D$ OWA is consistent with the baseline HabEx design with 64×64 actuators DMs.

The bandpass of the VVC is theoretically unlimited, but in practice is limited by the wavefront control system architecture. High-Contrast Imaging Testbed results indicate that surpassing a bandwidth of $\Delta\lambda/\lambda=0.2$ is challenging with a conventional dual DM coronagraph layout, thereby justifying the adopted bandwidth of 20%.

The total throughput of the system in **Table B.2-2** is evaluated at visible wavelengths and includes the reflectivity of all optical surfaces, the detector quantum efficiency (QE), IFS throughput, and a 5% contamination budget. This throughput metric does not include the core throughput of the coronagraph, which was taken into account separately via the off-axis PSF simulations discussed above. Detector parameters are discussed below.

### B.2.4.2 Starshade Assumptions

Starshade optical performance was estimated using a simple step function model for starlight suppression. This study adopted the performance metrics shown in **Table B.2-3**. Using the standardized yield metrics of Stark and Krist (2017), these performance metrics were then translated into the leaked starlight and off-axis PSFs for the starshade. Unlike the coronagraph, in this case the leaked starlight was assumed to be independent of stellar diameter.

**Table B.2-3** also lists the assumed starshade propulsion parameters. For yield calculations, HabEx assumed that 100 starshade slews were available for the overall exoplanet surveys (Section B.3.7). The surveys consist of a (3.5 year-long) "broad survey" of 111 stars (~770 coronagraph pointings, and ~70 starshade slews on the most interesting systems) and an additional 3 months of "deep survey" observations of a select group of nine stars (~30 starshade slews). The starshade fuel mass was then computed consistently, to allow a total of 100 over the 5 years of the mission, minus the starshade total observing time.

### B.2.4.3 Detector & Other Performance Assumptions

**Table B.2-4** lists the detector noise parameters that were assumed for yield calculations. The total detector noise count rate in the photometric aperture was calculated as

$$CR_{b,detector} \approx n_{pix}\left(\xi + RN^2/\tau_{expose} + 6.73 f CIC\right),$$

where $f$ is the photon counting rate and $n_{pix}$ is the number of pixels contributing to the signal and noise. The parameter $f$ was tuned to each individual target, such that the photon-counting detector time-resolves photons from sources 10 times as bright as an Earth-twin at quadrature.

**Table B.2-3.** Summary of adopted starshade performance.

| Parameter | Value | Description |
|-----------|-------|-------------|
| $\zeta$ | $10^{-10}$ | Uniform raw contrast level, relative to theoretical Airy pattern peak |
| $\Delta mag_{floor}$ | 26 | Systematic noise floor (faintest detectable point source) |
| $T_{core}$ | 0.69 | Starshade core throughput |
| $T$ | 0.20 | End-to-end facility throughput, excluding core throughput |
| $IWA$ | 60 mas | Inner working angle |
| $OWA$ | $\infty$ | Outer working angle |
| $\Delta\lambda$ | 0.7 µm | Instantaneous spectral Bandwidth |
| $D_{ss}$ | 72 m | Diameter of starshade |
| $z_{ss}$ | 124 Mm | Telescope-starshade separation |
| $m_{dry}$ | 6,394 kg | Dry mass of starshade spacecraft including contingency |
| $m_{fuel}$ | 7,007 kg | Total mass of starshade propellant |
| $I_{sk}$ | 308 s | Specific impulse of station keeping propellant (chemical) |
| $I_{slew}$ | 3,000 s | Specific impulse of slew propellant (electric) |
| $\epsilon_{sk}$ | 0.8 | Efficiency of station keeping fuel noise |
| $T$ | 1.04 N | Thrust |

The analysis assumed that the IFS splits the core of the PSF into 4 lenslets at the shortest wavelength, each of which are dispersed into 6 pixels per spectral channel for a total of 24 pixels per spectral channel at the shortest wavelength. For spectral characterization with the starshade, a larger average—$n_{pix} = 72$ per spectral channel—was adopted, assuming that the starshade IFS must Nyquist sample at its shortest wavelength. For broadband coronagraphic detections using the imager, the analysis assumed 4 pixels for the core of the planet. Note that the assumed detector noise is sufficiently low that small changes to the number of pixels have a negligible impact on yield.

## B.3 Operations Concepts

Yield is commonly thought of as the number of planets detected and/or characterized. As shown by Stark et al. (2016), the yield of a mission is very sensitive to precisely what measurements are required for "characterization," and how the mission goes about making those measurements. Thus, the yield depends on the science products desired and how the mission conducts the observations.

### B.3.1 Desired Science Products

HabEx is designed to obtain three primary data products on planets identified as exo-Earth candidates:

1. Photometry: to detect planets and measure brightness and color

2. Spectra: to assess chemical composition of atmospheres

3. Orbit measurement: to determine if planet resides in HZ and measure spectro-photometric phase variations

In the following sections, this appendix describes how HabEx would obtain these data products in an efficient manner to maximize the yield of the mission. While the mission observing strategy and scheduling are optimized for the exo-Earth candidate characterizations, many other planets would be observed in these systems and their yields are also calculated in the direct imaging planet yield analysis.

### B.3.2 Dealing with Confusion

Upon initial detection of a possible companion, the nature of the source may be unclear. The mission would have only photometry, possibly one color, and a stellocentric separation to determine the nature of the object. Color, brightness, and the fact the source is unresolved may allow us to discriminate between background galaxies and exoplanets. However, recent work has shown that other planets can mimic the color of exo-Earth candidates (e.g., Krissansen-Totton et al. 2016). Furthermore, planets that most easily mimic Earth are small, hot terrestrial planets that have even higher occurrence rates than exo-Earth candidates (van Gorkom & Stark, in prep), so planet-planet confusion may be common. However, performing costly characterizations on all planets mimicking an Earth could decrease the efficiency of the exoplanet survey and reduce the yield of exo-Earth candidates; there may be a need to disambiguate point sources to identify high priority planets.

**Table B.2-4.** Photon-counting CCD noise parameters adopted for yield modeling.

| Parameter | Value | Description |
|-----------|-------|-------------|
| $\zeta$ | $3\times10^{-5}$ counts pix$^{-1}$ sec$^{-1}$ | Dark current |
| $RN$ | 0 counts pix$^{-1}$ read$^{-1}$ | Read noise (N/A) |
| $\tau_{read}$ | 1,000 s | Read time |
| $CIC$ | $1.3\times10^{-3}$ counts pix$^{-1}$ clock$^{-1}$ | Clock induced charge |

HabEx, with its dual coronagraph and starshade design, is capable of dealing with these expected sources of confusion without significantly impacting the yield. As shown by Stark et al. (2016), coronagraphs excel at orbit determination, but take longer to provide a spectrum with broad wavelength coverage. Starshades on the other hand, excel at quickly providing spectra, but can only constrain the orbits for a handful of targets due to the cost of slewing the starshade. Combined, these two instruments provide HabEx with multiple and complementary ways to characterize a system.

### B.3.3    Order of Operations

The order in which observations are conducted and the instrument used to perform those observations would impact the final yield of the mission. For example, performing all initial detection and proper motion follow-up with the starshade would be far from ideal, as this requires many costly slews of the starshade. A more efficient order of operations would play to the strengths of each instrument, e.g., by using the coronagraph for initial detections and then to establish orbits, followed by using the starshade to obtain spectra of interesting systems when planets are known to be at advantageous phases.

Ultimately these decisions would depend on uncertain quantities, like $\eta_{Earth}$ for nearby FGK stars and the rate of confusion with background objects. In a low $\eta_{Earth}$ scenario (~0.1), and because of finite search completeness per visit, HabEx would have to search tens of stars to detect 1 exo-Earth candidate. Because of the fuel cost associated with slewing the starshade, initial detection and orbit determination with the coronagraph would likely be better in this scenario, especially if the confusion rate is high. If $\eta_{Earth}$ is high (~1) and the rate of confusion is low, or if precursor observations with other facilities have already revealed which stars host exo-Earth candidates, it may be better to just search with the starshade and immediately take spectra.

A precise operations concept will require further detailed study and will surely be adapted "on the fly" during mission operations. HabEx's dual instrument design would allow maximum flexibility to adapt to these unavoidable astrophysical uncertainties.

### B.3.4    Simulating Operations Concepts

To simulate a given operations concept, this study would need to generate a fictitious universe and model the execution of the mission one observation at a time, adapting to the detections, non-detections, and false positives as the simulated mission progresses, with decision-making logic. While current yield codes are capable of dynamically scheduling with realistic mission constraints to desired decision making logic(Morgan et al. 2017, Savransky and Garrett 2015)(Morgan et al. 2017, Savransky and Garrett 2015), and Section B.3.8), a static time-budgeting approach is more agile for exploration of a variety of operations concepts and is used here.

HabEx approximated the impact of different operations concepts with the AYO yield code by adopting general rules that define the observation plan. For example, to include orbit determination, a system was required to be observed at least six times to a depth consistent with detecting an exo-Earth. To include the effects of confusion, the problem was bounded by assuming that either all systems have a source of confusion in the HZ with the expected flux of an exo-Earth, or none of the systems had a source of confusion in the HZ.

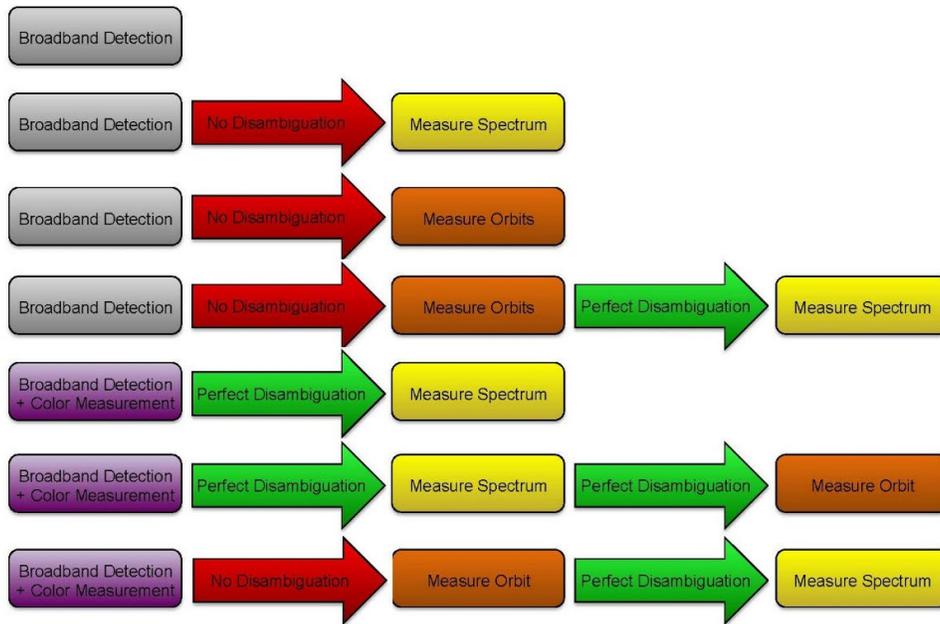

**Figure B.3-1.** A sample of the operations concepts studied for HabEx. Each row presents a simplified operations concept in which a measurement is made, which then provides either perfect disambiguation of exo-Earth candidates, or no disambiguation. The adopted HabEx operations concept is the bottom one, where planetary orbits are measured with the coronagraph before spectra are taken with the starshade, providing the highest science yield (see text for details).

Ten different operations concepts for a variety of HabEx mission designs, for both high and low $\eta_{Earth}$ scenarios, were studied. Each concept produced varying amounts and types of data, ranging from broadband detections only to orbits and spectra. The rows in **Figure B.3-1** show some of the operations concepts considered, with the order of operations for each concept proceeding from left to right. After each step in the operation plan, it is assumed that the information obtained up to that point allows us to perfectly disambiguate exo-Earth candidates (no confusion), or does not provide any disambiguation (100% confusion). Under the assumption of perfect disambiguation, subsequent measurements are only made on the expected yield of exo-Earth, i.e., a fraction of the target sample. Under the assumption of no disambiguation, subsequent measurements must be made on all systems.

### B.3.5 Adopted "Broad Survey" Operations Concept

By studying all of the operations concepts listed above, this study determined an operations concept that achieved all desired science products listed in **Figure B.3-1** and maximized the exo-Earth yield of the broad survey exoplanet search. Importantly, this operations concept is realistic, flexible, and does not require advanced autonomous decision making on board the spacecraft.

The following operations scenario for the "broad survey" 3.5-year exoplanet search was adopted:

1. Detect planets using the coronagraph in broad-band filter 1 (450–550 nm) and filter 4 (700–860 nm), providing color information for planets detected in both but only requiring detection in filter 1

2. Revisit *all* systems as necessary with the coronagraph until the orbits of high-priority planets are sufficiently constrained (likely more than 6 times each over the course of months to years)

3. Based on the color (in favorable cases), orbit, and brightness, identify high-priority targets for spectral characterization

4. Schedule and conduct starshade spectral characterization observations at an advantageous exo-Earth orbital phase, if possible

This operations scenario is both realistic and robust to error. By requiring orbit measurement regardless of what is detected, the operations concept is straightforward, does not rely on any confusion mitigation immediately after a detection, and proper motion would be established for free for all detected planets. Because the HabEx coronagraph's field of regard is nearly the full sky at any given time, the revisit schedule for each star can easily be optimized to maximize detections and constrain orbits without detailed consideration of whether or not the targets are inaccessible. Finally, with an expected yield of ~12 exo-Earth candidates and ~70 starshade slews over 5 years for the broad survey, HabEx expects to be able to measure the spectrum of every exo-Earth candidate (potentially multiple times) in addition to every other planetary system. In other words, the yield of characterized exo-Earth candidates is not expected to be limited by the starshade's fuel constraints.

### B.3.6    Adopted "Deep Survey" Operations Concept

Six months of mission time and ~30 starshade slews will be devoted to "deep survey" observations of ~10 high priority nearby stars. HabEx would perform spectroscopic observations of each of these systems on an average of 3 times over the course of the mission using the starshade and IFS instruments. Of these ~10 targets, more favorable targets may be observed up to 5 times while less favorable targets may be observed only twice.

For each deep survey observation, HabEx would use the starshade to obtain A) a deep broadband image limited by the systematic noise floor (~26.0 mags fainter than the host star), and B) a visible wavelength R = 140 IFS spectrum sufficient to obtain SNR = 10 per spectral channel on an Earth-twin at quadrature, if such a planet exists. These deep exposures and spectra would allow the first detailed understanding of Earth's nearest neighbors.

Deep survey targets are selected based on the high completeness of all planet types that can be achieved with relatively short exposure times. **Table B.3-1** lists the targets, which range from late G to K type stars. **Table B.3-1** also lists the total expected exposure time for each target and the expected exoplanet detection yields for each planet type.

**Table B.3-1.** HabEx's "deep survey" target list. Columns are, from left to right, star name, distance, and spectral type, followed by the total exposure time devoted to this target (sum of 3 visits, detection only), and the total expected exo-Earth candidate yield. For each of these stars the exo-Earth HZ search completeness is close to 100%, and the expected yield is then close to the assumed occurrence rate of exo-Earths around sunlike stars (0.24).

| Name | Dist (pc) | Type | $\Sigma_T$ (days) | $\Sigma Y_{EEC}$ |
|------|-----------|------|-------------------|------------------|
| τ Ceti | 3.65 | G8V | 0.27 | 0.24 |
| 82 Eri | 6.04 | G8V | 1.11 | 0.23 |
| ε Eri | 3.22 | K2V | 0.21 | 0.24 |
| 40 Eri | 4.98 | K1V | 0.55 | 0.24 |
| GJ 570 | 5.84 | K4V | 1.08 | 0.22 |
| σ Dra | 5.76 | K0V | 0.94 | 0.23 |
| 61 Cyg A | 3.49 | K5V | 0.33 | 0.21 |
| 61 Cyg B | 3.50 | K7V | 0.68 | 0.21 |
| ε Indi | 3.62 | K5V | 0.24 | 0.23 |

### B.3.7    Combined Exoplanet Survey and Overall Planet Yields

The overall exoplanet survey consists of:

- 3.5 years of broad survey operations, including coronagraph multi-epoch detections (1.95 year) followed by starshade spectra of all planetary systems with exo-Earths candidates detected (1.05 year) and starshade spectra of at least 50% of the other systems (0.5 year);

- 3 months of deep survey, using the starshade only

The characteristics of these broad and deep surveys are summarized in **Figure B.3-2.** The overall planet yield expected from the 2 surveys is summarized in **Table B.3-2**, using the nominal instrumental parameters but a variety of planet occurrence rate assumptions, ranging from pessimistic, to nominal to optimistic. For each planet type, the nominal case refers to the mean occurrence rate derived by SAG 13. The pessimistic and optimistic yield estimates assume

**Table B.3-2**. HabEx yield estimates for different planet types. As indicated in Figure B.2-1, planets are categorized by a range of surface temperatures (hot, warm, and cold) and planetary radii: small rocky planets (0.5–1 Re), large rocky planets (super-Earths 1–1.75 Re), sub-Neptune size (1.75–3.5 Re), Neptune-size (3.5–6 Re) and giant planets (6–14.3 Re). HZ exo-Earth candidates occupy a subset of the rocky planets bins, and their yield is given in the 2nd column. Planet yields are indicated for the broad survey, the deep survey and the combination of both. Assumed occurrence rates are consistent with estimates from the SAG 13 meta-analysis of Kepler data. "Nominal", "pessimistic" and "optimistic" planet yield estimates are given from top to bottom. They correspond to the nominal, +1σ and -1σ planet occurrence rates derived by SAG 13 (e.g., for Earth-like planets, $\eta_{Earth}$ = 0.24, 0.08 and 0.70, respectively) and also account for Poisson noise uncertainty.

| Planet Types | exo- Earths | Hot Small Rocky | Warm Small Rocky | Cold Small Rocky | Hot Super Earths | Warm Super Earths | Cold Super Earths | Hot Sub-Neptunes | Warm Sub-Neptunes | Cold Sub-Neptunes | Hot Neptune | Warm Neptune | Cold Neptune | Hot Jupiter | Warm Jupiter | Cold Jupiter |
|---|---|---|---|---|---|---|---|---|---|---|---|---|---|---|---|---|
| **Nominal Planet Occurrence Rates** | | | | | | | | | | | | | | | | |
| Planet Yields from Broad Survey | 9.7 | 10.6 | 5.1 | 1.2 | 29.4 | 19.1 | 14.8 | 37.9 | 24.5 | 45.5 | 7.2 | 8.1 | 21.5 | 4.5 | 5.0 | 13.7 |
| Planet Yields from Deep Survey | 2.0 | 2.8 | 2.1 | 0.6 | 2.0 | 1.9 | 2.1 | 1.9 | 2.0 | 4.5 | 0.4 | 0.6 | 2.1 | 0.2 | 0.4 | 1.3 |
| *Planet Yields from Both Surveys* | *11.7* | *13.4* | *7.2* | *1.8* | *31.4* | *21.0* | *16.9* | *39.8* | *26.5* | *50.0* | *7.6* | *8.7* | *23.6* | *4.7* | *5.4* | *15.0* |
| **Pessimistic Planet Occurrence Rates** | | | | | | | | | | | | | | | | |
| Planet Yields from Broad Survey | 3.5 | 3.9 | 1.7 | 0.4 | 14.0 | 8.1 | 5.7 | 23.7 | 13.7 | 22.7 | 4.6 | 4.5 | 9.8 | 2.3 | 2.2 | 4.9 |
| Planet Yields from Deep Survey | 0.7 | 0.99 | 0.6 | 0.2 | 0.9 | 0.8 | 0.7 | 1.2 | 1.07 | 2.2 | 0.2 | 0.3 | 0.9 | 0.1 | 0.2 | 0.4 |
| Planet Yields from Both Surveys | 4.2 | 4.8 | 2.3 | 0.6 | 14.9 | 8.9 | 6.4 | 24.9 | 14.8 | 24.9 | 4.8 | 4.8 | 10.7 | 2.4 | 2.4 | 5.3 |
| **Optimistic Planet Occurrence Rates** | | | | | | | | | | | | | | | | |
| Planet Yields from Broad Survey | 24.6 | 25.6 | 13.7 | 3.2 | 50.8 | 35.8 | 30.1 | 46.1 | 32.5 | 66.7 | 8.7 | 10.9 | 32.1 | 6.6 | 8.3 | 24.9 |
| Planet Yields from Deep Survey | 5.9 | 8.41 | 7.3 | 2.4 | 4.4 | 4.7 | 5.9 | 3.1 | 3.6 | 9.6 | 0.6 | 1.2 | 4.8 | 0.5 | 0.9 | 3.6 |
| Planet Yields from Both Surveys | 30.5 | 34.0 | 21.0 | 5.6 | 55.2 | 40.5 | 36.0 | 49.1 | 36.1 | 76.3 | 9.3 | 12.1 | 36.9 | 7.1 | 9.2 | 28.5 |

the +/-1σ SAG 13 limits on planet occurrence rates (**Figure B.2-1**).

### B.3.8    Scheduling of Exoplanet Surveys Observations

The adopted broad and deep surveys operation concepts were also simulated with EXOSIMS (Savransky, Delacroix, and Garrett 2017), a design reference mission code that uses a different approach to planet yield estimation than the AYO algorithm statistical completeness approach discussed in the previous sections. With EXOSIMS, many realizations of the universe are drawn, each with a different planet distribution around individual target stars, resulting in a different scheduling scheme and planet yield for each draw. As illustrated in

**Figure B.3-2**, EXOSIMS was also used to check that the target observations prioritized by the AYO algorithm were indeed schedulable using realistic mission factors such as solar keep-out (grey circle with yellow edge centered on "S" for Sun), starshade glint constraints (field of regard is the white region where the sun angle <83°), slew times and fuel use. The starshade slew path (black arrows) is scheduled with a three-step look-ahead Traveling Salesman Problem optimizer, and spectral characterization occurs at the end of each arrow, as prescribed in the broad and deep dive operations concepts.

During the starshade slews, coronagraph observations are scheduled and time allocated to other observatory science (i.e., using the HWC and UVS instruments). The synthetic planets are

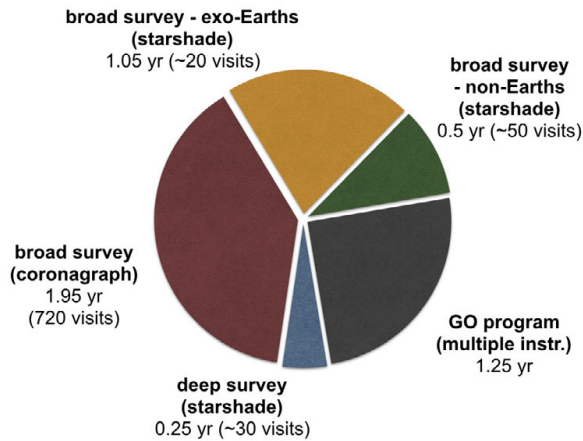

**Figure B.3-2.** HabEx time allocation for a nominal 5-year mission. The broad-survey uses both the coronagraph (for multi-epoch imaging) and the starshade (for spectroscopy). The deep survey only uses the starshade.

'observed' and considered detected or characterized if the goal SNR is reached: green for rocky planets in the HZ), purple for all other planets including rocky planets not in the HZ, red for insufficient SNR to detect any planets, of ~760 potential target stars. The size of the circle indicates the number of repeat detections or characterizations, with the case of 4 detections shown in the legend for scale. Spectral characterizations with the starshade are distinguished by a black edge to the circle and are at the tip of a black slew arrow. The simulated 5-year Design Reference Mission (DRM) shown here, one of a Monte Carlo ensemble of DRMs, performed the deep survey and the follow-up characterization of coronagraph-discovered planets (broad survey) with ~100 starshade slews and 990 kg of fuel remaining (**Figure B.3-3**). These current EXOSIMS results make us highly confident that the observations above are indeed schedulable when taking dynamics mission constraints into account. A more thorough cross-check of the yields predicted independently by the AYO and EXOSIMS algorithms will be the object of future analysis.

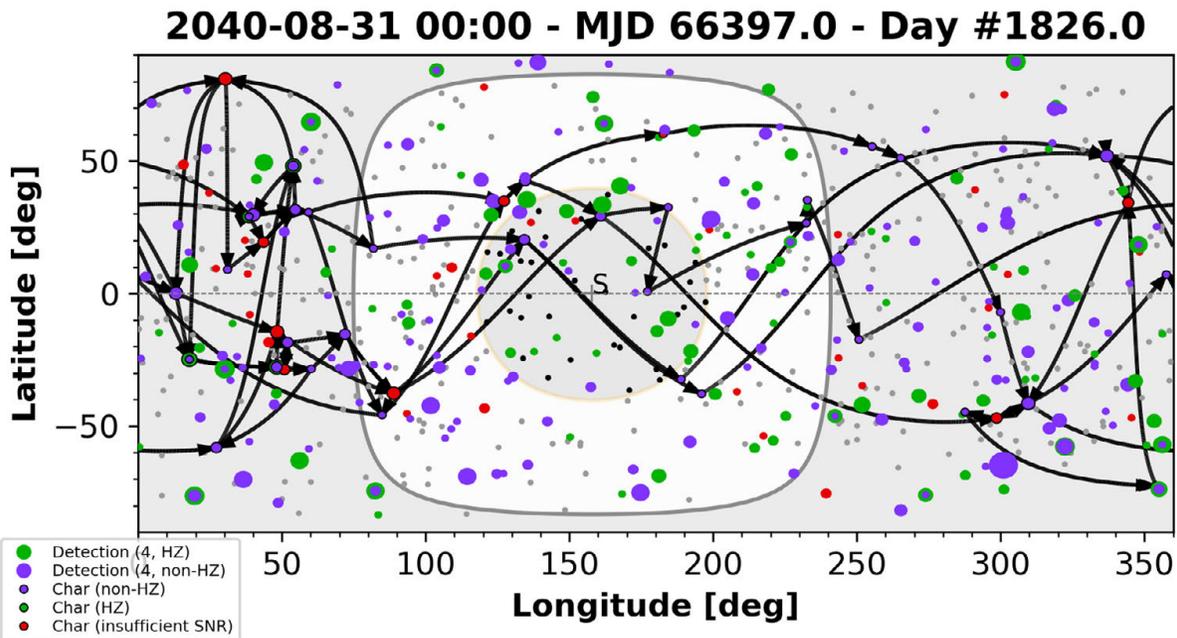

**Figure B.3-3.** EXOSIMS design reference mission simulation scheduling observations planned for the broad and deep survey operations concept. Starshade slews over a nominal 5-year mission (2035–2040) are indicated by black arrows. Using realistic mission factors such as solar field-of-regard, slew times and fuel usage, 100 slews can be accommodated with fuel margin.

grey for an unobserved star, all from a broad list

# C TARGET LIST FOR EXOPLANET SURVEYS

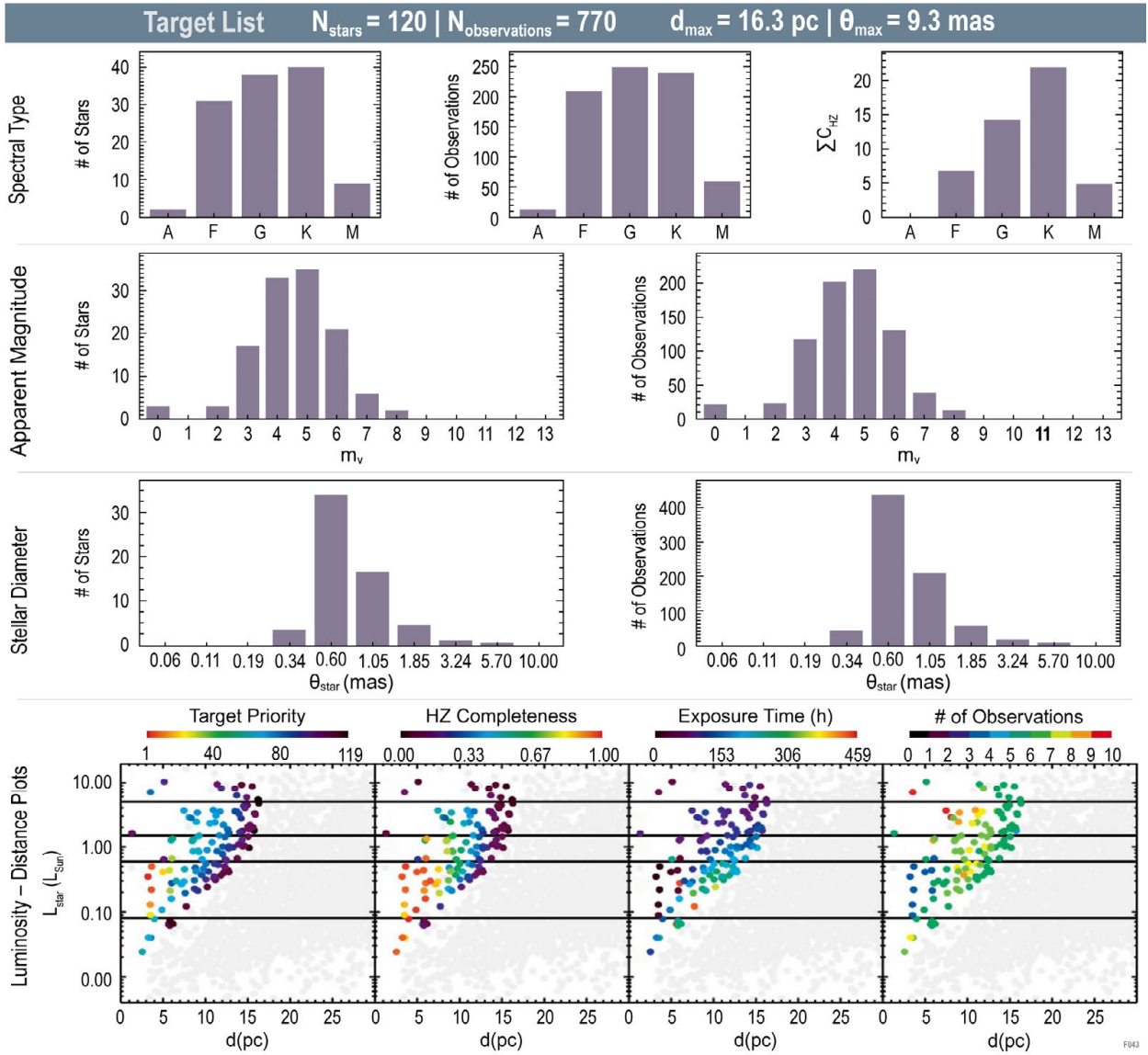

**Table C-1** shows the total list of 120 target stars considered for the exoplanet "broad survey" and for the exoplanet "deep survey." Stars are listed by increasing distance and the 9 deep survey targets are highlighted in *purple*.

*Column Headings:*

A. Star number, ranked by increasing distance
B. Star HIP number
C. Visible apparent magnitude
D. Distance in parsec
E. Spectral type
F. Earth equivalent insolation distance in milliarcsec (mas)
G. Inner edge of the habitable zone (in mas)
H. Outer edge of the habitable zone (in mas)
I. Total exposure time used for broadband imaging summed over all visits (in days)

J. Total completeness for exo-Earths in the habitable zone, summed over all visits.
K. Total number of exo-Earth candidates statistically detected, assuming an exo-Earth occurrence rate of 0.243
L. Cumulative completeness (CC) for exo-Earths in the habitable zone, summed overall all target stars up to the current one. For example, observing the nearest 67 stars in the list (targets within 11.4 pc) yields a cumulative completeness of 40.
M. Average completeness (AC) for exo-Earths in the habitable zone, computed for all target stars up to the current one. For example, the nearest 67 stars in the list (targets within 11.4 pc) are observed with an average completeness per star of 60%.

| A | B | C | D | E | F | G | H | I | J | K | L | M |
|---|---|---|---|---|---|---|---|---|---|---|---|---|
| Star Number | HIP | Vmag | Dist (pc) | Type | EEID (mas) | HZ Inner Edge (mas) | HZ Outer Edge (mas) | Total BB Imaging Time (days) | Total HZ Complete-ness | Total EEC Yield | CC | AC |
| 1 | 71683 | -0.01 | 1.32 | G2V | 964 | 916 | 1610 | 3.6 | 0.067 | 0.016 | 0.067 | 0.067 |
| 2 | 54035 | 7.49 | 2.55 | M2V | 61 | 58 | 102 | 4.98 | 0.961 | 0.234 | 1.028 | 0.514 |
| 3 | 16537 | 3.72 | 3.22 | K2V | 184 | 175 | 307 | 0.21 | 0.97 | 0.236 | 1.998 | 0.666 |
| 4 | 114046 | 7.35 | 3.28 | M2/M3V | 61 | 58 | 102 | 6.96 | 0.961 | 0.234 | 2.959 | 0.74 |
| 5 | 104214 | 5.2 | 3.49 | K5V | 103 | 98 | 172 | 0.33 | 0.865 | 0.211 | 3.824 | 0.765 |
| 6 | 104217 | 6.05 | 3.5 | K7V | 85 | 81 | 142 | 0.68 | 0.843 | 0.205 | 4.667 | 0.778 |
| 7 | 37279 | 0.4 | 3.51 | F5IV-V | 757 | 719 | 1264 | 2.3 | 0.148 | 0.036 | 4.815 | 0.688 |
| 8 | 1475 | 8.09 | 3.56 | M1V | 56 | 53 | 94 | 10.93 | 0.836 | 0.204 | 5.65 | 0.706 |
| 9 | 108870 | 4.69 | 3.62 | K5V | 129 | 123 | 215 | 0.24 | 0.951 | 0.232 | 6.602 | 0.734 |
| 10 | 8102 | 3.49 | 3.65 | G8V | 194 | 184 | 324 | 0.27 | 0.983 | 0.24 | 7.585 | 0.758 |
| 11 | 105090 | 6.69 | 3.98 | M1/M2V | 70 | 67 | 117 | 5.68 | 0.997 | 0.243 | 8.582 | 0.78 |
| 12 | 49908 | 6.6 | 4.87 | K8V | 67 | 64 | 112 | 6.73 | 0.985 | 0.24 | 9.567 | 0.797 |
| 13 | 19849 | 4.43 | 4.98 | K1V | 129 | 123 | 215 | 0.55 | 0.971 | 0.237 | 10.538 | 0.811 |
| 14 | 88601 | 4.03 | 5.08 | K0V | 158 | 150 | 264 | 10.56 | 0.244 | 0.06 | 10.783 | 0.77 |
| 15 | 97649 | 0.76 | 5.13 | A7IV-V | 629 | 598 | 1050 | 1.53 | 0.048 | 0.012 | 10.831 | 0.722 |
| 16 | 25878 | 7.97 | 5.65 | M1V | 45 | 43 | 75 | 19.11 | 0.338 | 0.082 | 11.17 | 0.698 |
| 17 | 96100 | 4.67 | 5.76 | K0V | 115 | 109 | 192 | 0.94 | 0.964 | 0.235 | 12.134 | 0.714 |
| 18 | 29295 | 8.15 | 5.79 | M1/M2V | 43 | 41 | 72 | 12.52 | 0.153 | 0.037 | 12.287 | 0.683 |
| 19 | 45343 | 7.64 | 5.81 | M0V | 46 | 44 | 77 | 10.05 | 0.09 | 0.022 | 12.377 | 0.651 |
| 20 | 73184 | 5.72 | 5.84 | K4V | 79 | 75 | 132 | 1.08 | 0.91 | 0.222 | 13.287 | 0.664 |
| 21 | 3821 | 3.46 | 5.95 | G0V | 189 | 180 | 316 | 3.35 | 0.621 | 0.151 | 13.908 | 0.662 |
| 22 | 84478 | 6.33 | 5.97 | K5V | 67 | 64 | 112 | 8.26 | 0.979 | 0.239 | 14.887 | 0.677 |
| 23 | 99461 | 5.32 | 6.02 | K2V | 88 | 84 | 147 | 18.7 | 0.952 | 0.232 | 15.839 | 0.689 |
| 24 | 15510 | 4.26 | 6.04 | G8V | 135 | 128 | 225 | 1.11 | 0.952 | 0.232 | 16.79 | 0.7 |
| 25 | 99240 | 3.55 | 6.11 | G5IV-Vvar | 189 | 180 | 316 | 3.32 | 0.881 | 0.215 | 17.672 | 0.707 |
| 26 | 99701 | 7.97 | 6.16 | M0V | 41 | 39 | 68 | 10.77 | 0.103 | 0.025 | 17.774 | 0.684 |
| 27 | 114622 | 5.57 | 6.55 | K3Vvar | 83 | 79 | 139 | 5.19 | 0.977 | 0.238 | 18.752 | 0.695 |

| A | B | C | D | E | F | G | H | I | J | K | L | M |
|---|---|---|---|---|---|---|---|---|---|---|---|---|
| Star Number | HIP | Vmag | Dist (pc) | Type | EEID (mas) | HZ Inner Edge (mas) | HZ Outer Edge (mas) | Total BB Imaging Time (days) | Total HZ Complete-ness | Total EEC Yield | CC | AC |
| 28 | 12114 | 5.79 | 7.18 | K3V | 72 | 68 | 120 | 7.58 | 0.869 | 0.212 | 19.62 | 0.701 |
| 29 | 3765 | 5.74 | 7.45 | K2V | 73 | 69 | 122 | 6.01 | 0.842 | 0.205 | 20.462 | 0.706 |
| 30 | 2021 | 2.82 | 7.46 | G2IV | 256 | 243 | 428 | 4.26 | 0.486 | 0.118 | 20.948 | 0.698 |
| 31 | 7981 | 5.24 | 7.53 | K1V | 89 | 85 | 149 | 4.59 | 0.848 | 0.207 | 21.796 | 0.703 |
| 32 | 113283 | 6.48 | 7.61 | K4Vp | 58 | 55 | 97 | 13.06 | 0.754 | 0.184 | 22.55 | 0.705 |
| 33 | 85295 | 7.54 | 7.74 | K7V | 45 | 43 | 75 | 18.37 | 0.225 | 0.055 | 22.775 | 0.69 |
| 34 | 22449 | 3.19 | 8.07 | F6V | 210 | 200 | 351 | 4.22 | 0.595 | 0.145 | 23.37 | 0.687 |
| 35 | 86974 | 3.42 | 8.31 | G5IV | 201 | 191 | 336 | 4.77 | 0.302 | 0.074 | 23.672 | 0.676 |
| 36 | 61317 | 4.24 | 8.44 | G0V | 132 | 125 | 220 | 3.51 | 0.743 | 0.181 | 24.414 | 0.678 |
| 37 | 64924 | 4.74 | 8.56 | G5V | 108 | 103 | 180 | 4.71 | 0.753 | 0.184 | 25.167 | 0.68 |
| 38 | 1599 | 4.23 | 8.59 | F9V | 133 | 126 | 222 | 3.54 | 0.731 | 0.178 | 25.898 | 0.682 |
| 39 | 23311 | 6.22 | 8.71 | K3V | 64 | 61 | 107 | 11.1 | 0.695 | 0.169 | 26.594 | 0.682 |
| 40 | 32984 | 6.58 | 8.71 | K3V | 55 | 52 | 92 | 11.99 | 0.451 | 0.11 | 27.045 | 0.676 |
| 41 | 84720 | 5.47 | 8.8 | M0V | 79 | 75 | 132 | 6.74 | 0.393 | 0.096 | 27.438 | 0.669 |
| 42 | 99825 | 5.73 | 8.91 | K3V | 73 | 69 | 122 | 7.9 | 0.712 | 0.174 | 28.15 | 0.67 |
| 43 | 27072 | 3.59 | 8.93 | F7V | 175 | 166 | 292 | 3.98 | 0.636 | 0.155 | 28.786 | 0.669 |
| 44 | 17378 | 3.52 | 9.04 | K0IV | 204 | 194 | 341 | 5.14 | 0.505 | 0.123 | 29.291 | 0.666 |
| 45 | 57939 | 6.42 | 9.09 | G8Vp | 51 | 48 | 85 | 9.81 | 0.379 | 0.092 | 29.67 | 0.659 |
| 46 | 64394 | 4.23 | 9.13 | G0V | 133 | 126 | 222 | 3.6 | 0.669 | 0.163 | 30.339 | 0.66 |
| 47 | 15457 | 4.84 | 9.14 | G5Vvar | 102 | 97 | 170 | 4.58 | 0.675 | 0.165 | 31.014 | 0.66 |
| 48 | 57443 | 4.89 | 9.22 | G3/G5V | 99 | 94 | 165 | 4.59 | 0.679 | 0.166 | 31.694 | 0.66 |
| 49 | 105858 | 4.21 | 9.26 | F6V | 132 | 125 | 220 | 3.49 | 0.652 | 0.159 | 32.346 | 0.66 |
| 50 | 56452 | 5.96 | 9.56 | K0V | 64 | 61 | 107 | 11.36 | 0.591 | 0.144 | 32.937 | 0.659 |
| 51 | 56997 | 5.31 | 9.61 | G8Vvar | 84 | 80 | 140 | 6 | 0.647 | 0.158 | 33.584 | 0.659 |
| 52 | 81300 | 5.77 | 9.75 | K2V | 70 | 67 | 117 | 8.16 | 0.605 | 0.147 | 34.189 | 0.657 |
| 53 | 68184 | 6.49 | 10.06 | K3V | 56 | 53 | 94 | 10.98 | 0.351 | 0.086 | 34.54 | 0.652 |
| 54 | 8362 | 5.63 | 10.07 | K0V | 74 | 70 | 124 | 8.6 | 0.596 | 0.145 | 35.136 | 0.651 |
| 55 | 29271 | 5.08 | 10.2 | G5V | 93 | 88 | 155 | 5.99 | 0.58 | 0.141 | 35.716 | 0.649 |
| 56 | 58345 | 6.99 | 10.22 | K4V | 48 | 46 | 80 | 14.62 | 0.179 | 0.044 | 35.895 | 0.641 |
| 57 | 13402 | 6.05 | 10.35 | K1V | 62 | 59 | 104 | 12.06 | 0.493 | 0.12 | 36.388 | 0.638 |
| 58 | 14632 | 4.05 | 10.54 | G0V | 144 | 137 | 240 | 4.16 | 0.484 | 0.118 | 36.872 | 0.636 |
| 59 | 10644 | 4.84 | 10.78 | G0V | 101 | 96 | 169 | 5.12 | 0.541 | 0.132 | 37.412 | 0.634 |
| 60 | 57757 | 3.59 | 10.93 | F8V | 176 | 167 | 294 | 4.3 | 0.403 | 0.098 | 37.816 | 0.63 |
| 61 | 86400 | 6.53 | 11 | K3V | 52 | 49 | 87 | 11.74 | 0.246 | 0.06 | 38.062 | 0.624 |
| 62 | 88972 | 6.38 | 11.02 | K2V | 54 | 51 | 90 | 9.74 | 0.281 | 0.068 | 38.343 | 0.618 |
| 63 | 3093 | 5.88 | 11.06 | K0V | 67 | 64 | 112 | 10.27 | 0.417 | 0.102 | 38.76 | 0.615 |
| 64 | 12777 | 4.1 | 11.13 | F7V | 139 | 132 | 232 | 3.6 | 0.377 | 0.092 | 39.138 | 0.612 |
| 65 | 42808 | 6.58 | 11.18 | K2V | 50 | 48 | 84 | 10.76 | 0.188 | 0.046 | 39.326 | 0.605 |
| 66 | 78072 | 3.85 | 11.25 | F6V | 155 | 147 | 259 | 3.93 | 0.408 | 0.099 | 39.734 | 0.602 |
| 67 | 47080 | 5.4 | 11.37 | G8IV-V | 82 | 78 | 137 | 6.51 | 0.386 | 0.094 | 40.119 | 0.599 |
| 68 | 67927 | 2.68 | 11.4 | G0IV | 271 | 257 | 453 | 3.04 | 0.057 | 0.014 | 40.176 | 0.591 |
| 69 | 72848 | 6 | 11.51 | K2V | 63 | 60 | 105 | 9.13 | 0.317 | 0.077 | 40.493 | 0.587 |
| 70 | 23693 | 4.71 | 11.65 | F7V | 105 | 100 | 175 | 4.77 | 0.403 | 0.098 | 40.896 | 0.584 |
| 71 | 109176 | 3.77 | 11.73 | F5V | 160 | 152 | 267 | 3.63 | 0.337 | 0.082 | 41.233 | 0.581 |
| 72 | 69972 | 6.66 | 11.82 | K3V | 51 | 48 | 85 | 10.9 | 0.142 | 0.035 | 41.375 | 0.575 |
| 73 | 107556 | 2.85 | 11.87 | A5mF2(IV) | 242 | 230 | 404 | 2.65 | 0.084 | 0.021 | 41.459 | 0.568 |
| 74 | 15330 | 5.53 | 12.01 | G2V | 74 | 70 | 124 | 6.92 | 0.356 | 0.087 | 41.815 | 0.565 |
| 75 | 15371 | 5.24 | 12.03 | G1V | 84 | 80 | 140 | 7.01 | 0.357 | 0.087 | 42.172 | 0.562 |

| A | B | C | D | E | F | G | H | I | J | K | L | M |
|---|---|---|---|---|---|---|---|---|---|---|---|---|
| Star Number | HIP | Vmag | Dist (pc) | Type | EEID (mas) | HZ Inner Edge (mas) | HZ Outer Edge (mas) | Total BB Imaging Time (days) | Total HZ Complete-ness | Total EEC Yield | CC | AC |
| 76 | 77257 | 4.42 | 12.12 | G0Vvar | 122 | 116 | 204 | 4.25 | 0.342 | 0.083 | 42.514 | 0.559 |
| 77 | 41926 | 6.38 | 12.17 | K0V | 52 | 49 | 87 | 8.12 | 0.122 | 0.03 | 42.636 | 0.554 |
| 78 | 26779 | 6.21 | 12.28 | K1V | 57 | 54 | 95 | 9.54 | 0.21 | 0.051 | 42.847 | 0.549 |
| 79 | 80686 | 4.9 | 12.31 | F9V | 97 | 92 | 162 | 4.86 | 0.3 | 0.073 | 43.146 | 0.546 |
| 80 | 43587 | 5.96 | 12.34 | G8V | 65 | 62 | 109 | 8.18 | 0.242 | 0.059 | 43.388 | 0.542 |
| 81 | 40693 | 5.95 | 12.49 | K0V | 63 | 60 | 105 | 7.28 | 0.203 | 0.05 | 43.591 | 0.538 |
| 82 | 24813 | 4.69 | 12.63 | G0V | 108 | 103 | 180 | 4.81 | 0.295 | 0.072 | 43.887 | 0.535 |
| 83 | 10798 | 6.33 | 12.67 | G8V | 52 | 49 | 87 | 8.51 | 0.118 | 0.029 | 44.005 | 0.53 |
| 84 | 58576 | 5.54 | 12.76 | K0IV | 76 | 72 | 127 | 7.34 | 0.26 | 0.063 | 44.265 | 0.527 |
| 85 | 85235 | 6.44 | 12.76 | K0V | 50 | 48 | 84 | 8.69 | 0.105 | 0.026 | 44.37 | 0.522 |
| 87 | 51459 | 4.82 | 12.78 | F8V | 100 | 95 | 167 | 4.48 | 0.243 | 0.059 | 44.873 | 0.516 |
| 86 | 80337 | 5.37 | 12.78 | G3/G5V | 79 | 75 | 132 | 6.32 | 0.259 | 0.063 | 44.63 | 0.519 |
| 88 | 22263 | 5.49 | 13.28 | G3V | 75 | 71 | 125 | 5.68 | 0.211 | 0.051 | 45.083 | 0.512 |
| 89 | 46853 | 3.17 | 13.48 | F6IV | 212 | 201 | 354 | 2.89 | 0.061 | 0.015 | 45.144 | 0.507 |
| 90 | 7513 | 4.1 | 13.49 | F8V | 140 | 133 | 234 | 3.71 | 0.183 | 0.044 | 45.326 | 0.504 |
| 91 | 98036 | 3.71 | 13.7 | G8IIvvar | 183 | 174 | 306 | 3.89 | 0.146 | 0.035 | 45.472 | 0.5 |
| 92 | 116771 | 4.13 | 13.71 | F7V | 137 | 130 | 229 | 3.19 | 0.143 | 0.035 | 45.615 | 0.496 |
| 93 | 544 | 6.07 | 13.77 | K0V | 59 | 56 | 99 | 8.82 | 0.124 | 0.03 | 45.738 | 0.492 |
| 94 | 79672 | 5.49 | 13.9 | G1V | 75 | 71 | 125 | 5.64 | 0.114 | 0.028 | 45.853 | 0.488 |
| 95 | 16852 | 4.29 | 13.96 | F9V | 129 | 123 | 215 | 3.4 | 0.156 | 0.038 | 46.009 | 0.484 |
| 96 | 53721 | 5.03 | 14.06 | G0V | 93 | 88 | 155 | 4.85 | 0.164 | 0.04 | 46.173 | 0.481 |
| 97 | 12843 | 4.47 | 14.22 | F5/F6V | 117 | 111 | 195 | 3.61 | 0.132 | 0.032 | 46.306 | 0.477 |
| 98 | 102422 | 3.41 | 14.27 | K0IV | 214 | 203 | 357 | 4.36 | 0.05 | 0.012 | 46.356 | 0.473 |
| 99 | 84862 | 5.38 | 14.33 | G0V | 79 | 75 | 132 | 5.63 | 0.162 | 0.039 | 46.518 | 0.47 |
| 100 | 25278 | 5 | 14.39 | F8VSB | 93 | 88 | 155 | 4.1 | 0.075 | 0.018 | 46.592 | 0.466 |
| 101 | 42438 | 5.63 | 14.41 | G1.5Vb | 70 | 67 | 117 | 6.4 | 0.121 | 0.03 | 46.713 | 0.463 |
| 102 | 70497 | 4.04 | 14.53 | F7V | 143 | 136 | 239 | 3.2 | 0.118 | 0.029 | 46.831 | 0.459 |
| 103 | 75181 | 5.65 | 14.64 | G2V | 70 | 67 | 117 | 7.06 | 0.112 | 0.027 | 46.943 | 0.456 |
| 104 | 102485 | 4.13 | 14.68 | F5V | 136 | 129 | 227 | 2.96 | 0.075 | 0.018 | 47.018 | 0.452 |
| 105 | 28103 | 3.71 | 14.88 | F1V | 163 | 155 | 272 | 2.61 | 0.055 | 0.013 | 47.074 | 0.448 |
| 106 | 59199 | 4.02 | 14.94 | F0IV/V | 141 | 134 | 235 | 2.78 | 0.042 | 0.01 | 47.116 | 0.444 |
| 107 | 47592 | 4.93 | 15.01 | G0V | 95 | 90 | 159 | 4.43 | 0.083 | 0.02 | 47.199 | 0.441 |
| 108 | 49081 | 5.37 | 15.05 | G1V | 80 | 76 | 134 | 6.44 | 0.096 | 0.023 | 47.295 | 0.438 |
| 109 | 5862 | 4.97 | 15.11 | F8V | 94 | 89 | 157 | 4.65 | 0.093 | 0.023 | 47.388 | 0.435 |
| 110 | 3583 | 5.8 | 15.16 | G5IV | 65 | 62 | 109 | 8.74 | 0.073 | 0.018 | 47.461 | 0.431 |
| 111 | 95447 | 5.17 | 15.18 | G8IVvar | 90 | 86 | 150 | 5.57 | 0.101 | 0.025 | 47.562 | 0.428 |
| 112 | 82860 | 4.88 | 15.26 | F6Vvar | 97 | 92 | 162 | 4.54 | 0.079 | 0.019 | 47.641 | 0.425 |
| 113 | 86796 | 5.12 | 15.51 | G5V | 90 | 86 | 150 | 5.45 | 0.066 | 0.016 | 47.707 | 0.422 |
| 114 | 95501 | 3.36 | 15.53 | F0IV | 191 | 181 | 319 | 2.38 | 0.027 | 0.006 | 47.734 | 0.419 |
| 115 | 88745 | 5.05 | 15.64 | F7V | 90 | 86 | 150 | 5.37 | 0.081 | 0.02 | 47.815 | 0.416 |
| 116 | 3909 | 5.17 | 15.75 | F7IV-V | 85 | 81 | 142 | 3.64 | 0.029 | 0.007 | 47.844 | 0.412 |
| 117 | 71284 | 4.47 | 15.83 | F3Vwvar | 115 | 109 | 192 | 3.58 | 0.037 | 0.009 | 47.881 | 0.409 |
| 118 | 77760 | 4.6 | 15.89 | F9V | 112 | 106 | 187 | 4.41 | 0.073 | 0.018 | 47.954 | 0.406 |
| 119 | 50954 | 3.99 | 16.22 | F2IV | 143 | 136 | 239 | 2.65 | 0.021 | 0.005 | 47.975 | 0.403 |
| 120 | 112447 | 4.2 | 16.3 | F7V | 133 | 126 | 222 | 3.69 | 0.031 | 0.007 | 48.006 | 0.4 |

# D  ARCHITECTURE TRADES

In late 2015, the Exoplanet Program Advisory Group (ExoPAG) released an advisory report to NASA's Astrophysics Division (APD) on the large mission concepts being considered for further study ahead of the upcoming Decadal Survey (ExoPAG 2015). The ExoPAG envisioned the HabEx mission carrying out direct imaging of Earth analogs and having the breakthrough capability to search for habitability on planets outside our solar system. In addition, ExoPAG advocated for an observatory capability aimed at cosmic origins science with a particular focus on ultraviolet instrumentation. The advisory group considered the likely HabEx design to include a telescope with an aperture larger than existing visible telescopes but smaller than 8 meters in diameter. This guidance defines a fairly large tradespace of possible architectures for the HabEx team to consider. How large a telescope? Is a starshade or a coronagraph included? Which instruments are needed? At the core of the tradespace evaluation is the question: which mission architecture will return the most science for the least cost and risk? Over the first year of the HabEx study, the STDT conducted a series of discussions and an architecture trade study to answer this question. This section summarizes the results of that work.

## D.1  Why 4 Meters?

Aperture size was the first architecture issue addressed by the HabEx Science and Technology Definition Team (STDT). Two sizes—4 m and 6.5 m—were selected as the cornerstones for two unique mission designs to be included in this study. The two diameters were seen as nicely spanning the tradespace aperture range (from the existing 2.4 m Hubble Space Telescope (HST) to the suggested 8 m limit given by the ExoPAG), and both had some specific advantages that could translate into lower cost and risk for the two mission designs. The 4 m choice represented the current industrial limit for a monolithic mirror fabricated out of low coefficient of thermal expansion (CTE) glass ceramic. An unobscured monolithic telescope is the preferred optical front-end for coronagraphs since there is no secondary mirror obscuration or segment edges to mask and no segment phasing to address. Since inner working angle (IWA) for coronagraphs is a function of diameter, 4 meters will greatly expand the number of habitable zones (HZs) that can be searched, in comparison to the Wide Field Infrared Survey Telescope (WFIRST). Finally, the 2010 Astrophysics Decadal Survey (NRC 2010) identified a 4 m telescope addressing both exoplanet imaging and observatory science as "compelling." The segmented 6.5 m aperture was selected for study since design work on the James Webb Space Telescope (JWST) may be leveraged to reduce development risk and possibly cost. Its increased aperture size also promises to improve coronagraph exoplanet yields over the 4 m design. This last point about science yield remains to be verified since more factors beyond aperture size will come into consideration when the yield estimates are calculated for the 6.5 m option.

## D.2  The 4-Meter Architecture Trade

Once the STDT agreed to take on two different aperture options, the architecture trade became two separate trade exercises. The first, the 4 m design, was addressed for this interim report. The 6.5 m architecture will be settled following this report.

Given a 4 m aperture, what is the best approach to detecting and characterizing exoplanets? Given a telescope primarily designed for exoplanet science, what are the most complimentary observatory instruments and what important science questions can they address? Even after reducing the architecture problem by deciding the telescope aperture diameter, the tradespace is still extensive.

To address observatory science, HabEx assigned the STDT members with backgrounds in non-exoplanet astrophysics to identify major observatory research areas that could be advanced with a 4 m space telescope. The members identified six important areas that could be used to build fiducial observatory science cases that would be compatible with the primary exoplanet science design constraints:

1. Stellar archaeology – constraining galaxy formation and evolution by measuring star formation histories in nearby galaxies

2. Constraining the nature of dark matter with the smallest galaxies

3. Measuring the local value of the Hubble constant

4. Tracing the life cycle of baryons

5. Understanding the nature of reionization and the escape fraction of ionizing radiation from star-forming galaxies

6. Understanding massive stars and their relationship to the flow of material from the intergalactic medium to galaxies

All of these areas were included in the HabEx baseline design and are described in detail in Section 3. The STDT recognized that the required measurements could be achieved with only two instruments: a near-ultraviolet, visible, near-infrared (NUV-Vis-NIR) camera with an internal spectrometer, and a high-resolution UV spectrograph. Neither placed major requirements on the telescope or flight system, or added new technologies requiring development. Both were adopted as part of the fiducial architectures and eventually, the baseline design. Instrument designs are detailed in Sections 5.5.7 and 5.5.8.

### D.2.1 Architecture Options

The architecture options considered by HabEx centered on the two primary methods for starlight suppression: external occulting (with a starshade) or internal occulting (with a coronagraph). Either, or both, or multiple starshades could be used, so a number of options needed to be assessed. In addition, a number of design trades also needed to be made to establish fiducial designs representative of each architecture option. Starshade sizes, IWA, and bandpass needed to be settled. An obscured or unobscured telescope choice was required. Coronagraph and starshade camera sensing bands, deformable mirror (DM) sizes, UV, visible and IR detector types, and notional coronagraph designs all needed to be selected to evaluate exoplanet science yield, which was fundamental to the architecture decision.

This section includes detailed descriptions about each of the four architecture options, along with the rationale behind some of the high-level design decisions, and some key characteristics of each are summarized in **Table D.2-1**. In brief, the coronagraph-only option (Option 1) and the coronagraph-and-starshade option (Option 3) utilize an unobscured telescope with a $f/2.5$ monolithic primary mirror. This design offers the best possible front-end optical system for coronagraphy. Without an obscuration or segmentation, the telescope will maximize throughput to the coronagraph and simplify wavefront control. The slow f-number helps minimize polarization effects that can reduce contrast performance (performance requirements, including contrast, are discussed in Section 5.2). The starshade-only options (Option

Table D.2-1. HabEx 4-m architecture trade option design characteristics.

| | Option 1 | Option 2 | Option 3 (baseline) | Option 4 |
|---|---|---|---|---|
| | Coronagraph Only | Starshade Only | Starshade and Coronagraph | Two Starshades |
| Telescope | $f/2.5$ PM, unobscured monolith Zerodur® | $f/1.25$ PM on-axis segmented ULE | $f/2.5$ PM, unobscured monolith Zerodur® | $f/1.0$ PM, on-axis segmented ULE |
| Starshade size(s) | N/A | 72 m | 72 m | 72 m and 32 m |
| IWA | 62 mas @ 500 nm 2.4 λ/D | 60 mas for 300–1,000 nm 108 mas for 1,000–1,800 nm | | |
| Observable bands | 450–1,800 nm | 200–1,800 nm | | |
| Detectors | Visible – EMCCDs; IR – HgCdTe APD detectors | | | |
| Launch vehicle(s) | SLS Block 1B | (2) Falcon H | SLS Block 1B co-launched | (2) Falcon H |

2 and Option 4) use a 4 m segmented on-axis telescope design since wavefront control is not an issue in these architectures. The $f/1$ segmented primary mirror is lighter and mirror segments are easier to manufacture than a large monolith. The on-axis configuration allows launching in a 5 m fairing without requiring the telescope to deploy. Telescope mass and cost are less with the segmented on-axis design. In contrast, the slow, unobscured telescope requires the new, NASA-built Space Launch System (SLS) with the 8.6 m fairing to launch due to both mass and volumetric design characteristics. The segmented on-axis design can be launched on any of the super heavy-lift launch vehicles (SHLLVs) but will require two launch vehicles since the SHLLVs do not have the volume or delivery mass to support launching the telescope with the starshade on a single launch vehicle

For this architecture trade, the primary starshade used in the option comparison was a 72 m perimeter truss design. An 80 m deployable boom design was also developed as part of this study but was not compatible with the co-launched option (Option 3) where both the telescope and starshade spacecraft needed to fit on a single launch vehicle to minimize costs. Both designs are described in Section 5. The starshade size was selected to allow observations with the full near-UV to visible detector band (300–1,000 nm) and IWA of 60 mas in a single pass. The IR band is accommodated by repositioning the starshade closer to the telescope but with a loss in IWA (108 mas). In the case of the two-starshade architecture (Option 4), the 72 m starshade was used with a 32 m starshade. The smaller starshade is used only at the blue end of the spectral band, at a closer separation distance, and only for planet detection. Half-scale was decided for commonality in design with the larger starshade.

These four architecture options represented a fairly broad examination of the architecture tradespace around one of the two apertures most likely to produce good science yield per unit cost and at reasonable technical development risk. As such, all were accepted by the STDT for inclusion in an architecture trade study using the Kepner-Tregoe (KT) rational decision methodology.

### D.2.2 The Kepner-Tregoe Rational Decision Method

In the 1950s, while working for the RAND Corporation, Charles Kepner and Benjamin Tregoe conducted research into decision-making processes within the Strategic Air Command. This research led to the identification of processes for effective decision-making and the founding of the Kepner-Tregoe management consulting company. The principles for the KT methodology are described in their book *The Rational Manager* (Kepner and Tregoe 1965).

The KT decision-making process consists of the following steps:

1. Identify the decision to be made
2. Identify the criteria to be used to make the decision
3. Separate the criteria into "musts" (requirements) and "wants" (desirements)
4. Define the alternative options to be compared
5. Evaluate the alternative options against the decision criteria
6. Identify any risks or opportunities that might follow with the selection each option
7. Select an option

Within the KT process, the failure of an option to meet a "must" results in rejection of that option from further consideration for selection. "Wants" are evaluated on a relative basis, i.e., which option does best and which does worst against a given "want" decision criteria. "Want" criteria can be weighted to elevate the importance of one "want" over another, which was the case with the HabEx 4 m architecture decision. To evaluate the different architectures against the criteria, some level of design definition was needed so fiducial designs and operations concepts were constructed for each architecture options. Consistency in design choices across the options was maintained where possible.

The KT approach was selected because it offers several advantages. It allows the whole STDT to decide which architecture options will be evaluated and against which criteria. All decision information is held in common by the deciding group: there is no private information. Weightings, risks, and opportunities are all discussed and agreed upon by the STDT. The process is completely transparent and allows all decision makers an opportunity to review and challenge the data being used to make the decision. KT has had a successful track record in use by the NASA Exoplanet Exploration Program, allowing stakeholders with initially divergent views on the best solution to a problem, to reach consensus.

### D.2.3  Decision Criteria

The HabEx architecture trade statement was to "recommend a 4 m exoplanet direct detection architecture for…study concept development." This statement limits the trade to direct imaging missions using a 4 m telescope, but also leaves the tradespace open enough for the STDT to assess the relative merits of the major architectures of interest to the exoplanet direct imaging science community. After considerable discussion, the STDT adopted the following criteria for evaluation of the different architectures under consideration:

#### "Musts"

The STDT identified a number of science-related criteria that were considered important enough so that all architectures under evaluation needed to be able to meet these "musts" to remain in consideration. Essentially, these criteria were floor science requirements to the architecture trade exercise. The "musts" include:

- The ability to detect Earth-sized planets in the habitable zones around other nearby stars at a minimum level of completeness
- A requirement to spectrally characterize exoplanets in the visible and near-infrared
- Minimum spectral resolution and signal-to-noise ratio (SNR) capabilities

- The ability of each architecture to determine orbits
- The ability of each architecture to confirm common proper motion of HZ planet candidates

Several programmatic "musts" were also identified:

- A minimum 5-year mission duration
- At least one observatory science instrument must be in the payload
- A minimum allocation of mission time to a Guest Observer (GO) program

These three criteria were taken as requirements on the fiducial designs. In addition, the "musts" included mission cost and schedule start date limitations aimed at making the concepts suitable for selection as the next large mission following WFIRST. Finally, the STDT placed a maximum number of new required technologies (i.e., technologies at or below Technology Readiness Level [TRL] 3) to limit the choice to concepts that would appear technologically ready to start after the launch of WFIRST.

#### "Wants"

The "wants" used in the evaluation were more extensive than the "must" criteria and were often open-ended restatements of the floor "must" requirements. For example, where the "must" might say "no more than N new technologies," the "want" would say, "minimize the number of new technologies." In addition to this relative performance comparison of the important criteria for the "must" list, the "wants" also expanded the planet characterization spectral range, added some observatory science criteria and a few new exoplanet science criteria. These additions include:

- The ability to search binary systems for planets
- Maximizing the outer working angle
- The capability for multi-object spectroscopy
- The capability for host star UV spectral characterization

The STDT placed a heavier weighting on science "wants" over cost, technology, and other programmatic "wants." This was done because the programmatic "musts" were set at levels intended to make the mission selectable, so the "wants" were more focused on how to get an architecture with the most science given that all architectures able to reach the "want" evaluation level are likely to be seen as selectable from a cost and risk point of view. Still, cost, schedule and the number of new technologies remained in the "want" evaluation but at a lower weighting than science.

Essentially, the dimensions of the tradespace being evaluated are performance, cost, and risk. Accordingly, the "musts" can be seen as limits on these dimensions while the "wants" as comparative measures along these dimensions.

The science-related "musts" are largely focused on establishing the floor performance on detecting and characterizing HZ planets and are intended to ensure that the architecture selected will have the capability to detect and characterize Earth-sized planets within the HZ. Additional science "musts" are included to guarantee compelling observatory science with the selected architecture. Risk-related technology development "musts" are aimed at keeping the selected architecture's number of new technologies at or below the number seen in past prioritized Decadal Survey concepts. The cost limit ($7B FY17) was based on HabEx's assessment of available funding in the Astrophysics Division budget over a 10-year period, assuming that funding levels for the JWST returned to the APD budget upon launch of JWST. This annual budget "free energy" estimate is based on a 20-year budget forecast released by the NASA Astrophysics Division Director in October 2015 at the X-ray Surveyor Workshop in Washington, DC. While a larger number could be justified with a longer development period, the HabEx STDT was comfortable with this limit since the number is in keeping with previous, large, facility-type observatories such as HST and Chandra.

The "wants" looked to broaden the exoplanet science performance beyond floor detection and characterization of HZ Earth-sized planets, and valued not only strong direct imaging performance in the HZ, but also the detection and characterization of a broader range of planet types as well as circumstellar disks. In addition, greater spectral range in the observatory science performance and more time for the observatory program were seen as strengths in the architecture evaluation. Architectures that minimized cost and technical risk were also favored in the "wants." The "wants" were weighted heavily in favor of science performance. This choice was made for two reasons. First, the mission was seen as a world-class space-science facility on par with HST so science performance must be the first priority. And second, the constraints on cost and risk levied in the "musts" were aimed at keeping the concept implementable with current funding levels and a manageable number of new, required technologies—further emphasis on cost and risk minimization were seen as desirable but not critical.

To assess these "musts" and "wants," the STDT divided up into Working Groups (WGs) and the evaluation of the different "musts" and "wants" was then divided among the WGs. Engineering support provided assessments of the number of new technologies, rough mass estimates, and preliminary cost estimates associated with each candidate architecture. Yield estimates were assessed using the widely accepted Altruistic Yield Optimization (AYO) yield modeling tool.

### D.2.4   The Trade Results

In a facilitated face-to-face meeting, the trade "wants" were weighted by the STDT and the assessments of the trade criteria were compiled into a trade matrix. For each "want," the best performing option was awarded 10 points. Options with high performance but not best-performing received eight points; good performance received five and low performance received two. For a few criteria, some options could not deliver performance at all; these cases received zero points. The points were then multiplied by the criteria's weighting factor and then points were totaled across all "wants" for each architecture option.

The resulting evaluation favored the combined starshade and coronagraph option (Option 3) slightly ahead of the two-starshade option (Option 4). Both options benefit from the high throughput and broad spectral range offered by the large starshade, but the coronagraph is more efficient at detecting planets (and, consequently, determining orbits) than the small starshade due to the coronagraph's faster target acquisition. Cost advantages with the starshade-only option and the coronagraph-only option (Options 1 and 2) were muted by the weightings. The impact of alternative weightings were examined by the STDT but the STDT remained in support of their original science-heavy weighting.

After establishing the assessments for all the "wants," the STDT identified a number of risks and opportunities connected to different architecture options. KT uses this step to make sure that any additional, relevant information that did not get into the "must"/"want" evaluation is brought into the decision process. If compelling, these added facts can motivate the deciding group to select a lower scoring option as their final choice. In the case of the 4 m architecture trade, the risks and opportunities largely still favored the starshade/coronagraph architecture. The HabEx STDT reached consensus and decided to advance the combined starshade and coronagraph architecture for further development. Details of that baseline architecture are discussed in Section 5.

## D.3    Alternative 4-Meter Architectures

The HabEx 4 m architecture trade evaluated four different architectures against science return, cost, and risk. These four options were selected for the trade because they represented highly varied, yet feasible, mission configurations within the tradespace, and as such, promised to give a good understanding of the design sensitivities of exoplanet direct imaging missions built around a 4 m telescope. None of the fiducial designs were optimized; HabEx itself is just a proof-of-concept design intended to make the case that compelling exoplanet science can be achieved at an affordable cost and with low technical risk. Optimizations will be handled by any future mission. Instead, the trade is intended to establish a rationale for why the STDT selected the architecture that they did for this report, and to communicate the constraints and sensitivities involved in a direct imaging mission using a 4 m telescope.

The architecture trade required early assessments of rough cost and risk to compare the options. A rough cost estimation tool was assembled using:

1. Analogue instrument and spacecraft cost estimates from the NASA Instrument Cost Model (NICM) and past Team X estimates for similar concepts;

2. Telescope costs form the "Update to single-variable parametric cost models for space telescopes" model (Stahl et al. 2013);

3. Starshade costs from scaling an earlier model estimate that had been reviewed by the Exo-Starshade CATE team;

4. Analogue annual telescope spacecraft operations costs from historic data, and starshade operations costs form a previous CATE estimate; and

5. Other ancillary project costs using percentages taken from past CATE estimates.

Assessing the risk of a mission is less straightforward. As a surrogate for risk, the architecture study used a count of the number of new technologies (i.e., technologies at or below TRL 3 at the time of the HabEx final report's release) for each option. A concept requiring more technologies to be developed before launch will likely run into more development setbacks than one that has fewer new technologies. HabEx has no technologies below TRL 3. At the time of the release of the HabEx final report, three areas of technology development are likely to remain.

### D.3.1 The 4-Meter Architectures

The four options selected by the STDT for evaluation in the initial architecture trade are: the coronagraph-only architecture (Option 1), the starshade-only architecture (Option 2), the combined starshade and coronagraph architecture (Option 3), and the starshade-only option with a second, smaller starshade (Option 4). The options were summarized earlier in **Table D.2-1**. Option 3 was selected as the HabEx report 4 m baseline and is covered in detail in Section 5. The other three alternative architectures are covered here.

### D.3.2 Common Design Elements within All Architectures

Prior to conducting the trade evaluation, the STDT needed to settle on fiducial designs to represent the competing architectures in the trade. These designs were not detailed; they only defined what was necessary to assess science yield, cost and technical maturity (a surrogate for risk) at a very course level. Early engineering trades and operational scenario discussions provided enough insight to allow specification of several design requirements that would be needed no matter which option was selected so all four fiducials needed to include these elements.

**Orbit Location.** The first consideration in the architecture trade concepts was where to locate the observatory. Various Earth orbits were unattractive due to the thermal variability of the orbits and its detrimental impact on coronagraph measurements. In addition, starshade operations were not possible due to the need for large-separation, formation flying and long-period, target tracking. Heliocentric Earth-trailing and -leading orbits were not possible due to the need to make the observatory serviceable—a future servicing mission could not practically reach a telescope in such an orbit years after the initial launch. The ideal location for an exoplanet direct imaging mission would be at the Earth-Sun L2 point as for JWST and WFIRST. This location provides a low disturbance environment, simplifies starshade formation flying, and allows the possibility of future observatory servicing.

Other Lagrange points are not as advantageous due to their distance from Earth (reduced data volume and more difficult servicing) or, in the case of L1, inferior observing field of regard.

**Upper Limit on the Direct Imaging Observing Band.** Determining where to set the direct imaging spectral limit at the long wavelength end of the observing band is a trade of access to desired molecular spectral features and operating temperature for the telescope. 1,800 nm was adopted as the upper limit since it permits the telescope to operate near room temperature. Operation at wavelengths longer than 1,800 nm would require cooling the telescope well below room temperature, which adds cost and complexity to the integration and test of the payload. More concerning, a cold telescope primary mirror would condense contaminants on the mirror surface, which would have detrimental consequences for the UV observatory science.

**Direct Imaging Detectors.** The coronagraph and starshade camera need to have visible and IR detectors to cover the full spectral range. The chosen detectors are a mercury-cadmium-telluride (HgCdTe) avalanche photo diode (APD) device to cover the near-infrared to 1,800 nm, and an electron multiplying charge coupled device (EMCCD) to cover the visible spectrum. The EMCCD can be modified by a delta doping process to extend its sensitivity into the near-UV imaging. Both device types are in production with performance meeting HabEx's direct imaging requirements; the EMCCD is baselined for the WFIRST coronagraph so it will have been flown in space before the HabEx mission. EMCCDs, HgCdTe APDs, and delta doping are all discussed in more detail in Section 6.

**Mirror Coatings.** Primary and secondary telescope mirrors are used by both the direct imaging instruments and the observatory science instruments. The STDT held extensive discussions on what coating material should be used on the two mirrors. Aluminum was traded against silver for the reflecting material. Silver did not permit observing in the UV and early simulations indicated that silver introduced polarization errors more readily than aluminum leading to inferior coronagraph contrast performance. Although silver offered slightly better reflectivity across the visible band, the STDT gave priority to contrast and UV performance and chose aluminum for the primary and secondary mirrors. UV observatory science preferred an aluminum mirror coating with a protective overcoat that extended the UV spectral cut-off as far into the UV as possible. A number of overcoats were discussed including magnesium-fluoride, lithium-fluoride and lithium-fluoride/magnesium-fluoride. Magnesium-fluoride is the overcoat used on HST so it has a proven operational life approaching 30 years, but it also has a sharp observational cut-off at about 115 nm. The other coatings promise useable reflectance below 110 nm with some reaching down to 100 nm, but none have yet demonstrated the desired lifetime stability. For this study, the STDT elected not to add a new technological development with the mirror coating, and chose to use aluminum with a magnesium-fluoride overcoat like HST. More details on the mirror coatings can be found in Section 6.

**Observatory Science Instruments.** As noted earlier, the STDT identified a number of additional astrophysics science goals that could be realized with a 4 m telescope. The six highest-priority goals were associated with two different instrument types: an ultraviolet spectrograph and a general purpose camera with a spectrometer, operating in a spectral band from the near-UV to the near-IR. These instruments were compatible with each of the four architecture options.

### D.3.3    Other Fiducial Design Choices

**Mirror Material.** Early design trades included a look at the best material for a 4 m mirror. The options were Corning ULE® or Schott Zerodur®. The Zerodur® had an advantage in better thermal stability and homogeneity but the closed-back ULE® design made for a stiffer mirror, which would be better for rejecting mechanical disturbances. The possibility of using micro-thrusters and eliminating the major mechanical disturbance shifted the decision toward Zerodur®.

**Lower Limit on Direct Imaging Observing Bands.** In the architecture trade, the blue-end limits for the coronagraph and starshade camera were set at different wavelengths. The starshade camera was set at 300 nm since such a limit would allow good characterization of exoplanet atmospheric Raleigh scattering. Starshade camera mirrors in one channel could be coated in aluminum like the telescope primary and secondary mirrors. The coronagraph limit was complicated by high throughput losses due its greater number of mirrors. The STDT saw throughput as a significant factor and decided on adopting silver coating for all mirrors within the coronagraph. Silver has a low-end reflectance drop-off starting around 450 nm so going down to 300 nm is not possible with the coronagraph.

**Direct Imaging Instrumentation.** The two types of direct imaging instruments are coronagraphs using the internal occulting method of starlight suppression, and the camera supporting the external occulting starshade. Within the coronagraphs there are a number of different internal occulting methods that can be used. The final coronagraph trade will be carried out after this interim report but for the interim report the vortex charge-6 and vortex charge-8 designs are considered along with the hybrid Lyot coronagraph (HLC). Early simulations showed that the vortex charge design would be less sensitive to telescope thermal and mechanical disturbances than the HLC. The vortex charge-6 design was far less sensitive to telescope mirror rigid body motion but at the cost of a larger IWA (about 2.4 $\lambda$/D). The vortex charge-8 is even more immune to telescope disturbances than the vortex charge-6 but again, the IWA will increase and there would be some decrease in the number of reachable habitable zones. The vortex charge-6 was adopted for the interim report coronagraph; a more detailed structural, thermal, and optical performance (STOP) analysis will be performed to verify which coronagraph actually produces the best science yield. The starshade camera was not traded; it was designed to support a specific starshade size (~72 m) over a 300–1,800 nm band.

**Starshade Size.** Starshade sizing is fundamentally a trade between diameter, IWA, and contrast level for fixed observational bands. HabEx set a primary objective of being able to characterize from 300–1,800 nm on a single target visit. Capturing such a broad spectrum reduces the number of starshade visits needed to complete characterization of a planetary system, and greatly increases the chance of finding evidence of atmospheric gases associated with life during the baseline 5-year mission. Early yield trade studies suggested an IWA of about 60 nm would produce enough habitable zones to deliver a total HZ completeness greater than 40 over the mission; a number considered compelling to the STDT. Coupled with the need for $10^{-10}$ starlight suppression needed to detect and characterize Earth-sized planets in the habitable zone, the architecture fiducial designs were driven toward starshades in the 70–80 m range. The JPL starshade designers developed a 72 m design suitable for co-launch with a telescope spacecraft on the SLS or as a stand-alone launch on an SHLLV, and an NGAS starshade team created an 80 m design for SHLLV launch only. For simplicity in the architecture trade, the 72 m design was used for all starshade fiducial designs. In the case of the two starshade options, a 32 m starshade was assumed as the second starshade, for design efficiency.

**Launch Vehicles.** In the timeframe of a future HabEx mission, the likely launch vehicles that will be available to the mission would be the SLS and several new SHLLVs. None exist today. Only the Delta IV H is available today and that launch vehicle will be replaced by ULA's Vulcan launch vehicle long before any HabEx mission. The different fiducial designs require either the SLS or an SHLLV for all configurations. For the architecture trade, the two launch vehicles closest to launch, the Falcon Heavy and the SLS Block 1B, were adopted as the fiducial launch vehicles and set the mass and volume constraints for the trade.

## D.4  Coronagraph without a Starshade

Perhaps the least complicated architecture of the four examined is the coronagraph-only architecture (Option 1). This option avoids some of the programmatic complexity of the dual-launch starshade-only architectures (Options 2 and 4) and cost and technical complexity of the starshade plus coronagraph baseline architecture (Option 3) but at the price of diminished science yield when compared to the baseline, and limited spectral range and spectral characterization capability when compared to the starshade-only options.

### D.4.1  Concept Overview

For the architecture trade, the coronagraph-only option is essentially the baseline without the starshade spacecraft. The orbit and launch vehicle remain the same. The payload differences are small. Operations are less complicated since there is only one spacecraft and no need for formation flying.

### D.4.2  Payload Differences from Baseline Option

The payload (i.e., the telescope and its associated instruments) is only slightly simplified from the payload in the baseline option. The telescope remains the same $f/2.5$ PM, unobscured design with a 4 m monolithic primary as used in the baseline design. The coronagraph, UV spectrograph, and workhorse camera instruments remain unchanged, but the starshade camera is eliminated. This means that the direct imaging spectral coverage is reduced from 300–1,800 nm to 450–1,800 nm. This change also requires the coronagraph to handle all spectral characterization science whereas the baseline design largely handled spectral measurements through the starshade camera's integral field spectrograph (IFS) in one 300–1,000 nm broadband observation. The coronagraph would cover the 450–1,000 nm band in 20% increments requiring longer duration observations than the starshade. Since the baseline design used the coronagraph for planet detection, there is no change in the number of planets detected with Option 1, but the number of planets characterized is reduced by about two-thirds.

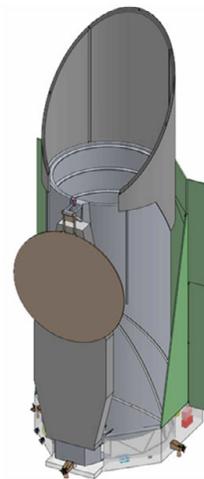

**Figure D.4-1.** Telescope spacecraft used with Options 1 and 3.

### D.4.3  Other Differences from Baseline Option

While the removal of the starshade eliminates considerable mass from the baseline option, the telescope still exceeds the expected launch capabilities of the SHLLVs to L2 and the off-axis design does not fit in a 5 m fairing, so the coronagraph-only option will still require the SLS Block 1B launch vehicle like the baseline design.

### D.4.4  Mission Cost

The Option 1 cost savings over the baseline option largely stem from the elimination of the starshade and the starshade camera. There is also some small additional cost savings from simplified operations. Option 1 was the lowest cost option of the four evaluated, running about 20% lower than the baseline option.

### D.4.5  Required Technology Development

In addition to being the lowest cost option of the four examined, the coronagraph-only option was also the lowest risk option since it only required closing the one technology gap on coronagraph contrast performance.

## D.5  Starshade without a Coronagraph

The starshade with no coronagraph option (Option 2) looks at using the starshade for both exoplanet detection and characterization. While easing requirements on telescope performance, this option is limited by the amount of propellant on the starshade, and the speed at which it can slew from target to target. Additionally, the long-duration slew maneuvers also limit the number of targets that can be reached within the 5-year mission.

### D.5.1  Concept Overview

While the starshade and starshade spacecraft remain the same as in the baseline option, the concept takes advantage of the relaxed telescope wavefront stability requirements to reduce telescope size, mass, and cost. Orbit and mission duration remain the same as in the baseline option.

### D.5.2  Payload Differences from Baseline Option

Unlike the baseline option, the starshade-only options (Option 2 and Option 4) use an on-axis telescope design with a segmented primary mirror. The notional segmented telescope design and mass estimate were developed at JPL by optical engineers having experience with segmented telescopes in the visible spectrum, including ground testbeds for Next Generation Space Telescope (NGST)/JWST, the Space Interferometry Mission (SIM), and ground observatories, such as the Palomar Testbed Interferometer, the Keck Telescopes, the Thirty Meter Telescope, and the Keck Interferometer. The telescope is a non-deployed, on-axis design with six petal segments.

The petals are 60° pie-shaped segments made from ULE® closed back glass with 10 cm thickness and a maximum dimension of 1.7 m. Their proportions are well within the current glass and mirror manufacturing capabilities for space telescopes. Petal tip, tilt, and defocus are controlled with six rigid body actuators and a laser-metrology truss for position sensing. Petal surfaces are also actively controlled using figure control actuators.

This telescope was used in the architecture trade because:

- The on-axis design and faster $f/\#$ (an $f/\#$ of 1.25 for the segmented telescope verses 2.5 for the baseline monolith) allow the telescope to fit in an SHLLV launch fairing, which will save cost and reduce programmatic risk by eliminating the need for the SLS.

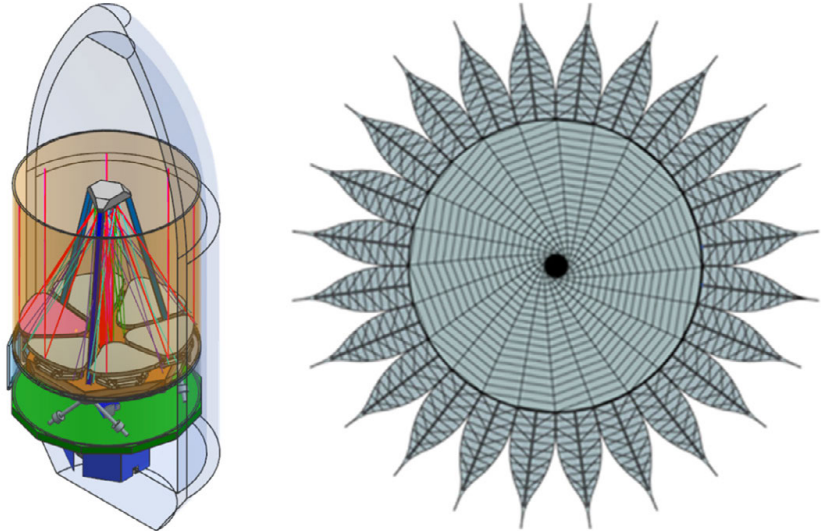

**Figure D.5-1.** The starshade-only options (Options 2 and 4) used a segmented on-axis telescope with the starshades.

- The telescope and overall flight system masses are reduced significantly. Again, this enables the segmented telescope to fit in the SHHLV's deliverable mass to an L2 orbit.

- The smaller mirrors reduce primary mirror fabrication problems.

- Experience from previous mirror development projects (AMSD, MMSD, and other projects) suggests that the telescope can be fabricated for significantly less than the off-axis monolith design.

Without the demanding diffraction suppression and contrast requirements of a coronagraph, the segmented telescope design is a more attractive option.

Like the baseline option, the starshade-only telescope instrumentation includes a starshade camera, UV spectrograph, and near-UV/visible/near-IR camera. No coronagraph was included in the architecture trade. With the more common on-axis design, the instruments were located behind the primary mirror.

Without the coronagraph, operations would be very different for the starshade-only architecture in comparison to the baseline. In the baseline, the coronagraph handled most of the planet detections and orbit determinations. These require blind searches of target systems to detect new exoplanets and repeat visits to establish planet orbits. While this work is possible with the starshade-only architecture, the time and fuel required to reposition the starshade for each visit and revisit would limit their numbers. Orbit determination in particular is a challenge for this architecture option. Modeled number of orbits determined by a single starshade in the architecture trade, were less than a third of the number of orbits captured with the coronagraph options (i.e., the baseline and Option 1).

As noted above, the starshade-only telescope flight system has more launch options than the baseline telescope spacecraft. The telescope and starshade can be launched together on an SLS, with considerable launch mass margin, or they can be launched separately on SHLLVs. For the architecture trade, the latter was assumed.

### D.5.3   Mission Cost

The Option 2 cost savings over the baseline option came from the elimination of the coronagraph and the reduction in the size and cost of the starshade telescope and telescope spacecraft. Option 2 was higher but close in cost to Option 1—the coronagraph-only architecture.

### D.5.4   Required Technology Development

Option 2 had two areas requiring technology development: starshade deployment and petal shape stability. Since it does not carry a coronagraph, Option 2 carries less risk than the baseline, but more than Option 1.

## D.6 Two Starshades without a Coronagraph

With the addition of a second, smaller starshade, the two-starshade option (Option 4) looks at mitigating the orbit determination shortcoming of the single starshade option (Option 2).

### D.6.1 Concept Overview

Option 4 has the same hardware as in Option 2, but with the addition of a 32 m starshade. This smaller starshade maintains the same IWA as the 72 m but operates over a spectral band of 300–500 nm. This smaller size and smaller blue-end bandpass allow the starshade to operate at a much closer separation distance from the telescope than that of the 72 m (55,000 km verses 124,000 km) while still preserving the same deep shadow at the telescope. The smaller size also brings a significantly lower mass. With its lower mass and closer separation, the smaller starshade is able to perform target-to-target repositioning maneuvers in much less time and using much less fuel than the large starshade. While the narrow bandpass will make the small starshade a poor platform for spectral characterization, the band is sufficient for planet detection. Therefore, the greater agility of the 32 m starshade is used for exoplanet searches to find planetary systems and exoplanet orbit determination. The larger starshade will follow up with broadband exoplanet spectral characterization, much like in the baseline architecture case.

Orbit and launch vehicles remain as in Option 2. A single launch option with both starshades and the telescope on the SLS does not exist due to volumetric constraints. For the architecture trade, the telescope would launch on one SHLLV and the two starshades would share a launch on another SHLLV.

### D.6.2 Mission Cost

With three separate spacecraft, Option 4 was close in cost to the baseline option. The small additional starshade was slightly less in cost than the savings realized by the lighter telescope meeting less demanding requirements, and the elimination of the coronagraph.

### D.6.3 Required Technology Development

Like Option 2, Option 4 had the same two areas requiring technology development: starshade deployment and petal shape stability. Since it had the same technologies requiring development, Option 4 was evaluated as having the same level of technology risk as Option 2.

## D.7 Alternative Weightings of the "Wants"

While a consensus was reached within the STDT to support Option 3 as the baseline architecture for further design development, several STDT members noted that the adopted outcome of the architecture trade was greatly influenced by the weightings assigned to the "want" criteria. The Kepner-Tregoe "musts"— minimum acceptable performance and programmatic thresholds—were not a factor since the four options being evaluated all met these gate requirements in the trade. Performance and programmatic desirements captured in the KT "wants" were largely what established the baseline choice. These "wants" were weighted by agreement within the STDT since some criteria were seen as more critical to the selection than others. In particular, the "wants" were weighted in favor of science-based criteria over cost and schedule. The rationale for this decision was that as a major space observatory, HabEx needed to give science the priority. Minimum successful cost and schedule criteria were established and met in the "musts."

A subsequent weighting sensitivity study by the STDT showed that a small decrease in the science criteria weightings, accompanied with a corresponding increase in the cost criteria weighting would shift the KT trade in favor of the coronagraph-only option. With this awareness, the STDT reaffirmed the original weightings and original decision to develop the coronagraph and starshade architecture.

# E  TECHNOLOGY ROADMAPS TO TRL 5

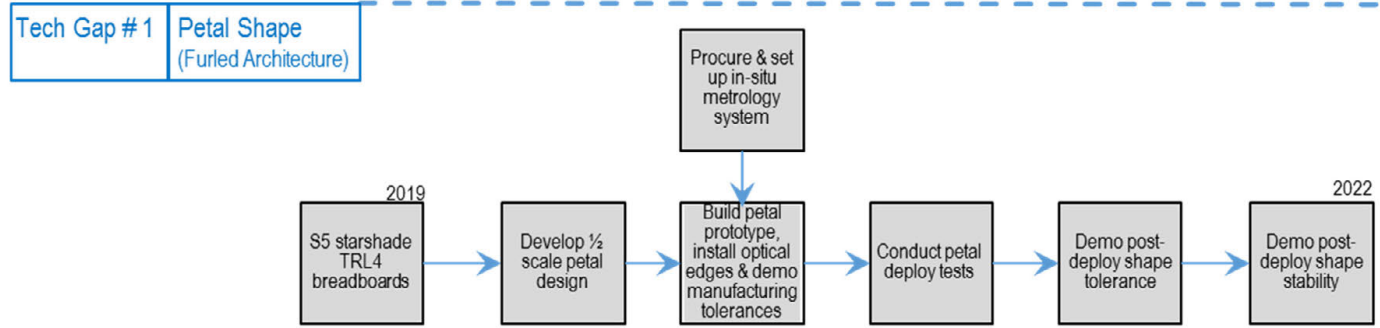

S5 will mature an 8m petal to TRL 5 which will also qualify the HabEx petal for TRL 5 as a 1/2 scale demonstration.

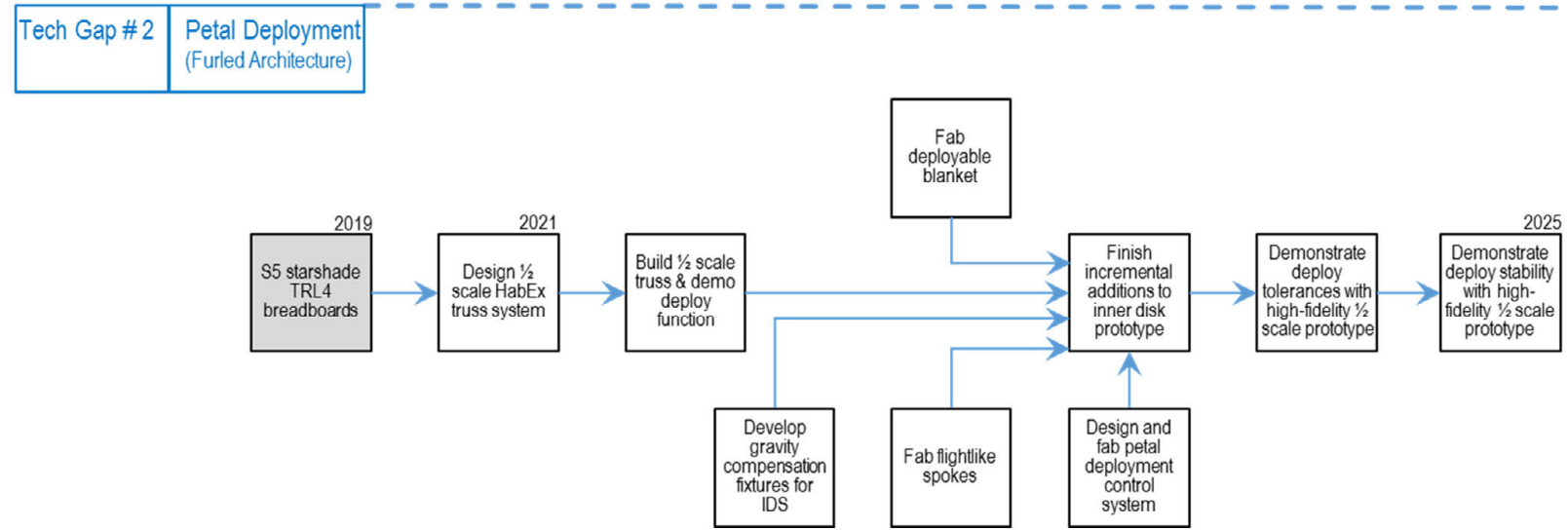

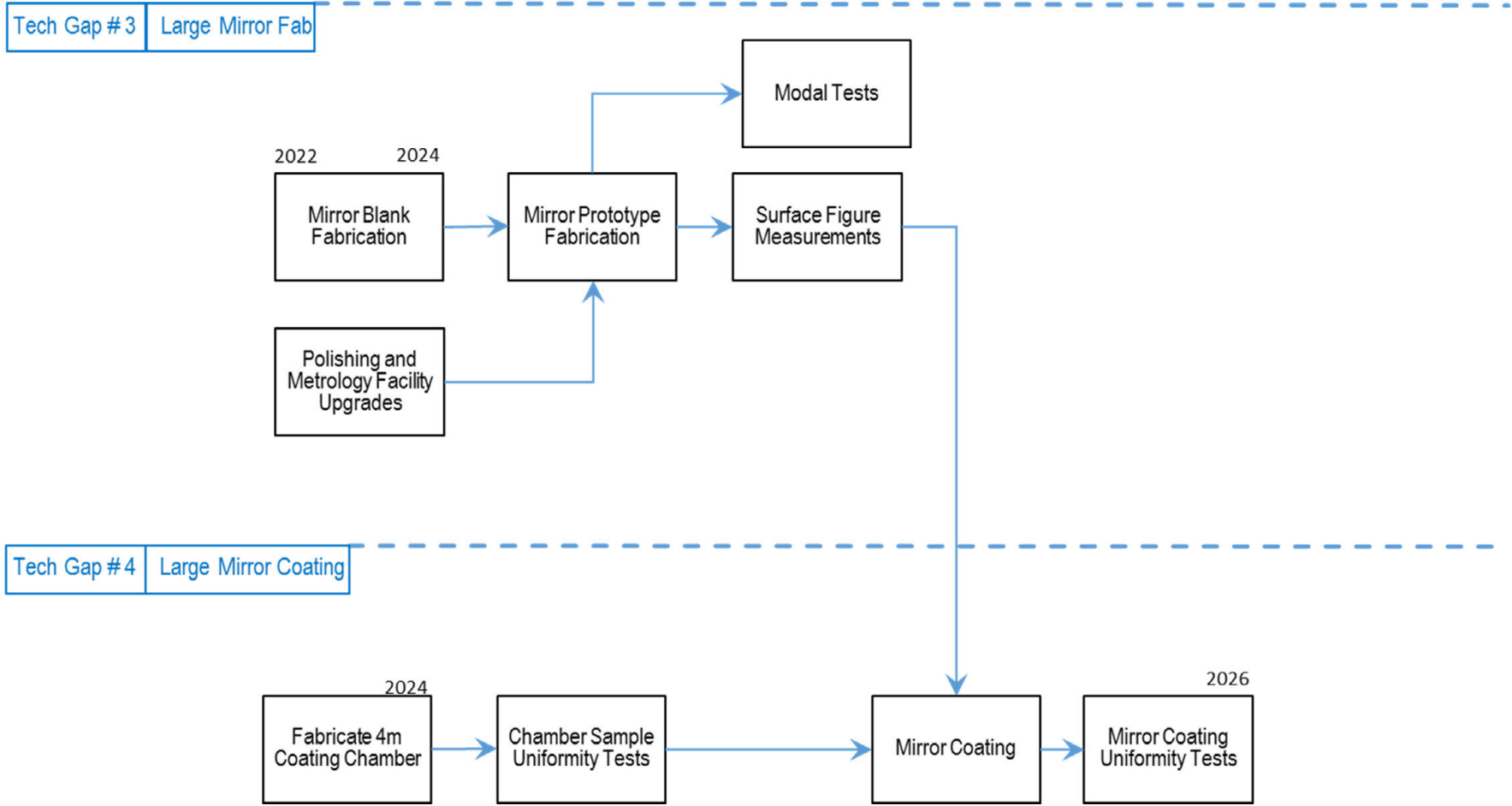

| Tech Gap #5 | LOFWS |
|---|---|

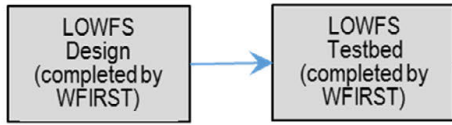

LOWFS design is based on CGI. Only requires verification in HabEx coronagraph to reach TRL 5.

| Tech Gap #6 | VV6 Coronagraph |
|---|---|

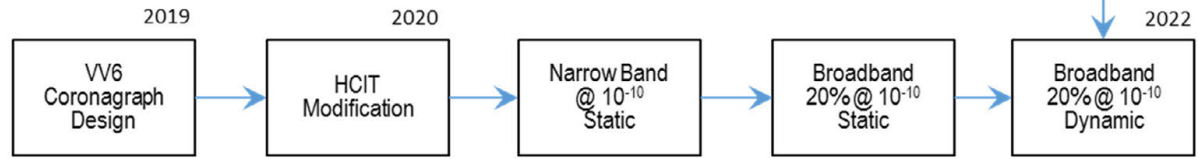

| Tech Gap #7 | DMs |
|---|---|

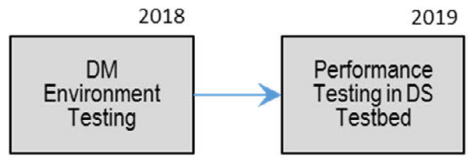

DM advancement to TRL 5 is covered under existing TDEMs

| Tech Gap # 8 | Starshade Edge Scatter |
|---|---|

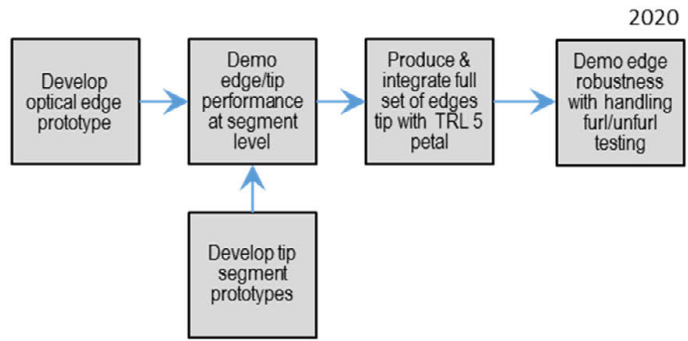

Edge scatter gap will be closed through the S5 task

**Table E-1.** Technology cost estimate.

| Gap# | Title | Tasks to Resolve | Cost Estimate ($FY20) | Basis of Estimate |
|---|---|---|---|---|
| #1 | Demonstration of flight-like petal fabrication and deployment | a) Develop HabEx ½ petal design<br>b) Procure and set up in-situ metrology system<br>c) Build petal prototype, install optical edges and demo manufacturing tolerances<br>d) Conduct petal deployment tests<br>e) Demo postdeployment shape tolerances<br>f) Demo postdeployment shape stability | n/a | Covered by S5 |
| #2 | Demonstration of inner disk system deployment | a) Design ½ scale HabEx truss system<br>b) Build ½ scale truss and demo deployment<br>c) Build ½ scale deployable blanket<br>d) Develop gravity compensation fixtures for inner disk system (IDS)<br>e) Demo postdeployment shape tolerances<br>f) Fabricate ½ scale spokes<br>g) Design and fab petal deployment control system<br>h) Add components to truss build IDS<br>i) Demo deployment tolerances with ½ scale IDS prototype<br>j) Demo deployment stability with ½ scale prototype | TBD | Scaled up from S5 costs based on inner disk sizes |
| #3 | Large mirror fabrication | a) 4 m mirror blank fabrication<br>b) Mirror fabrication and metrology facilities upgrades<br>c) Prototype mirror fabrication<br>d) Mirror modal tests<br>e) Surface figure measurements | TBD | Quote from industry |
| #4 | Large mirror coating uniformity demonstration | a) Fabricate 4 m mirror coating chamber<br>b) Demo chamber uniformity<br>c) Coat prototype mirror<br>d) Demo mirror coating uniformity | | |
| #5 | LOWFS performance demonstration | a) LOWFS performance demo with VVC 6 | n/a | Included in #6 |
| #6 | Coronagraph performance demonstration | a) VVC 6 coronagraph design<br>b) HCIT modification needed<br>c) Coronagraph narrow band static tests<br>d) Coronagraph broadband static tests<br>e) Coronagraph broadband dynamic tests | TBD | Based on WFIRST CGI actual costs |
| #7 | Deformable mirror performance demonstration | a) DM environmental testing<br>b) DM performance in the Decadal Survey testbed | n/a | Covered by TDEMs |
| #8 | Starshade edge scatter control demonstrations | a) Develop optical edge prototypes<br>b) Develop petal tip prototypes<br>c) Demo edge and tip performance at the segment level<br>d) Produce and integrate a full set of edges and tips for a TRL 5 petal<br>e) Demo robustness with handling and furling/unfurling tests | n/a | Covered by S5 |